\documentclass[traditabstract]{aa} 

\usepackage{amsmath}
\usepackage{amssymb}
\usepackage{graphicx}
\usepackage{txfonts}
\usepackage{color}
\usepackage{natbib}
\usepackage{verbatim}
\usepackage{multirow}
\usepackage{float}

\usepackage{color,pstcol}
\newcommand{\ytextmodifhershelvtwostepfiftyone}[1]
{{#1}}
\newcommand{\ytextmodifhershelvonestepfive}[1]
{{\color{black}#1}}
\usepackage[algoruled,linesnumbered]{algorithm2e}

\begin{document}
\title{Unmixing methods 
\ytextmodifhershelvonestepfive{based on}
nonnegativity and weakly mixed pixels for astronomical hyperspectral datasets}
\author{A. Boulais, 
O. Bern\'e, G. Faury, and Y. Deville}
\institute{IRAP, Universit\'e de Toulouse, UPS, CNRS, CNES, \\
9 Av. colonel Roche, BP 44346, F-31028 Toulouse cedex 4, France}
\titlerunning{Unmixing methods for astronomical hyperspectral datasets}
\authorrunning{}
\date{Received ??? ; accepted ???}
\abstract{%
\ytextmodifhershelvonestepfive{An increasing number of astronomical
instruments (on Earth and space-based) provide hyperspectral images, that is three-dimensional
data cubes with two spatial dimensions and one spectral dimension. The intrinsic limitation in  spatial resolution of these instruments implies that
the spectra associated with pixels of such images are
most often mixtures of the spectra of the ``pure'' components
that exist in the considered region.
In order to estimate the spectra and spatial abundances of these pure
components, we here propose an original blind signal separation (BSS),
that is to say an unsupervised unmixing method.}
Our approach is  
\ytextmodifhershelvonestepfive{based on extensions
and combinations of
linear BSS methods that belong to two major
classes of methods, namely
nonnegative matrix factorization (NMF) and}
Sparse 
Component Analysis (SCA). The former performs the decomposition 
\ytextmodifhershelvonestepfive{of hyperspectral images,
as a set of pure spectra and abundance maps, by using}
nonnegativity 
\ytextmodifhershelvonestepfive{constraints,}
but the estimated solution is not unique%
\ytextmodifhershelvonestepfive{:}
It highly depends on 
\ytextmodifhershelvonestepfive{the}
initialization of the algorithm. 
\ytextmodifhershelvonestepfive{The considered SCA methods are}
based on the assumption of the existence 
of 
\ytextmodifhershelvonestepfive{points or}
tiny spatial zones where only one {source} is active
\ytextmodifhershelvonestepfive{(i.e., one pure component is present)}%
. These 
\ytextmodifhershelvonestepfive{points or}
zones are 
then used to estimate the mixture and perform the decomposition. In real 
conditions, the assumption of perfect single-source 
\ytextmodifhershelvonestepfive{points or}
zones is not always 
realistic. 
\ytextmodifhershelvonestepfive{In such conditions, SCA yields approximate
versions of the unknown sources and mixing coefficients.}
%
We 
\ytextmodifhershelvonestepfive{ propose}
to use 
\ytextmodifhershelvonestepfive{part of}
these preliminary estimates from the SCA to initialize
\ytextmodifhershelvonestepfive{several runs of}
the NMF in order to refine 
\ytextmodifhershelvonestepfive{these estimates}
and further constrain the convergence of the 
\ytextmodifhershelvonestepfive{NMF}
algorithm. 
%
\ytextmodifhershelvonestepfive{The proposed methods also estimate the
number of pure components involved in the data and they provide error
bars associated with the obtained solution.}
%
\ytextmodifhershelvonestepfive{Detailed}
tests 
\ytextmodifhershelvonestepfive{with synthetic data}
show that the decomposition achieved with 
\ytextmodifhershelvonestepfive{such hybrid methods}
is nearly unique 
and provides good 
\ytextmodifhershelvonestepfive{performance, illustrating the potential
of applications to real data.}
%
}
\keywords{astrochemistry - ISM: molecules - molecular processes - Methods: 
numerical}

\maketitle

\section{Introduction}
Telescopes keep growing in diameter, and detectors are more and more sensitive 
and made up of an increasing number of pixels. Hence, the number of photons that 
can be captured by astronomical instruments, in a given amount of time and at a 
given wavelength, has increased significantly, thus allowing astronomy to go { 
hyperspectral}. More and more, astronomers do not deal with 2D images or 1D 
spectra, but with a combination of these media resulting in three-dimensional (3D) 
\ytextmodifhershelvonestepfive{data cubes}
(two spatial 
dimensions, one spectral dimension). We 
\ytextmodifhershelvonestepfive{hereafter provide an overview}
of the instruments that provide 
\ytextmodifhershelvonestepfive{hyperspectral data in astronomy, mentioning}
specific examples without any objective to be exhaustive. Several integral field 
unit spectrographs (e.g., MUSE on the Very Large Telescope) 
\ytextmodifhershelvonestepfive{provide}
spectral cubes at visible 
wavelengths, yielding access to the optical tracers of ionized gas (see for 
instance \citealt{wei15}). Infrared missions such as the Infrared Space 
Observatory (ISO) and {\it Spitzer} performed spectral mapping in the 
mid-infrared, a domain that is particularly suited to observe the emission of UV 
heated polycyclic aromatic hydrocarbon (e.g., \citealt{ces96, wer04}). In the 
millimeter wavelengths, large spectral maps in rotational lines of abundant 
molecules (typically CO) have been used for several decades to trace the 
dynamics of molecular clouds (e.g., \citet{bal87, mie94, fal09}). The PACS, SPIRE, 
 and HIFI instruments, on board  {\it Herschel} all have a mode that allows for 
spectral 
mapping (e.g. \citealt{vke10, hab10, job10}) in atomic and molecular lines. 
Owing to its high spectral resolution, HIFI allows 
\ytextmodifhershelvonestepfive{one}
to resolve the profiles of 
these lines, enabling 
\ytextmodifhershelvonestepfive{one}
to study the kinematics of, for example, the immediate surroundings 
of protostars (\cite{kri11}) or of star-forming regions (\cite{pil12}) using 
radiative transfer models. Similarly, the GREAT instrument on board the 
Stratospheric Observatory For Infrared Astronomy (SOFIA) now provides large-scale
 spectral maps in the C$^+$ line at 1.9 THz (\citet{pab17}). The Atacama 
Large Millimeter Array (ALMA) also provides final products that are spectral 
cubes (see e.g., \citet{goi16}).
A majority of astronomical spectrographs to be employed at large observatories 
in the future will provide spectral 
\ytextmodifhershelvonestepfive{maps. This}
is the case for the MIRI and NISPEC instruments on the James Webb Space 
Telescope (JWST) and the METIS instrument on the Extremely Large Telescope 
(ELT). 

Although such 3D datasets have become common, 
\ytextmodifhershelvonestepfive{few methods have been}
developed by 
astronomers to analyze the outstanding amount of information they contain. 
Classical analysis methods tend to decompose the spectra by fitting them with 
simple functions (typically 
\ytextmodifhershelvonestepfive{mixtures}
of Gaussians) but this has several 
disadvantages: 1) the {\it a priori} 
\ytextmodifhershelvonestepfive{assumption}
made by the use of a given function is 
usually not founded physically%
\ytextmodifhershelvonestepfive{,}
2) if the number of parameters is high, the 
result of the fit may be degenerate%
\ytextmodifhershelvonestepfive{,}
3) for large datasets and fitting with 
nonlinear functions, the fitting may be very time consuming%
\ytextmodifhershelvonestepfive{,}
4) initial guesses 
must be provided, and%
\ytextmodifhershelvonestepfive{,}
5) the spectral fitting is usually performed on a (spatial) 
pixel by pixel basis, so that the extracted components are spatially independent, 
whereas physical components are often present at large scales on the image.  An 
alternative is to 
\ytextmodifhershelvonestepfive{analyze}
the data by means of principal component analysis 
(e.g., \citealt{neu07,gra17}), which provides a representation of the data in 
\ytextmodifhershelvonestepfive{an orthogonal basis of a subspace, thus}
allowing interpretation. However, this may be limited by the fact that the 
principal components are orthogonal, and hence they are not easily interpretable 
in physical terms. An alternative analysis was proposed by \cite{juv96}, { which 
is based on a Blind Signal Separation (BSS) approach. It  
consists in decomposing spectral cubes (in their case, CO spectral maps) into 
the product of a small number of spectral components, or ``end members'', and 
spatial ``abundance'' maps.}

\ytextmodifhershelvonestepfive{This requires}
no {\it a priori} on spectral properties of 
the components, and hence this can provide deeper insights into the physical 
structure represented in the data, as demonstrated in this pioneering paper. 
This method uses the positivity constraint 
\ytextmodifhershelvonestepfive{for}
the maps and spectra (all their 
points must be positive) combined with the minimization of a statistical 
criterion to derive the maps and spectral components. This method is referred to 
as positive matrix factorization (PMF, \cite{paa94}). Although it contained the 
original idea of using positivity as a constraint to estimate a matrix product, 
this work used a classical optimization algorithm. Several years later, 
\cite{lee99} introduced a novel algorithm to perform PMF using simple 
multiplicative iterative rules, making the PMF algorithm extremely fast. This 
algorithm is usually referred to as Lee and Seung's Non Negative Matrix 
Factorization (NMF hereafter) and has been widely used in a vast number of 
applications outside astronomy. This algorithm has 
\ytextmodifhershelvonestepfive{proved to be efficient}
including in astrophysical applications \citep{ber07}. However, NMF has several 
disadvantages: 1)
the number of spectra to be extracted must be given by the user%
\ytextmodifhershelvonestepfive{,}
2) the 
\ytextmodifhershelvonestepfive{error bars}
related to the procedure are not 
\ytextmodifhershelvonestepfive{derived}
automatically%
\ytextmodifhershelvonestepfive{,}
3) convergence 
to a unique point is not guaranteed and may depend on initialization (see 
\citealt{don04} on these latter aspects). When applying NMF to astronomical 
hyperspectral data, the above drawbacks become critical and can jeopardize the 
integrity of the results.

In this paper, we  
{ evaluate possibilities to improve application of BSS to hyperspectral
positive data by hybridizing NMF with sparsity-based algorithms}.
{ Here, we focus on synthetic data, so as 
to perform a detailed comparison of the performances 
of the proposed approaches.}
{ A first application on real data of { one of} the methods presented here is provided in \citet{fos19}.} 
The 
\ytextmodifhershelvonestepfive{proposed methods}
should be applicable to any hyperspectral dataset fulfilling the properties that 
we will describe hereafter. 
The paper is organized as follows. In the next section we present the 
{ adopted mathematical model for hyperspectral astronomical data, 
using tow possible conventions, spatial or spectral. We describe the 
mixing model and associated ``blind signal separation'' (BSS) problem.
In Section \ref{sec_preprocessing}, we describe the preliminary steps
(preprocessing steps) that are required before applying the proposed algorithms.
In Section \ref{seq_Methods} we describe in details the three methods 
that are used in this paper, that is, NMF (with an extension using a Monte-Carlo
approach referred to as MC-NMF) and two methods based on sparsity 
(Space-CORR and Maximum Angle Source Separation, MASS). We then detail
how MC-NMF can be hybridized with the latter two methods.} 

%
%
In Section \ref{sec_exp},
a comparative performance analysis of studied methods is performed.
We conclude in Section 
\ytextmodifhershelvonestepfive{\ref{sec_concl}.}

\section{Data model and blind source separation
\ytextmodifhershelvonestepfive{problem}%
}\label{sec_model}

The observed data consist of a spectral cube $\underline{C}(p_x,p_y,f)$ of 
dimension $P_x \times P_y \times N$ where ($p_x,p_y$) define the spatial 
coordinates and $f$ 
\ytextmodifhershelvonestepfive{is}
the spectral index. To 
\ytextmodifhershelvonestepfive{help one interpret}
the results, the spectral index is 
\ytextmodifhershelvonestepfive{hereafter}
expressed as a Doppler-shift velocity in 
km/s, using $v = c \times (f-f_0) / f_0$, with $f$ the observed frequency, $f_0$ 
the emitted frequency and $c$ the light speed. We assume that all observed 
values in $\underline{C}$ are nonnegative. We call 
each vector 
$\underline{C}(p_x,p_y,.)$ recorded at a position ($p_x,p_y$) 
\ytextmodifhershelvonestepfive{``}%
spectrum%
\ytextmodifhershelvonestepfive{''}
and we call each matrix $\underline{C}(.,.,v)$ recorded at a given velocity
\ytextmodifhershelvonestepfive{``}%
spectral band.%
\ytextmodifhershelvonestepfive{''}
\ytextmodifhershelvonestepfive{Each}
observed spectrum corresponding to a given pixel results from a 
mixture of different kinematic components that are present on the line of sight 
of the instrument. Mathematically, the observed spectrum obtained for one pixel 
is then a combination (which will be assumed 
\ytextmodifhershelvonestepfive{to be}
linear and instantaneous) of 
elementary spectra. 

In order to recover these elementary spectra, one can use methods known as Blind 
Source Separation (BSS). BSS consists in estimating a set of unknown source 
signals from a set of observed signals that are mixtures of these source 
signals. The linear mixing coefficients are unknown and are also to be 
estimated. The observed spectral cube is then decomposed 
\ytextmodifhershelvonestepfive{as}
a set of elementary 
spectra and a set of abundance maps (the contributions of elementary spectra in 
each pixel).

Considering BSS terminology and a linear mixing model, the matrix containing all 
observations is expressed as the product of a mixing matrix and a 
{source} 
matrix. Therefore, it is necessary here to restructure the 
hyperspectral cube $\underline{C}$ into a matrix and
\ytextmodifhershelvonestepfive{to}
identify what we call 
``observations'', ``samples'', ``mixing coefficients'', and ``%
{sources}''.
A spectral cube can be modeled in two different ways: a spectral model where we 
consider the cube as a set of spectra and a spatial model where we consider the 
cube as a set of images (spectral bands)%
\ytextmodifhershelvonestepfive{, as detailed hereafter.}

\subsection{Spectral model}\label{subseq_spectral_model}

\begin{table}
\caption{ List of major variables \label{tab}}
\begin{center}
\begin{tabular}{cc}
\hline \hline
 Variable 			& Description  \\
 \hline
 $spec (l,v)$		& Value of elementary spectrum $l$ at velocity index $v$ \\
 $Spec $			& Matrix of values $spec (l,v)$ of elementary spectra \\
 \hline
 $map (m, l)$		& Scale factor of elementary spectrum $p$ in pixel $m$\\
 $Map$				& Matrix of scale factors $map (m, l)$\\
 \hline
 $obs (m,v)$		& Observed value of pixel $m$ at velocity $v$ \\
 $Obs$				& Matrix of observed values $obs (m,v)$\\
 \hline

\end{tabular}
\end{center}
\end{table}

For 
\ytextmodifhershelvonestepfive{the spectral}
data model, we define the observations as being the spectra 
$\underline{C}(p_x,p_y,.)$. The data cube $\underline{C}$ is reshaped into a new 
matrix of observations $Obs$ { (variables defined in this section are summarized in Table 1)}, where 
\ytextmodifhershelvonestepfive{the}
rows contain the $P_x \times P_y = M$ 
observed spectra of $\underline{C}$ arranged in any order and indexed by $m$. 
Each column of $Obs$ corresponds to a given spectral sample 
\ytextmodifhershelvonestepfive{with an integer-valued index also denoted
as}
$v\in\{1,\ldots,N\}$ for all observations. Each observed spectrum $obs(m,.)$ is 
a linear combination of $L$ ($L \ll M$) unknown elementary spectra and yields a 
different mixture of the same elementary spectra:
\begin{gather}
obs(m,v) = \sum_{\ell=1}^L map(m,\ell)~spec(\ell,v)\\
m\in\{1,\ldots,M\}, ~ v\in\{1,\ldots,N\}, \ell\in\{1,\ldots,L\}, \nonumber
\end{gather}
where $obs(m,v)$ is the $v^{th}$ sample of the $m^{th}$ observation, 
$spec(\ell,v)$ is the $v^{th}$ sample of the $\ell^{th}$ elementary spectrum and 
$map(m,\ell)$ defines the contribution scale of elementary spectrum $\ell$ in 
observation $m$. Using 
\ytextmodifhershelvonestepfive{the}
BSS terminology, $map$ stands for the mixing coefficients 
and $spec$ stands for the sources. This model can be 
\ytextmodifhershelvonestepfive{rewritten}
in matrix form:
\begin{equation}\label{eq_spectral_model}
Obs = Map\times Spec,
\end{equation}
where $Map$ is an $M \times L$ mixing matrix
\ytextmodifhershelvonestepfive{and}
$Spec$ is an $L \times N$ 
{source} 
\ytextmodifhershelvonestepfive{matrix.}

\subsection{Spatial model}\label{subseq_spatial_model}
For 
\ytextmodifhershelvonestepfive{the spatial}
data model, we define the observations as being the spectral bands 
$\underline{C}(.,.,v)$. The construction of the spatial model is performed by 
transposing the spectral model (\ref{eq_spectral_model}). In this configuration, 
the rows of the observation matrix $Obs^T$ (the transpose of the original matrix 
of observations $Obs$) contain the $N$ spectral bands
\ytextmodifhershelvonestepfive{with a one-dimensional structure.}
Each column of $Obs^T$ corresponds to a given spatial sample index 
$m\in\{1,\ldots,M\}$ for all observations (i.e., each column corresponds to a 
pixel). Each spectral band $Obs^T(v,.)$ is a linear combination of $L$ ($L \ll 
N$) unknown abundance maps and yields a different mixture of the same abundance 
maps:
\begin{equation}\label{eq_spacial_model}
Obs^T=Spec^T\times Map^T,
\end{equation}
where $Obs^T$ is the 
\ytextmodifhershelvonestepfive{transpose}
of the original observation matrix, $Spec^T$ is 
the $N \times L$ mixing matrix and $Map^T$
\ytextmodifhershelvonestepfive{is}
the $L \times M$ source matrix. In 
this alternative data model, the elementary spectra in $Spec^T$ stand for the 
mixing coefficients and the abundance maps in $Map^T$ stand for the 
\ytextmodifhershelvonestepfive{sources.}
%

\subsection{Problem statement}
\label{Sec_prob-stat}
In this section, we
\ytextmodifhershelvonestepfive{denote the mixing matrix as $A$ and
the source matrix as $S$, whatever}
the nature of the adopted model 
(spatial or spectral) to simplify notations, the following remarks being valid 
in both cases.

The goal of BSS methods is to find estimates of a mixing matrix $A$ and a 
{source} matrix $S$, respectively 
\ytextmodifhershelvonestepfive{denoted as}
$\hat{A}$ and $\hat{S}$%
\ytextmodifhershelvonestepfive{, and}
such that:
\begin{equation}
X \approx \hat{A}\hat{S}.
\end{equation}
However this problem is ill-posed. Indeed, if $\{\hat{A},\hat{S}\}$ is a 
solution, then $\{\hat{A}P^{-1},P\hat{S}\}$ is also a solution for any 
invertible matrix $P$. To achieve the decomposition, we must add two extra 
constraints. 
The first one is a constraint on the 
\ytextmodifhershelvonestepfive{properties}
of the unknown matrices $\hat{A}$ and/or $\hat{S}$. The type of constraint 
(independence of 
{sources}, nonnegative matrices, sparsity) leads 
directly to the class of methods that will be used for the decomposition. The 
case of linear instantaneous mixtures was first studied in the 1980s, then three 
classes of methods became important:
\begin{itemize}
\item Independent component analysis (ICA) (\cite{car98,hyv99}): It is based on 
a probabilistic formalism and requires the source signals to be mutually 
statistically independent. Until the early 2000s, ICA was the only class of 
methods available to achieve BSS.
\item Nonnegative matrix factorization (NMF) (\cite{lee99}): It requires the 
source signals and mixing coefficients values to be nonnegative.
\item Sparse component analysis (SCA) (\cite{gri06}): It requires the source 
signals to be sparse in the considered representation domain (time, 
time-frequency, time-scale, wavelet...).
\end{itemize}

The second constraint is to determine the dimensions 
\ytextmodifhershelvonestepfive{of}
$\hat{A}$ and $\hat{S}$. Two of these dimensions are obtained directly from 
observations $X$ ($M$ and $N$). The third dimension, common to both $\hat{A}$ 
and $\hat{S}$ matrices, is the number of 
{sources} $L$, which must be 
estimated.
\\

Here, we consider astrophysical hyperspectral data 
that have the properties listed below. These are relatively general properties
that are applicable to a number of cases with Herschel-HIFI, ALMA, Spitzer,
JWST, etc:  
\begin{itemize}
\item They do not satisfy the condition of independence of the 
{sources}. 
In our simulated 
 data, elementary spectra have, by 
construction, similar variations (Gaussian spectra with different means, see 
Section \ref{subseq_Data}). Likewise, abundance maps associated with each 
elementary spectrum have similar shapes. Such data involve nonzero correlation 
coefficients between elementary spectra and between abundance maps. Hence ICA 
methods will not be discussed in this paper.
\item 
\ytextmodifhershelvonestepfive{These data}
are nonnegative if we disregard noise. Each pixel provides an emission 
spectrum, hence 
\ytextmodifhershelvonestepfive{composed of}
positive or zero values. Such data thus correspond to the 
conditions of use of NMF that we detail in Section \ref{subseq_NMF}. 
\item If we consider the data in a spatial framework (spatial model), the cube 
provides a set of images. We can then formulate the hypothesis that there are 
regions in these images where only one source is present. This is detailed in 
Section \ref{subseq_SCA}. This hypothesis then refers to a ``sparsity'' 
assumption in the data and SCA methods are then applicable to hyperspectral 
cubes.
%
%
%
\ytextmodifhershelvonestepfive{On the contrary, 
sparsity properties do not exist
in the spectral framework in our case, as discussed below.}
%
%
%
\item If the data 
\ytextmodifhershelvonestepfive{have some sparsity properties,%
%
%
}
%
%
%
adding the nonnegativity assumption enables the use of geometric methods. The 
geometric methods are a subclass of 
\ytextmodifhershelvonestepfive{BSS}
methods based on the identification of the 
convex hull containing the mixed data.
However, the majority of geometric methods%
\ytextmodifhershelvonestepfive{, which are}
used in hyperspectral unmixing in Earth 
\ytextmodifhershelvonestepfive{%
observation, are not 
applicable to Astrophysics because
they set}
an additional constraint on the data 
\ytextmodifhershelvonestepfive{model: they
require all abundance coefficients to sum to one in each pixel,}
which changes the geometrical representation of the mixed data. 
\ytextmodifhershelvonestepfive{On the contrary, in}
Section \ref{subsec_MASS}, we introduce a geometric method called MASS,
for Maximum Angle Source Separation
\ytextmodifhershelvonestepfive{(\cite{bou15}), which may be used}
in an astrophysical context (i.e., for data respecting the models 
presented 
\ytextmodifhershelvonestepfive{above}%
).
\end{itemize}

The sparsity constraint required for SCA and geometric methods is carried by the 
source matrix $S$. 
\ytextmodifhershelvonestepfive{These methods may therefore potentially be 
applied
in two ways to the above-defined data: either}
we suppose that there exist 
spectral indices for which a unique 
\ytextmodifhershelvonestepfive{spectral source}
is nonzero, 
\ytextmodifhershelvonestepfive{or}
we suppose that there exist some regions in the image for which a unique 
\ytextmodifhershelvonestepfive{spatial source}
is zero. In our context of studying the properties of photodissociation 
regions, only the second case is realistic. Thus only the mixing model 
(\ref{eq_spacial_model}) is relevant.
\ytextmodifhershelvonestepfive{Therefore,}
throughout the rest of this paper, we will only use 
\ytextmodifhershelvonestepfive{that}
spatial data model
\ytextmodifhershelvonestepfive{(\ref{eq_spacial_model}),
so that we here define the associated final}
notations and vocabulary: let $X = Obs^T$ 
\ytextmodifhershelvonestepfive{be}
the $(N \times M)$ 
observation matrix, $A=Spec^T$ the $(N \times L)$ mixing matrix containing the 
elementary spectra and $S=Map^T$ the $(L \times M)$ source matrix
containing the spatial abundance maps%
\ytextmodifhershelvonestepfive{, each associated with an}
elementary spectrum.

Moreover, we note that in the case of the NMF, the 
\ytextmodifhershelvonestepfive{spectral and spatial}
models are equivalent but the community generally prefers the more intuitive 
spectral model. 
\\

Before thoroughly describing the algorithms used for the aforementioned BSS 
methods, we present 
%
\ytextmodifhershelvonestepfive{preprocessing stages required 
for}
%
the decomposition of 
\ytextmodifhershelvonestepfive{data}
cubes.

\section{Data preprocessing}
\label{sec_preprocessing}
\subsection{Estimation of number of sources}\label{subseq_nb_S}
An inherent problem in BSS is to estimate the number $L$ of 
{sources} 
(the 
\ytextmodifhershelvonestepfive{dimension shared by the}
$\hat{A}$ and $\hat{S}$ matrices). This parameter 
should be fixed before performing the decomposition in the majority of cases. 
Here, this estimate is based on the eigen-decomposition of the covariance matrix 
of the data. As in Principal Component Analysis (PCA), we look for the minimum 
number of components that most contribute to the total variance of the data. 
Thus the number of high eigenvalues is the number of 
{sources} in the 
data. Let $\Sigma_X$ be the $(N \times N)$ covariance matrix of observations 
$X$:

\begin{equation}
\Sigma_X = \frac{1}{M} X_c X_c^T = \sum_{i=1}^N \lambda_i e_i e_i^T,
\end{equation}
where $\lambda_i$ is the $i^{th}$ eigenvalue associated with eigenvector $e_i$ 
and $X_c$ is the matrix of centered data (i.e. each observation has 
zero mean: $x_c(n,.)=x(n,.)-\bar{x}(n,.)$ ).  

The eigenvalues of $\Sigma_X$ have the following properties (their proofs are 
available in \cite{dev14a}):
\\~~

\textbf{Property 1:} For noiseless data ($X_0=AS$), 
\ytextmodifhershelvonestepfive{the}
$\Sigma_X$ 
\ytextmodifhershelvonestepfive{matrix}
has $L$ positive 
eigenvalues and $N-L$ eigenvalues equal to zero.
\\~~

The number $L$ of 
{sources} is therefore simply inferred from this 
property. Now, we consider the data with an additive spatially white noise $E$, 
with standard deviation $\sigma_E$, 
\ytextmodifhershelvonestepfive{i.e.,}
$X=X_0+E$. The relation between the 
covariance matrix $\Sigma_{X_0}$ of noiseless data and the covariance matrix 
$\Sigma_X$ of noisy data is then:
\begin{equation}
\Sigma_X = \Sigma_{X_0} + \sigma_E^2 I_N,
\end{equation}
where $I_N$ is the identity matrix.
\\~~

\textbf{Property 2:} The eigenvalues $\lambda$ of $\Sigma_X$ and the eigenvalues 
$\lambda_0$ of 
\ytextmodifhershelvonestepfive{$\Sigma_{X_0}$}
are linked by:
\begin{equation}
\lambda = \lambda_0 + \sigma_E^2.
\end{equation}
These two properties then 
\ytextmodifhershelvonestepfive{show}
that the ordered eigenvalues $\lambda_{(i)}$ 
of $\Sigma_X$ for a mixture of $L$ 
{sources} are such that: 
\begin{equation}
\lambda_{(1)} \geq \ldots \geq \lambda_{(L)} > \lambda_{(L+1)} = \ldots = 
\lambda_{(N)} = \sigma_E^2.
\end{equation}
But in practice, because 
\ytextmodifhershelvonestepfive{of the limited number of samples and
since}
the strong assumption of a white noise with the same standard deviation in all 
pixels
\ytextmodifhershelvonestepfive{is not fulfilled}%
, the equality $\lambda_{(L+1)} = \ldots = \lambda_{(N)} = \sigma_E^2$ is 
not met. However, the differences between the eigenvalues $\lambda_{(L+1)}, 
\ldots, 
\ytextmodifhershelvonestepfive{\lambda_{(N)}}
$ are small compared to the differences between the eigenvalues $\lambda_{(1)}, 
\ldots, \lambda_{(L)}$. The curve of the ordered eigenvalues is therefore 
constituted of two parts. The first part%
\ytextmodifhershelvonestepfive{,}
$\Omega_S$, contains the first 
\ytextmodifhershelvonestepfive{$L$}
eigenvalues associated with a strong contribution in the total variance. In this 
part, eigenvalues are significantly different. The second part%
\ytextmodifhershelvonestepfive{,}
$\Omega_E$, 
contains the other eigenvalues, associated with noise. In this part, eigenvalues 
are similar. 

The aim is then to identify from which rank $r=L+1$ eigenvalues no longer vary 
significantly. To this end, we use a method based on the gradient of the curve  
of ordered eigenvalues 
(\cite{luo00})
in order to identify a break in this curve (see Fig.~\ref{fig_VP_example}).

Moreover, a precaution must be taken 
\ytextmodifhershelvonestepfive{concerning}
the difference between $\lambda_{(L)}$ and $\lambda_{(L+1)}$. In simulations, we 
found that in the noiseless case, it is possible that the last eigenvalues of 
$\Omega_S$ are close to zero. Thus, for very noisy mixtures, the differences 
between these eigenvalues become negligible relative to the noise variance 
$\sigma_E^2$. These eigenvalues are then associated with $\Omega_E$ and 
therefore rank 
\ytextmodifhershelvonestepfive{$r$}
where a ``break'' appears will be underestimated.

The procedure described by \cite{luo00} is as follows:
\begin{enumerate}
\item Compute the eigen-decomposition of the covariance matrix $\Sigma_X$ and 
arrange the eigenvalues in decreasing order.
\item Compute the gradient of the curve of the logarithm of the $L$ first 
(typically $L = 20$) ordered eigenvalues:
\begin{equation}
\nabla \lambda_{(i)} = \ln(\lambda_{(i)} / \lambda_{(i+1)}) \qquad 
i\in{1,\ldots,L}.
\end{equation}
\item Compute the average gradient of all these eigenvalues:
\begin{equation}
\overline{\nabla \lambda} = \frac{1}{(L-1)} \ln(\lambda_{(1)} / \lambda_{(L)}).
\end{equation}
\item Find all $i$ satisfying $\nabla\lambda_{(i)} < \overline{\nabla \lambda}$ 
to construct the set $\{\mathcal{I}\} = \{i~|~\nabla\lambda_{(i)} <
\overline{\nabla \lambda}\}$.
\item Select the index $r$, such that 
\ytextmodifhershelvonestepfive{it}
is the first one of the last continuous 
block of $i$ in the set $\{\mathcal{I}\}$.
\item The number of 
{sources} is then $L=r-1$.
\end{enumerate}

\subsection{Noise reduction}\label{subseq_NoiseRed}
The observed spectra are contaminated by noise. In synthetic data, this noise is 
added assuming it is white and Gaussian. Noise in real data may have different 
properties, however the aforementioned assumptions are made here in order to 
evaluate the sensitivity of the method to noise in the general case. To improve 
the performance of the above BSS methods, we propose different 
\ytextmodifhershelvonestepfive{preprocessing}
%
stages to reduce the influence of noise on the results.

The first 
\ytextmodifhershelvonestepfive{preprocessing}
stage consists of applying a spectral 
\ytextmodifhershelvonestepfive{thresholding,}
i.e., only the continuous range 
of $v$ containing signal is preserved. Typically many first and last channels 
contain only noise and are therefore unnecessary for the BSS. This is done for 
all BSS methods presented in the next section.  

The second 
\ytextmodifhershelvonestepfive{preprocessing}
stage consists of  applying a spatial thresholding. Here, we must distinguish the 
case of each BSS method because the SCA method requires to retain the spatial 
structure of data. For NMF, the observed spectra (columns of $X$) whose 
\ytextmodifhershelvonestepfive{``normalized power''}
is lower than a threshold $\alpha_e$ are discarded. Typically some spectra 
contain only noise and are therefore unnecessary for the spectra estimation step 
(Section \ref{subseq_NMF}). In our application, we 
\ytextmodifhershelvonestepfive{set the threshold to}
$\alpha_e=\max\limits_i \|X
\ytextmodifhershelvonestepfive{(.,i)}
\| \times 0.2 ~~ (\forall i\in\{1,\ldots,
\ytextmodifhershelvonestepfive{M}
\})$. 
For the SCA method, some definitions are necessary to 
\ytextmodifhershelvonestepfive{describe}
this 
spatial 
\ytextmodifhershelvonestepfive{thresholding}
step. This procedure is therefore 
\ytextmodifhershelvonestepfive{presented}
in the section 
regarding the method itself (Section \ref{subseq_SCA}).

Finally, synthetic and actual data from 
\ytextmodifhershelvonestepfive{the HIFI instrument}
contain some negative values due to 
noise. To stay in the assumption of NMF, these values are 
\ytextmodifhershelvonestepfive{reset}
to $\epsilon = 
10^{-16}$.

\section{%
\ytextmodifhershelvonestepfive{Blind signal separation}
methods}\label{seq_Methods}
\subsection{\ytextmodifhershelvonestepfive{nonnegative matrix factorization and
our extension}}
\label{subseq_NMF}

NMF is a class of methods introduced by \cite{lee99}. The standard algorithm 
 iteratively and simultaneously computes $\hat{A}$ and $\hat{S}$, minimizing an 
objective function of 
\ytextmodifhershelvonestepfive{the}
initial $X$ matrix and 
\ytextmodifhershelvonestepfive{the}
$\hat{A}\hat{S}$ product. In our 
case, we use the minimization of the Euclidean distance $\delta = 
\frac{1}{2}\|X-\hat{A}\hat{S}\|^{2}_F$, using multiplicative update rules:
\begin{gather}
\hat{A} \leftarrow \hat{A} \odot (X \hat{S}^T) \oslash (\hat{A} \hat{S} 
\hat{S}^T) \label{eq_up_A}
\\
\hat{S} \leftarrow \hat{S} \odot (\hat{A}^T X) \oslash (\hat{A}^T \hat{A} 
\hat{S}), \label{eq_up_S}
\end{gather}
where $\odot$ and $\oslash$ 
\ytextmodifhershelvonestepfive{are}
%
respectively the element-wise product and division.

Lee and Seung show that the Euclidean distance $\delta$ is non increasing under 
these update rules (\cite{lee01}), so that starting from random $\hat{A}$ and 
$\hat{S}$ matrices, the algorithm will converge toward a minimum for $\delta$. 
We estimate that the convergence is reached when:
\begin{equation}\label{eq_conv_crit}
1-\frac{\delta ^{i+1}}{\delta ^{i}} < \kappa,
\end{equation}
where $i$ corresponds to the iteration and $\kappa$ is a threshold typically set 
to $10^{-4}$.
\\~~

The main drawback of standard NMF is the uniqueness of the decomposition. 
The algorithm is sensitive to the initialization due to the existence of local 
minima of the objective function (\cite{cic09}). The convergence point highly 
depends on the distance between the initial point and 
\ytextmodifhershelvonestepfive{a}
global minimum. A 
random initialization without additional constraint is generally not 
satisfactory. To improve the quality of the decomposition, several solutions are 
possible: 
\begin{itemize}
\item 
\ytextmodifhershelvonestepfive{Use}
a Monte-Carlo analysis to estimate the elementary spectra and then 
rebuild the abundance maps (\cite{ber12}).
\item Further constrain the convergence by altering the initialization 
(\cite{lan06}).
\item 
\ytextmodifhershelvonestepfive{Use}
additional constraints on the 
{sources} and/or mixing 
coefficients, such as sparsity constraints (\cite{cic09}), or geometric 
constraints (\cite{mia07}).
\end{itemize}

The addition of geometric constraints is usually based on the sum-to-one of the 
abundance coefficients for each 
\ytextmodifhershelvonestepfive{pixel}
($\sum\limits_{\ell=1}^L s_m(\ell) = 1$). This condition is not realistic in an 
astrophysical context, where the total power received by the detectors vary from a pixel to another. 
Therefore, this type of constraints cannot be applied
\ytextmodifhershelvonestepfive{here. A standard type of
sparsity constraints imposes}
a sparse representation of the estimated matrices 
$\hat{A}$ and/or $\hat{S}$ 
\ytextmodifhershelvonestepfive{in the following sense:}
the spectra and/or the abundance maps 
have a large number of coefficients equal to zero or negligible.
Once again, 
this property is not verified in the data that we consider and so this type of 
constraint cannot be applied. However, the 
\ytextmodifhershelvonestepfive{above type of sparsity}
must be 
distinguished from 
\ytextmodifhershelvonestepfive{the sparsity 
properties exploited in the SCA methods used}
in this paper. 
This 
\ytextmodifhershelvonestepfive{is}
discussed in 
\ytextmodifhershelvonestepfive{Sections
\ref{subseq_SCA} 
and
\ref{subsec_MASS}
dedicated to these methods.}

Moreover, well-known indeterminacies of BSS appear 
\ytextmodifhershelvonestepfive{in}
the $\hat{A}$ and 
$\hat{S}$ estimated matrices. The first one is a possible permutation of 
{sources} in $\hat{S}$. The second one is the presence of a scale factor 
per estimated source. To offset these scale factors, the estimated 
{source} spectra are normalized 
\ytextmodifhershelvonestepfive{so}
that: 
\begin{equation}\label{eq_norm}
\int a_\ell(v) ~ dv = 1 \qquad \ell \in\{1,\ldots,L\}
\end{equation}
where $a_\ell$ is the $\ell^{th}$ column of $A$. This normalization allows the 
abundance maps to be expressed in physical 
\ytextmodifhershelvonestepfive{units.}

To improve the results of standard NMF, 
\ytextmodifhershelvonestepfive{we extend it as follows.}
First, the NMF is amended to take into account the normalization constraint 
(\ref{eq_norm}). At each iteration (i.e. each update of $\hat{A}$ 
according to (\ref{eq_up_A})), 
\ytextmodifhershelvonestepfive{the}
spectra are normalized in order to avoid the 
scale indeterminacies. Then NMF is complemented by a Monte-Carlo analysis 
described hereafter. Finally, we propose an alternative to initialize NMF with 
results from 
\ytextmodifhershelvonestepfive{one of the SCA methods described in Sections
\ref{subseq_SCA} 
and
\ref{subsec_MASS}.}

The NMF-based method used here (called MC-NMF hereafter), combining standard 
NMF, normalization and Monte-Carlo analysis, has the following structure:
\begin{itemize}
\item The Monte-Carlo analysis stage gives the most probable samples of 
elementary spectra and error bars associated with these 
\ytextmodifhershelvonestepfive{estimates provided by}
the 
normalized NMF.
\item The combination stage recovers abundance map 
{sources} from the 
above estimated elementary spectra and observations.
\end{itemize}
These two stages are described hereafter:

\paragraph{1. Monte-Carlo analysis:}Assuming that the number of 
{sources} 
$L$ is known (refer to Section \ref{subseq_nb_S} for its estimation), NMF is ran 
$p$ times, with different initial random matrices for each trial ($p$ is 
typically 
\ytextmodifhershelvonestepfive{equal to}
100). In each run, a set of $L$ elementary spectra are identified. The 
total number of obtained spectra at the end of this process is $p \times L$. 
These spectra are then grouped into $L$ sets 
$\{\omega_1,\omega_2,\ldots,\omega_L\}$, each set representing the same column 
of $\hat{A}$. To achieve this clustering, the method uses the K-means algorithm 
(\cite{the09}) with a correlation 
\ytextmodifhershelvonestepfive{criterion,}
provided in Matlab 
(\textit{kmeans}). More details about the K-means algorithm are provided in 
Appendix A.

To then derive the estimated value $\hat{a}_\ell(v)$ of each elementary 
\ytextmodifhershelvonestepfive{spectrum,}
at each velocity $v$ in a set $\omega_\ell$, we estimate the probability density 
function 
\ytextmodifhershelvonestepfive{(pdf)}
$f_{\omega_\ell,v}$ from the available $p$ intensities with 
\ytextmodifhershelvonestepfive{the Parzen kernel}
method provided in Matlab (\textit{ksdensity}). Parzen kernel (\cite{the09}) is 
a parametric method to estimate the
\ytextmodifhershelvonestepfive{pdf}
of a random variable at any point of 
its support. For more details about this method, refer to Appendix A.

Each estimated elementary spectrum $\hat{a}_\ell$ is obtained by selecting the 
intensity $u$ that has the highest probability at a given wavelength:
\begin{equation}\label{eq_probaMax}
\hat{a}_\ell(v) = \operatorname*{argmax}_{u}  f_{\omega_\ell,v}(u) \qquad 
\ell\in\{1,\ldots,L\}.
\end{equation}

The estimation error at each wavelength $v$ for a given elementary spectrum 
$\hat{a}_\ell$ is obtained by selecting the intensities whose 
\ytextmodifhershelvonestepfive{pdf values}
are equal to $max(f_{\omega_\ell,v})/2$. Let 
$\left[\alpha_\ell(n),\beta_\ell(n)\right]$ be the error interval of 
$\hat{a}_\ell(v)$ such that:
\begin{equation}\label{eq_errorBar}
f_{\omega_\ell,n}(\hat{a}_\ell(n)-\alpha_\ell(n)) = 
f_{\omega_\ell,n}(\hat{a}_\ell(n)+\beta_\ell(n)) = \frac{1}{2} 
\operatorname*{max}\left(f_{\omega_\ell,n}\right).
\end{equation} 

\begin{figure}[htb]
\begin{center}
\includegraphics[width=0.4\textwidth]{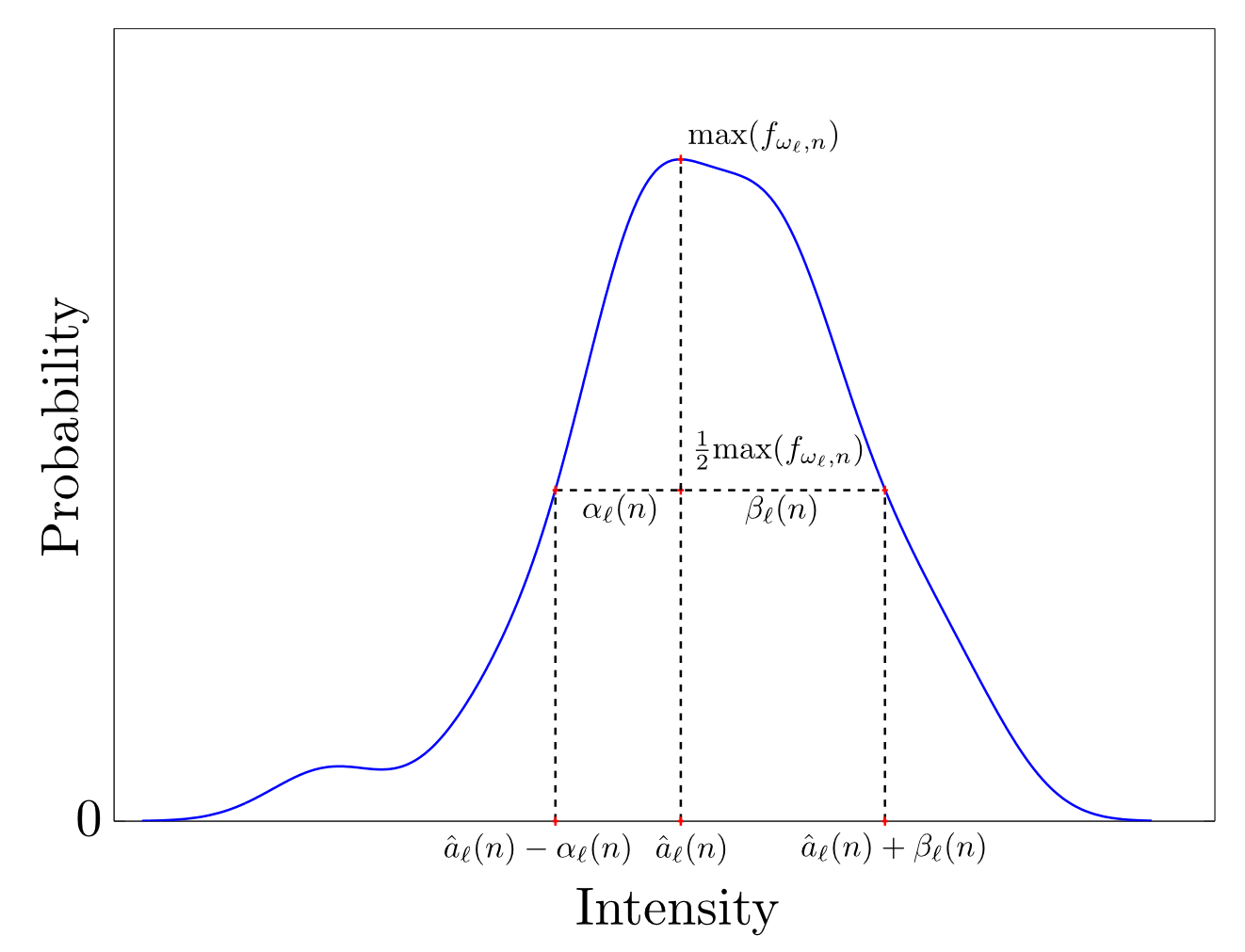}
\end{center}
\caption{Probability density function $f_{\omega_\ell,n}$ of intensities of the 
set $\omega_\ell$ at a given velocity $v$.}
\label{fig_pdf}
\end{figure}

The two endpoints $\alpha_\ell(n)$ and $\beta_\ell(n)$ are respectively the 
lower and upper error 
\ytextmodifhershelvonestepfive{bounds}
for each velocity. We illustrate this procedure in Figure \ref{fig_pdf} showing 
an example of 
\ytextmodifhershelvonestepfive{pdf}
annotated with the different characteristic points 
\ytextmodifhershelvonestepfive{defined above.}

\paragraph{2. Combination stage:}This final step consists of estimating the $L$ 
spatial 
{sources} from the estimation of elementary spectra and 
observations, under the 
\ytextmodifhershelvonestepfive{ nonnegativity constraint.}
Thus for each observed 
spectrum of index $m \in \{1,\ldots,M\}$, the 
{sources} are estimated by 
minimizing the objective function:
\begin{equation}\label{eq_NNLS}
\mathcal{J}(\hat{s}_m) = \frac{1}{2}\|x_m-\hat{A}\hat{s}_m\|^2_2. \qquad 
\hat{s}_m \geqslant 0,
\end{equation}
where $x_m$ is the $m^{th}$ observed spectrum (i.e., the $m^{th}$ column 
of $X$) and $\hat{s}_m$ the estimation of spatial contributions associated with 
each elementary spectrum (i.e., the $m^{th}$ column of $\hat{S}$). This 
is done by using the classical nonnegative least square algorithm 
(\cite{law74}). 
\ytextmodifhershelvonestepfive{We here}
used the version of this algorithm 
provided in Matlab (\textit{lsqnonneg}). The abundance maps are obtained by 
resizing the 
\ytextmodifhershelvonestepfive{columns}
of $\hat{S}$ into $P_x \times P_y$ matrices (reverse process 
\ytextmodifhershelvonestepfive{as compared with}
resizing 
$\underline{C}$).
\\~~

\noindent\rule{\linewidth}{.5pt}
\vspace{-0.7cm}
\begin{center}
Summary of MC-NMF method
\end{center}
\vspace{-0.5cm}
\noindent\rule{\linewidth}{.5pt}
\textbf{Requirement:} All points in $\underline{C}$ are nonnegative.
\begin{enumerate}
\item Identification of the number of 
{sources} $L$ (Section 
\ref{subseq_nb_S}).
\item Noise 
\ytextmodifhershelvonestepfive{reduction}
(Section \ref{subseq_NoiseRed}).
\item NMF:
	\begin{itemize}
	\item Random initialization of $\hat{A}$ and $\hat{S}$.
	\item Update $\hat{A}$ and $\hat{S}$ using (\ref{eq_up_A}) and 
(\ref{eq_up_S}). At each iteration, the column of $\hat{A}$ are normalized 
according to (\ref{eq_norm}).
	\item Stop updating when the convergence criterion (\ref{eq_conv_crit}) 
is reached.
	\end{itemize} 
\item Repeat Step 3. $p$ times for Monte-Carlo analysis.
\item Cluster normalized estimated spectra to form $L$ sets.
\item In each set, compute the 
\ytextmodifhershelvonestepfive{pdf}
of $p$ intensities at each velocity 
\ytextmodifhershelvonestepfive{and use}
(\ref{eq_probaMax}) to estimate the elementary spectra $\hat{A}$. The error bars 
of this estimate are deduced from the 
\ytextmodifhershelvonestepfive{pdf}
using (\ref{eq_errorBar}).
\item Reconstruct the spatial 
{sources} $\hat{S}$ 
\ytextmodifhershelvonestepfive{with a}
nonnegative least 
square 
\ytextmodifhershelvonestepfive{algorithm: see}
(\ref{eq_NNLS}).
\end{enumerate}
\vspace{-0.2cm}
\noindent\rule{\linewidth}{.5pt}

\subsection{\ytextmodifhershelvonestepfive{Sparse component analysis
based on single-source zones}}
\label{subseq_SCA}
SCA is another class of BSS 
\ytextmodifhershelvonestepfive{methods,}
based on the sparsity of 
{sources} in 
a given representation domain (time, space, frequency, time-frequency, 
time-scale). It became popular during the 2000s and several methods 
\ytextmodifhershelvonestepfive{then}
emerged. The 
\ytextmodifhershelvonestepfive{first SCA method used in this paper}
is derived from TIFCORR introduced by 
\cite{dev07}. In the original version, the method is used to separate 
one-dimensional signals, but an extension for images has been proposed by 
\cite{meg10}. This type of method is based on the assumption that there are some 
small zones in the considered domain of analysis where only one 
{source} 
is active, i.e.,  
\ytextmodifhershelvonestepfive{it}
has zero mean power in these zones called 
single-source zones. 
\ytextmodifhershelvonestepfive{We here use a spatial framework
(see model (\ref{eq_spacial_model})), so that we
assume}
that spatial single-source zones exist in the cube $\underline{C}$.

The sparsity considered here does not 
\ytextmodifhershelvonestepfive{correspond to the same property}
as the 
sparsity mentioned 
\ytextmodifhershelvonestepfive{in}
Section \ref{subseq_NMF}. In order to clarify this 
distinction, we introduce the notion of degree of sparsity. 
\ytextmodifhershelvonestepfive{Sparse signals may have different
numbers of coefficients equal to zero.}
If nearly all the coefficients 
are zero, we 
\ytextmodifhershelvonestepfive{define}
the signal as highly sparse. On the contrary, if only a 
few coefficients are zero, we 
\ytextmodifhershelvonestepfive{define}
the signal as weakly sparse.

\ytextmodifhershelvonestepfive{The sparsity assumption considered in
Section \ref{subseq_NMF} corresponds to the case when
the considered signal (spectrum 
or abundance map) contains}
a large number of negligible coefficients. 
\ytextmodifhershelvonestepfive{This therefore assumes a high sparsity,
which is not realistic in our context.}
%
%
%
\ytextmodifhershelvonestepfive{On the contrary,
the sparsity assumption used in the BSS method derived from TIFCORR
considered here only consists of requiring the existence of \emph{a
few tiny zones} in the considered domain (spatial 
domain in our case) where 
only one 
{source} is active. 
{ More precisely, separately for each source, that BSS method only requires the existence of at least one tiny zone (typically $5\times5$ pixels) where this source is active, and this corresponds to Assumption 1 defined below.}
We thus only require}
a weak 
spatial sparsity. More precisely, we use the joint sparsity (\cite{dev14b}) of 
the 
{sources} since we do not consider the sparsity of one 
{source} signal alone (i.e.,  the inactivity of this signal on a 
number of coefficients) but we consider the spatial zones where only one source 
signal is active%
\ytextmodifhershelvonestepfive{, whereas the others are}
simultaneously inactive. This constraint of joint sparsity is 
\ytextmodifhershelvonestepfive{weaker}
than a constraint of sparsity in the 
\ytextmodifhershelvonestepfive{sense of Section \ref{subseq_NMF},}
since it concerns a very 
small number of zones (at least one for each source).
The ``sparse component analysis method'' used hereafter might therefore be 
called a ``quasi-sparse component analysis method''.
\\~~

The method used here, called LI-2D-SpaceCorr-NC 
%
\ytextmodifhershelvonestepfive{and}
proposed by \cite{meg10} (which 
we just call SpaceCorr hereafter), is based on correlation parameters and has the 
following structure:
\begin{itemize}
\item The detection stage finds the single-%
{source} zones.
\item The estimation stage identifies the 
\ytextmodifhershelvonestepfive{columns}
of the mixing matrix 
corresponding to these single-%
{source} zones.
\item The combination stage recovers the 
{sources} from the estimated 
mixing matrix and the observations.
\end{itemize}

Before detailing these steps, some assumptions and definitions are to be 
specified. The spectral cube $\underline{C}$ is divided into small spatial zones 
(typically $5 \times 5$ pixels), denoted $Z$. These zones consist of adjacent 
pixels and the spectral cube is scanned spatially using adjacent or overlapping 
zones. We denote $X(Z)$ the matrix of observed spectra in $Z$ (each column of 
$X(Z)$ contains an observed spectrum).

First of all, as explained in Section \ref{subseq_NoiseRed}, preprocessing is 
necessary to minimize the 
\ytextmodifhershelvonestepfive{impact of noise}
on the results. For this particular 
method, we must keep the spatial data consistency. The aforementioned spatial 
thresholding is achieved by retaining only zones $Z$ whose power is greater than 
a threshold. Typically some zones contain only noise and are therefore 
unnecessary for the spectra estimation step (detection and estimation stages of 
SpaceCorr). As for the NMF, we set the threshold 
\ytextmodifhershelvonestepfive{to}
$\alpha_n=\max\limits_{Z} \|X(Z)\|_F \times 0.2$.
\\~~

\textbf{Definition 1:} A 
{source} is ``active'' in an analysis zone $Z$ 
if its mean power is zero in $Z$.
\\~~

\textbf{Definition 2:} A {source} is ``isolated'' in an analysis zone $Z$ 
if only this {source} is active in $Z$.
\\~~

\textbf{Definition 3:} A {source} is ``accessible'' in the representation 
domain if at least one analysis zone $Z$ where it is isolated exists.
\\~~

\textbf{Assumption 1:} Each {source} is spatially accessible.
\\~~

If the data satisfy this spatial sparsity assumption, then we can achieve the 
decomposition as follows:

\paragraph{1. Detection stage:}From expression (\ref{eq_spacial_model}) of $X$ 
and considering a single-source zone $Z$ where only the {source} 
$s_{\ell_0}$ is present, the observed signals become restricted to:
\begin{equation}
x_v(m) = a_{v\ell_0}~s_{\ell_0}(m) \qquad m\in Z,
\end{equation}
where $x_v$ is the $v^{th}$ row of $X$ and $s_{\ell_0}$ the $\ell_0^{th}$ row of 
$S$. We note that all the observed signals $x_v$ in $Z$ are proportional to each 
other. They all contain the same {source} $s_{\ell_0}$ weighted by 
a different factor $a_{v\ell_0}$ for each observation whatever the considered 
velocity $v$.
Thus, to detect the single-source zones, 
\ytextmodifhershelvonestepfive{the considered approach consists of using}
the 
correlation coefficients in order to quantify the observed signals 
proportionality. 
Let $R\{x_i,x_j\}(Z)$ 
\ytextmodifhershelvonestepfive{denote}
%
the centered 
\ytextmodifhershelvonestepfive{cross-correlation}
of the two observations 
$x_i$ and $x_j$ in $Z$:
\begin{equation}
R\{x_i,x_j\}(Z) = \frac{1}{Card(Z)} \sum_{m\in Z} x_i(m)x_j(m) \qquad \forall 
i,j\in\{1,\ldots,N\},
\end{equation}
where $Card(Z)$ is the number of samples (i.e., pixels) in $Z$. 
\ytextmodifhershelvonestepfive{On each analysis zone $Z$, we estimate}
the centered correlation coefficients 
$\rho\{x_i,x_j\}(Z)$ between all pairs of observations: 
\begin{equation}\label{eq_rho}
\rho\{x_i,x_j\}(Z) = \frac{R\{x_i,x_j\}(Z)}{\sqrt{R\{x_i,x_i\}(Z) \times 
R\{x_j,x_j\}(Z)}} \qquad \forall i,j\in\{1,\ldots,N\}.
\end{equation}
We note that these coefficients are undefined if all {sources} are equal to 
zero. So we add the following condition:
\\~~

\textbf{Assumption 2:} On each analysis zone $Z$, at least one {source} 
is active.
\\~~

For each zone $Z$ we obtain a correlation matrix $\rho$. In \cite{dev14b}, the 
authors show that for linearly independent {sources}, a necessary and 
sufficient condition for a {source} to be isolated in a zone $Z$ is:
\begin{equation}
|\rho\{x_i,x_j\}(Z)| = 1 \qquad i,j\in\{1,\ldots,N\}, ~ i<j.
\end{equation} 
To measure the single-source quality $q_Z$ of an analysis zone, the matrix 
$\rho$ is aggregated by calculating the mean $q_Z = 
\overline{|\rho\{x_i,x_j\}(Z)|}$, over $i$ and $j$ indices%
\ytextmodifhershelvonestepfive{, with
$
i<j
$}%
. The best single-source zones are the zones where the quality coefficient $q_Z$ is the 
highest. To ensure the detection of 
single-source 
\ytextmodifhershelvonestepfive{zones,}
the coefficient $q_Z$ 
must be less than 1 for multi-source zones. We then set the following 
constraint:
\\~~

\textbf{Assumption 3:} Over each analysis zone, all active {sources} are 
linearly independent if at least two active {sources} exist in this zone.
\\~~

The detection stage therefore consists in keeping the zones for which the 
quality coefficient is above a threshold defined by the user. 

\paragraph{2. Estimation stage:}
\ytextmodifhershelvonestepfive{Successively considering each
previously selected single-source zone,}
the correlation parameters $R\{x_i,x_j\}(Z)$ between pairs 
of bands allow one to estimate a column of the mixing matrix $A$ up to a scale 
factor:
\begin{equation}\label{eq_est_A}
\frac{R\{x_1,x_v\}(Z)}{R\{x_1,x_1\}(Z)} = \frac{a_{v\ell_0}}{a_{1\ell_0}} \qquad 
v\in\{1,\ldots,N\}.
\end{equation}

\ytextmodifhershelvonestepfive{The}
choice of the observed signal of index 1 as 
\ytextmodifhershelvonestepfive{a}
reference is 
arbitrary%
\ytextmodifhershelvonestepfive{:}
it can be replaced by any other 
\ytextmodifhershelvonestepfive{observation.}
In practice, the 
observation with the greatest power will be chosen as the reference in order to 
limit the risk of 
\ytextmodifhershelvonestepfive{using}
a highly noisy signal as 
\ytextmodifhershelvonestepfive{the}
reference.

\ytextmodifhershelvonestepfive{Moreover,}
to avoid any division by zero, we assume that:
\\~~

\textbf{Assumption 4:} All mixing coefficient $a_{1\ell}$ are zero.
\\~~

As for MC-NMF, the scale factor $\frac{1}{a_{1\ell_0}}$ of the estimated 
spectrum is then compensated 
\ytextmodifhershelvonestepfive{for,}
%
by normalizing each estimated spectrum so that 
$\int a_\ell(v) ~ dv = 1$. We thus obtain a set of potential columns of 
$\hat{A}$. We apply clustering (K-means  with a correlation criterion) to these 
best columns in order to regroup  the estimates corresponding to 
the same column of the mixing matrix in $L$ clusters. The mean of each cluster is retained to 
form a column of the matrix $\hat{A}$.

\paragraph{3. Combination stage:}The {source} matrix estimation step is 
identical to 
\ytextmodifhershelvonestepfive{that}
used for the NMF method (see previous section). It is 
performed by minimizing the cost function 
(\ref{eq_NNLS}) with a nonnegative 
least square algorithm.
\\~~

\noindent\rule{\linewidth}{.5pt}
\vspace{-0.7cm}
\begin{center}
Summary of 
\ytextmodifhershelvonestepfive{SpaceCorr}
method
\end{center}
\vspace{-0.5cm}
\noindent\rule{\linewidth}{.5pt}
\textbf{Requirements:} Each {source} is spatially accessible. On each 
zone $Z$, at least one {source} is active and all active {sources} 
are linearly independent. All mixing coefficient $a_{1\ell}$ are zero.
\begin{enumerate}
\item Identification of the number of {sources} $L$ (Section 
\ref{subseq_nb_S}).
\item Noise 
\ytextmodifhershelvonestepfive{reduction}
(Section \ref{subseq_NoiseRed}).
\item Compute the single-source quality coefficients $q_Z = 
\overline{|\rho\{x_i,x_j\}(Z)|}$ for all analysis zones $Z$.
\item Keep the zones where the quality coefficient is above a threshold.
\item For each above zone, estimate the potential column of $\hat{A}$ with 
(\ref{eq_est_A}) and normalize 
\ytextmodifhershelvonestepfive{it}
so that $\int a_\ell(v) ~ dv = 1$.
\item 
\ytextmodifhershelvonestepfive{Cluster}
potential columns to form $L$ sets. The mean of each cluster 
forms a final column of $\hat{A}$.
\item Reconstruct {sources} $\hat{S}$ 
\ytextmodifhershelvonestepfive{with a}
nonnegative least square
\ytextmodifhershelvonestepfive{algorithm: see (\ref{eq_NNLS}).}
\end{enumerate}
\vspace{-0.2cm}
\noindent\rule{\linewidth}{.5pt}
\\~~

The efficiency of SpaceCorr significantly depends on the size of the analysis 
zones $Z$. Too little zones do not allow one to reliably evaluate the 
correlation parameter $\rho\{x_i,x_j\}(Z)$, hence to reliably evaluate the 
single-source quality of the zones. Conversely, too large zones do not ensure 
the presence of single-source zones. Furthermore, the size of the zones must be 
compatible with the data. A large number of source signals or a low number of 
pixels in the data can jeopardize the presence of single-source zones for each 
source. 

Thus, it is necessary to relax the sparsity condition in order to separate such 
data. The size of these single-source zones being a limiting factor, we suggest 
to reduce them to a minimum, i.e., 
\ytextmodifhershelvonestepfive{to one pixel:}
we assume that there exists at least 
one single-source pixel per source in the data. 
\ytextmodifhershelvonestepfive{To exploit this property,}
we developed a 
\ytextmodifhershelvonestepfive{geometric BSS}
method called MASS (for Maximum Angle Source Separation) 
(\cite{bou15})%
\ytextmodifhershelvonestepfive{, which applies to}
data that do not meet the SpaceCorr assumptions. We note 
however that MASS does not make 
SpaceCorr 
obsolete. SpaceCore generally 
yields better results than MASS 
\ytextmodifhershelvonestepfive{for}
data with single-source zones. This will be 
detailed in Section \ref{subseq_unmixing} devoted to experimentations.

\subsection{\ytextmodifhershelvonestepfive{Sparse component analysis
basd on single-source pixels}}
\label{subsec_MASS}
The MASS method (\cite{bou15}) is a BSS method based on the geometrical 
representation of data and a sparsity assumption 
\ytextmodifhershelvonestepfive{on}
sources. For this method, we 
assume that there are at least one pure pixel per source. The spectrum 
associated with a pure pixel contains the contribution of only one elementary 
spectrum. This sparsity assumption is of the same nature as the one introduced 
for SpaceCorr (i.e., spatial sparsity), 
\ytextmodifhershelvonestepfive{but}
the size of the zones Z is
\ytextmodifhershelvonestepfive{here}
%
reduced to a single pixel. Once again, we use the spatial model described in 
Section \ref{subseq_spatial_model}. 
\ytextmodifhershelvonestepfive{With the terminology
introduced in Section
\ref{subseq_SCA} for 
the SpaceCorr method,
we here use}
the 
following assumption:
\\~~

\textbf{Assumption 
\ytextmodifhershelvonestepfive{$ 1^{\prime}$}
:} For each source, there exist at least one pixel (spatial 
sample) where this source is isolated (i.e., each source is spatially 
accessible).
\\~~

Before detailing the MASS algorithm, we provide a geometrical framework 
\ytextmodifhershelvonestepfive{for}
the 
BSS problem. Each observed spectrum (each column of $X$) is represented as an 
element of 
\ytextmodifhershelvonestepfive{the}
$\mathbb{R}^N$ vector space:
\begin{equation}
x_m = As_m,
\end{equation}
where $x_m$ is a nonnegative linear combination of columns of $A$. The set of 
all 
\ytextmodifhershelvonestepfive{possible 
(i.e., not necessarily present in the measured data matrix $X$)
nonnegative combinations $x_*$
of the $L$
columns of $A$ is}
\begin{equation}
\mathcal{C}(A) = \{x_*~|~x_*=As_*,~s_*\in \mathbb{R}^L_+\}.
\end{equation}
%
\ytextmodifhershelvonestepfive{This defines}
a simplicial cone whose 
$L$ edges 
are spanned by the $L$ column vectors 
\ytextmodifhershelvonestepfive{$a_\ell$}
of $A$:
\begin{equation}\label{eq_edge}
\mathcal{E}_\ell = \{x_*~|~x_*=\alpha a_\ell,~\alpha\in \mathbb{R}_+\},
\end{equation}
where $\mathcal{E}_\ell$ is the $\ell^{th}$ edge of the simplicial cone 
$\mathcal{C}(A)$. We notice that the simplicial cone $\mathcal{C}(A)$ is a convex 
hull, each nonnegative linear combination of columns of $A$ is contained within 
$\mathcal{C}(A)$.

\ytextmodifhershelvonestepfive{Here,}
the 
mixing coefficients and the sources are nonnegative. 
\ytextmodifhershelvonestepfive{The 
observed spectra are therefore}
contained in the simplicial cone spanned by the column of 
$A$, i.e., by the elementary spectra. If we add the 
\ytextmodifhershelvonestepfive{above-defined sparsity assumption}
(Assumption 
\ytextmodifhershelvonestepfive{$1^{\prime}$}%
), the observed data matrix contains at least 
one pure pixel (i.e., 
\ytextmodifhershelvonestepfive{a pixel}
containing the contribution of a unique column of 
$A$) for each source.

The expression of such a pure observed spectrum, where only the source of index 
$\ell_0\in\{1,\ldots,L\}$ is nonzero, is restricted to:
\begin{equation}
x_m = a_{\ell_0} s_{\ell_0m} 
\label{eq-x-m}
\end{equation}
where $a_{\ell_0}$ is the $\ell_0^{th}$ column of $A$. 
\ytextmodifhershelvonestepfive{Since $s_{\ell_0m}$ is a 
nonnegative scalar,  
(\ref{eq-x-m})
corresponds to}
an edge vector of 
\ytextmodifhershelvonestepfive{the}
simplicial 
cone $\mathcal{C} (A)$ 
\ytextmodifhershelvonestepfive{defined}
(\ref{eq_edge}). Therefore, the edge vectors are 
\ytextmodifhershelvonestepfive{actually present in the observed}
data. 

\begin{figure}[htb]
\begin{center}	
\includegraphics[width=0.5\textwidth]{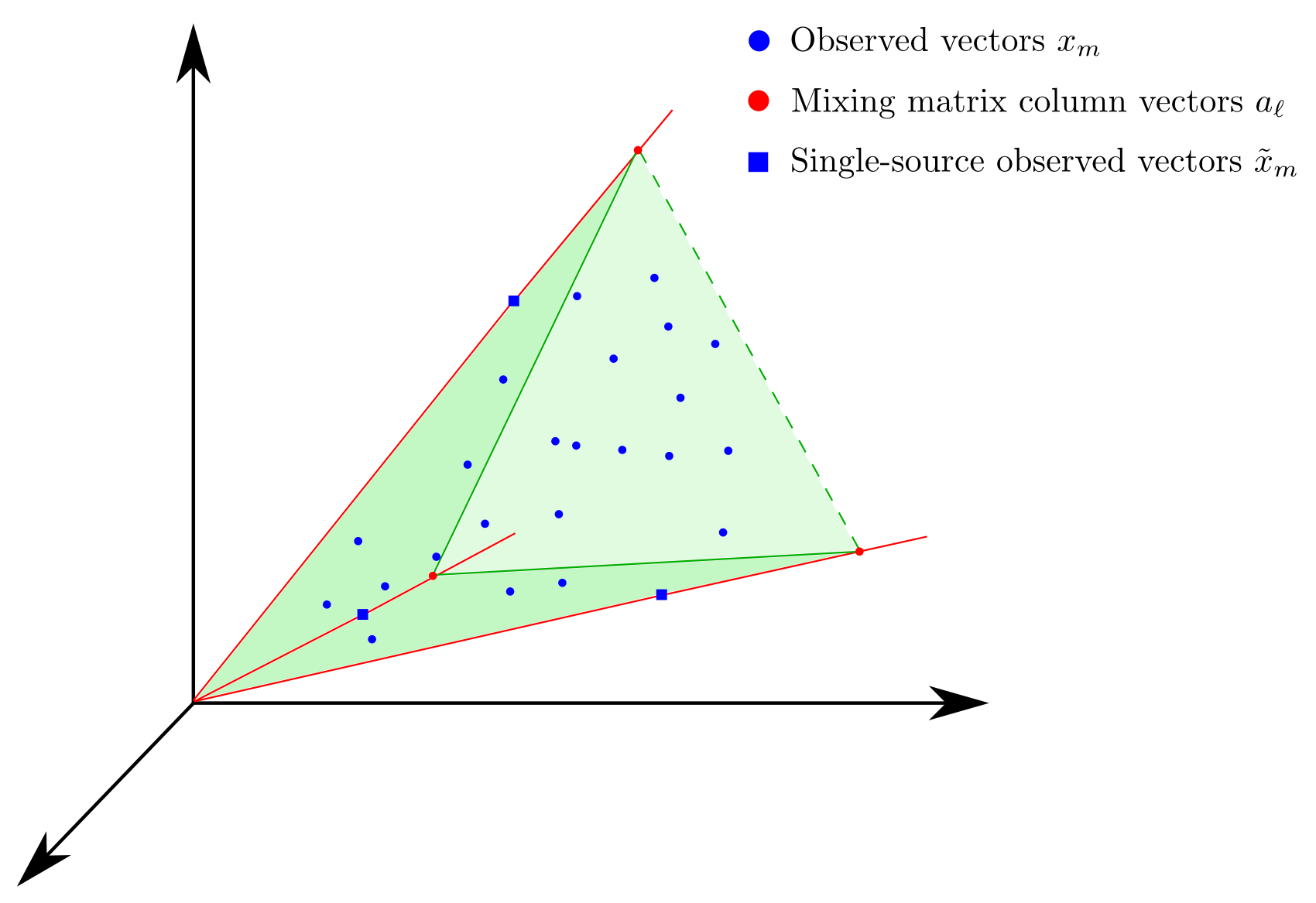}
\end{center}
\caption{Scatter plot of mixed data and edges $\mathcal{E}_\ell$ of the 
simplicial cone in 
\ytextmodifhershelvonestepfive{the three-dimensional}
case. The columns of $X$ are shown in blue 
and those of $A$ in red.}
\label{fig_example_cone}
\end{figure}

To illustrate these properties, we create a scatter plot of data in three 
dimensions (Fig.~\ref{fig_example_cone}). These points are generated from 
nonnegative linear combinations of 3 sources. On the scatter plot, the blue 
points represent the mixed data (i.e., the columns of $X$), the red 
points represent the generators of data (i.e., the columns of $A$). As 
previously mentioned, the observations $x_m$ are contained in the simplicial 
cone spanned by the columns of the mixing matrix $A$. Moreover, if the red 
points are among the observed vectors 
\ytextmodifhershelvonestepfive{(i.e., if}
Assumption 
\ytextmodifhershelvonestepfive{$1^{\prime}$}
is verified), the 
simplicial cone spanned by $A$ is the same as the simplicial cone spanned by 
$X$.

From these 
\ytextmodifhershelvonestepfive{properties,}
we develop the MASS method%
\ytextmodifhershelvonestepfive{, which}
aims to unmix the 
hyperspectral data. 
%
It 
operates in two stages.
The first 
\ytextmodifhershelvonestepfive{one}
%
is the identification 
of the mixing matrix $A$
\ytextmodifhershelvonestepfive{and}
the second 
\ytextmodifhershelvonestepfive{one}
is the reconstruction of source matrix $S$. 
If the data satisfy the spatial sparsity assumption, then we can achieve the 
decomposition as follows:

\paragraph{1. Mixing matrix identification:}Identifying the columns of the 
matrix $A$ (up to scale indeterminacies) is equivalent to identifying each edge 
vector of the simplicial cone $\mathcal{C} (A)$ spanned by the data matrix $X$. 
The observed vectors being nonnegative, the identification of the edge vectors 
reduces to identifying the observed vectors which are furthest apart in the 
angular sense. 

First of all, the columns of $X$ are normalized to unit length (i.e. $\|x_m\| = 
1$) to simplify the following equations. The 
\ytextmodifhershelvonestepfive{identification algorithm operates}
in $L-1$ steps.
The first step 
\ytextmodifhershelvonestepfive{identifies}
two columns of $\hat{A}$ by selecting the two 
columns of $X$ that have the largest angle. We denote $x_{m_1}$ and $x_{m_2}$ this 
pair of observed spectra. We have:
\begin{equation}\label{eq_AngleCos}
(m_1,m_2) = \operatorname*{argmax}_{i,j} \cos^{-1}({x_i}^T x_j) \qquad \forall 
i,j\in\{1,\ldots,M\}.
\end{equation}
Moreover, the $\cos^{-1}$ function being monotonically decreasing on $[0,1]$, 
Equation (\ref{eq_AngleCos}) can be simplified to:
\begin{equation}\label{eq_AnglePS}
(m_1,m_2) = \operatorname*{argmin}_{i,j} {x_i}^T x_j \qquad \forall 
i,j\in\{1,\ldots,M\}.
\end{equation}
We denote $\tilde{A}$
the sub-matrix of $\hat{A}$
formed by these two columns:
\begin{equation}
\tilde{A} = [x_{m_1},x_{m_2}]
.
\end{equation}

The next step 
\ytextmodifhershelvonestepfive{consists of identifying}
the column which has the largest angle with 
$x_{m_1}$ and $x_{m_2}$. This column is defined as the one which is furthest in 
the angular sense from its orthogonal projection on the simplicial cone spanned 
by $x_{m_1}$ and $x_{m_2}$. Let $\Pi_{\tilde{A}}(X)$ be the projection of 
columns of $X$ on the simplicial cone spanned by the columns of $\tilde{A}$:
\begin{equation}\label{eq_Proj}
\Pi_{\tilde{A}}(X) = \tilde{A} (\tilde{A}^T\tilde{A})^{-1} \tilde{A}^T X.
\end{equation}
To find the column of $X$ which is the furthest from its projection, we proceed 
in the same way as to identify the first two columns. Let $m_3$ be the index of 
this column:
\begin{equation}\label{eq_AngleProj}
m_3 = \operatorname*{argmin}_{i} {x_i}^T \pi_i \qquad \forall i\in\{1,\ldots,M\},
\end{equation}
where $\pi_i$ is the $i^{th}$ column of $\Pi_{\tilde{A}}(X)$. The new estimate 
of the mixing matrix is then $\tilde{A} = [x_{m_1},x_{m_2},x_{m_3}]$. This 
projection and identification procedure is then repeated to identify the $L$ 
columns of the mixing matrix. For example, the index $m_4$ can be identified by 
searching the column of $X$ which forms the largest angle with its projection on 
the simplicial cone spanned by the columns of $\tilde{A} = 
[x_{m1},x_{m2},x_{m3}]$. Finally, the mixing matrix is completely estimated:
\begin{equation}\label{eq_Atild}
\hat{A} = [x_{m_1},\ldots,x_{m_L}]
.
\end{equation}

However, this mixing matrix estimation is very sensitive to noise since 
$\hat{A}$ is constructed directly from observations. In order to make the 
estimate more robust to noise and to consider the case when several 
single-source vectors, relating to the same source, are present in the observed 
data, we introduce a tolerance margin upon selection of the columns. Instead of 
selecting the column that has the largest angle with its projection (or both 
first columns which are furthest apart), we select all columns which are nearly 
collinear 
\ytextmodifhershelvonestepfive{to}
the identified column. For each column $x_{m_\ell}$ previously 
identified according to Equation (\ref{eq_Atild}), we construct the set 
$\mathcal{A}_\ell$:
\begin{equation}
\mathcal{A}_\ell = \{{x_i}~|~x_{m_\ell}^T x_i \geq \kappa \} \qquad 
i\in\{1,\ldots,M\},~ \ell\in\{1,\ldots,L\},
\end{equation}
where $\kappa$ is the tolerance threshold of an inner product (thus included in 
$[0,1]$). It must be chosen close to $1$ to avoid selecting mixed observations 
(typically $\kappa=0.99$). The column of the new mixing matrix $\hat{A}$ is 
obtained by averaging the columns in each set $\mathcal{A}_\ell$%
\ytextmodifhershelvonestepfive{,}
which reduces 
the influence of noise:
\begin{equation}
\hat{A} = [\bar{\mathcal{A}_1},\ldots,\bar{\mathcal{A}_L}],
\end{equation}
where $\bar{\mathcal{A}_\ell}$ is the average column of the set 
$\mathcal{A}_\ell$. Thus, we obtain an estimate of the mixing matrix $A$ up to 
permutation and scale factor indeterminacies.

\paragraph{2. Source matrix reconstruction:}The {source} matrix 
estimation step is identical to those used for the NMF and SpaceCorr methods. It 
is performed by minimizing the cost function 
(\ref{eq_NNLS}) with a 
nonnegative least square algorithm.
\\~~

\noindent\rule{\linewidth}{.5pt}
\vspace{-0.7cm}
\begin{center}
Summary of MASS method
\end{center}
\vspace{-0.5cm}
\noindent\rule{\linewidth}{.5pt}
\textbf{Requirements:} 
\ytextmodifhershelvonestepfive{For}
each 
source, there exist at least one pixel (spatial sample) where this source is 
isolated. All points in $\underline{C}$ are nonnegative.

\begin{enumerate}
\item Identification of the number of {sources} $L$ (Section 
\ref{subseq_nb_S}).
\item Noise 
\ytextmodifhershelvonestepfive{reduction}
(Section \ref{subseq_NoiseRed}).
\item Normalization 
\ytextmodifhershelvonestepfive{of the observed spectra $x_m$ 
to unit length.}
\item Selection of the two columns of $X$ that have the largest angle according 
to (\ref{eq_AnglePS}). 
\item Repeat $L-
\ytextmodifhershelvonestepfive{2}
$ times the procedure of projection (\ref{eq_Proj}) and 
identification (\ref{eq_AngleProj}) to obtain the whole mixing matrix $\hat{A}$.
\item Normalization of the columns of $\hat{A}$ so that $\int a_\ell(v) ~ dv = 
1$.
\item Reconstruct the {sources} $\hat{S}$ using 
\ytextmodifhershelvonestepfive{a nonnegative least square
algorithm: see (\ref{eq_NNLS}).}
\end{enumerate}
\vspace{-0.2cm}
\noindent\rule{\linewidth}{.5pt}
\\~~

\subsection{Hybrid 
\ytextmodifhershelvonestepfive{methods}%
}
The BSS methods presented above have advantages and drawbacks. 
\ytextmodifhershelvonestepfive{NMF and its extended version, MC-NMF, are
attractive because they explicitly
request only the nonnegativity of the considered data
(as opposed, e.g., to sparsity). However, without additional assumptions,
they e.g. do not
provide a unique decomposition, as mentioned above.}
The 
SpaceCorr method is influenced by the degree of spatial sparsity present in the 
data. Indeed, in practice, the assumption of perfectly single-source zones ($q_Z 
= 1$) may not be realistic. In such conditions, the zones $Z$ retained for the 
unmixing are contaminated by the presence, small but not negligible, of other 
{sources}. However, SpaceCorr provides a unique decomposition and the 
algorithm does not require
initialization.
\ytextmodifhershelvonestepfive{MASS then allows one to reduce the
required size of single-source zones to a single pixel, but 
possibly at the expense of a higher sensitivity to noise.}

In order to take advantage of the benefits and reduce the drawbacks specific to 
each of these methods, we 
\ytextmodifhershelvonestepfive{hereafter}
combine them. The spectra and abundance maps 
estimated with SpaceCorr may not be perfectly 
\ytextmodifhershelvonestepfive{unmixed, i.e.}
elementary, but provide a good 
approximation of the actual components. To improve the decomposition, these 
approximations are then refined by initializing MC-NMF with these estimates of 
elementary spectra or abundance maps from SpaceCorr (the choice of $\hat{A}$ or 
$\hat{S}$ initialized in this way will be discussed in Section 
\ref{subseq_unmixing}). Thus the starting point of 
\ytextmodifhershelvonestepfive{MC-NMF}
is close to 
\ytextmodifhershelvonestepfive{a}
global 
minimum of the objective function%
%
\ytextmodifhershelvonestepfive{,}
which reduces the possibility 
%
\ytextmodifhershelvonestepfive{for MC-NMF}
to 
converge to a local minimum. The variability of results is greatly reduced%
%
\ytextmodifhershelvonestepfive{,}
which 
leads to 
\ytextmodifhershelvonestepfive{low-amplitude error bars.}

Thus we obtain two new%
\ytextmodifhershelvonestepfive{, hybrid,}
methods: 
MC-NMF initialized with the spectra obtained 
from SpaceCorr, which we call 
\ytextmodifhershelvonestepfive{SC-NMF-Spec},
and 
%
\ytextmodifhershelvonestepfive{MC-NMF}
initialized with the 
abundance maps obtained from SpaceCorr, which we call 
\ytextmodifhershelvonestepfive{SC-NMF-Map.}

\ytextmodifhershelvonestepfive{Similarly, two other new hybrid methods
are obtained by using the MASS method, instead of
SpaceCorr, to initialize MC-NMF:
initializing MC-NMF with the spectra obtained with MASS yields the
MASS-NMF-Spec method, whereas
initializing MC-NMF with the maps obtained with MASS yields the
MASS-NMF-Map method.}
%

\section{Experimental results}
\label{sec_exp}
\subsection{Synthetic data}\label{subseq_Data}

\begin{figure*}[htb]
\begin{center}	
\includegraphics[width=1\textwidth]{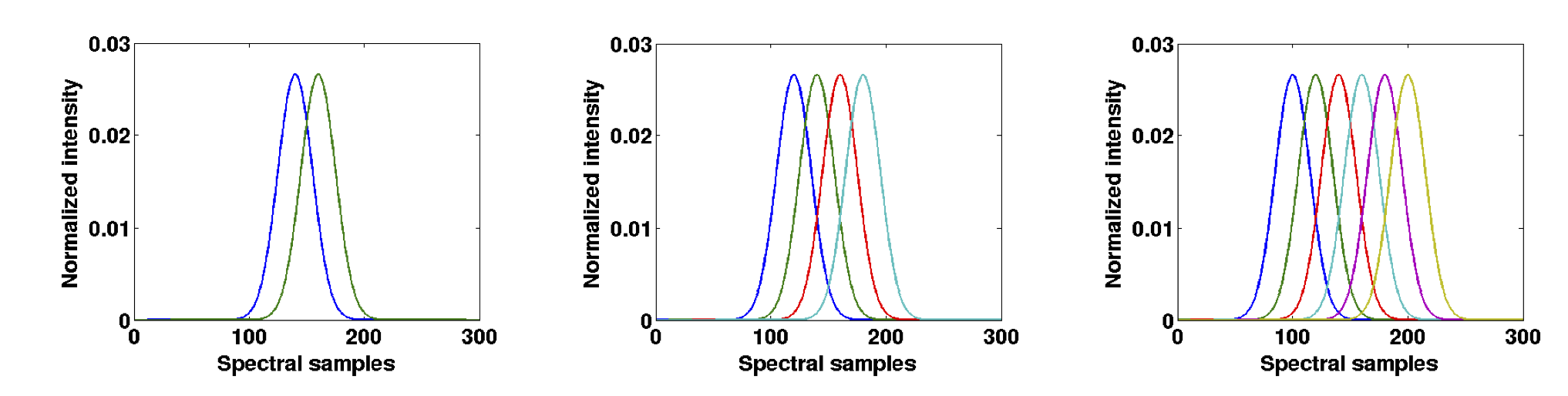}
\caption{\label{Fig_spectral_sources} From left to right, two, four, or six elementary spectra used to create 
the synthetic data.}
\end{center}
\end{figure*}

To evaluate the performance of 
\ytextmodifhershelvonestepfive{all considered}
methods, we generate data cubes containing 2, 
4 or 6 elementary spectra (Fig.~\ref{Fig_spectral_sources})
\ytextmodifhershelvonestepfive{which have}
300 samples. The 
\ytextmodifhershelvonestepfive{spectra}
are simulated using 
\ytextmodifhershelvonestepfive{Gaussian}
functions 
\ytextmodifhershelvonestepfive{(}%
that integrate to one) with same 
standard deviation 
\ytextmodifhershelvonestepfive{$\sigma_{Spec}$}
(cases with different standard deviations were 
also studied and yield similar results). To obtain the different elementary 
spectra of a mixture, we vary the mean of the Gaussian functions. Thus we 
simulate the Doppler-Shift specific to each {source}. 

\begin{figure}[htb]
\begin{center}	
\begin{tabular}{cc}
\includegraphics[width=0.23\textwidth]{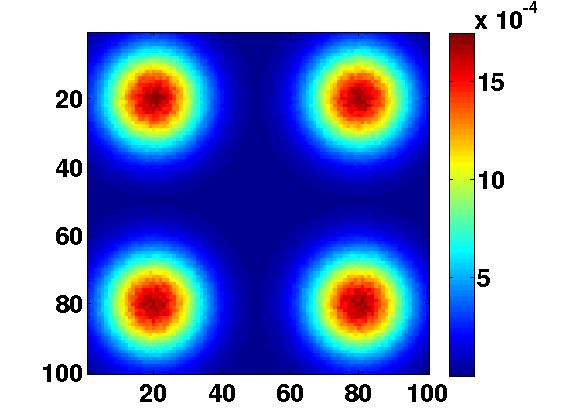}&
\includegraphics[width=0.23\textwidth]{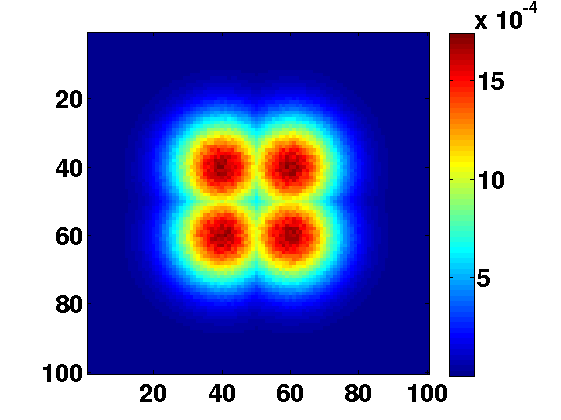}
\end{tabular}
\end{center}
\caption{Spatial positions of different 2D Gaussian functions for four 
{sources}. The left map shows the case without overlap 
\ytextmodifhershelvonestepfive{($d=6 \sigma_{Map}$)}
and the right map shows the case with maximum overlap 
\ytextmodifhershelvonestepfive{($d=2 \sigma_{Map}$).}
The 
intermediate cases 
\ytextmodifhershelvonestepfive{($d=5 \sigma_{Map}$,
$d=4 \sigma_{Map}$
and
$d=3 \sigma_{Map}$)}
are not 
represented.} 
\label{fig_Maps}
\end{figure}

The spatial abundance 
\ytextmodifhershelvonestepfive{maps}
are simulated using 
\ytextmodifhershelvonestepfive{2D}
Gaussian 
functions, each map having the same standard deviation 
\ytextmodifhershelvonestepfive{$\sigma_{Map}$}
on the $x$ and 
$y$ axes. For each 2D Gaussian, we define its influence zone as the 
\ytextmodifhershelvonestepfive{pixel}
locations between its peak and a distance of 
\ytextmodifhershelvonestepfive{$3 \sigma_{Map}$}.
Beyond this distance, 
the corresponding spatial contributions will be assumed 
\ytextmodifhershelvonestepfive{to be}
negligible. To add 
spatial sparsity, we vary the spatial position of each 2D Gaussian to get more 
or less overlap between them (see Fig.~\ref{fig_Maps}). The distance $d$ between 
two peaks is varied from 
\ytextmodifhershelvonestepfive{$6 \sigma_{Map}$ down}
to 
\ytextmodifhershelvonestepfive{$2 \sigma_{Map}$}
with a 
\ytextmodifhershelvonestepfive{$1 \sigma_{Map}$}
step. The 
extreme case 
\ytextmodifhershelvonestepfive{$2 \sigma_{Map}$}
still yields single-source zones to meet the 
assumptions of SpaceCorr. Thus we build 5 different mixtures of the same 
{sources}, each 
\ytextmodifhershelvonestepfive{involving}
more or less sparsity.

Moreover, to ensure the assumption of linear independence of {sources} 
(i.e., abundance maps from the point of view of SpaceCorr), each map is 
slightly disturbed by a uniform multiplicative noise. Thus the symmetry of 
synthetic scenes does not introduce linear relationships between the different 
maps. Finally%
\ytextmodifhershelvonestepfive{, we add white Gaussian noise to each cube,}
to get a signal to noise 
ratio (SNR) of 10, 20 or 30 dB%
\ytextmodifhershelvtwostepfiftyone{, unless otherwise stated (see in particular Appendix \ref{sec-appendix-Results-low-SNR} where the case of low SNRs is considered).}

It is important to note, finally, that the objective here is not to produce 
a realistic simulated astrophysical scene or simulated dataset,
but rather to have synthetic data that fulfill the 
statistical properties listed in Sect.~\ref{Sec_prob-stat}, and in which
we can vary simple parameters to test their effect on the performances of the method. 
We also note that \citet{gui19} are currently developing a model of scene
and instruments to provide a realistic synthetic 
JWST hyperspectral datasets, and the present method could be tested 
on these upcoming data.

\subsection{Estimation of number of sources}
\begin{figure}[htb]
\begin{center}	
\includegraphics[width=0.35\textwidth]{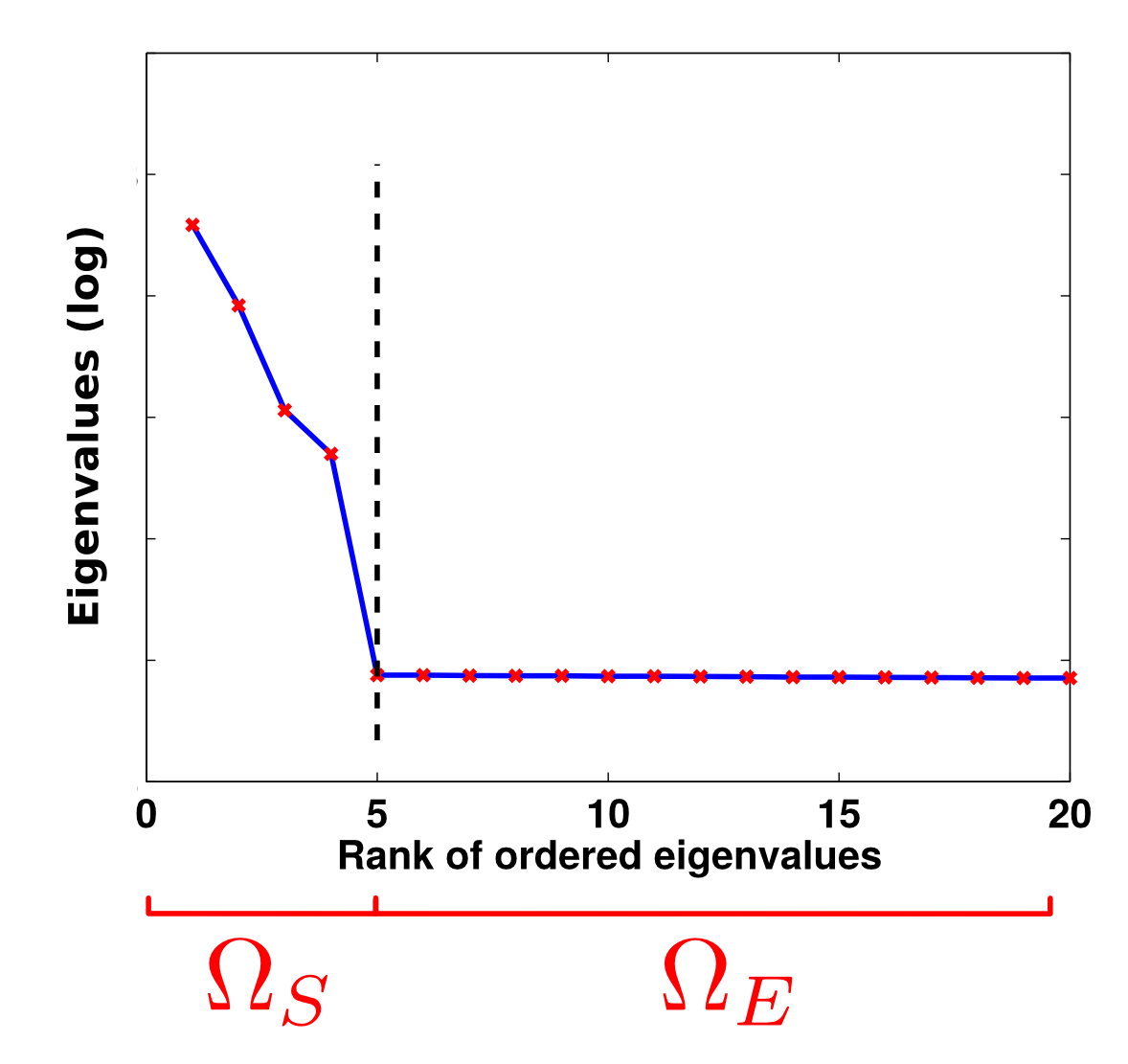}
\end{center}
\caption{Example of identification of number 
\ytextmodifhershelvonestepfive{of}
%
{sources} for a synthetic 
mixture of four {sources} with 
SNR = 10 dB
and 
\ytextmodifhershelvonestepfive{$d=2 \sigma_{Map}$}%
.}
\label{fig_VP_example}
\end{figure}

We tested the method 
\ytextmodifhershelvonestepfive{used}
to estimate the number of {sources} (Section 
\ref{subseq_nb_S}) on our 45 synthetic data cubes. For each of them, we found 
the true number of {sources} of the mixture. These results are 
unambiguous because the difference between $\lambda_{(L)}$ and $\lambda_{(L+1)}$ 
clearly appears. We can easily differenciate the two parts $\Omega_S$ and 
$\Omega_E$ on each curve of ordered eigenvalues. We illustrate the method for a 
mixture of 4 {sources} in Fig.~\ref{fig_VP_example}. A ``break'' is 
clearly observed in the curve of ordered eigenvalues at index $r=5$. The number 
of {sources} identified by the method is correct, $L=r-1=4$.

\subsection{Unmixing}\label{subseq_unmixing}
\subsubsection{Quality measures}
We now present the performance of the different BSS methods introduced in 
Section \ref{seq_Methods}: MC-NMF, SpaceCorr, MASS and their hybrid versions. To 
study the behavior of these methods, we apply them to the 45 synthetic cubes. 
We use two measures of error as performance criteria, one for maps and the other 
for spectra. The Normalized Root Mean Square Error (NRMSE) 
\ytextmodifhershelvonestepfive{defines}
the error of 
estimated maps:
\begin{equation}
NRMSE_\ell = \frac{\| s_\ell - \hat{s}_\ell\|}{\|  s_\ell \|}.
\label{eq-def-NRMSE}
\end{equation}
The spectral angle mapper (SAM) normalized root mean square error

\ytextmodifhershelvonestepfive{defines}
the 
error of estimated spectra. This usual measurement in hyperspectral imaging for 
Earth observation is defined as the angle formed by two spectra:
\begin{equation}
SAM_\ell = \operatorname*{arccos} \left( \frac{a_\ell^T 
\hat{a}_\ell}{\|a_\ell\|.\|\hat{a}_\ell\|} \right).
\label{eq-def-SAM}
\end{equation}

The Monte Carlo analysis associated with the NMF makes it possible to define the 
\ytextmodifhershelvonestepfive{spread}
of the solutions given by each of the $K$ 
\ytextmodifhershelvonestepfive{runs}
of the NMF. For 
each estimated {source}, we construct the envelope giving the 
\ytextmodifhershelvonestepfive{spread}
of the solutions around the most probable solution according to 
(\ref{eq_errorBar}). The amplitude of the envelope is normalized by the maximum 
intensity in order to obtain the error bars as a percentage of the maximum 
intensity. This normalization is arbitrary and makes it possible to express the 
\ytextmodifhershelvonestepfive{spread}
of the MC-NMF independently 
\ytextmodifhershelvonestepfive{from}
the spectral intensity. 
%
\ytextmodifhershelvonestepfive{We first denote as}
$\text{NMCEB}_\ell$ (for Normalized Monte Carlo Error Bar) the 
normalized error associated with the $\ell^{th}$ elementary spectrum:
\begin{equation}
\text{NMCEB}_\ell(n) = \frac{\alpha_\ell(n) + \beta_\ell(n)}{U_{\ell}} \qquad 
\forall n \in \{1,\ldots,N\},
\end{equation}
where $U_{\ell} = \underset{n}{\operatorname*{max}} \{a_\ell(n)\}$ is the 
maximal intensity of the $\ell^{th}$ elementary spectrum.
To quantify the total 
\ytextmodifhershelvonestepfive{spread}
of MC-NMF solutions for a data cube, the 
\ytextmodifhershelvonestepfive{above parameter}
is 
\ytextmodifhershelvonestepfive{then}
maximized along the spectral axis:
\begin{equation}
\text{NMCEB}_\ell^{\text{max}} = \operatorname*{max}_{n} 
\{\text{NMCEB}_\ell(n)\}.
\end{equation}

For clarity, we 
\ytextmodifhershelvonestepfive{hereafter detail}
two examples of 
\ytextmodifhershelvonestepfive{mixtures}
of four 
{sources} with 
SNR = 20 dB%
. The results for other mixtures 
\ytextmodifhershelvtwostepfiftyone{with an SNR of 10, 20, or 30 dB}
lead to the 
same conclusions and are available in Appendix B. 
\ytextmodifhershelvtwostepfiftyone{More specifically, some 
additional tests
with a very low SNR (1, 3, or 5 dB)
are also reported in Appendix
\ref{sec-appendix-Results-low-SNR}
for the preferred two methods.
Their relative merits are then modified as expected,
as compared to the above cases involving significantly higher
SNRs.}
\\

\subsubsection{Results}

The first example is a case of a highly sparse mixture 
\ytextmodifhershelvonestepfive{($d = 6 \sigma_{Map}$),}
whose map 
is 
\ytextmodifhershelvonestepfive{shown in the leftmost part}
of Fig. \ref{fig_Maps}. The second 
\ytextmodifhershelvonestepfive{example concerns a weakly sparse mixture}
\ytextmodifhershelvonestepfive{($d = 2 \sigma_{Map}$),}
whose map is 
\ytextmodifhershelvonestepfive{shown in the rightmost part}
of Fig. \ref{fig_Maps}. Again 
to simplify the figures, we present only one component but the results are 
similar for the remaining 3 components.

\begin{figure*}[htb]
\begin{center}	
\begin{tabular}{cc}
\includegraphics[width=0.40\textwidth]{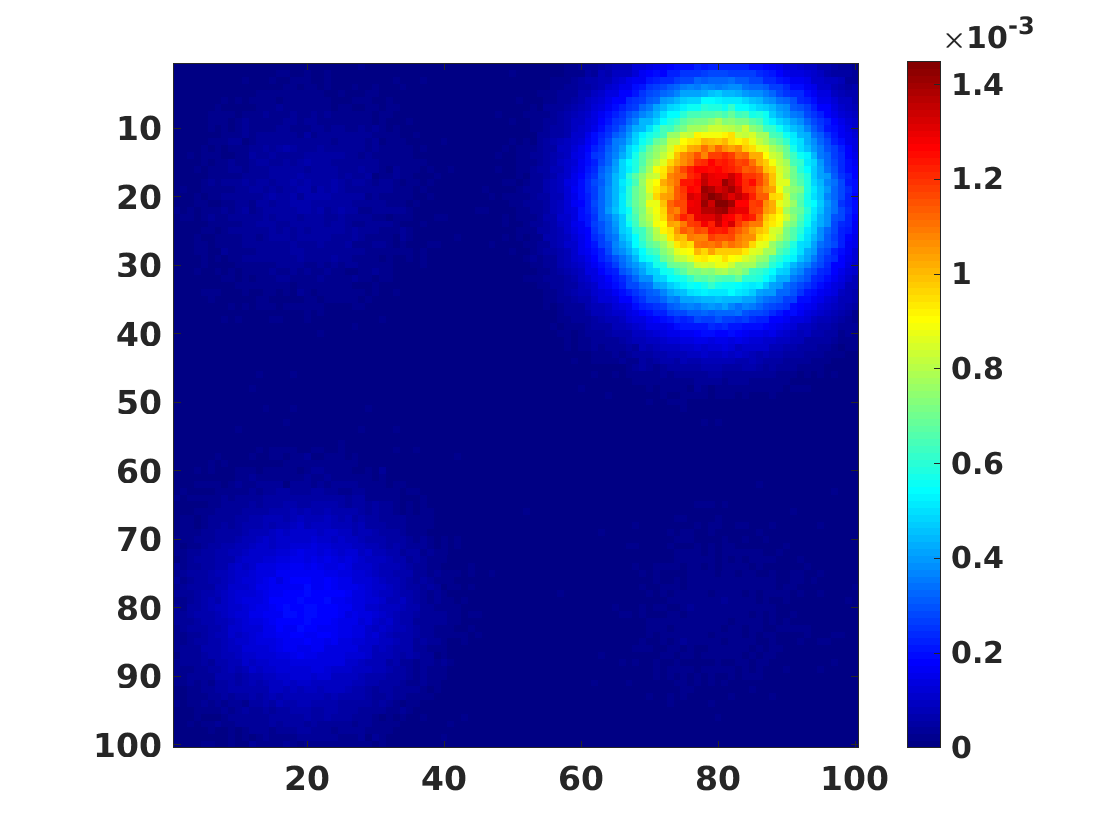}
&
\includegraphics[width=0.40\textwidth]{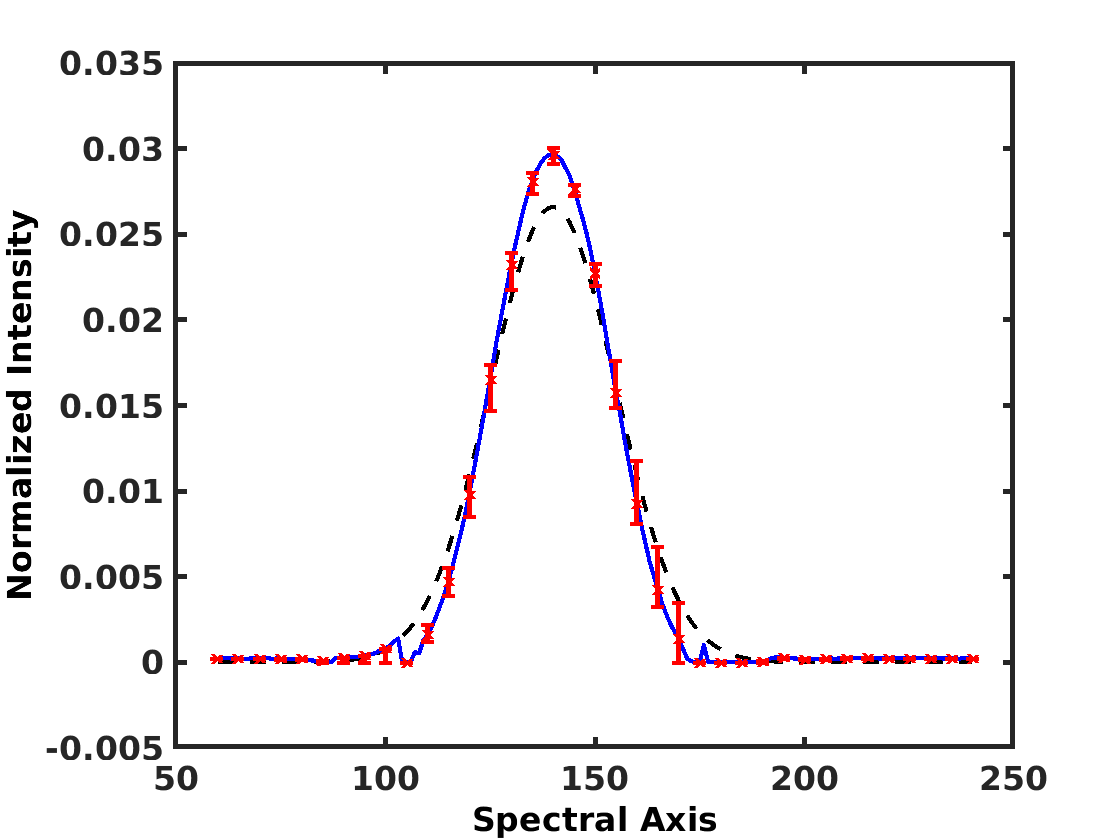}\\
\multicolumn{2}{c}{(a) 
\ytextmodifhershelvonestepfive{MC-NMF}
%
\ytextmodifhershelvtwostepfiftyone{(20.81\%, 0.107 rad)}}\\
\includegraphics[width=0.40\textwidth]{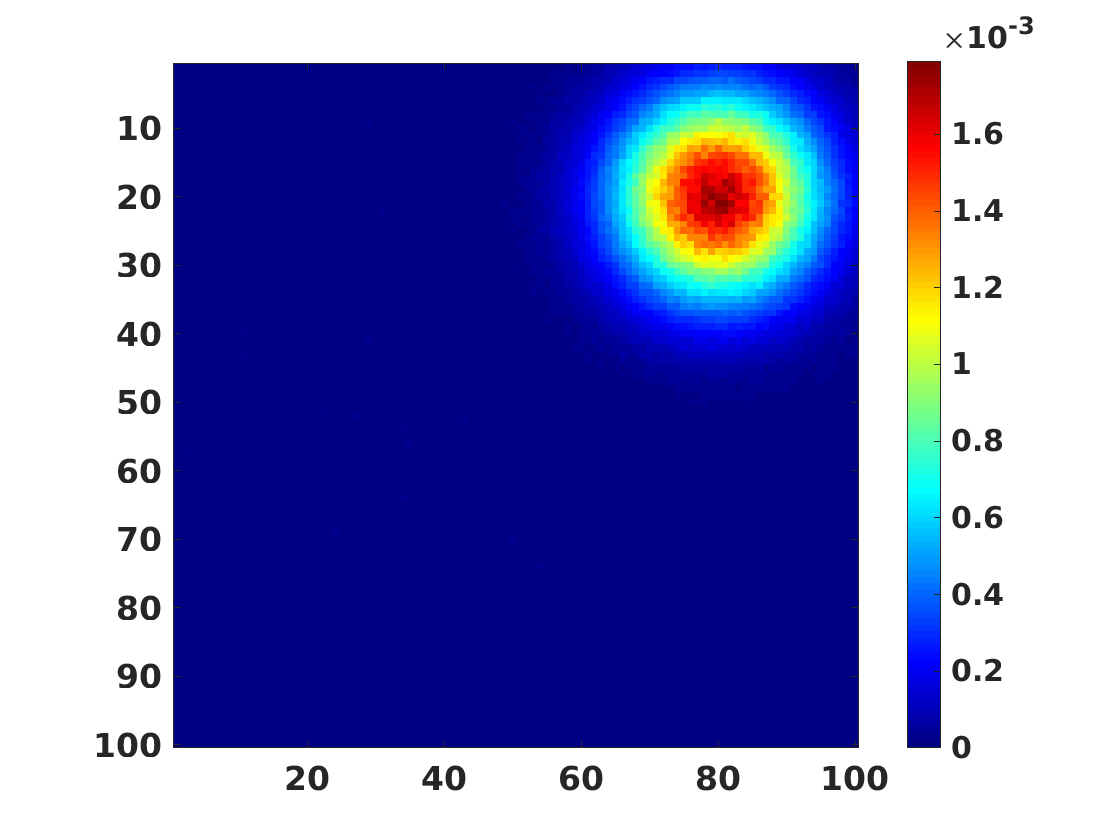}&
\includegraphics[width=0.40\textwidth]{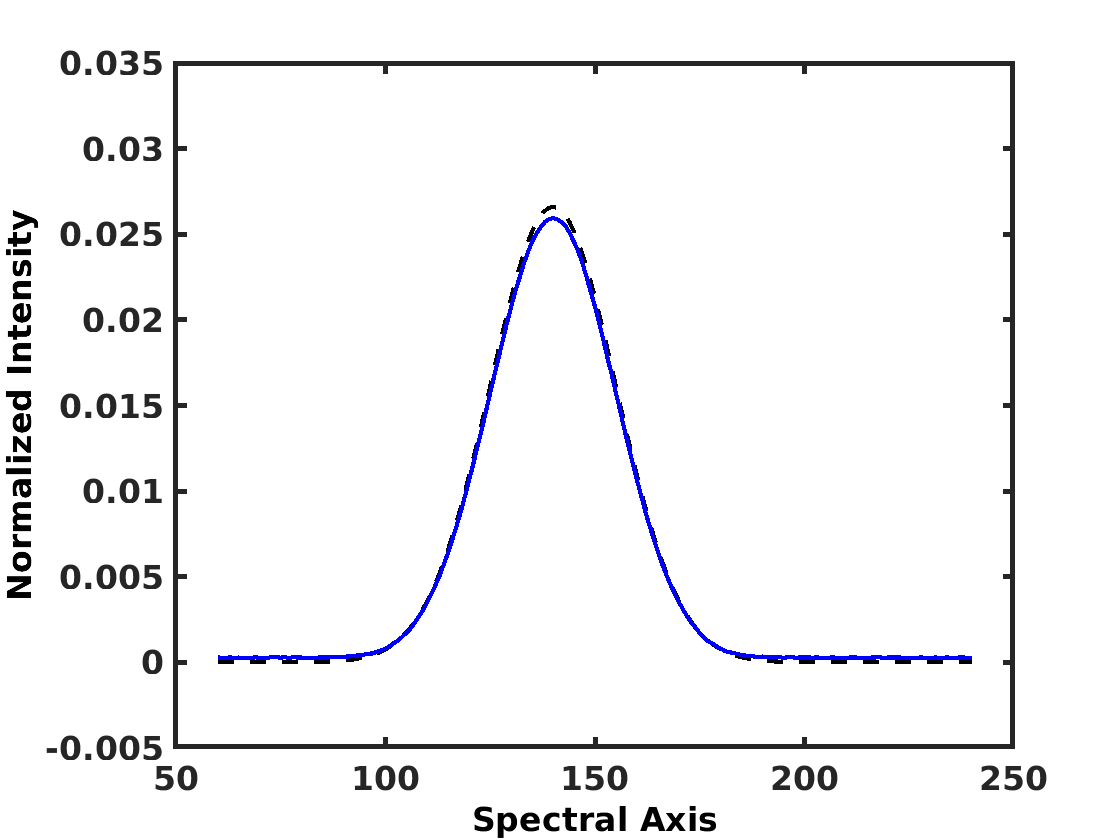}\\
\multicolumn{2}{c}{(b) SpaceCorr 
%
\ytextmodifhershelvtwostepfiftyone{(2.69\%, 0.019 rad)}}\\
\includegraphics[width=0.40\textwidth]{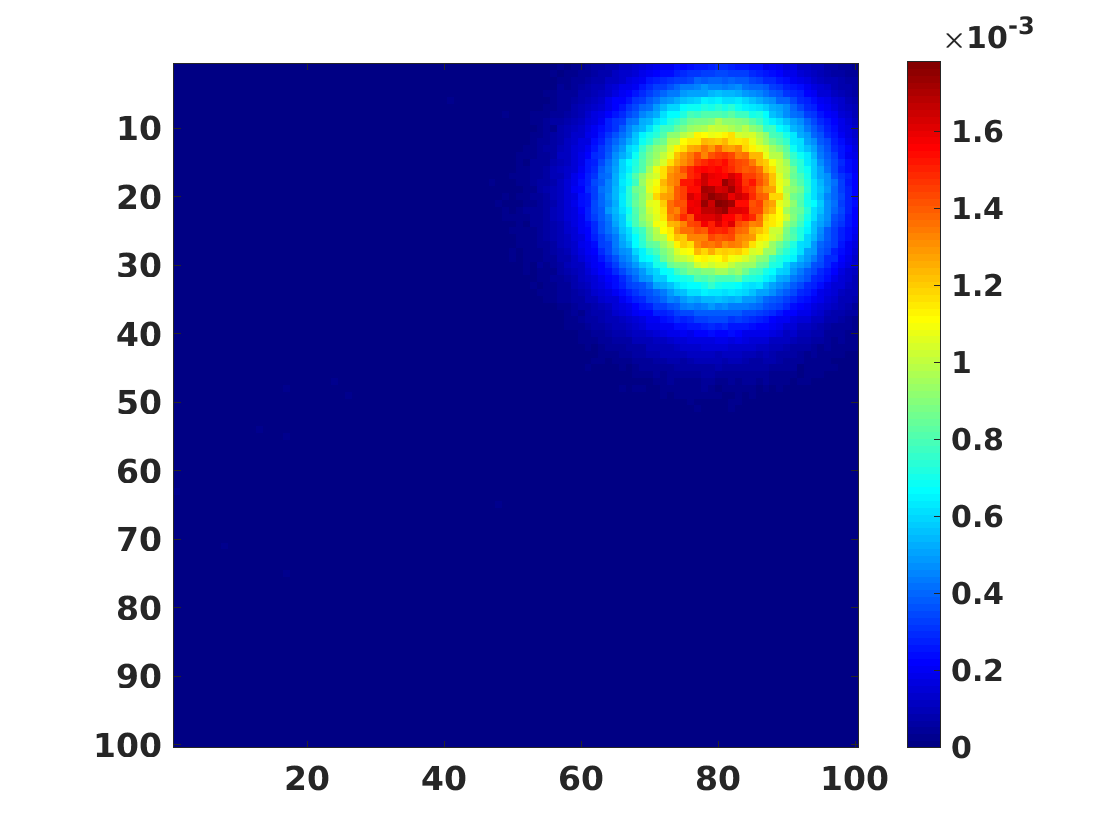}&
\includegraphics[width=0.40\textwidth]{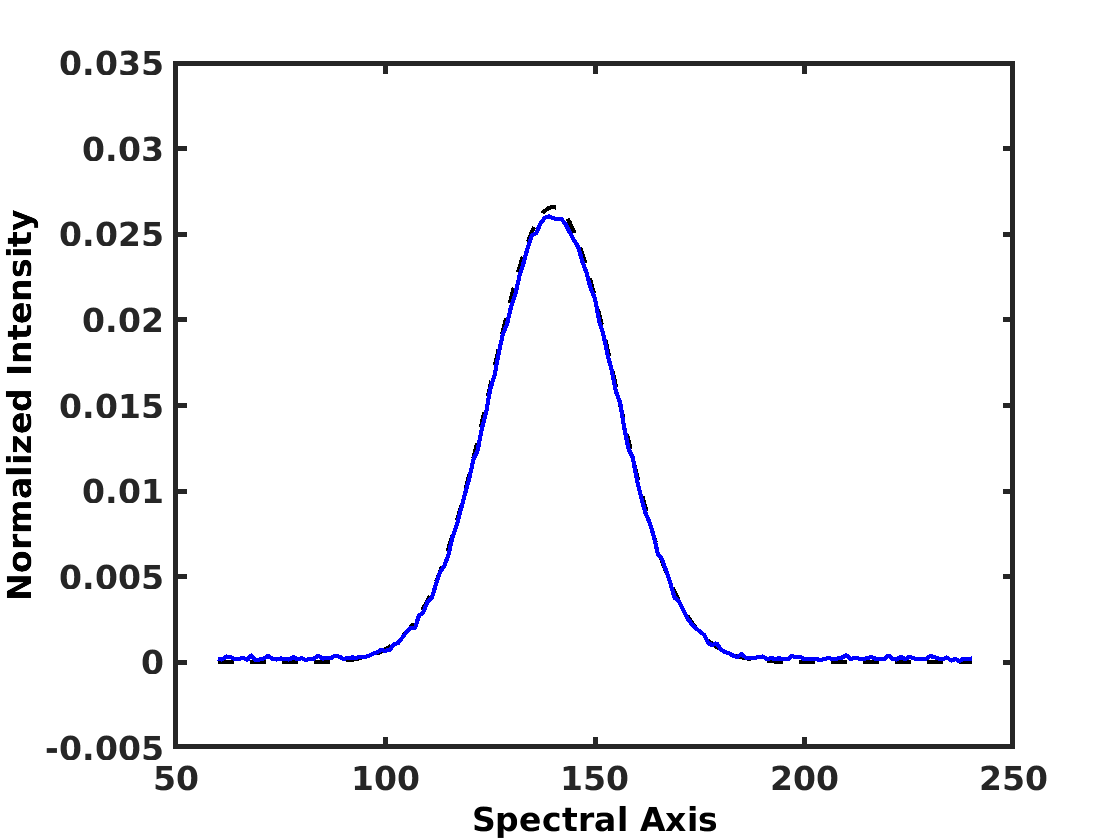}\\
\multicolumn{2}{c}{(c) MASS 
%
\ytextmodifhershelvtwostepfiftyone{(2.07\%, 0.019 rad)}}\\
\end{tabular}
\end{center}
\caption{Results of the decomposition for the MC-NMF, SpaceCorr, and MASS methods in 
the most sparse case 
\ytextmodifhershelvonestepfive{($d= 6 \sigma_{Map}$)}.
Estimated spectrum is in blue, actual 
spectrum is in black dashes, and red error bars give the 
\ytextmodifhershelvonestepfive{spread}
of the 
solutions of 
MC-NMF.
\ytextmodifhershelvonestepfive{Each subfigure caption contains the
name of the considered BSS method, followed by the
NRMSE of the estimated abundance map and the SAM of the estimated spectrum
(see
(\ref{eq-def-NRMSE}) and
(\ref{eq-def-SAM})). This also applies to the subsequent figures.}
} 
\label{fig_sparse_seule}
\end{figure*}

\begin{figure*}[htb]
\begin{center}	
\begin{tabular}{cc}

\includegraphics[width=0.40\textwidth]{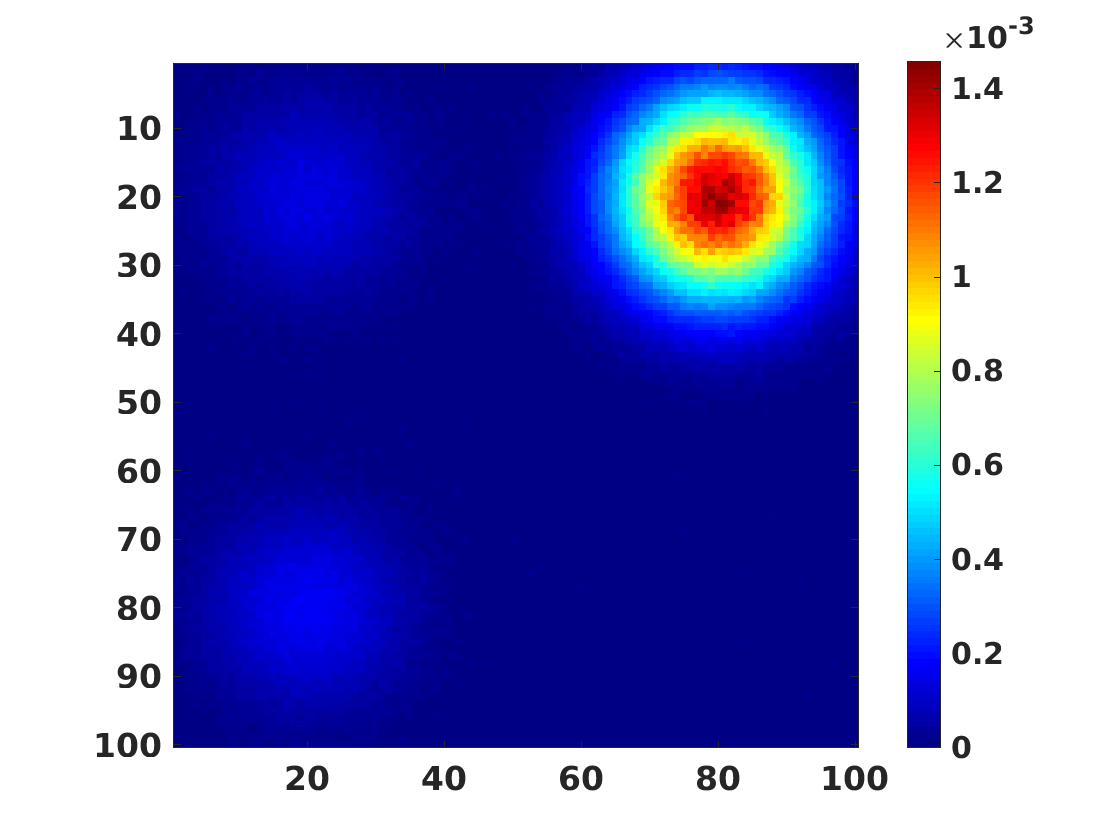}&
\includegraphics[width=0.40\textwidth]{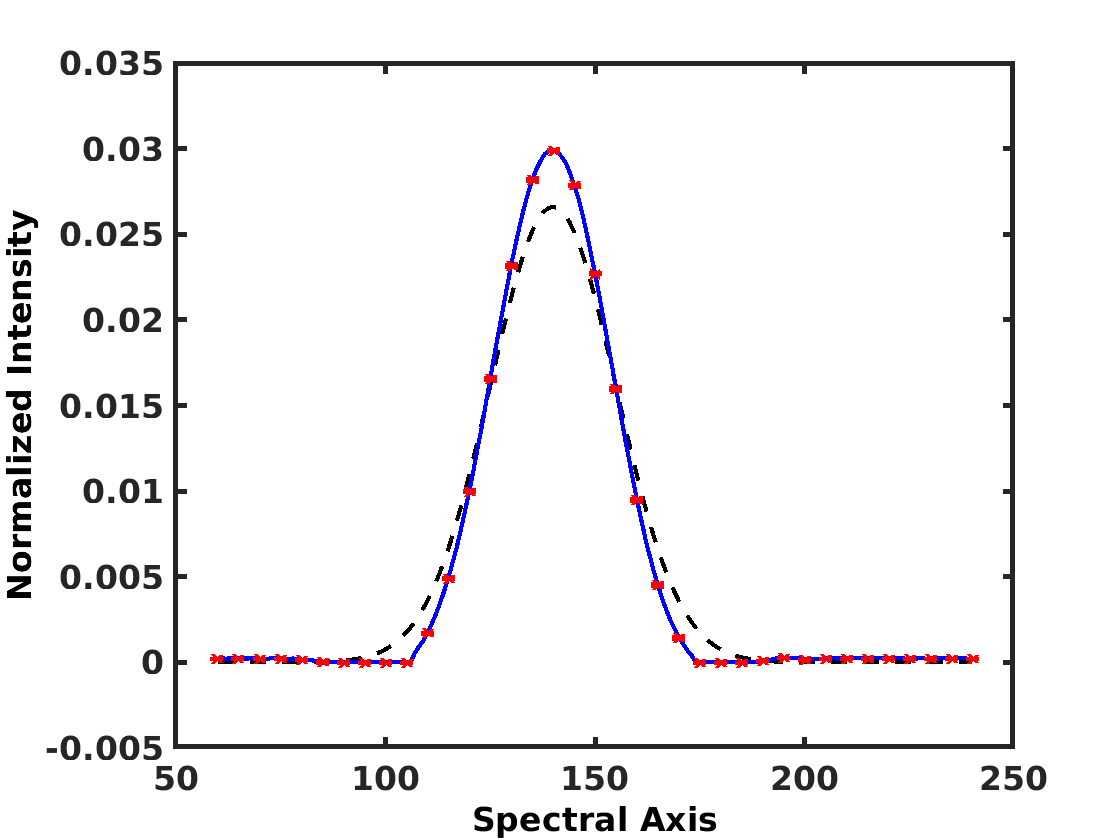}\\
\multicolumn{2}{c}{(a) 
\ytextmodifhershelvonestepfive{SC-NMF-Spec}
%
\ytextmodifhershelvtwostepfiftyone{(21.47\%, 0.102 rad)}}\\

%
\includegraphics[width=0.40\textwidth]{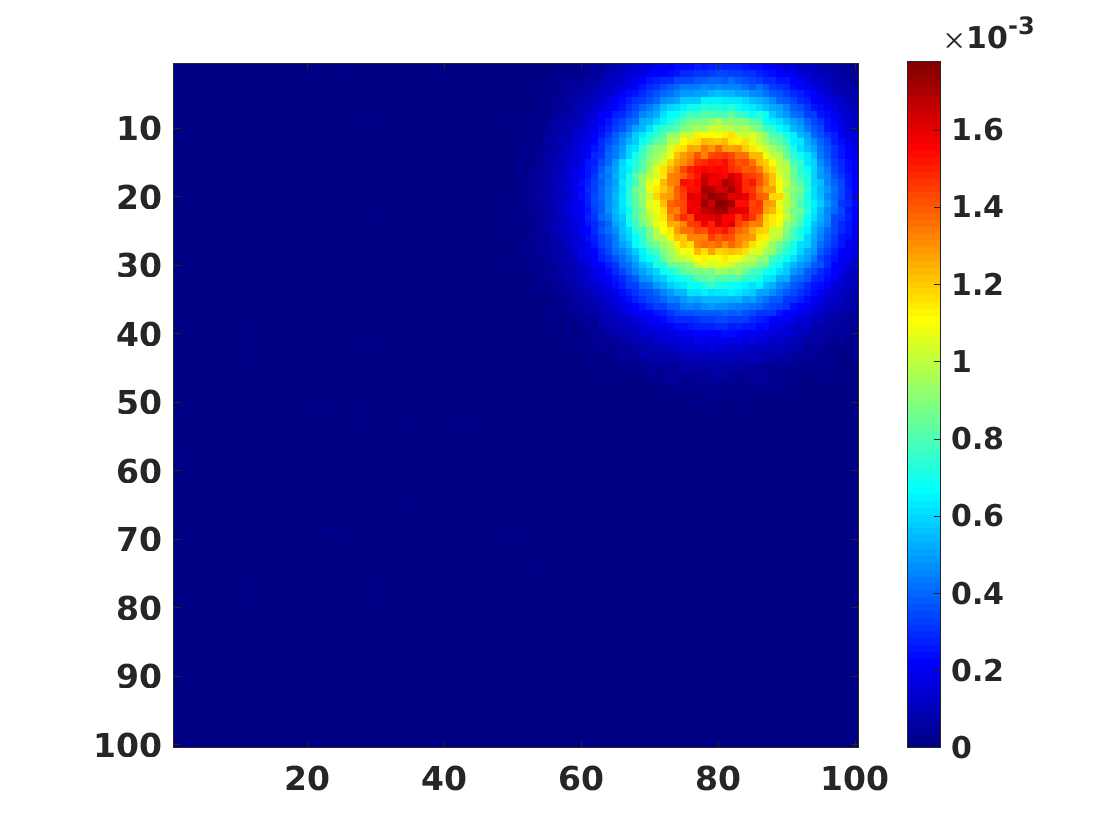}&
\includegraphics[width=0.40\textwidth]{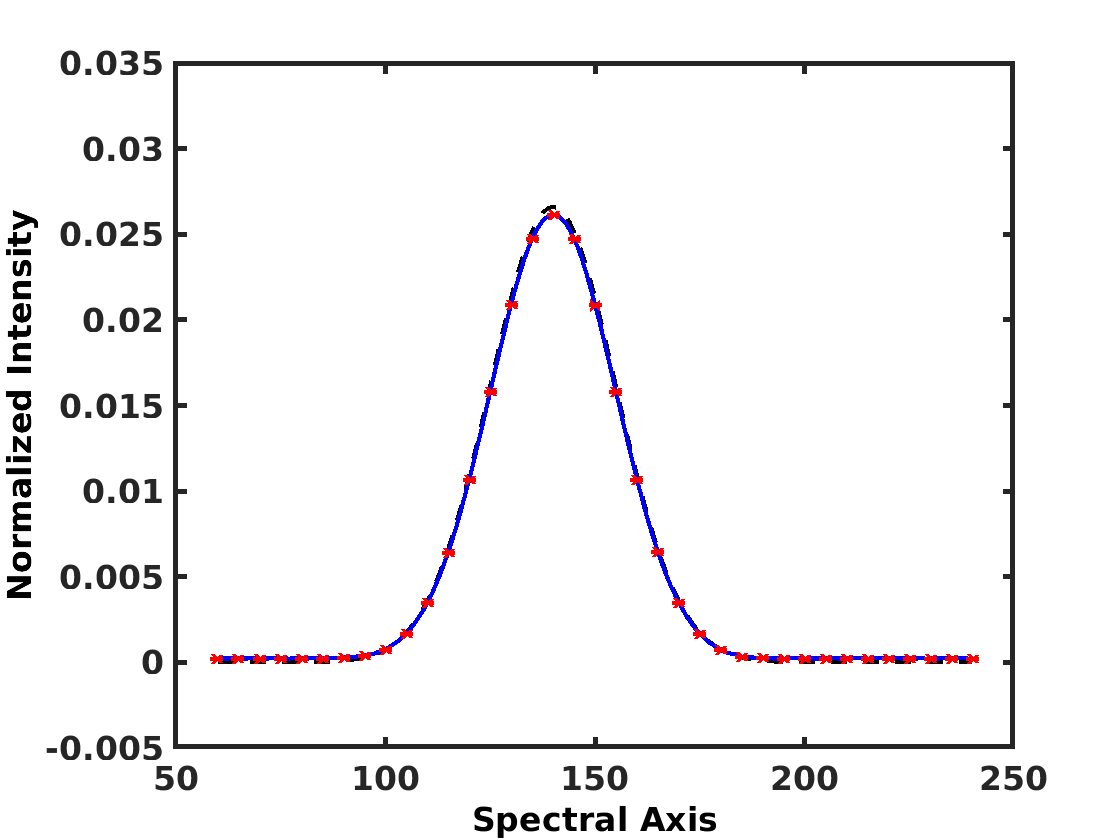}\\
\multicolumn{2}{c}{(b) 
\ytextmodifhershelvonestepfive{SC-NMF-Map}
%
\ytextmodifhershelvtwostepfiftyone{(2.13\%, 0.016 rad)}}\\
\includegraphics[width=0.40\textwidth]{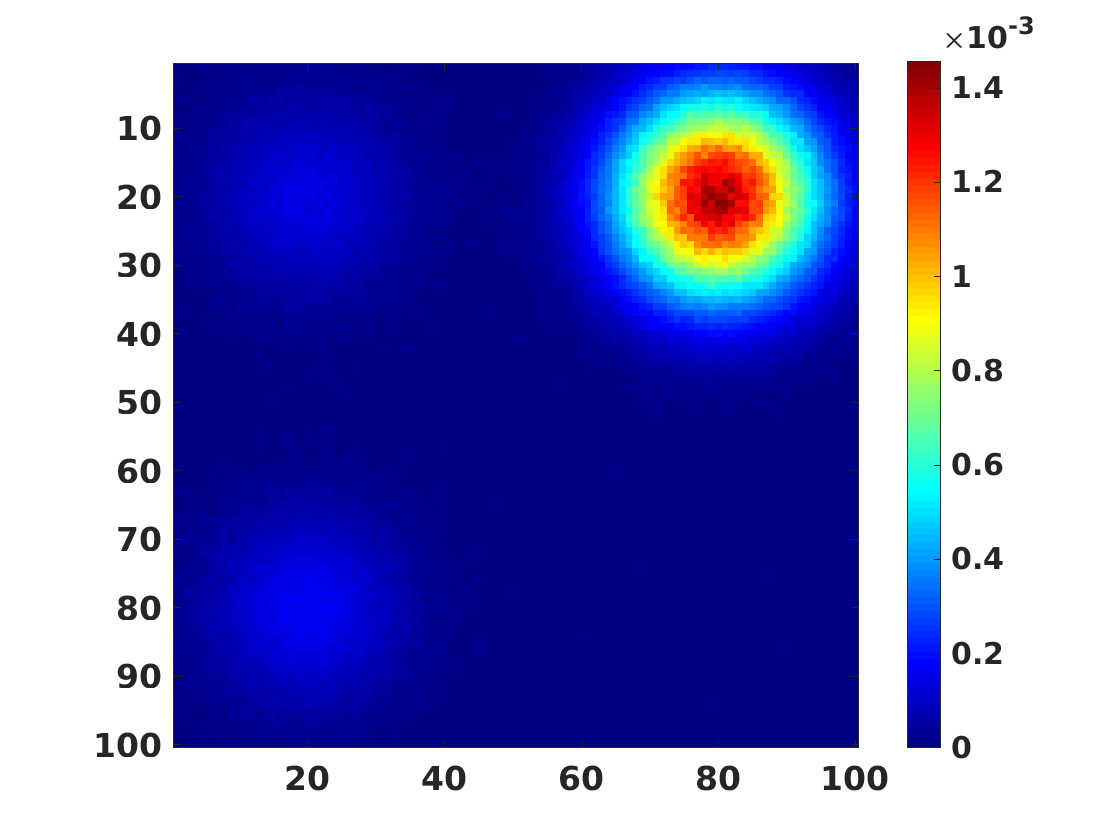}&
\includegraphics[width=0.40\textwidth]{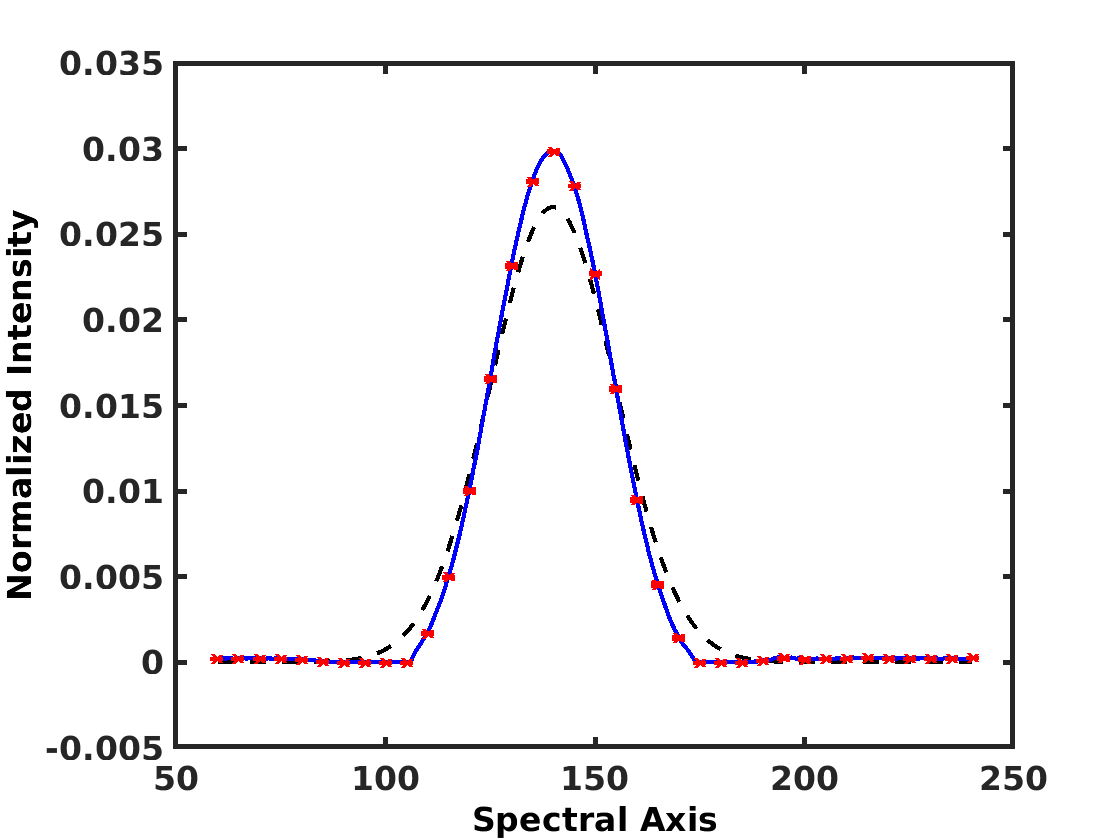}\\
\multicolumn{2}{c}{(c) MASS-NMF-Spec 
%
\ytextmodifhershelvtwostepfiftyone{(21.32\%, 0.101 rad)}}\\
\includegraphics[width=0.40\textwidth]{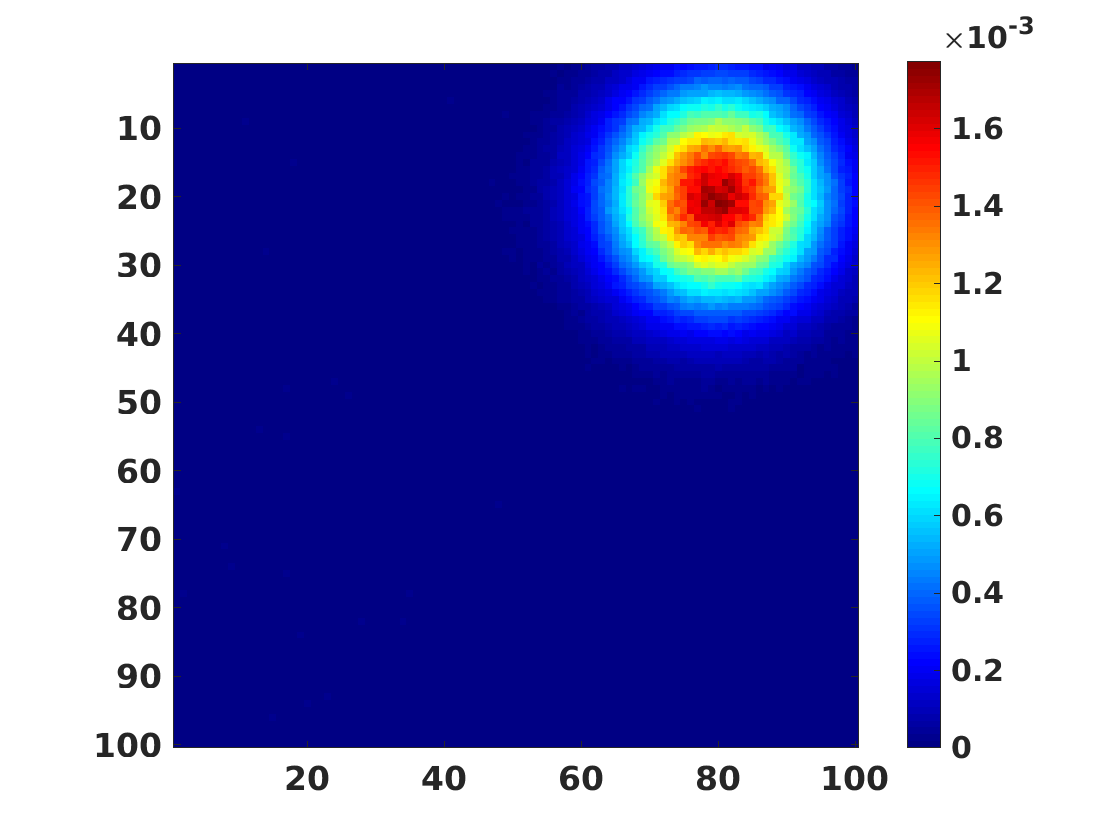}&
\includegraphics[width=0.40\textwidth]{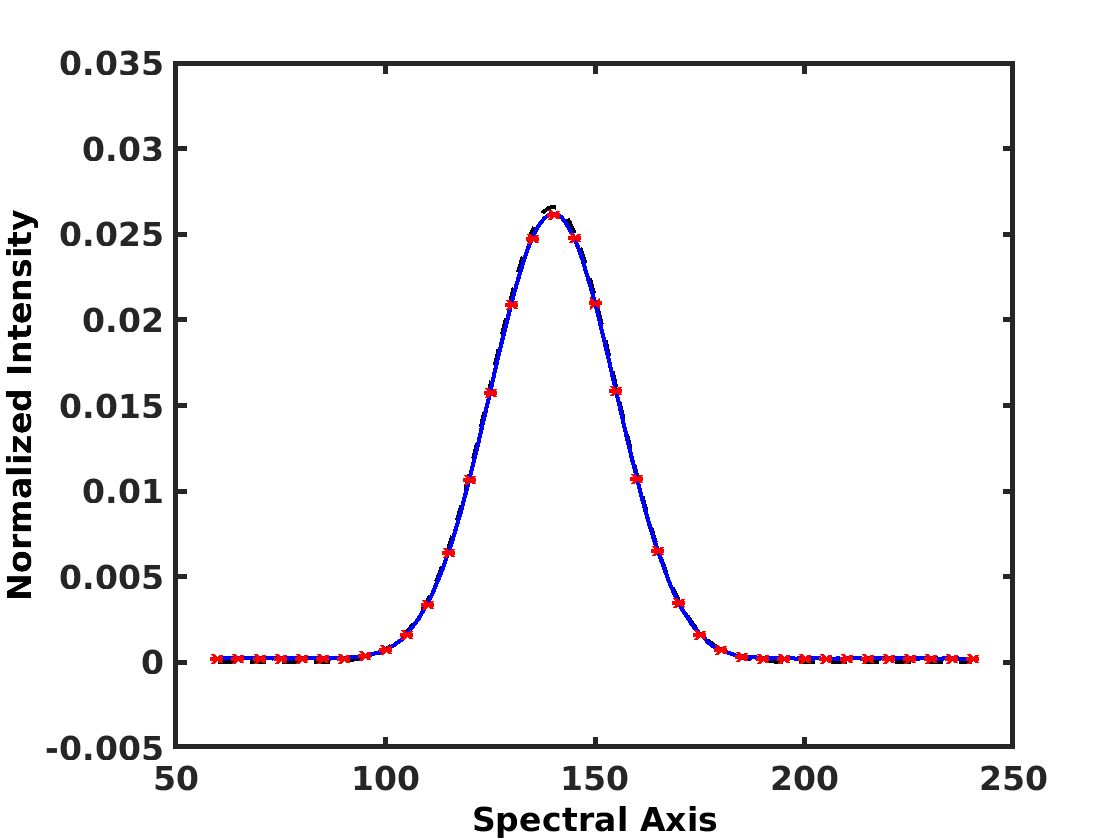}\\
\multicolumn{2}{c}{(d) MASS-NMF-Map
%
\ytextmodifhershelvtwostepfiftyone{(1.95\%, 0.016 rad)}%
}
\end{tabular}
\end{center}
\caption{Results of the decomposition for hybrid methods in the most sparse case 
\ytextmodifhershelvonestepfive{($d= 6 \sigma_{Map}$)}.
Estimated spectrum is in blue, actual spectrum is in black 
dashes, and red error bars give the 
\ytextmodifhershelvonestepfive{spread}
of the 
\ytextmodifhershelvonestepfive{solutions.}%
} 
\label{fig_sparse_hybride}
\end{figure*}

\begin{figure*}[htb]
\begin{center}	
\begin{tabular}{cc}
\includegraphics[width=0.40\textwidth]{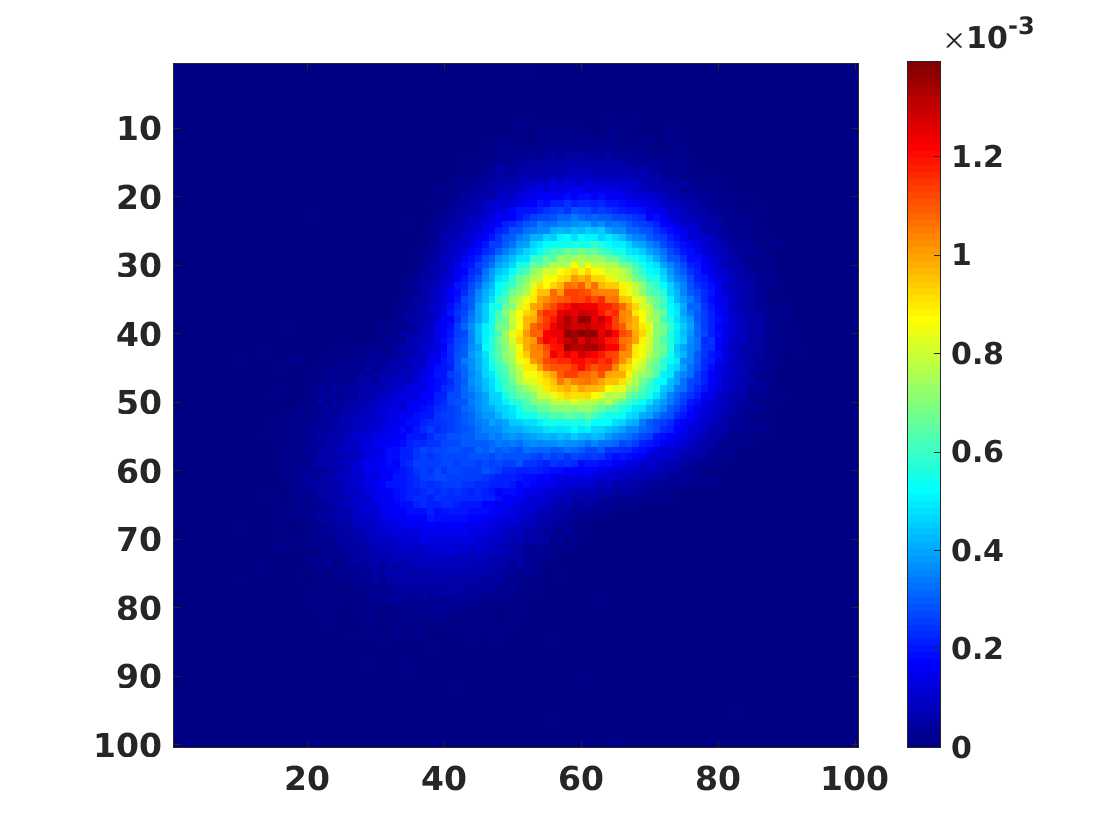}&
\includegraphics[width=0.40\textwidth]{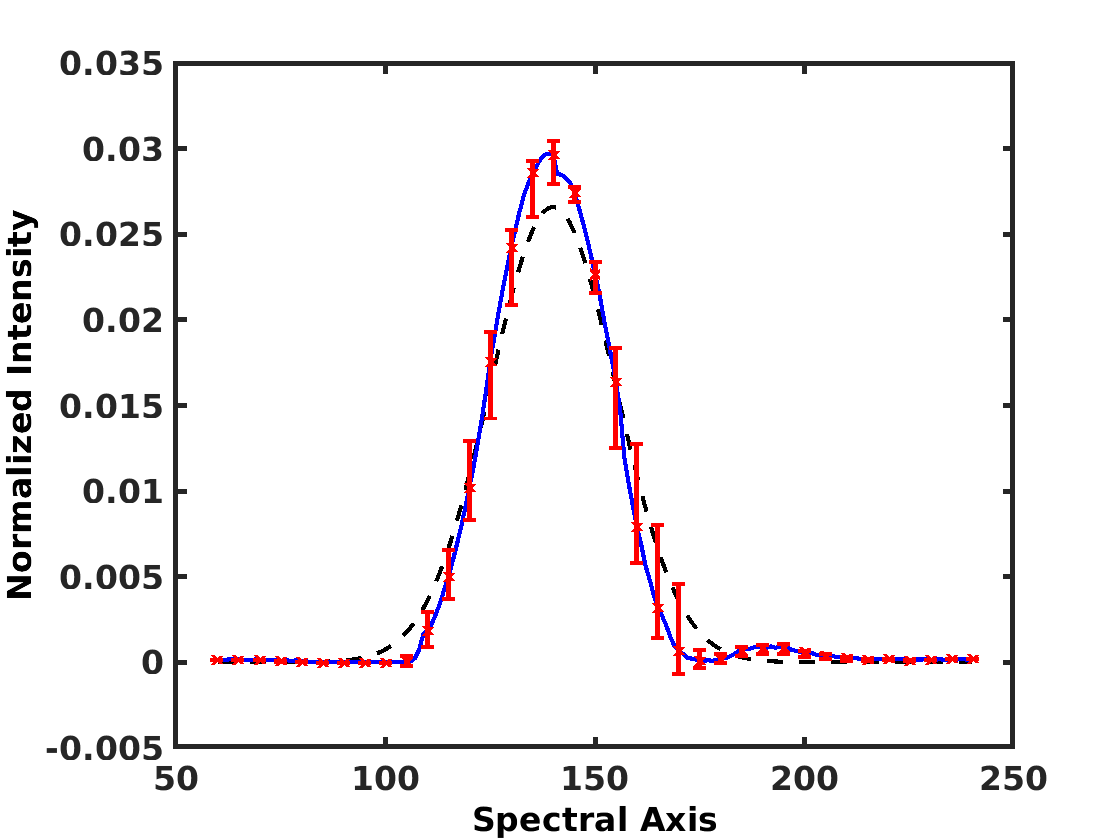}\\
\multicolumn{2}{c}{(a) 
\ytextmodifhershelvonestepfive{MC-NMF}
%
%
\ytextmodifhershelvtwostepfiftyone{(23.63\%, 0.126 rad)}}\\
\includegraphics[width=0.40\textwidth]{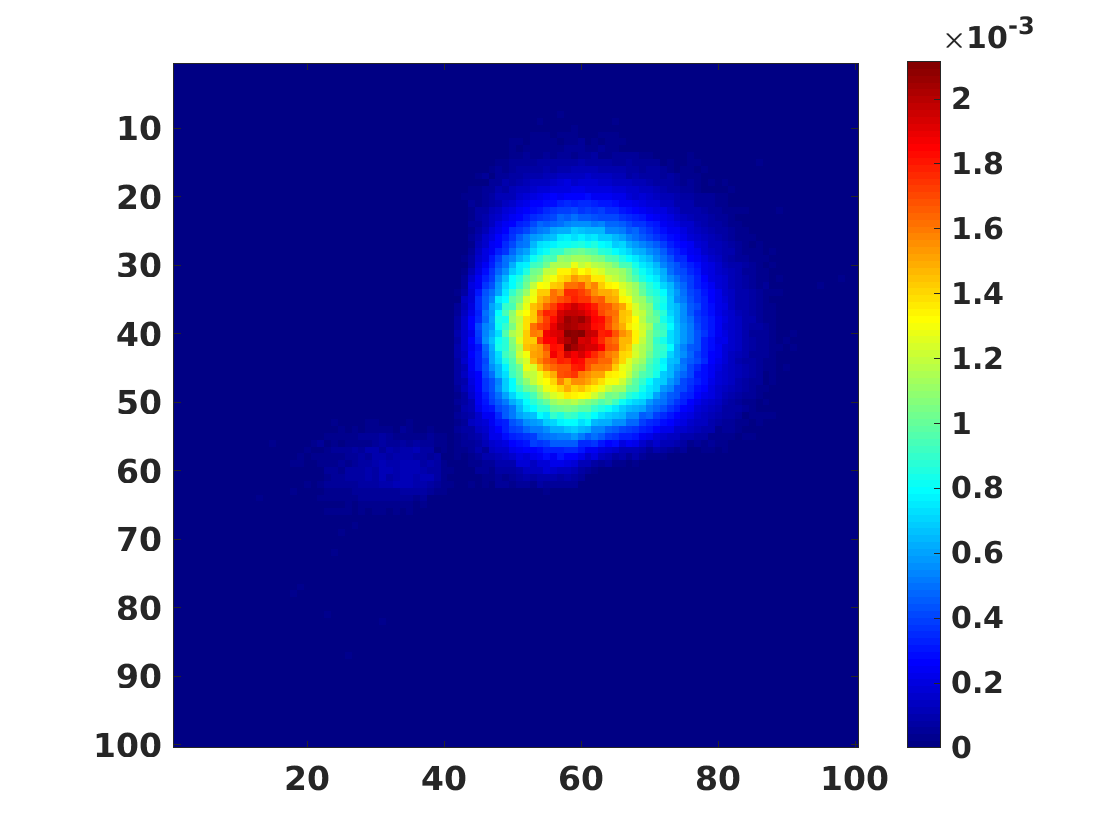}&
\includegraphics[width=0.40\textwidth]{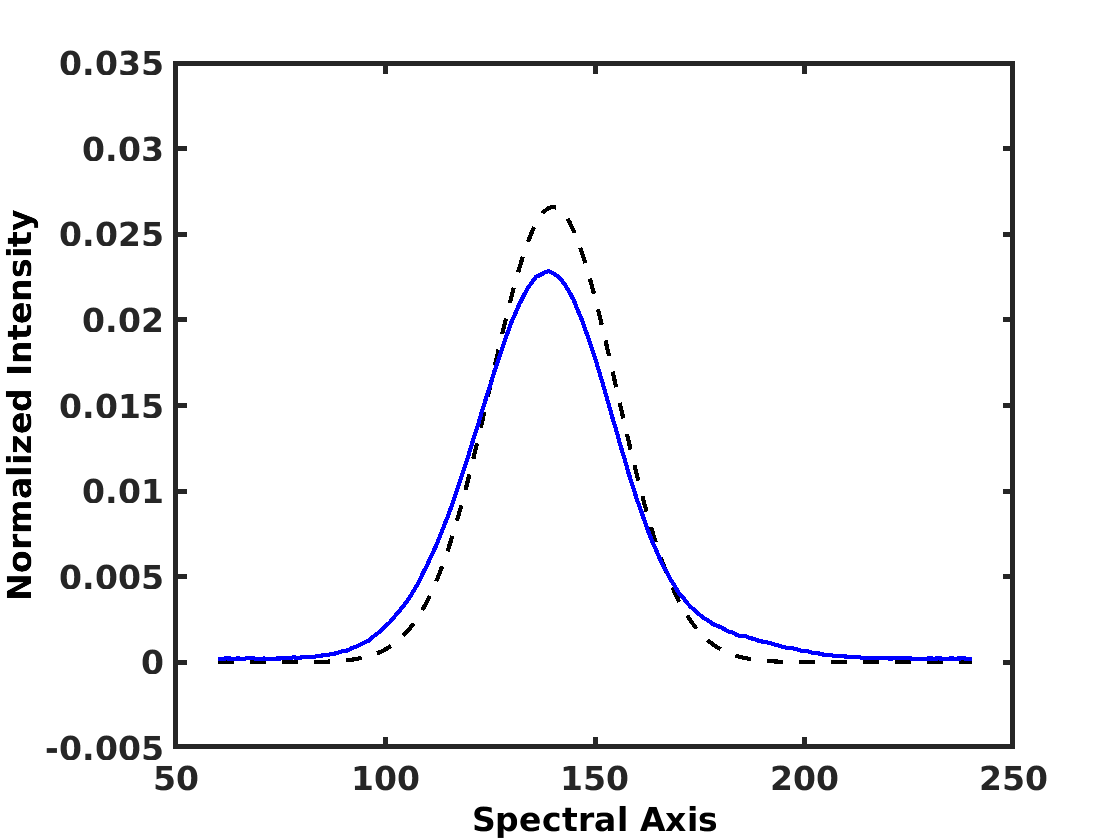}\\
\multicolumn{2}{c}{(b) SpaceCorr 
%
\ytextmodifhershelvtwostepfiftyone{(19.67\%, 0.122 rad)}}\\
\includegraphics[width=0.40\textwidth]{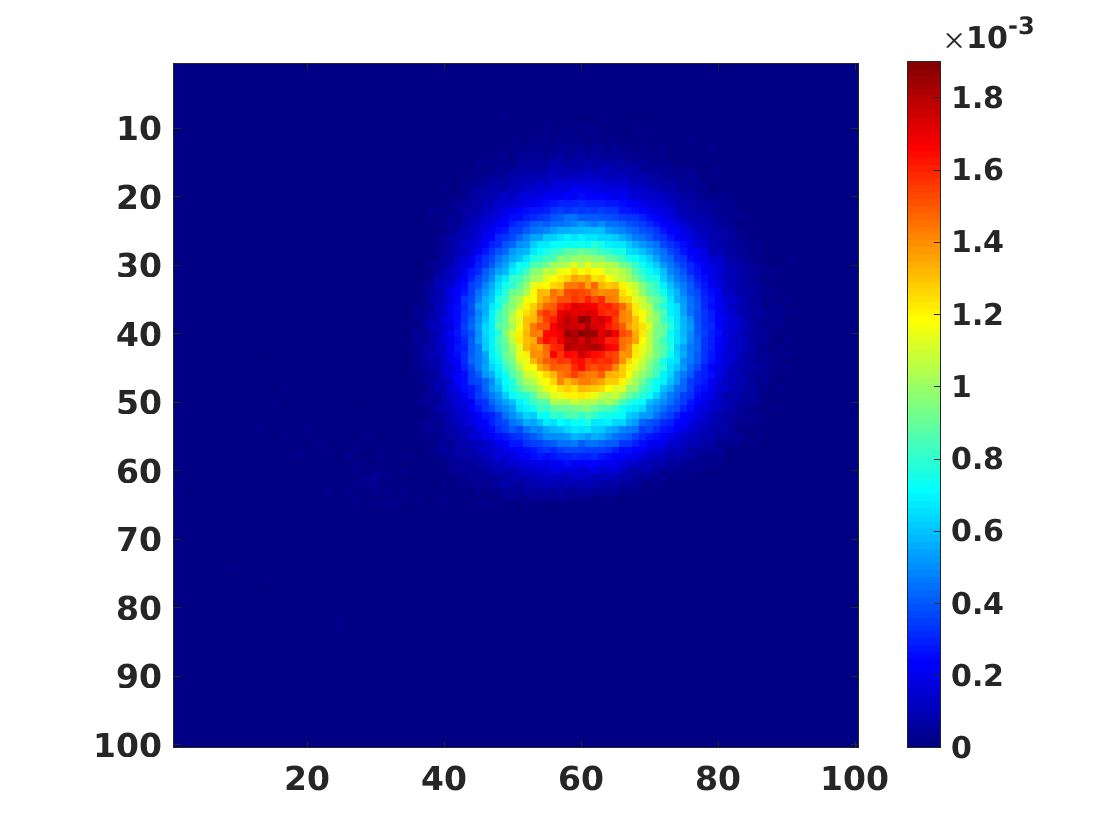}&
\includegraphics[width=0.40\textwidth]{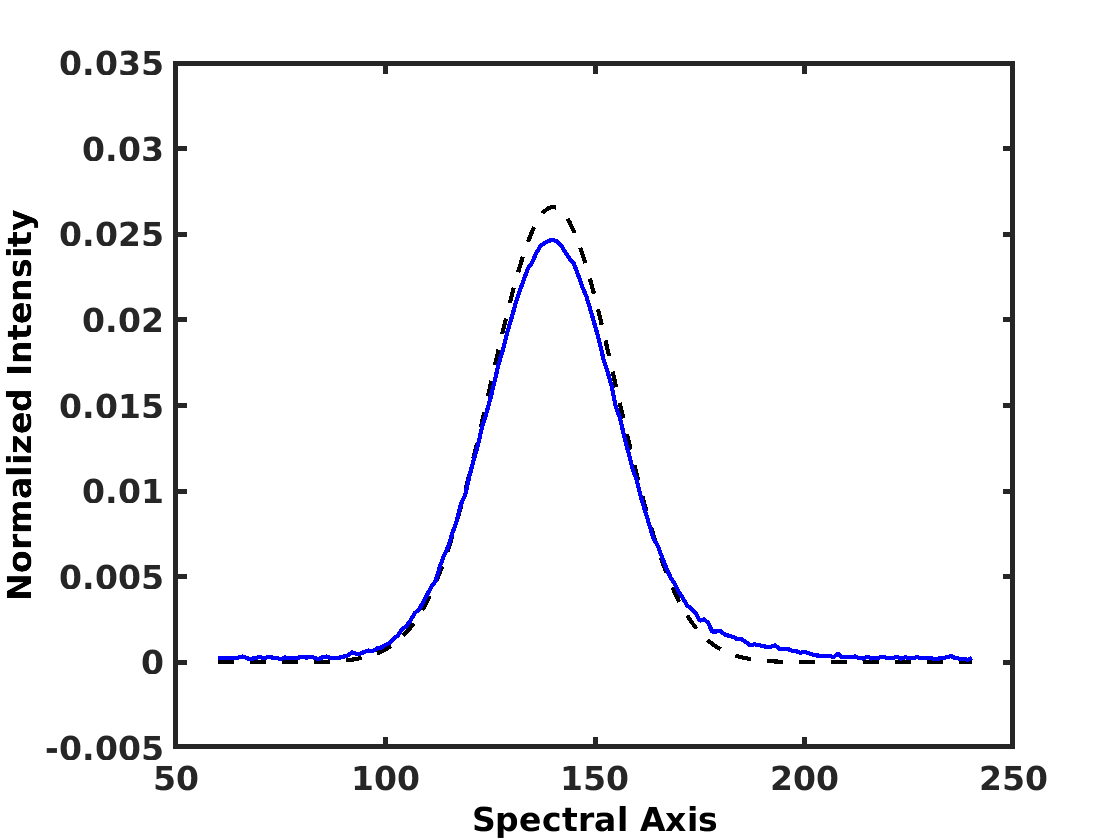}\\
\multicolumn{2}{c}{(c) MASS
%
\ytextmodifhershelvtwostepfiftyone{(8.84\%, 0.051 rad)}}\\
\end{tabular}
\end{center}
\caption{Results of the decomposition for the MC-NMF, SpaceCorr and MASS methods in 
the least sparse case 
\ytextmodifhershelvonestepfive{($d= 2 \sigma_{Map}$)}.
Estimated spectrum is in blue, actual 
spectrum is in black dashes, and red error bars give the 
\ytextmodifhershelvonestepfive{spread}
of the 
solutions of 
MC-NMF.} 
\label{fig_nonsparse_seule}
\end{figure*}

\begin{figure*}[htb]
\begin{center}	
\begin{tabular}{cc}
\includegraphics[width=0.4\textwidth]{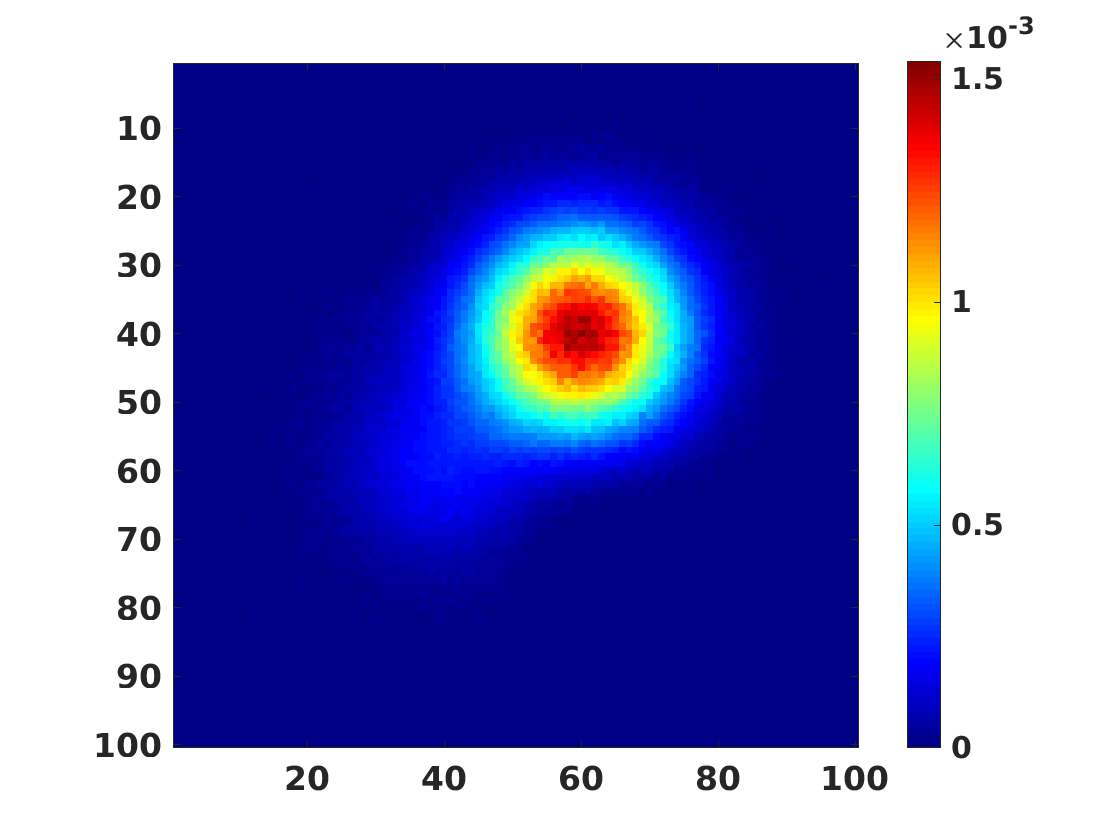}&
\includegraphics[width=0.4\textwidth]{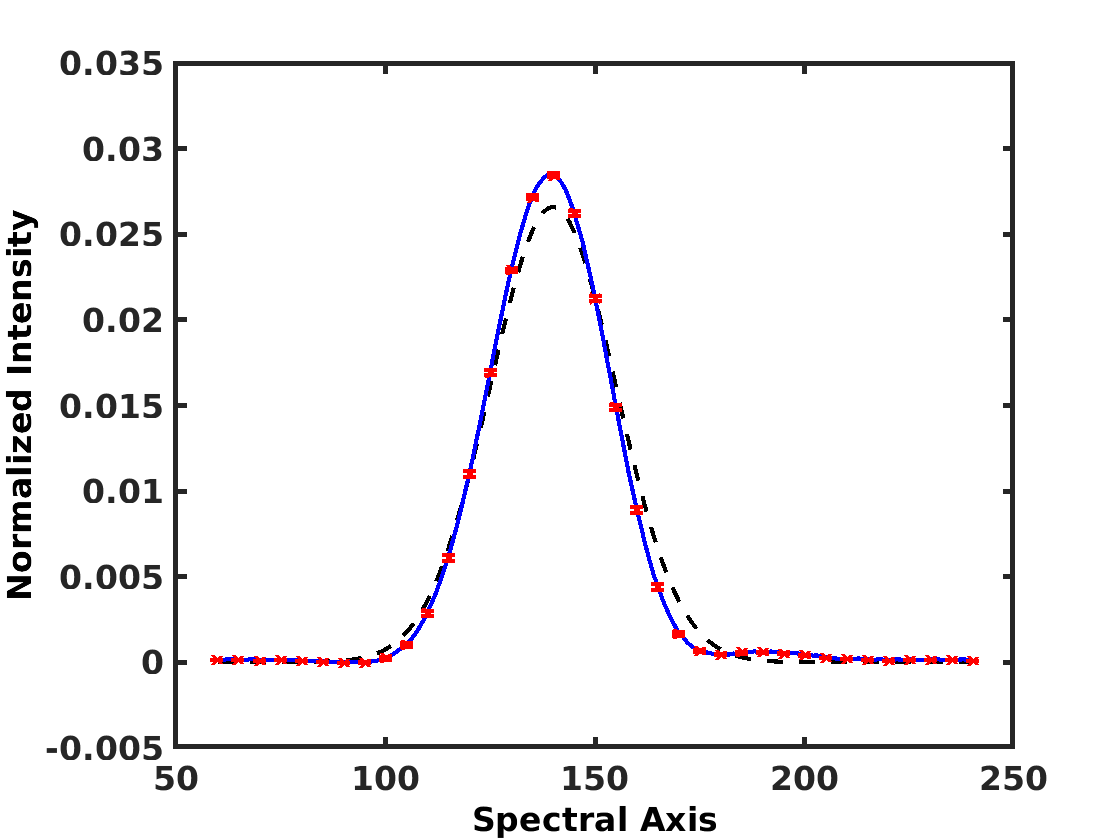}\\
\multicolumn{2}{c}{(a) 
\ytextmodifhershelvonestepfive{SC-NMF-Spec}
%
\ytextmodifhershelvtwostepfiftyone{(16.20\%, 0.082 rad)}}\\
%
\includegraphics[width=0.4\textwidth]{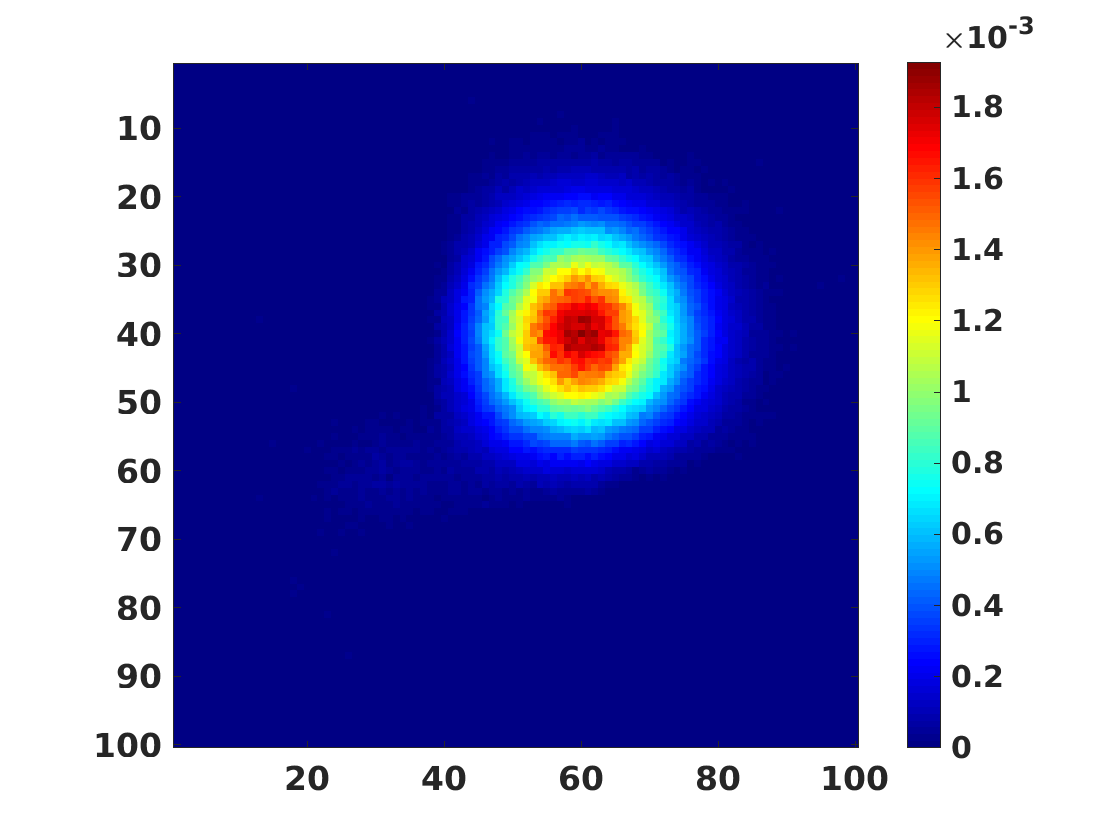}&
\includegraphics[width=0.4\textwidth]{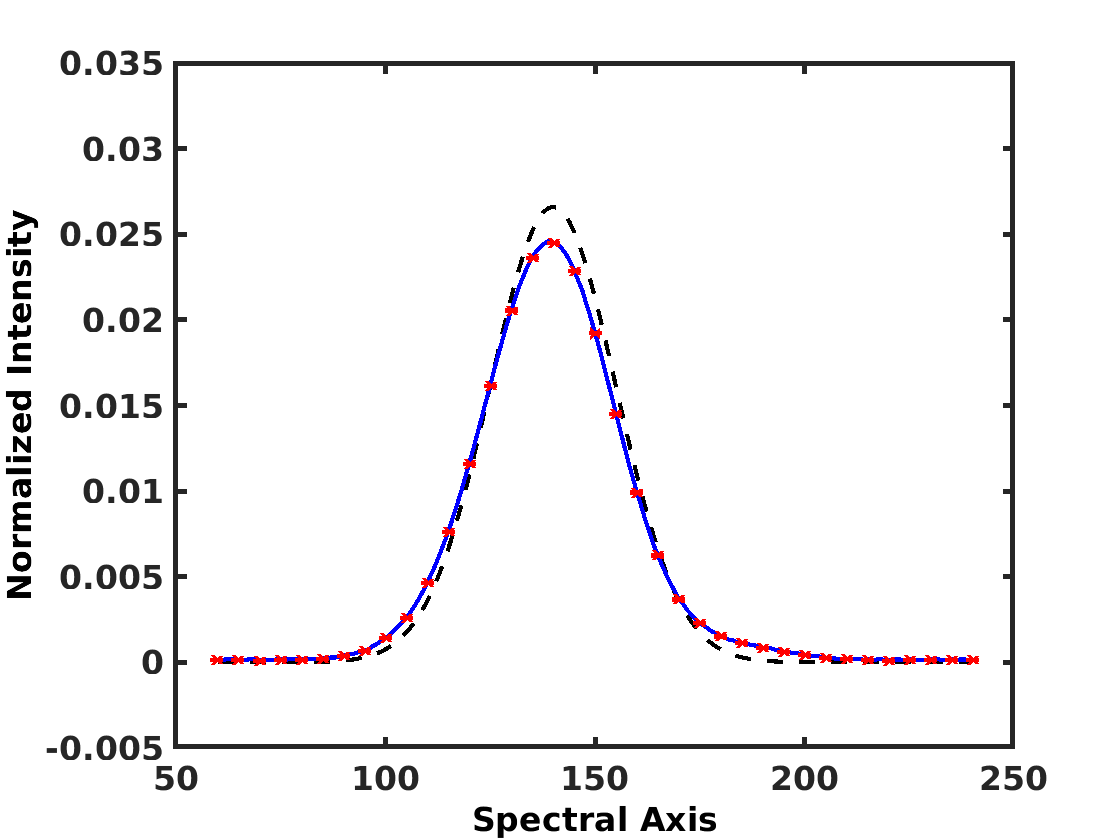}\\
\multicolumn{2}{c}{(b) 
\ytextmodifhershelvonestepfive{SC-NMF-Map}
%
\ytextmodifhershelvtwostepfiftyone{(10.37\%, 0.065 rad)}}\\
%
\includegraphics[width=0.4\textwidth]{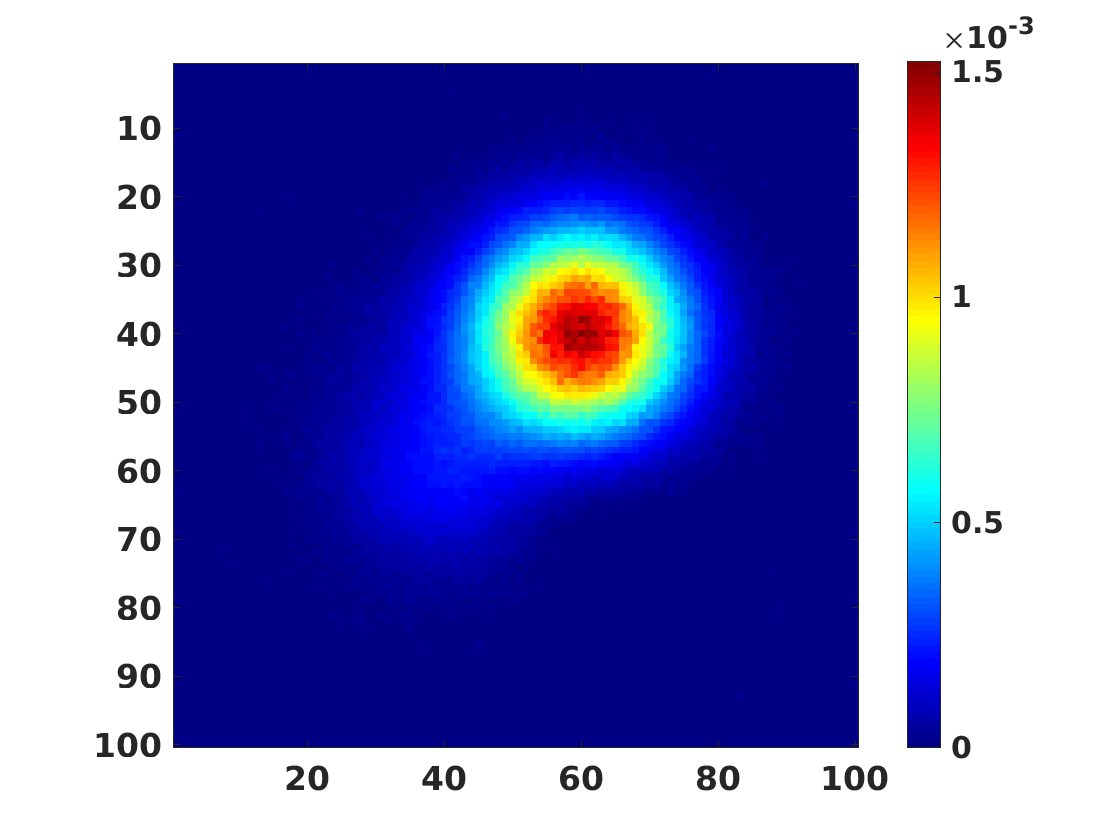}&
\includegraphics[width=0.4\textwidth]{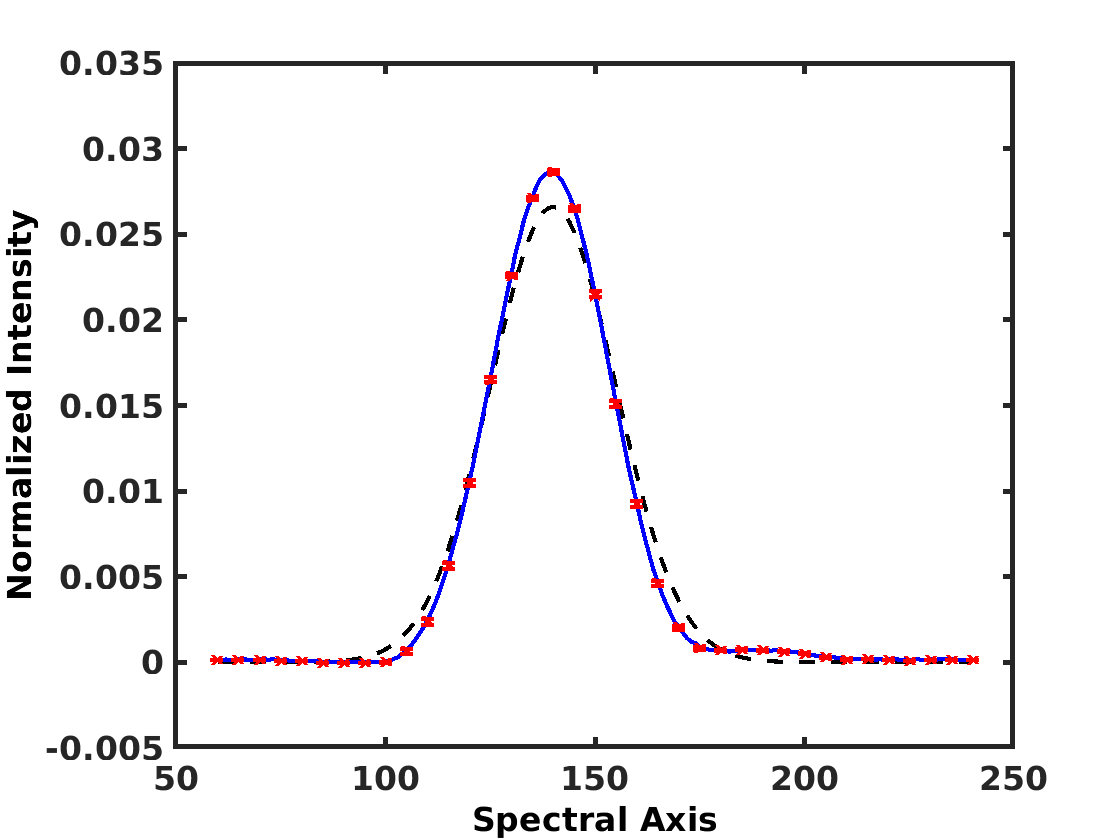}\\
\multicolumn{2}{c}{(c) MASS-NMF-Spec 
%
\ytextmodifhershelvtwostepfiftyone{(16.69\%, 0.080 rad)}}\\
%
\includegraphics[width=0.4\textwidth]{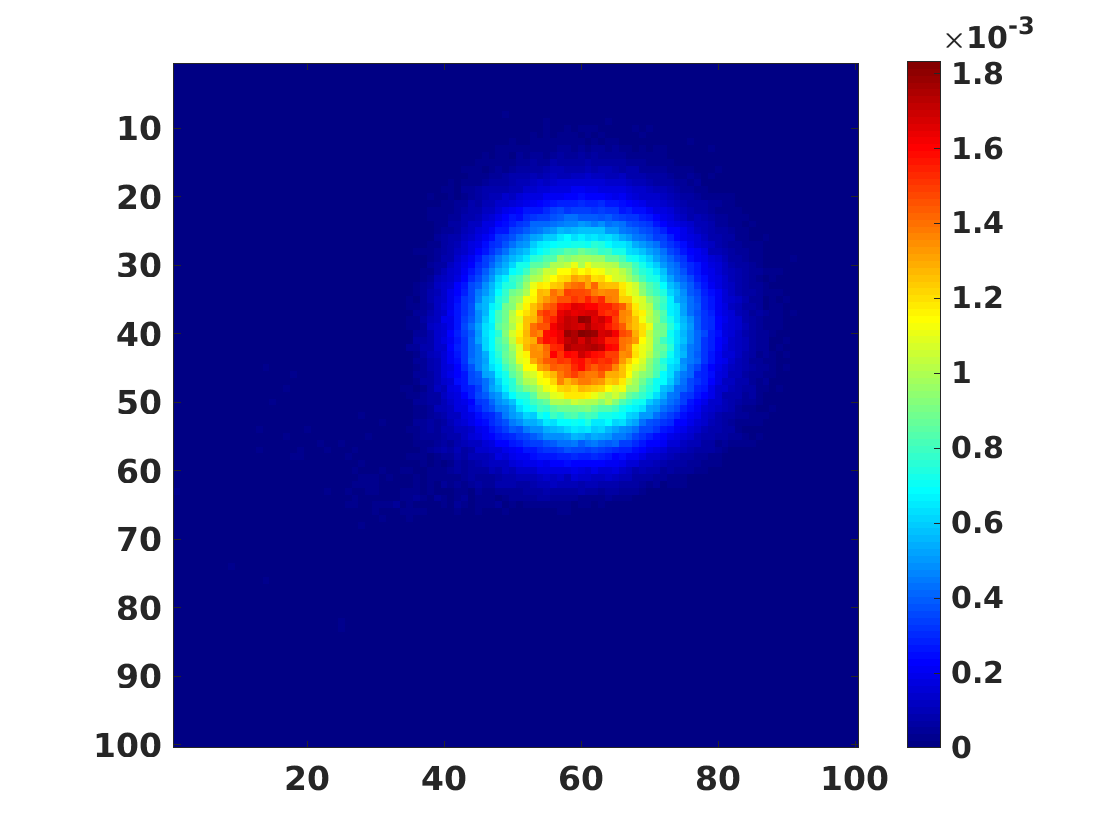}&
\includegraphics[width=0.4\textwidth]{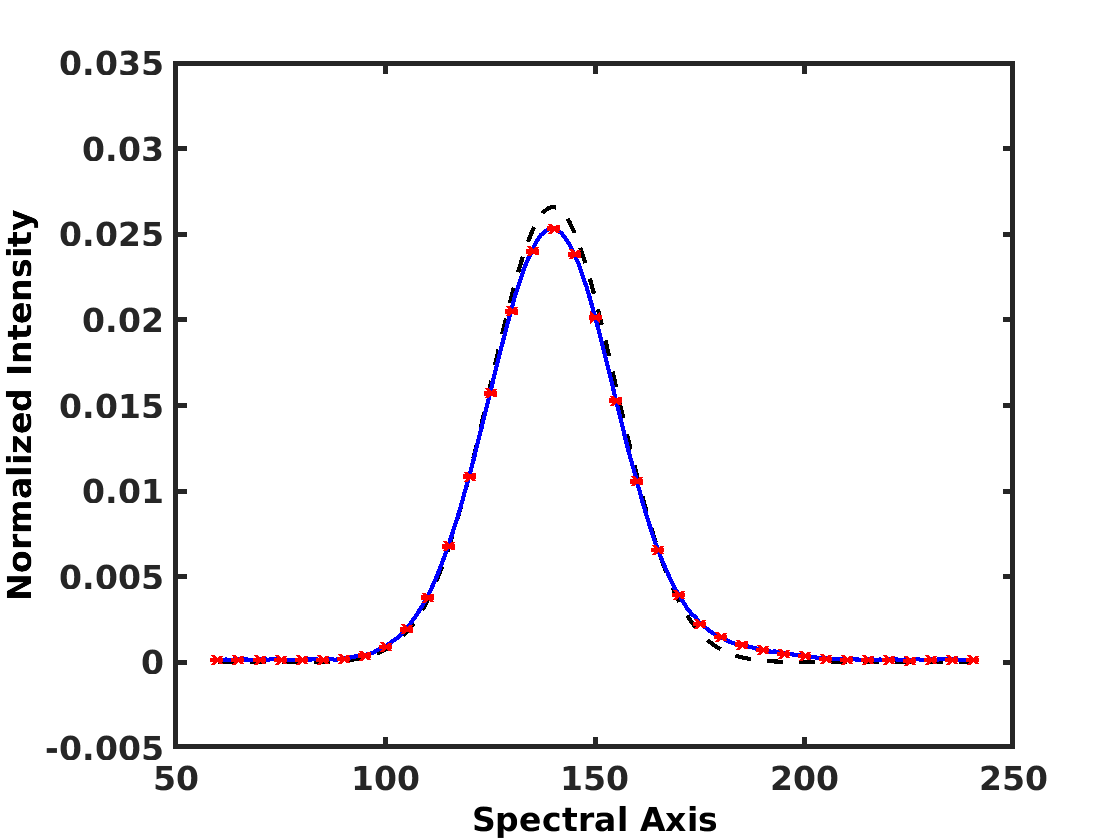}\\
\multicolumn{2}{c}{(d) MASS-NMF-Map 
%
\ytextmodifhershelvtwostepfiftyone{(6.08\%, 0.034 rad)}%
}
\end{tabular}
\end{center}
\caption{Results of the decomposition for hybrid methods in the least sparse 
case 
\ytextmodifhershelvonestepfive{($d= 2 \sigma_{Map}$)}.
Estimated spectrum is in blue, actual spectrum is in black 
dashes and red error bars give the 
\ytextmodifhershelvonestepfive{spread}
of the 
\ytextmodifhershelvonestepfive{solutions}.
} 
\label{fig_nonsparse_hybride}
\end{figure*}

The results 
\ytextmodifhershelvonestepfive{in}
the most sparse case are given in Fig. \ref{fig_sparse_seule} and 
Fig. \ref{fig_sparse_hybride}. The first 
\ytextmodifhershelvonestepfive{figure}
illustrates the performance of the 
\ytextmodifhershelvonestepfive{MC-NMF, SpaceCorr or MASS}
methods used alone and the second 
\ytextmodifhershelvonestepfive{figure shows the results}
of the 
\ytextmodifhershelvonestepfive{resulting four}
hybrid methods. Similarly, the results 
\ytextmodifhershelvonestepfive{in}
the least sparse case are given in Fig. \ref{fig_nonsparse_seule} and Fig. 
\ref{fig_nonsparse_hybride}.
\\

In the most sparse case, the results of MC-NMF are consistent with the 
\ytextmodifhershelvonestepfive{previously highlighted drawbacks.}
On the one hand, we note a variability of 
results which leads to significant error bars.
On the other hand, 
\ytextmodifhershelvonestepfive{the estimated spectrum yields a
significant error.}
This corresponds to an overestimation of the maximum intensity of the spectrum 
and an underestimation of the width of the beam. Furthermore, we observe the 
residual presence of close {sources}, visible on the map
\ytextmodifhershelvonestepfive{of}
Fig. 
\ref{fig_sparse_seule} (a). 

The SpaceCorr and MASS methods provided excellent results 
\ytextmodifhershelvonestepfive{(see}
Fig. 
\ref{fig_sparse_seule} (b-c)%
\ytextmodifhershelvonestepfive{),}
which is consistent with the theory. This first 
case is in the optimal 
\ytextmodifhershelvonestepfive{conditions of use}
of the method, 
\ytextmodifhershelvonestepfive{since}
many adjacent observed 
pixels are single-source.

Regarding hybrid methods, we observe a significant reduction of the error bars 
in agreement with the objective of these methods.
However when MC-NMF is initialized with the
\ytextmodifhershelvonestepfive{previously estimated}
spectra 
\ytextmodifhershelvonestepfive{(see}
Fig.~\ref{fig_sparse_hybride} (a-c)), we find on the estimated spectra
the same inaccuracy as 
\ytextmodifhershelvonestepfive{with}
MC-NMF used alone (overestimation of the maximum 
intensity and underestimation of the width of the beam). Initialization with 
\ytextmodifhershelvonestepfive{previously estimated}
abundance maps gives the best performance%
\ytextmodifhershelvtwostepfiftyone{, with very similar results for
the two algorithms based on this approach}
(Fig.~\ref{fig_sparse_hybride} (b-d)). 
For the 
\ytextmodifhershelvonestepfive{SC-NMF-Map}
and MASS-NMF-Map methods, there is  
\ytextmodifhershelvonestepfive{performance improvement, as compared}
respectively with SpaceCorr and MASS used alone, although 
\ytextmodifhershelvonestepfive{the latter two methods}
are already excellent.
   
~

In the least sparse case, MC-NMF provides 
\ytextmodifhershelvtwostepfiftyone{estimated spectra which
have almost the same accuracy}
as in the 
\ytextmodifhershelvonestepfive{most sparse
case (see}
Fig. \ref{fig_nonsparse_seule} (a)). We observe the same deformation 
of the estimated beam, a large 
\ytextmodifhershelvonestepfive{spread}
of the solutions 
and a residual source on the abundance map.

This time, SpaceCorr does not provide satisfactory results. Indeed, abundance 
maps seem to give a good approximation of ground truth but estimated spectra are 
contaminated by the presence of the other spectral 
\ytextmodifhershelvonestepfive{components}
%
\ytextmodifhershelvonestepfive{(see}
Fig. 
\ref{fig_nonsparse_seule} (b)). This contamination leads to an underestimation of 
the peak of intensity, the loss of the symmetry of the beam as well as a 
positioning error 
\ytextmodifhershelvonestepfive{for the maximum of intensity
on the spectral axis.}
This 
perturbation is explained by the fact that there are few single-source zones in 
the cube. Furthermore, the detection step is sensitive to the fixed threshold 
for selection of the best single-source zones. Depending on the choice of the 
threshold, some ``quasi-single-source'' zones 
\ytextmodifhershelvonestepfive{may turn out to}
be used to estimate the 
columns of the mixing matrix $A$. 

\ytextmodifhershelvonestepfive{In this case,
the MASS method yields a better estimate than SpaceCorr
(see Fig. \ref{fig_nonsparse_seule} (c)), thanks to its 
ability to operate with single-source pixels,
instead of complete single-source spatial zones.}
The 
\ytextmodifhershelvonestepfive{obtained spatial source is correctly located}
and is circular (unlike 
\ytextmodifhershelvonestepfive{with}
the SpaceCorr method, where it was slightly 
deformed). The estimated spectrum is better than that estimated by SpaceCorr, 
however it is slightly noisy because of the sensitivity of MASS to the high 
noise level 
(see Appendix B).

\ytextmodifhershelvonestepfive{Here again, all four hybrid methods}
significantly reduce the error bars%
\ytextmodifhershelvonestepfive{, as compared with applying
MC-NMF alone. Initializations}
with SpaceCorr results 
\ytextmodifhershelvonestepfive{(}%
Fig. \ref{fig_nonsparse_hybride} (a-b))
improve the results of SpaceCorr without completely removing the residue of 
other spectral components
\ytextmodifhershelvonestepfive{(i.e., the estimated spectrum is still
somewhat asymmetric)}%
%
. In addition, we observe again that when MC-NMF is 
initialized with the spectra (Fig. \ref{fig_nonsparse_hybride} (a-c)), we 
\ytextmodifhershelvonestepfive{obtain the estimated spectra with}
the same inaccuracy as 
\ytextmodifhershelvonestepfive{with}
the MC-NMF used alone. The 
initialization with MASS results 
(Fig. \ref{fig_nonsparse_hybride} (c-d)) improves 
the results of MASS by removing the residual noise of the estimated spectrum. 
\ytextmodifhershelvonestepfive{As an overall result,}
%
the initialization of MC-NMF with 
\ytextmodifhershelvonestepfive{the abundance maps provided by MASS (see}
Fig. 
\ref{fig_nonsparse_hybride} (d)%
\ytextmodifhershelvtwostepfiftyone{, including a 6.08\% NRSME
 and a 0.034 rad SAM}%
)
gives the best performance in this difficult 
case of weakly sparse and highly noisy data.

\renewcommand\arraystretch{1.5} 
\begin{table*}[ht]
\begin{center}
\begin{tabular}{|c|c||c|c||c|c||c||c|c|}
\hline
\ytextmodifhershelvonestepfive{$d$}
&
\ytextmodifhershelvonestepfive{$criterion$}
%
& MC-NMF & SpaceCorr & 
\ytextmodifhershelvonestepfive{SC-NMF-Spec}
& 
\ytextmodifhershelvonestepfive{SC-NMF-Map}
& MASS 
& MASS-NMF-Spec & MASS-NMF-Map\\
\hline
%
\ytextmodifhershelvonestepfive{$6\sigma_{Map}$}
& NRMSE & 15.59 \% & 3.05 \% & 16.52 \% & \textbf{\underline{2.58 \%}} & 3.58 \% 
& 16.60 \% & 3.30 \% \\
\cline{2-9}
& SAM (rad) & 0.077 & 0.019 & 0.089 & \textbf{\underline{0.015}} & 0.030 & 0.090 
& 0.018\\
\cline{2-9}
& $\text{NMCEB}^{\text{max}}$ & 13.65 \% & - & 0.69 \% & 2.58e-3 \% & - & 0.70 \% 
& 4.92e-3 \%\\
\hline	
\hline
\ytextmodifhershelvonestepfive{$2\sigma_{Map}$}
& NRMSE & 18.25 \% & 20.59 \% & 11.76 \% & 11.67 \% & 10.37 \% 
& 12.82 \% & \textbf{\underline{8.36 \%}}\\
\cline{2-9}
& SAM (rad) & 0.084 & 0.145 & 0.061 & 0.082 & 0.077 & 
0.057 & \textbf{\underline{0.050}}\\
\cline{2-9}
& $\text{NMCEB}^{\text{max}}$ & 22.25 \% & - & 1.51 \% & 8.93e-2 \% & - & 1.24 \% 
& 0.72 \%\\
\hline
\end{tabular}
\caption{Performance 
\ytextmodifhershelvonestepfive{obtained with the considered}
methods for a cube 
containing 4 sources with a 
\ytextmodifhershelvonestepfive{20 dB SNR.}
Results in bold identify the cases 
when hybrid methods improve 
\ytextmodifhershelvonestepfive{ performance, as compared
with the MC-NMF, SpaceCorr or MASS}
methods used alone. The 
underlined results identify the best results obtained for each of the two cubes 
\ytextmodifhershelvonestepfive{respectively corresponding to}
$d = 6
\ytextmodifhershelvonestepfive{\sigma_{Map}}
$ 
\ytextmodifhershelvonestepfive{and}
$d = 2
\ytextmodifhershelvonestepfive{\sigma_{Map}}
$.}
\label{tab_exemple_All_chap3}
\end{center}
\end{table*} 
\renewcommand\arraystretch{1} 

\subsubsection{Summary of the results}
\label{sec-results-summary}

To conclude on the synthetic tests, we group in Table 
\ref{tab_exemple_All_chap3} the performances obtained by the different methods 
for the cube containing four sources with 
\ytextmodifhershelvonestepfive{an}
SNR of 20 dB.
\ytextmodifhershelvtwostepfiftyone{The reported performance values
are averaged over all four components and 100 noise
realizations.}

First, the four hybrid methods presented here 
\ytextmodifhershelvonestepfive{highly}
improve the 
\ytextmodifhershelvonestepfive{spread}
of 
the solutions given by 
MC-NMF used alone. This point is the first interest 
to use hybrid methods.

With regard to the decomposition quality achieved by the 
\ytextmodifhershelvtwostepfiftyone{SC-NMF}
and MASS-NMF 
hybrid methods, the synthetic data tests show that
\ytextmodifhershelvonestepfive{the Map initialization 
provides
better results than the
``Spec'' one in almost all cases, whatever the considered source sparsity.
Therefore,
as an overall result, Map initialization is the preferred option.
The only exception to that trend is that
\ytextmodifhershelvonestepfive{SC-NMF-Spec}
yields a better result than 
\ytextmodifhershelvonestepfive{SC-NMF-Map}
in terms of SAM
for low-sparsity sources.%
}

The last point to be specified in these tests is the choice of the method 
\ytextmodifhershelvonestepfive{used}
to 
initialize 
%
\ytextmodifhershelvonestepfive{MC-NMF
with the ``Map'' initialization
selected above,
namely
SpaceCorr or MASS.
The results of Table
\ref{tab_exemple_All_chap3}
%
%
%
\ytextmodifhershelvtwostepfiftyone{refine those
derived above from Fig.
\ref{fig_sparse_seule}
to
\ref{fig_nonsparse_hybride},
with a much
better statistical confidence,
because they are here averaged over all sources and
100 noise realizations,
instead of considering
only one source and one noise realization in the above figures.
The results of Table
\ref{tab_exemple_All_chap3},
in terms of NRSME and SAM,
thus
show that
SC-NMF-Map yields slightly better performance than 
MASS-NMF-Map in the most sparse case,
whereas MASS-NMF-Map provides significantly better performance
in the least sparse case. This result is not surprising, because
SC-NMF-Map sets more stringent constraints on source sparsity, but
is expected to yield somewhat better performance when these
constraints are made (thanks to the data averaging that it peforms
over analysis zones). In a ``blind configuration'', i.e. when
the degree of sparsity of the sources
is not known, the preferred method for the overall set of data
considered here is MASS-NMF-Map because, as compared with
SC-NMF-Map, it is globally better in the sense that it may yield
significantly better or only slightly worse
performance, depending on sparsity.}

It should be noted at this stage that we suspect 
that what favors MASS-NMF-Map is related to 
the dimension of the data cube, i.e. that this latter method
performs better here because there are, in the synthetic 
data, more points spatially than spectrally (i.e. 10$^4$ spatial points
vs 300 spectral points). Hence, 
by initializing with SCA results the matrix that contains the largest 
number of points, MASS-NMF-Map provides a better solution. 
On the contrary, in the recent study by \citet{fos19}, the authors have found that
MASS-NMF-Spec performs better. In their specific case case, the (real) data 
contain only 31 spatial positions and 6799 spectral points. This suggests that
the general recommendation is to use the version of the method (Spec or Map) 
that initializes the largest number of points in the NMF with SCA results. }

%
%
%
%
\section{Conclusion and future work}
\label{sec_concl}
In this paper, we proposed different versions of Blind Source Separation 
methods for astronomical hyperspectral data. Our approach was to combine two 
well-known classes of methods, 
\ytextmodifhershelvonestepfive{namely}
NMF and Sparse 
SCA, in order to leverage their respective advantages while 
compensating their disadvantages. 
\ytextmodifhershelvonestepfive{We developed several 
hybrid methods based
on this principle, depending on the considered SCA algorithm
(SpaceCorr or MASS) and depending whether that SCA algorithm 
is used to set the
initial values of spectra or abundances then updated by our 
Monte-Carlo version of NMF, called MC-NMF.
In particular, our MASS-NMF-Map
hybrid method,
based on
initializing MC-NMF with the abundance maps provided by MASS,
yields}
a quasi-unique 
solution to the decomposition of a synthetic hyperspectral data cube, 
with an 
\ytextmodifhershelvonestepfive{average error 
(summarized in Table 
\ref{tab_exemple_All_chap3})
which is always better, and often
much better, than that of the MC-NMF, SpaceCorr}
and MASS 
methods used separately. 
Our various tests on simulated data also show
robustness to additive white noise. 
%
%
%
\ytextmodifhershelvonestepfive{Since the initialization of NMF 
with SCA methods was here shown to yield
encouraging results, our future work will especially aim at
developing SCA methods with lower sparsity constraints, in order
to further extend the domain where the resulting hybrid
SCA-NMF 
methods apply.
A first application of the MASS-NMF-Spec method presented in this paper 
on real data is presented in \citet{fos19} and shows the potential
of such methods for current and future hyperspectral datasets. 
}

\bibliographystyle{aa}
\bibliography{biblio.bib}

\newpage
\onecolumn

\appendix

\section{\ytextmodifhershelvonestepfive{Data processing methods}}
\subsection{\ytextmodifhershelvonestepfive{K-means method}}
K-means is a standard unsupervised classification method (\cite{the09}) which 
aims to partition a set of $N$ vectors $x$ into $k$ sets 
$\{\sigma_1,\sigma_2,\ldots,\sigma_k\}$. Within each set $\sigma_i$, a measure 
of dissimilarity $d$ between vectors $x \in \sigma_i$ and cluster representative 
$c_i$ is minimized. The cluster representative $c_i$ (or centroid) is the mean 
vector in the cluster $\sigma_i$. In our case, the measure of dissimilarity is 1 
minus the correlation between $x \in \sigma_i$ and $c_i$:
\begin{equation}\label{eq_d}
d(x,
\ytextmodifhershelvonestepfive{c_i}
) = 1-corr(x,
\ytextmodifhershelvonestepfive{c_i}
).
\end{equation}
Performing clustering then amounts to minimizing the following objective 
function:
\begin{equation}
J(U,C) = \sum\limits_{i=1}^k \sum\limits_{j=1}^N U_{ij}d(x_j,c_i),
\end{equation}
where $C$ is the matrix of centroids 
\ytextmodifhershelvonestepfive{$c_i$}
and $U$ is a partition matrix such that 
$U_{ij}=1$ if $x_j \in \sigma_i$ and $0$ otherwise. Moreover, 
\ytextmodifhershelvonestepfive{each vector belongs to a single cluster}
(hard clustering):
\begin{equation}
\sum\limits_{i=1}^k U_{ij} = 1 \qquad j\in\{1,\ldots,N\}.
\end{equation}
The K-means algorithm is defined as follows:

\begin{algorithm}[H]
\SetKwInOut{Input}{Input}\SetKwInOut{Output}{Output}
\Input{A set of vectors $x$, the number of cluster $k$.}
\Output{A partition matrix $U$, a set of centroids $C$.}
\BlankLine
\Begin{
	Choose arbitrary initial $C_0$ for the centroids\tcc*[r]{In practice, 
$k$ vectors $x$%
\ytextmodifhershelvonestepfive{,}
randomly selected}
	$U \leftarrow 0$\;
	\Repeat{no change in $c_i$'s occurs between two successive iterations}{
		\tcc{Update clusters}
		\For{$j=1$ \KwTo $N$}
		{
			Determine the closest representative, say $c_i$, for 
$x_j$ in the sense of $d$\;
			$U_{ij} \leftarrow 1$\;
		}
		\tcc{Update centroids}
		\For{$i=1$ \KwTo $k$}
		{
			Determine $c_i$ as the mean of the vectors 
$x\in\sigma_i$\;
		}
	}
}
\caption{K-means}
\end{algorithm}

However, a well-known drawback of this algorithm is its sensitivity to the 
initialization, the final centroids depending on the choice of initial cluster 
\ytextmodifhershelvonestepfive{representatives}
$c_i$. But for our two clustering steps, a manual initialization 
is possible and avoids this drawback: 
\begin{itemize}
\item As part of Monte-Carlo analysis of NMF (Section \ref{subseq_NMF}), we use 
as initial centroids $C_0$ the 
\ytextmodifhershelvonestepfive{$L$}
spectra estimated by the first Monte-Carlo 
trial.
\item As part of the estimation step of SpaceCorr (Section \ref{subseq_SCA}), we 
use as initial centroids $C_0$ the 
\ytextmodifhershelvonestepfive{$L$}
farthest spectra in the sense of the 
measure $d$ (%
\ytextmodifhershelvonestepfive{see Eq. (}%
\ref{eq_d})), among the potential columns of the matrix 
\ytextmodifhershelvonestepfive{$\hat{A}$.}
\end{itemize}

\subsection{\ytextmodifhershelvonestepfive{Parzen kernel method}}
The estimation by Parzen kernel or Parzen windows (\cite{the09,sil98}) is a 
parametric method for estimating the probability density function (pdf) of a 
random variable at any point of its support $\Omega$. Let $(x_1,x_2,\ldots,x_N)$ 
be an independent and identically distributed sample of a random variable $X$ 
with an unknown pdf $f$. Its kernel density estimator is:
\begin{equation}\label{eq_density}
\hat{f}(x) = \frac{1}{Nh} \sum\limits_{i=1}^N K\left(\frac{x_i-x}{h}\right) 
\qquad x\in\Omega,
\end{equation}
where $K$ is a kernel, i.e., a smooth nonnegative function that 
integrates to one, and $h>0$ is a smoothing parameter called the bandwidth (or 
width of window). In our case, we use as the kernel the standard normalized 
Gaussian function (zero expectation and unit standard deviation):
\begin{equation}
K(x) = \frac{1}{\sqrt{2\pi}}e^{\left(-\frac{1}{2}x^2 \right)}.
\end{equation}
The bandwidth $h$ of the kernel is a free parameter which exhibits a strong 
influence on the resulting estimate. In the case of a Gaussian kernel, it can be 
shown (\cite{sil98}) that the optimal choice for $h$ is:
\begin{equation}\label{eq_h}
h = \left(\frac{4\hat{\sigma}^5}{3N}\right)^{\frac{1}{5}} \approx 
1.06\hat{\sigma}N^{-\frac{1}{5}},
\end{equation}
where $\hat{\sigma}$ is the sample standard deviation. The Parzen kernel 
algorithm is defined as follows:

\begin{algorithm}[H]
\SetKwInOut{Input}{Input}\SetKwInOut{Output}{Output}
\Input{Sample points $X$.}
\Output{The estimated pdf $\hat{f}$, a support $\Omega$ covering the range of 
the data.}
\BlankLine
\Begin{
	Compute $h$\tcc*[r]{According to (\ref{eq_h})}
	Define the support such that $\Omega=[min(X),max(X)]$\tcc*[r]{In 
practice, Card $(\Omega)=100$}
	\For{$k=1$ \KwTo Card$(\Omega)$}
	{
		Compute the 
\ytextmodifhershelvonestepfive{estimate}
		of $\hat{f}(\Omega_k)$\tcc*[r]{According 
to (\ref{eq_density})}
	}
}
\caption{Parzen Kernel}
\end{algorithm}

\section{\ytextmodifhershelvonestepfive{Additional test results}}
The next figures show the performance of the methods applied to all cubes. 
\ytextmodifhershelvonestepfive{The}
SAM and the NRMSE 
\ytextmodifhershelvonestepfive{values}
presented in the following figures
\ytextmodifhershelvonestepfive{are}
obtained by averaging these measurements for each method 
\ytextmodifhershelvonestepfive{over all 
\ytextmodifhershelvtwostepfiftyone{100}
noise realizations.}
We 
\ytextmodifhershelvonestepfive{also provide error bars which define}
the standard deviation 
\ytextmodifhershelvonestepfive{over all noise realizations.}
%
\ytextmodifhershelvonestepfive{Each such}
NMCEB 
\ytextmodifhershelvonestepfive{spread}
measurement
\ytextmodifhershelvonestepfive{is}
obtained in two stages. Firstly, for each source, the average spectrum and the 
average envelope are defined 
\ytextmodifhershelvonestepfive{over all noise realizations.}
In a second 
step, we identify the maximum%
\ytextmodifhershelvonestepfive{, first}
along the spectral axis ($n$)
\ytextmodifhershelvonestepfive{for a given source ($\ell$),
and then with respect to all sources.}
%

To summarize, we associate with each method a 
\ytextmodifhershelvonestepfive{first}
figure composed of six frames 
arranged in two 
\ytextmodifhershelvonestepfive{rows}
and three columns. The first 
\ytextmodifhershelvonestepfive{row}
gives the SAM and the 
second the NRMSE. The first column concerns cubes containing 2 sources, the 
middle one containing 4 sources and the last one containing 6 sources. Within 
each frame, the $x$ axis gives the distance $d$ defining the 
\ytextmodifhershelvonestepfive{overlap between}
the 
spatial sources, from $d=6
\ytextmodifhershelvonestepfive{\sigma_{Map}}
$ 
\ytextmodifhershelvonestepfive{down}
to $d=2
\ytextmodifhershelvonestepfive{\sigma_{Map}}
$. 
\ytextmodifhershelvtwostepfiftyone{In Sections
\ref{sec-appendix-Results-MC-NMF}
to
\ref{sec-appendix-Results-MASS-NMF-Map},}
%
\ytextmodifhershelvonestepfive{inside each such frame,}
the three curves 
\ytextmodifhershelvtwostepfiftyone{respectively correspond to}
each of the 
three 
\ytextmodifhershelvtwostepfiftyone{main}
noise levels tested in this study: blue for 30 dB
\ytextmodifhershelvonestepfive{Signal to Noise Ratio (SNR)}%
, red for 20 dB and black 
for 10 dB. 
%
%
%
\ytextmodifhershelvtwostepfiftyone{Similarly,
in Section
\ref{sec-appendix-Results-low-SNR},
inside each frame,
the three curves 
respectively correspond to
each of the 
three 
additional tested
low SNRs:
blue for 5 dB,
red for 3 dB and black 
for 1 dB.}
%
%
%
For the MC-NMF method and the hybrid methods, 
\ytextmodifhershelvonestepfive{we also provide a second figure,}
consisting of 
3 frames giving the NMCEB maximum for each of the 45 cubes.
%

In conclusion, all of these tests and illustrations aim to quantify several 
phenomena. First, evaluate the impact of the noise level and the distance 
between the sources on the performance of the methods. Next, quantify the 
contribution of hybrid methods compared to 
\ytextmodifhershelvonestepfive{the}
MC-NMF, SpaceCorr and MASS 
\ytextmodifhershelvonestepfive{methods}
used alone. The study of the error bars of the Monte-Carlo analysis associated 
with the NMF makes it possible to evaluate the 
\ytextmodifhershelvonestepfive{spread}
of the solutions as a 
function of the noise level, the sparsity of the sources and the initialization.

\subsection{\ytextmodifhershelvonestepfive{Results with MC-NMF}}
\begin{figure}[htb]
\begin{center}	
\begin{tabular}{ccc}
\includegraphics[width=0.3\textwidth]{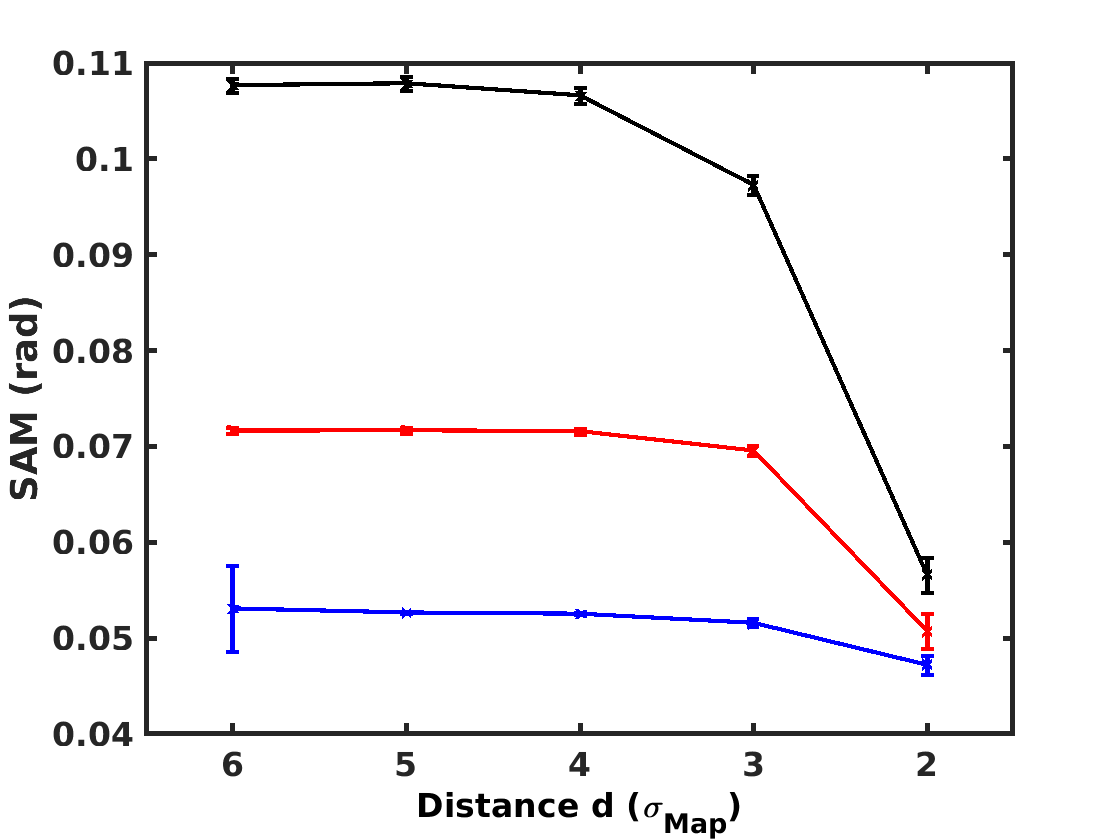}&
\includegraphics[width=0.3\textwidth]{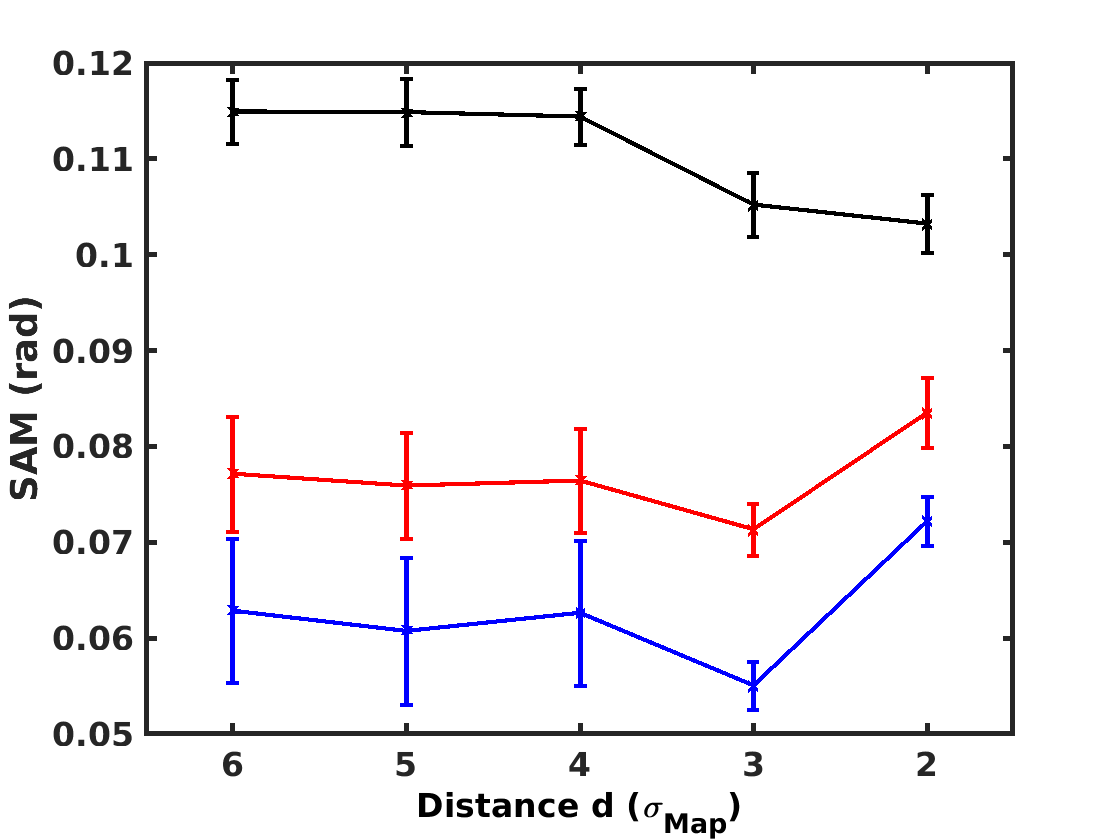}&
\includegraphics[width=0.3\textwidth]{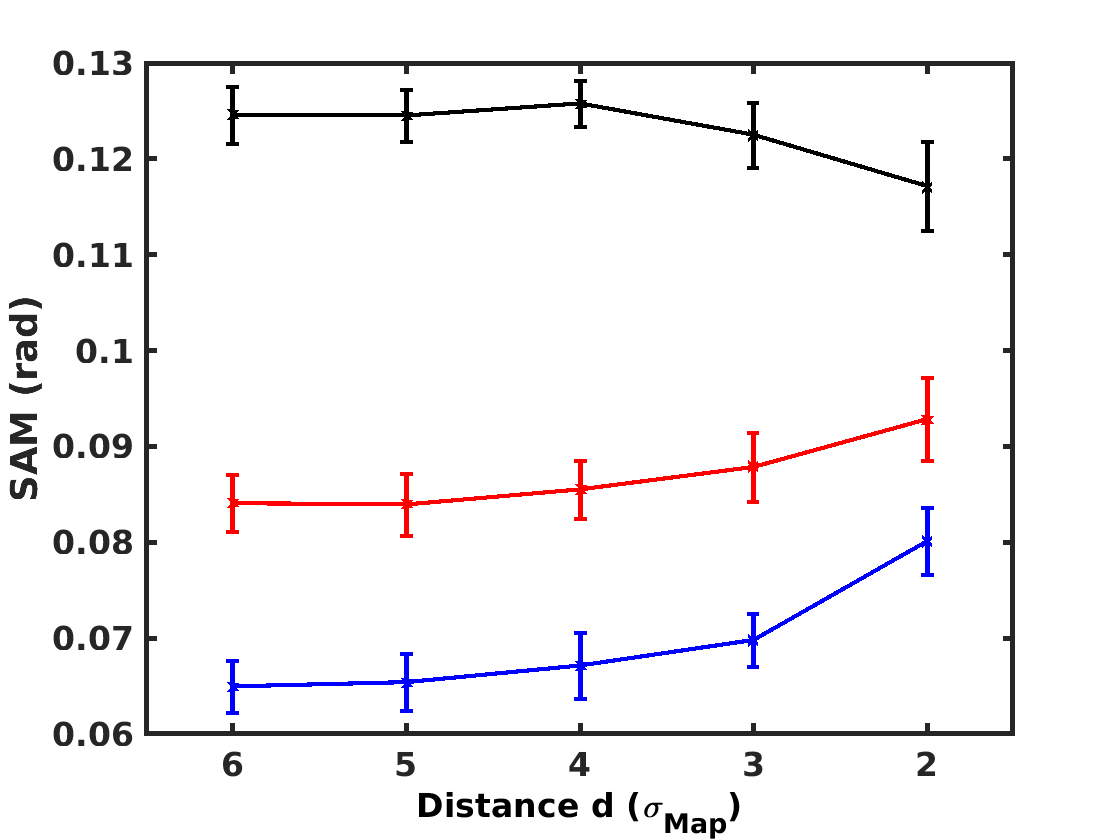}\\
\includegraphics[width=0.3\textwidth]{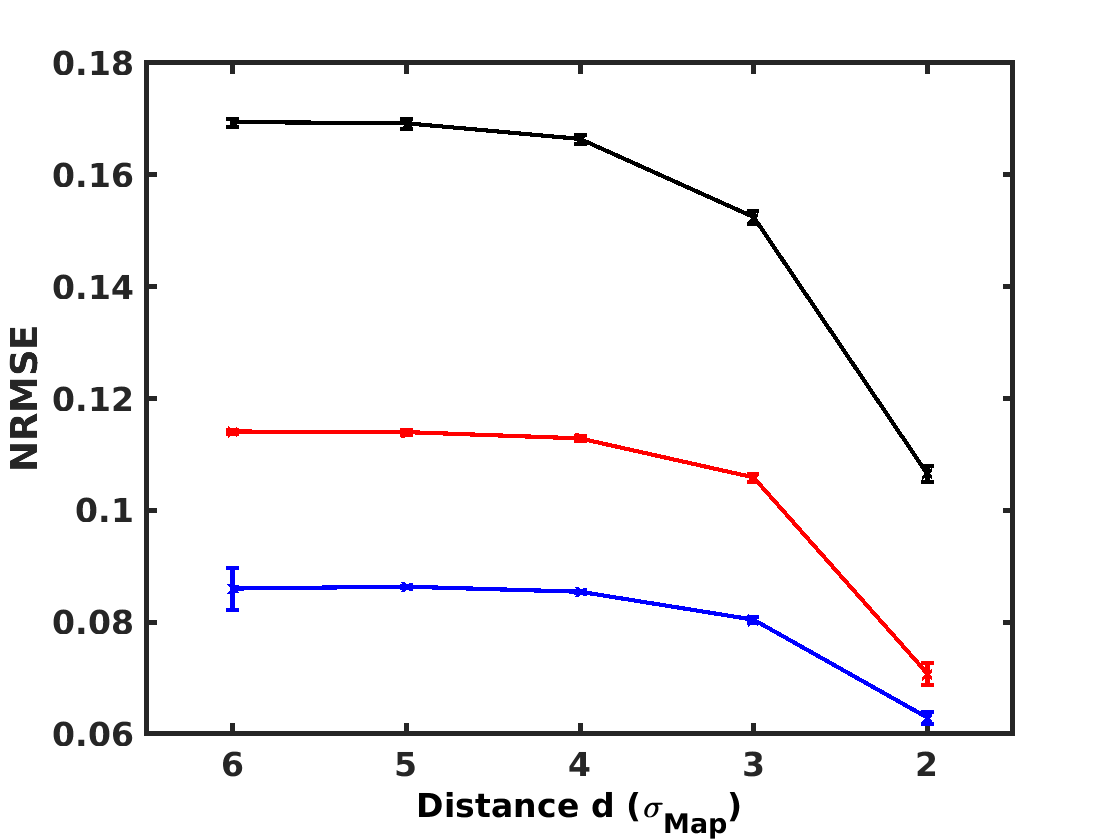}&
\includegraphics[width=0.3\textwidth]{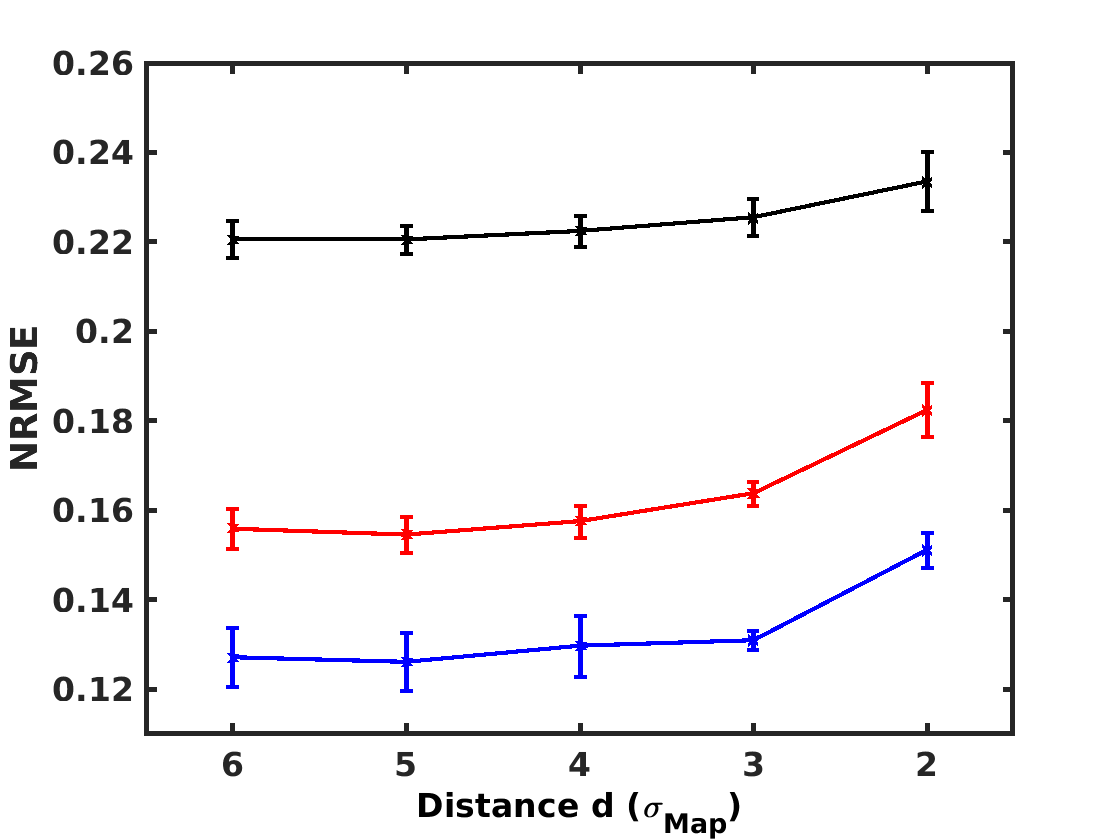}&
\includegraphics[width=0.3\textwidth]{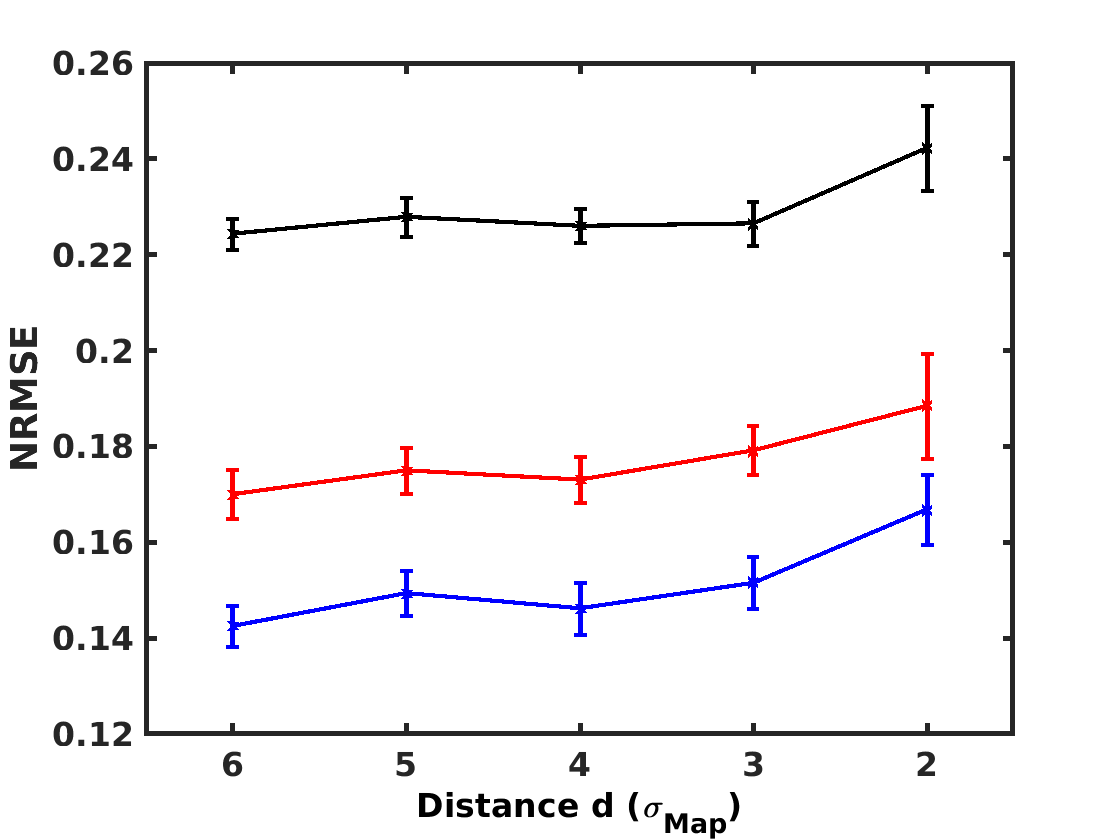}\\
2 sources & 4 sources & 6 sources
\end{tabular}
\end{center}
\caption{Performances achieved by MC-NMF on the 45 synthetic cubes for 100 
realizations of noise with 
\ytextmodifhershelvonestepfive{an}
SNR of 30 dB (in blue), 20 dB (in red), and 10 dB 
(in black). The error bars give the standard deviation 
\ytextmodifhershelvonestepfive{over}
the 100 
realizations of noise.} 
\label{fig_perf_NMF_chap3}
\end{figure}

\begin{figure}[htb]
\begin{center}	
\begin{tabular}{ccc}
\includegraphics[width=0.3\textwidth]{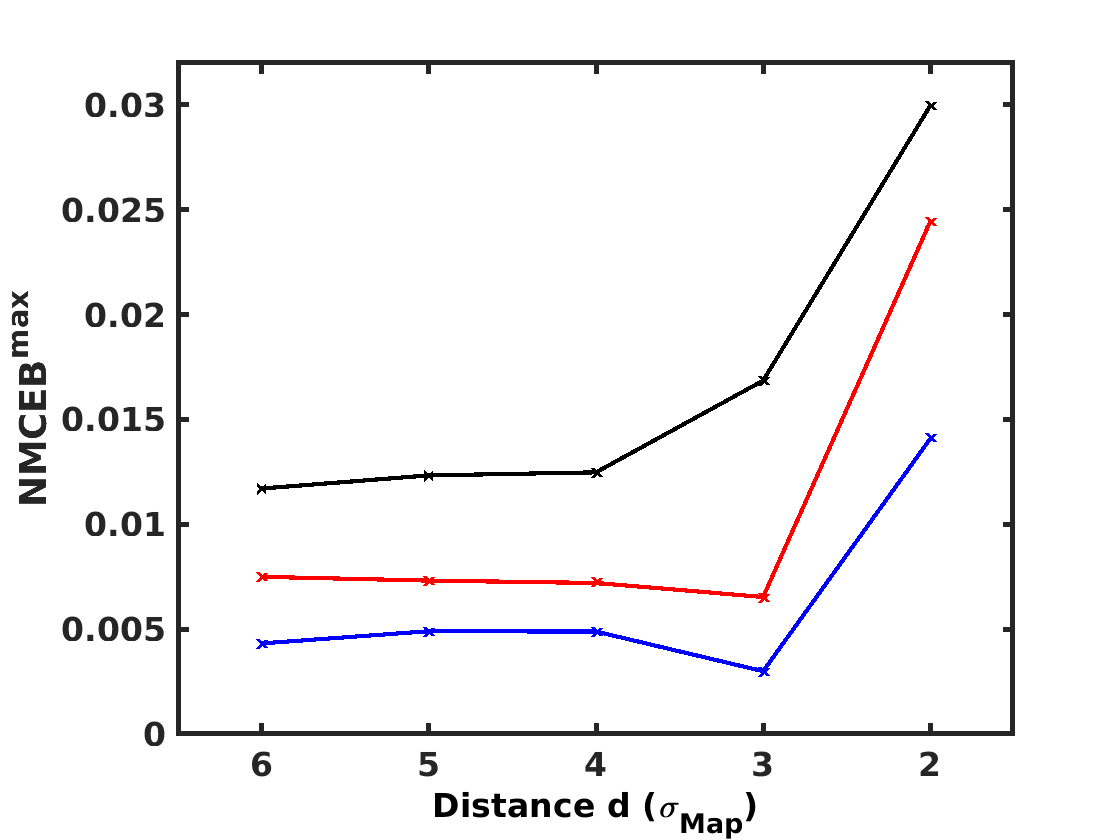}&
\includegraphics[width=0.3\textwidth]{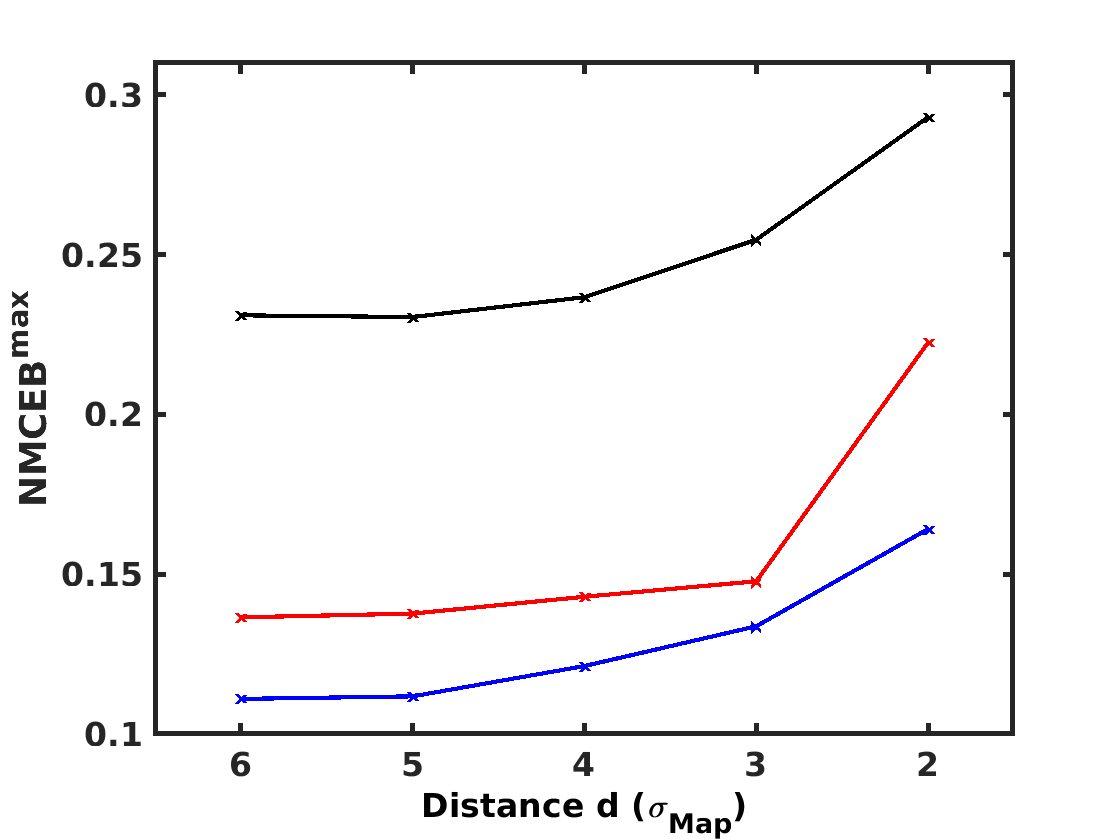}&
\includegraphics[width=0.3\textwidth]{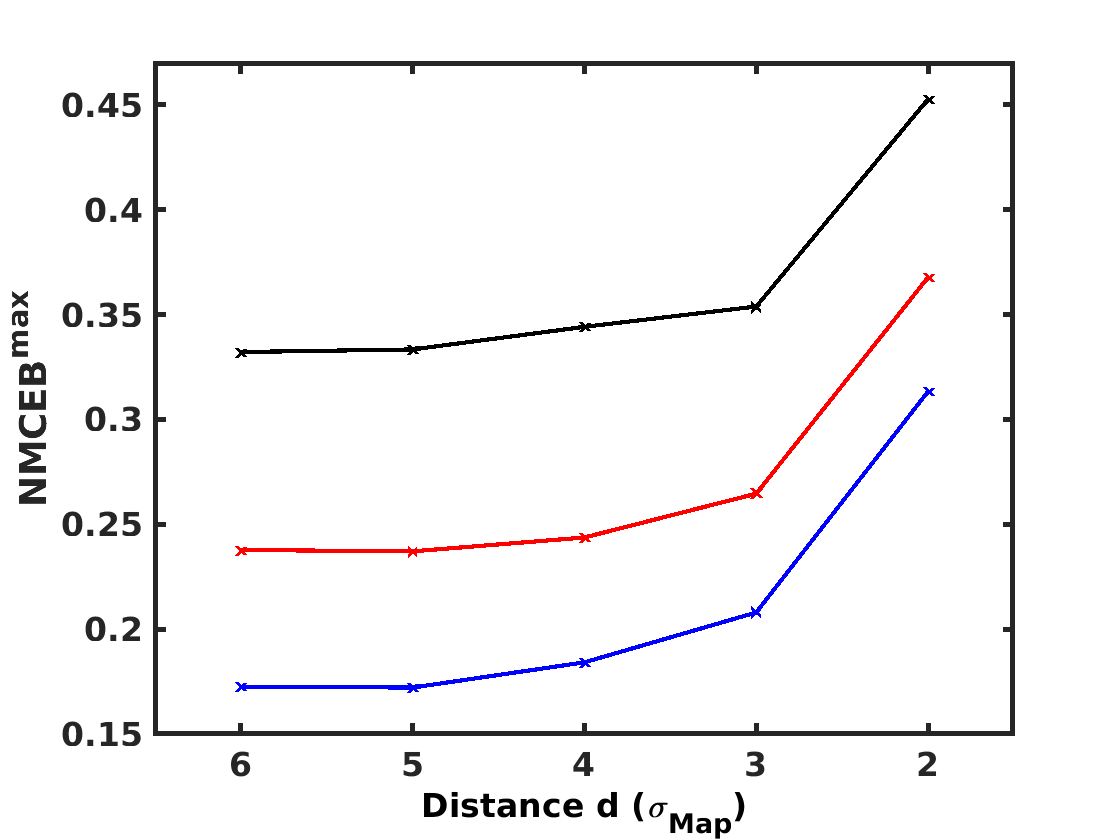}\\
2 sources & 4 sources & 6 sources
\end{tabular}
\end{center}
\caption{%
\ytextmodifhershelvonestepfive{Spread (NMCEB)}
of the solutions of MC-NMF obtained on the 45 synthetic 
cubes with 100 realizations of noise.} 
\label{fig_errbar_NMF_chap3}
\end{figure}

%
\label{sec-appendix-Results-MC-NMF}
%
%
\ytextmodifhershelvonestepfive{%
The performance of the MC-NMF method is shown in Fig. 
\ref{fig_perf_NMF_chap3}
and
\ref{fig_errbar_NMF_chap3}.
We will first consider the cases
involving four or six sources. The performance of MC-NMF then
has a low sensitivity}
%
%
%
to 
\ytextmodifhershelvonestepfive{the distance $d$, i.e. to the source joint}
%
%
%
sparsity. 
%
%
Similarly, the number of 
sources is a criterion having a relatively 
\ytextmodifhershelvonestepfive{limited}
effect on the performance 
of MC-NMF
\ytextmodifhershelvonestepfive{in}
our simulations. 
%
\ytextmodifhershelvonestepfive{On the contrary, the noise level has}
%
a significant effect on the quality of the solutions given by  
MC-NMF (although the noiseless case does not give ideal results).
It should be noted that the amplitude of the error bars of MC-NMF depends on 
all the tested parameters (degree of sparsity, number of sources and noise 
level). 
The variability of MC-NMF solutions is often substantial and the goal of 
hybrid methods is to reduce this variability.

\ytextmodifhershelvonestepfive{We will now focus on the case of 2 sources.
For distances ranging from $d=6 \sigma_{Map}$ down to $d=3 \sigma_{Map}$ 
(or at least 
$
4
\ytextmodifhershelvonestepfive{\sigma_{Map}}
$%
), the same comments as those provided above
for 4 or 6 sources still apply, as expected.
On the contrary, further decreasing $d$ to
$d=2
\ytextmodifhershelvonestepfive{\sigma_{Map}}
$ results in an unexpected behavior:
the SAM and NRMSE values then strongly decrease.}
In other words, in this situation 
\ytextmodifhershelvonestepfive{when the sources become more mixed,
performance improves.}
%
Several causes can lead to this result, such as the presence of noise, the 
symmetry of the maps, the great similarity of the spectra or the number of 
iterations of the NMF, although all these 
\ytextmodifhershelvonestepfive{features also apply to}
the cubes 
containing 4 or 6 {sources}. 
\ytextmodifhershelvonestepfive{To analyze the influence of such
features
on the shape of the cost function to be minimized
and thus on performance, we performed the following additional tests
%
%
%
for 2 sources:}

\begin{itemize}
\item No noise in the mixtures.
\item Deleting 
the normalization of the spectra at each iteration of the NMF.
\item Random switching of columns of $X$.
\item 
\ytextmodifhershelvonestepfive{Avoiding}
the symmetry of abundance maps by deforming a 2D Gaussian.
\item 
\ytextmodifhershelvonestepfive{Avoiding}
the similarity of the spectra by changing the nature of the 
function simulating the line (triangle or square function).
\item Fixed number of iterations of the NMF.
\item Deleting 
the observed spectra having a power lower than 90\% of the 
maximal 
\ytextmodifhershelvonestepfive{observed power.}%
\end{itemize}
%
\ytextmodifhershelvonestepfive{Each of these tests led to
the same, unexpected, trend as that 
observed in the first column of Fig \ref{fig_perf_NMF_chap3}.}
%
%
%
%
%
%
\ytextmodifhershelvonestepfive{On the contrary, the expected behavior
was observed in the following additional test, where the sparsity of
the sources was varied.}
%
The abundance maps simulated by 2D Gaussians are replaced by a matrix $S$ of 
dimension $2\times 100 $ whose elements are drawn 
\ytextmodifhershelvonestepfive{with}
a uniform distribution 
between 0.5 and 1. The first 25 elements of the first row are multiplied by a 
coefficient $\alpha$. The elements 26 to 50 of the second 
\ytextmodifhershelvonestepfive{row}
are also 
multiplied by the same coefficient $\alpha$. Thus, 
\ytextmodifhershelvonestepfive{depending on}
the value of 
$\alpha $, we simulate more or less sparsity in the data. In this test, we 
observe a decrease of the performance of  MC-NMF when $\alpha$ increases%
\ytextmodifhershelvonestepfive{, i.e. when the source sparsity decreases,
as expected.}
%
%
%
\ytextmodifhershelvonestepfive{Finally, we performed tests
for all the cubes (as well as for the hybrid methods),
where we
further decreased the distance
$d$ to $d=1
\ytextmodifhershelvonestepfive{\sigma_{Map}}
$, which corresponds to very low sparsity.
The SAM 
and NMRSE then considerably increase, which corresponds
to the expected behavior of BSS methods.}
%
%
%
%
%
%
\ytextmodifhershelvonestepfive{The above-defined unexpected behavior
is therefore only observed in the specific
case involving 2 sources and the intermediate distance
$d = 2
\sigma_{Map}
$, and it could be further analyzed in future work.}

\subsection{\ytextmodifhershelvonestepfive{Results with SpaceCorr}}
\begin{figure}[H]
\begin{center}	
\begin{tabular}{ccc}
\includegraphics[width=0.3\textwidth]{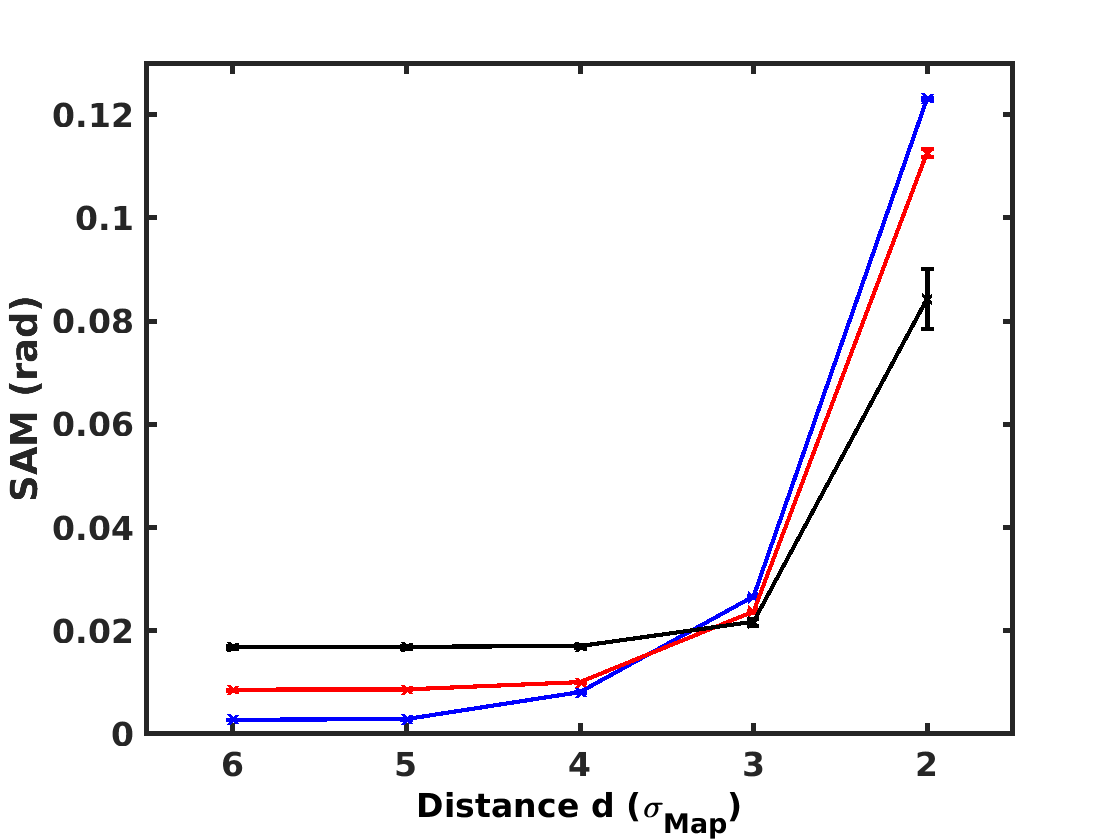}&
\includegraphics[width=0.3\textwidth]{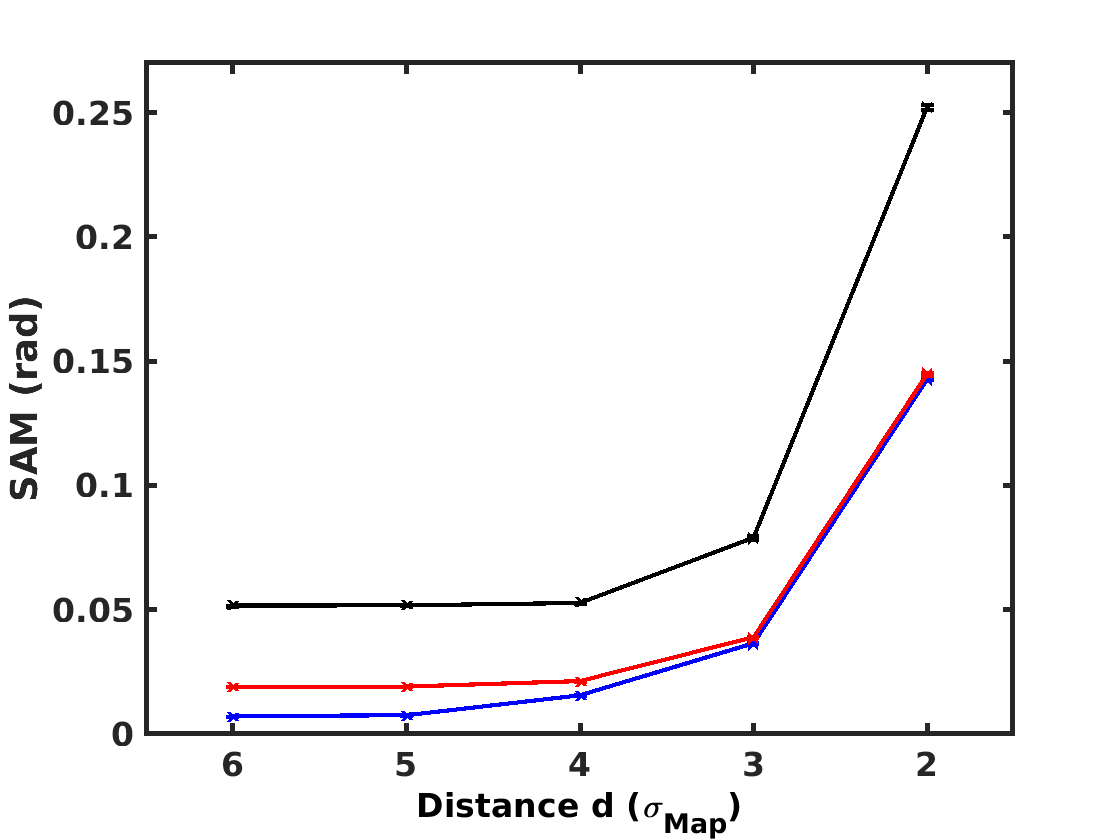}&
\includegraphics[width=0.3\textwidth]{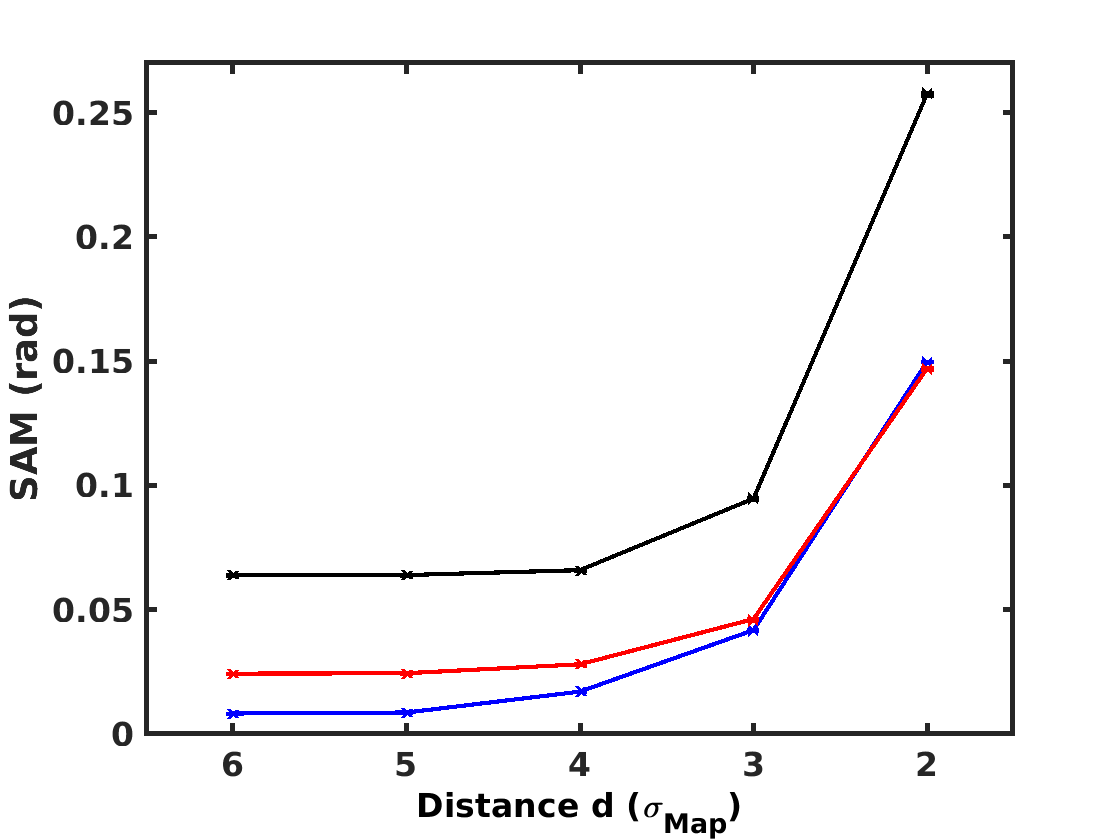}\\
\includegraphics[width=0.3\textwidth]{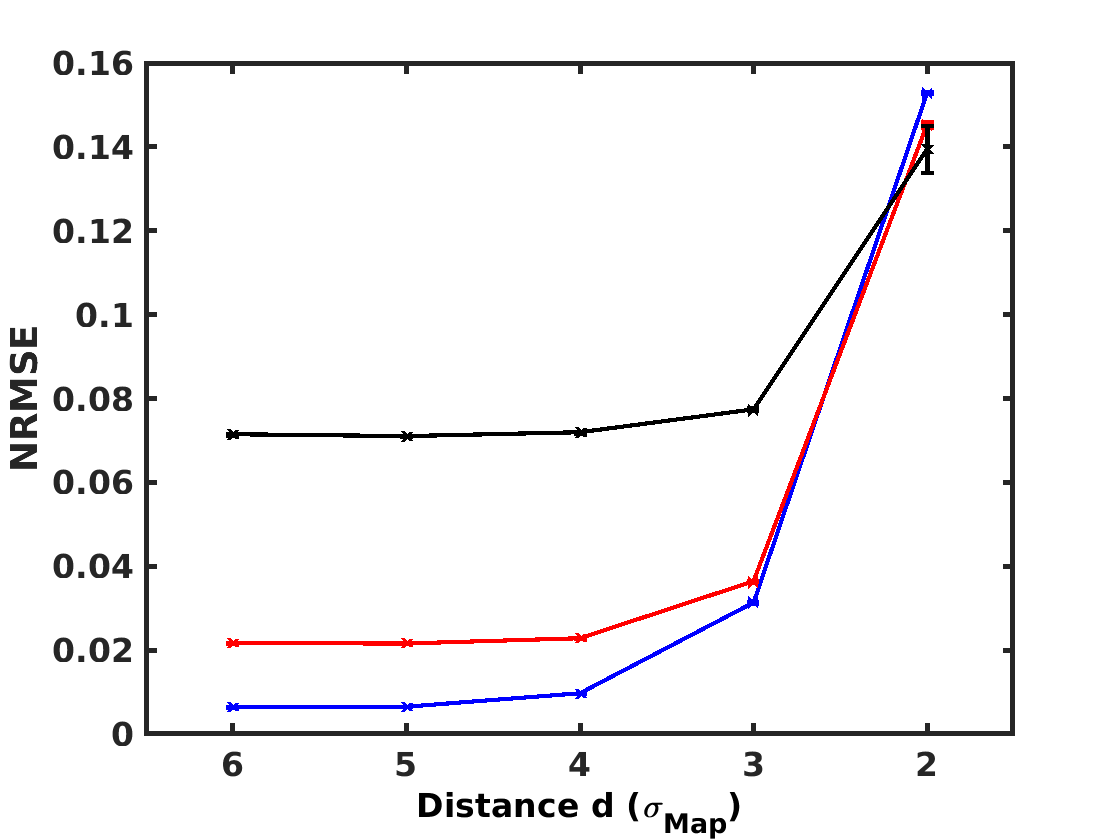}&
\includegraphics[width=0.3\textwidth]{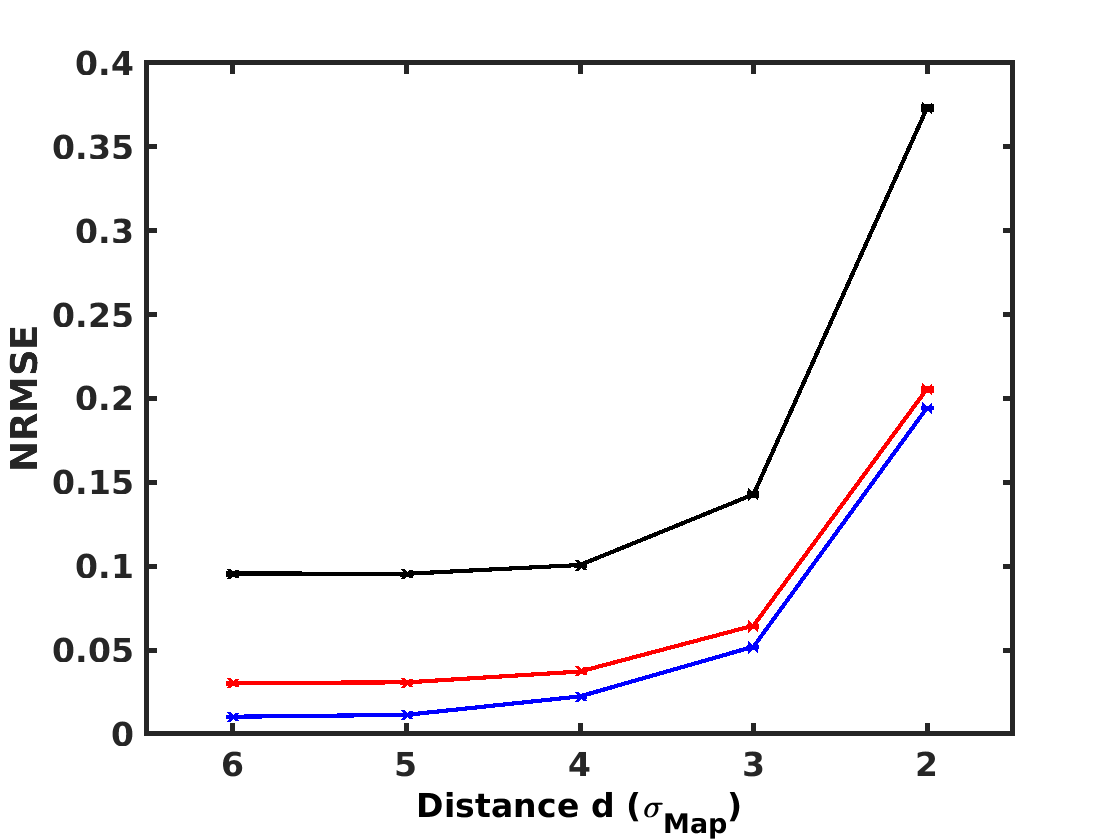}&
\includegraphics[width=0.3\textwidth]{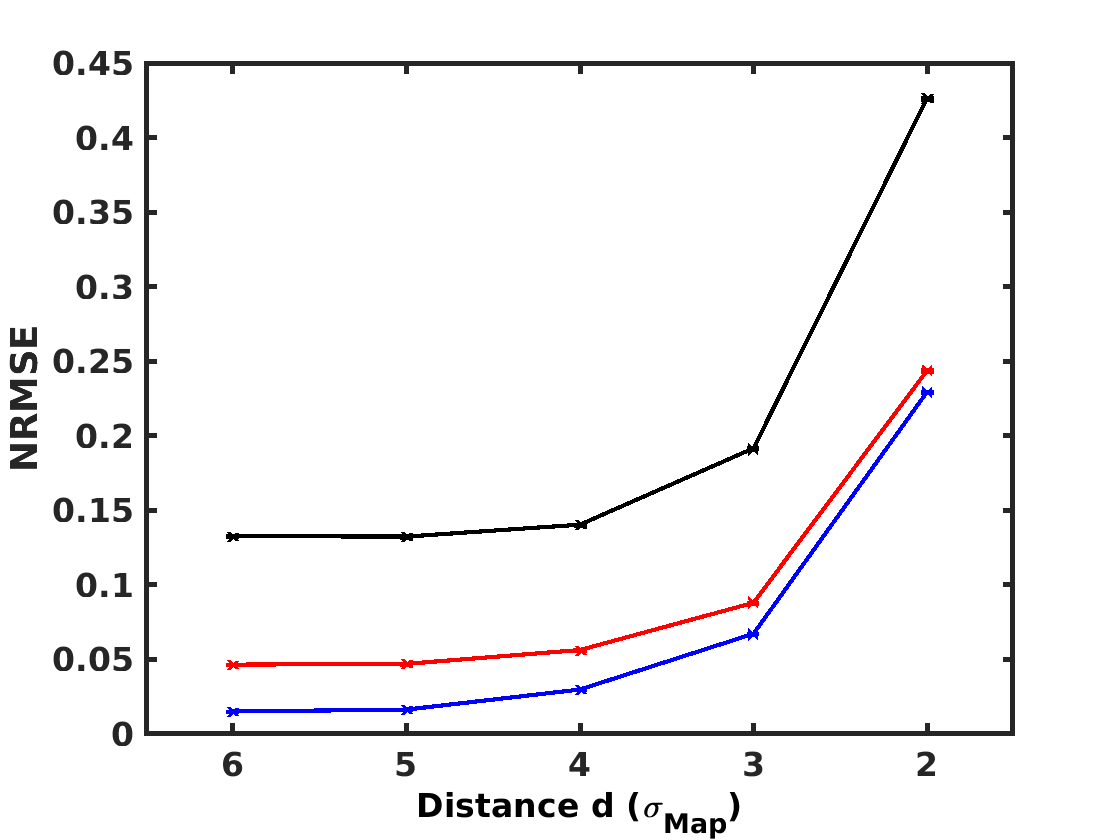}\\
2 sources & 4 sources & 6 sources
\end{tabular}
\end{center}
\caption{Performances achieved by SpaceCorr on the 45 synthetic cubes for 100 
realizations of noise with 
\ytextmodifhershelvonestepfive{an}
SNR of 30 dB (in blue), 20 dB (in red), and 10 dB 
(in black). The error bars give the standard deviation 
\ytextmodifhershelvonestepfive{over}
the 100 
realizations of noise.} 
\label{fig_perf_SpaceCORR_chap3}
\end{figure}

\ytextmodifhershelvonestepfive{The performance of the SpaceCorr method 
is shown in Fig. 
\ref{fig_perf_SpaceCORR_chap3}.
This method}
%
gives excellent results if the data are sparse enough, i.e., 
if there 
\ytextmodifhershelvonestepfive{are}
a sufficient number of single-source zones%
%
%
%
\ytextmodifhershelvonestepfive{, which here corresponds to a large 
enough distance $d$.}
%
%
%
We also note that 
SpaceCorr is not very sensitive to the number of sources.
\ytextmodifhershelvonestepfive{Its (limited) sensitivity is at least
due to the fact that the number of sources
%
%
%
over the considered
fixed spatial area 
may have an influence on the degree of source sparsity in
terms of available single-source zones.}
Finally, we emphasize that SpaceCorr is relatively robust to noise in data.
The presence of residuals in the estimates of the least sparse cases is due to 
the small number of single-source zones 
\ytextmodifhershelvonestepfive{per}
source in the data. In addition, the 
step of detection of the single-source zones is sensitive to the 
\ytextmodifhershelvonestepfive{choice}
of the 
threshold used to select the best 
\ytextmodifhershelvonestepfive{single-source}
zones. 
\ytextmodifhershelvonestepfive{Depending on}
this 
threshold, almost single-source zones 
\ytextmodifhershelvonestepfive{may}
be used to estimate the columns of the 
mixing matrix%
\ytextmodifhershelvonestepfive{, which yields}
contributions from other sources. 
However, the sensitivity of the method to the choice of this parameter will be 
attenuated by hybridization with MC-NMF.

\subsection{\ytextmodifhershelvonestepfive{Results with SC-NMF-Spec}}

\begin{figure}[H]
\begin{center}	
\begin{tabular}{ccc}
\includegraphics[width=0.3\textwidth]{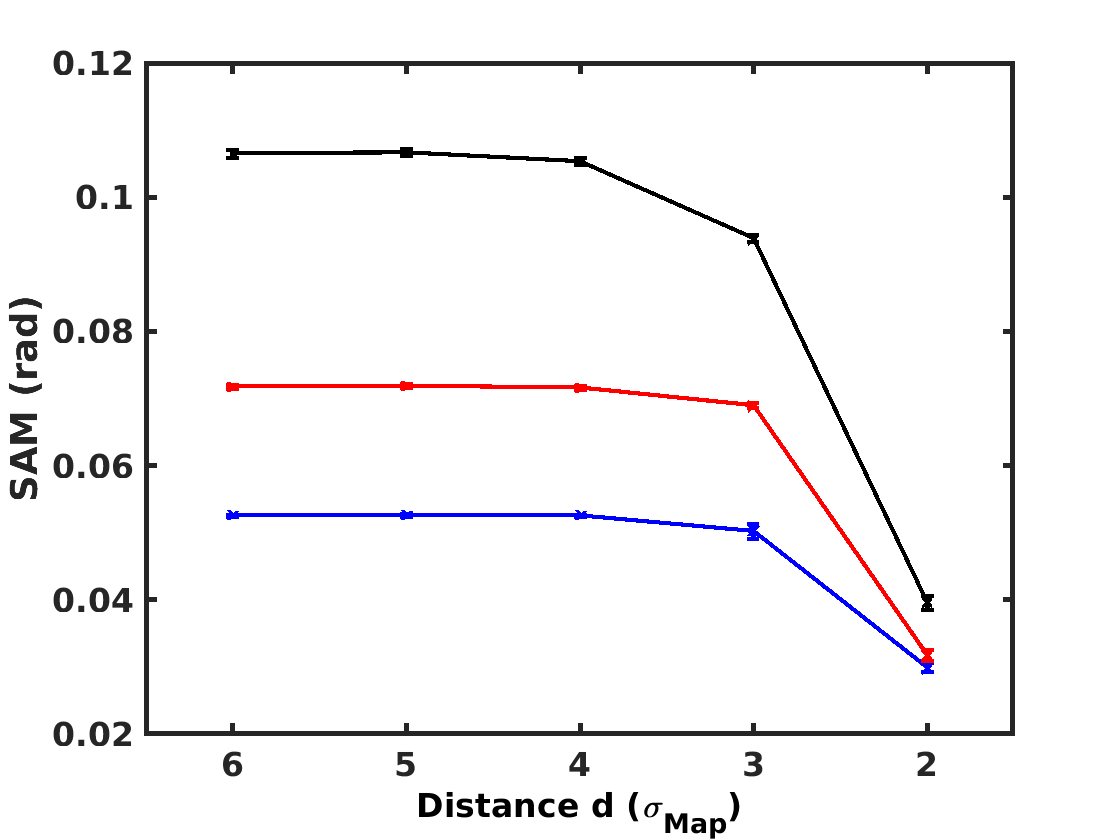}&
\includegraphics[width=0.3\textwidth]{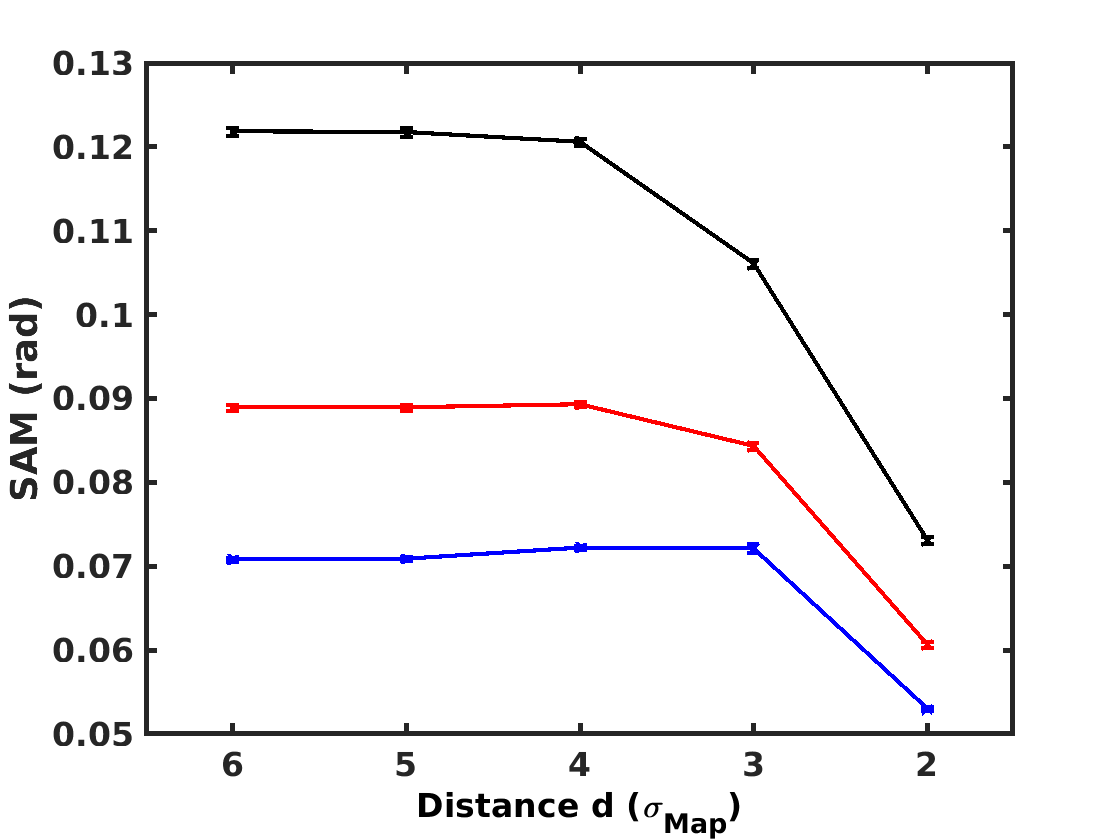}&
\includegraphics[width=0.3\textwidth]{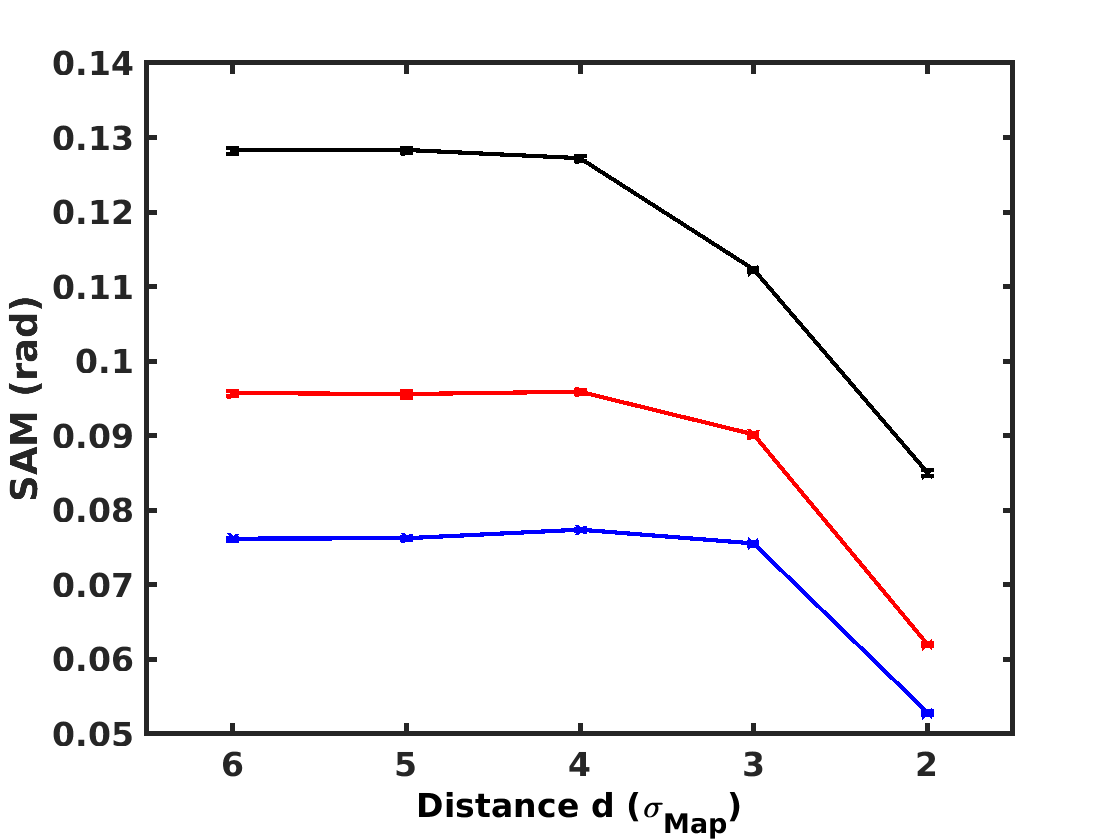}\\
\includegraphics[width=0.3\textwidth]{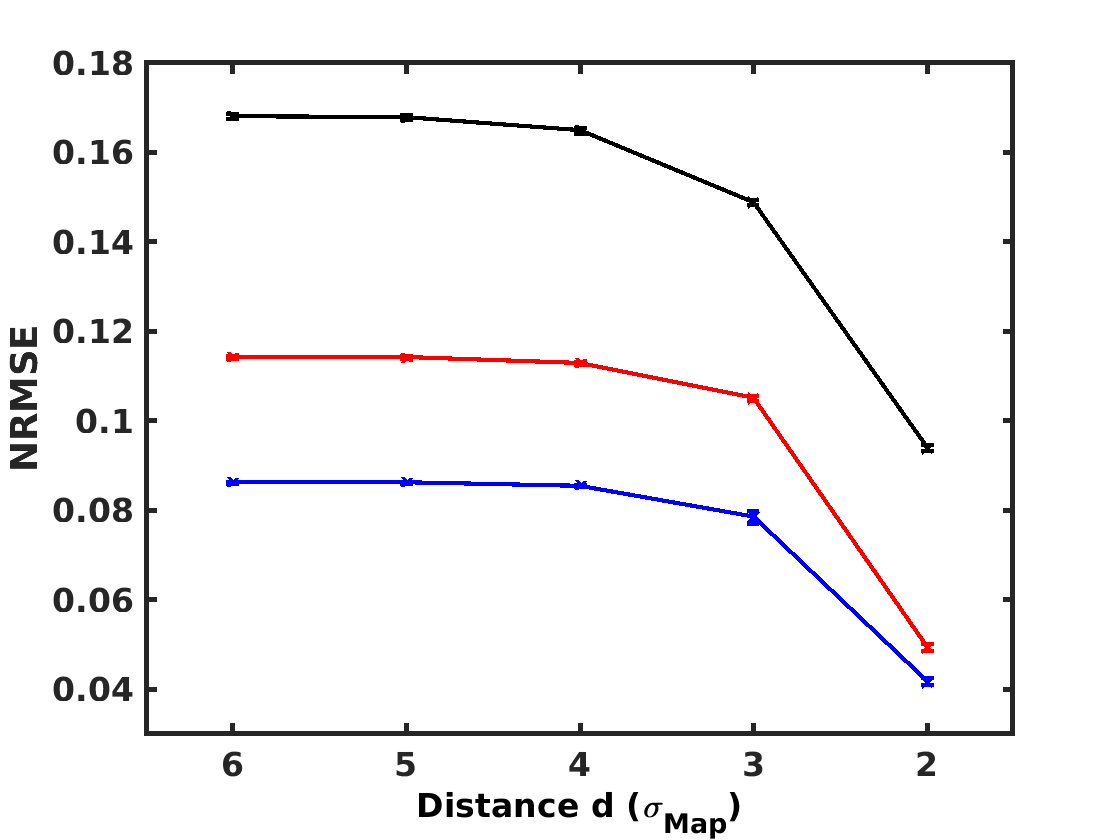}&
\includegraphics[width=0.3\textwidth]{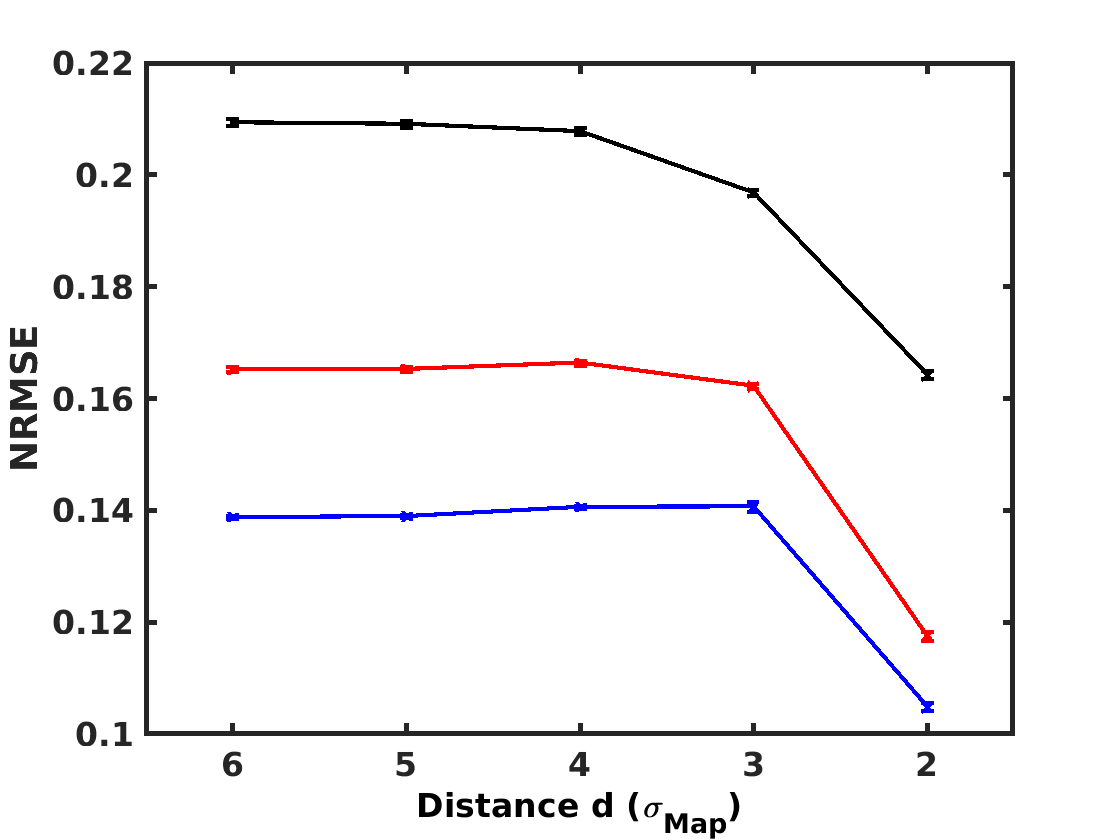}&
\includegraphics[width=0.3\textwidth]{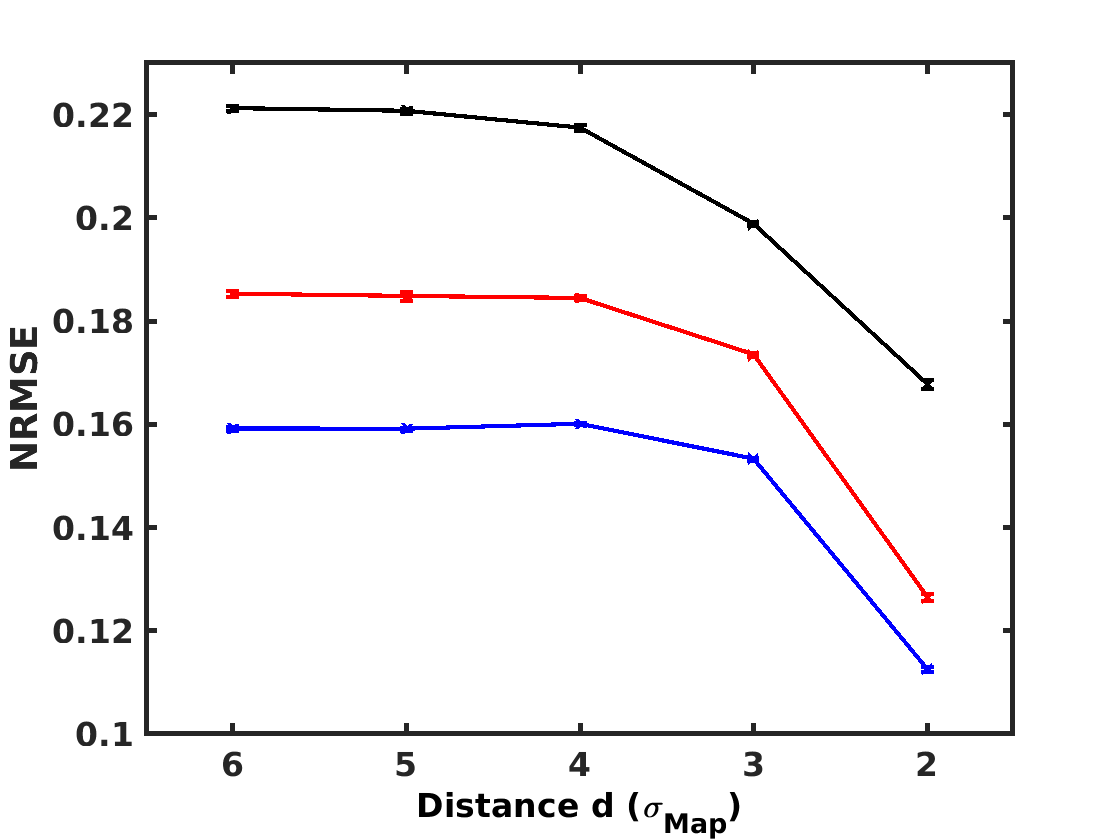}\\
2 sources & 4 sources & 6 sources
\end{tabular}
\end{center}
\caption{Performances achieved by 
\ytextmodifhershelvonestepfive{SC-NMF-Spec}
on the 45 synthetic cubes for 100 
realizations of noise with 
\ytextmodifhershelvonestepfive{an}
SNR of 30 dB (in blue), 20 dB (in red), and 10 dB 
(in black). The error bars give the standard deviation 
\ytextmodifhershelvonestepfive{over}
the 100 
realizations of noise.} 
\label{fig_perf_HybridSpec_chap3}
\end{figure}

\begin{figure}[h]
\begin{center}	
\begin{tabular}{ccc}
\includegraphics[width=0.3\textwidth]{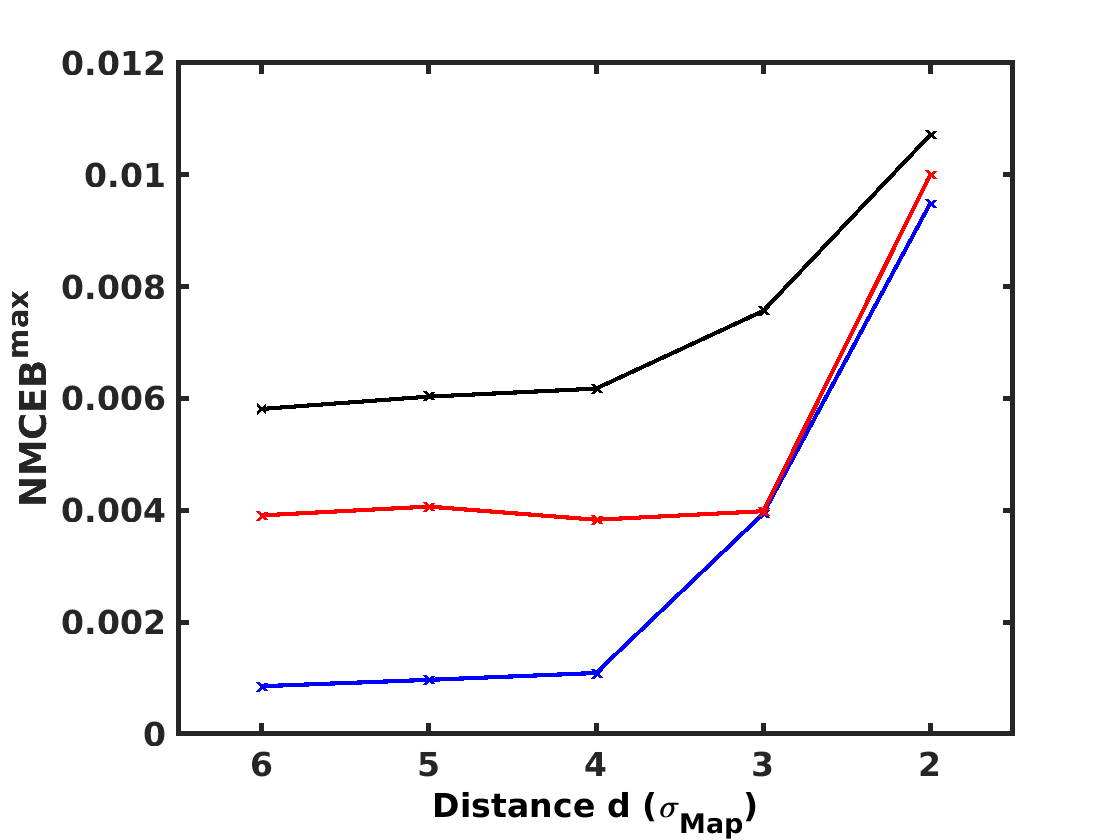}&
\includegraphics[width=0.3\textwidth]{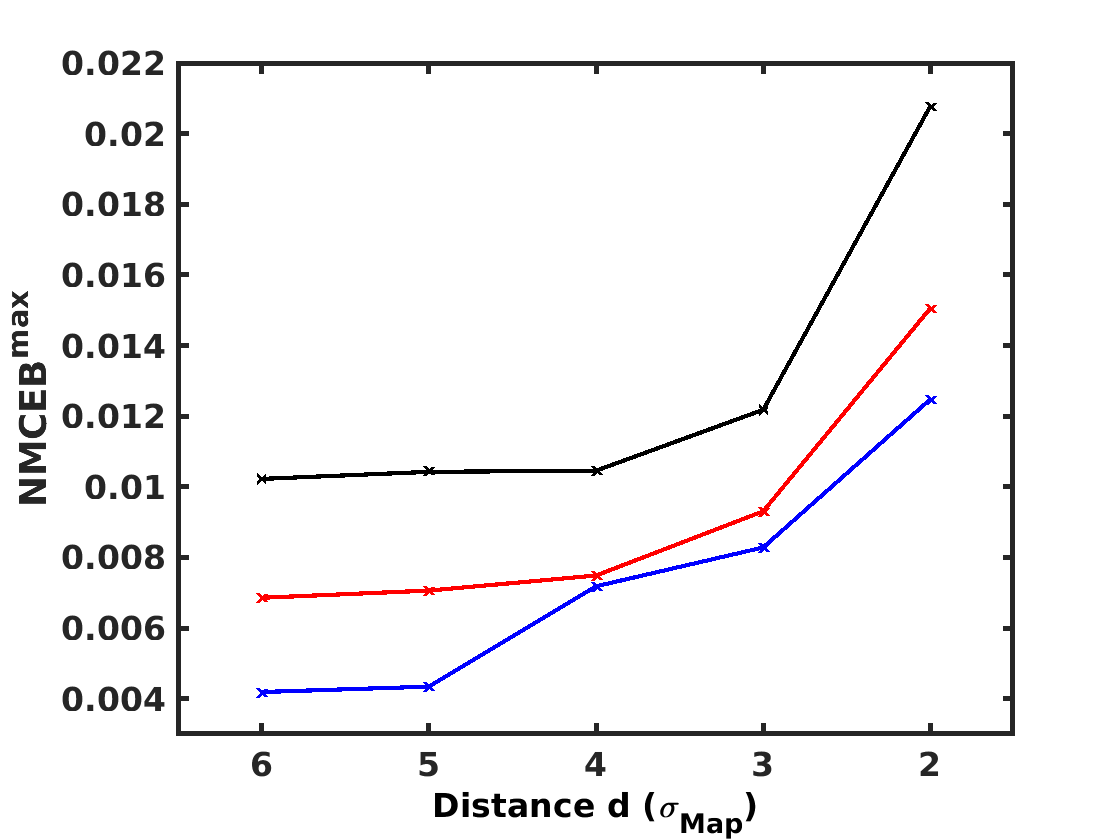}&
\includegraphics[width=0.3\textwidth]{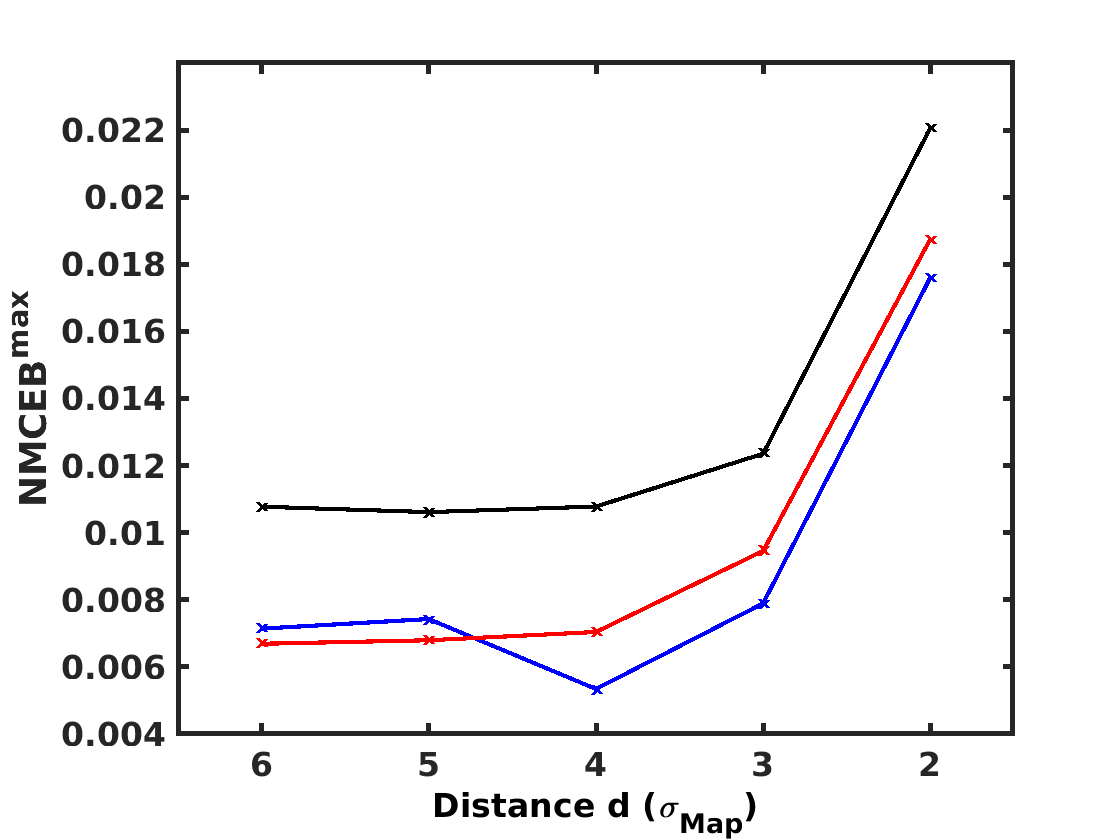}\\
2 sources & 4 sources & 6 sources
\end{tabular}
\end{center}
\caption{%
\ytextmodifhershelvonestepfive{Spread (NMCEB)}
of the solutions of 
\ytextmodifhershelvonestepfive{SC-NMF-Spec}
obtained on the 45 
synthetic cubes with 100 realizations of noise.} 
\label{fig_errbar_HybridSpec_chap3}
\end{figure}

\ytextmodifhershelvonestepfive{
Whatever the number of sources,
the 
\ytextmodifhershelvonestepfive{SC-NMF-Spec}
hybrid version 
(Fig.
\ref{fig_perf_HybridSpec_chap3})
yields results with the same trend as that 
obtained by MC-NMF used alone for
two sources (Fig. \ref{fig_perf_NMF_chap3}, leftmost column).
Moreover, when considering two sources for both methods, they yield 
similar 
estimation errors
for given sparsity and
noise level.}
%
%
%
\ytextmodifhershelvonestepfive{Besides, SC-NMF-Spec
results in a much lower 
\ytextmodifhershelvonestepfive{spread}
(Fig.
\ref{fig_errbar_HybridSpec_chap3})
than
MC-NMF
(Fig. 
\ref{fig_errbar_NMF_chap3}),
especially for four or six sources.}
This property is the main goal of 
hybridization. 
%
%
%


\subsection{\ytextmodifhershelvonestepfive{Results with SC-NMF-Map}}

\begin{figure}[H]
\begin{center}	
\begin{tabular}{ccc}
\includegraphics[width=0.3\textwidth]{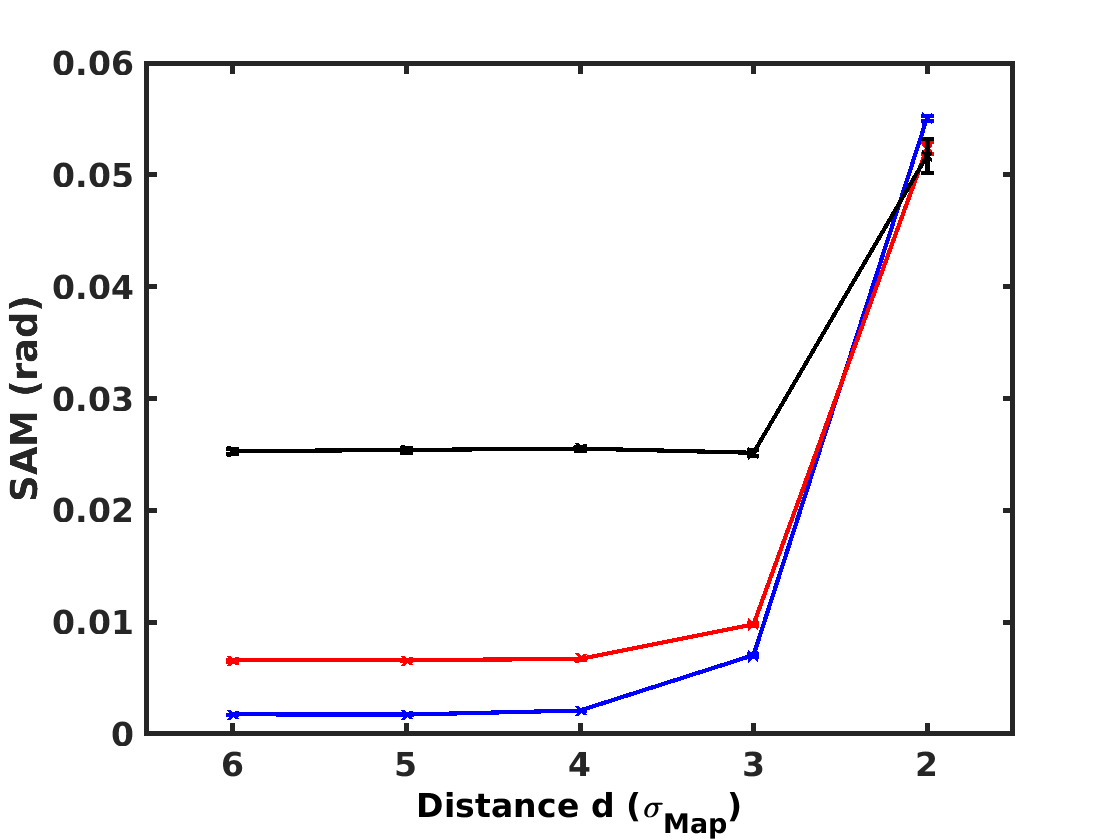}&
\includegraphics[width=0.3\textwidth]{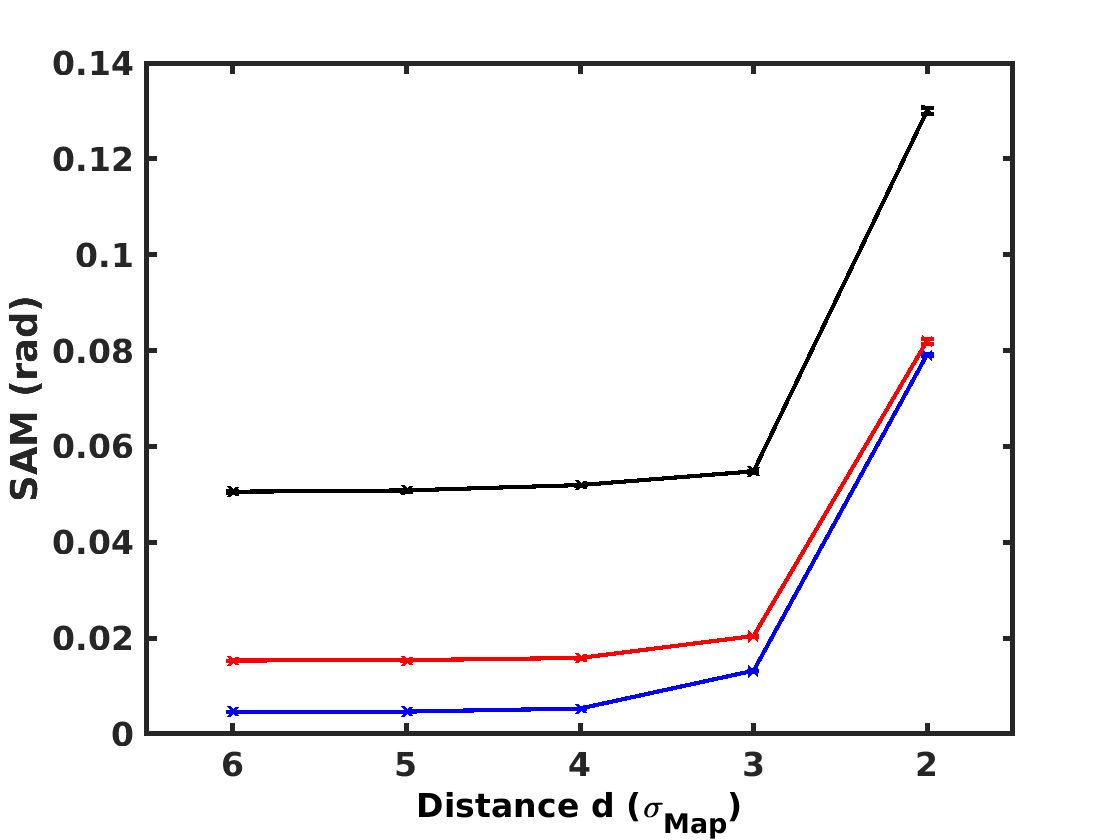}&
\includegraphics[width=0.3\textwidth]{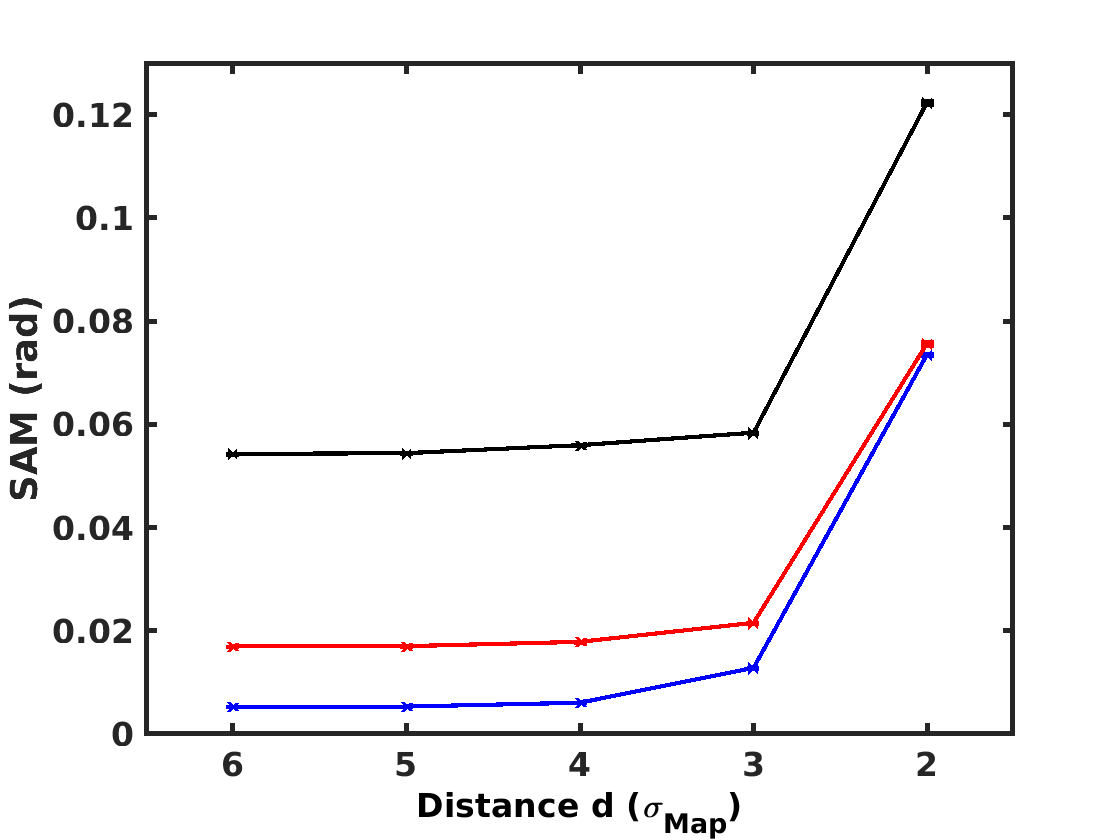}\\
\includegraphics[width=0.3\textwidth]{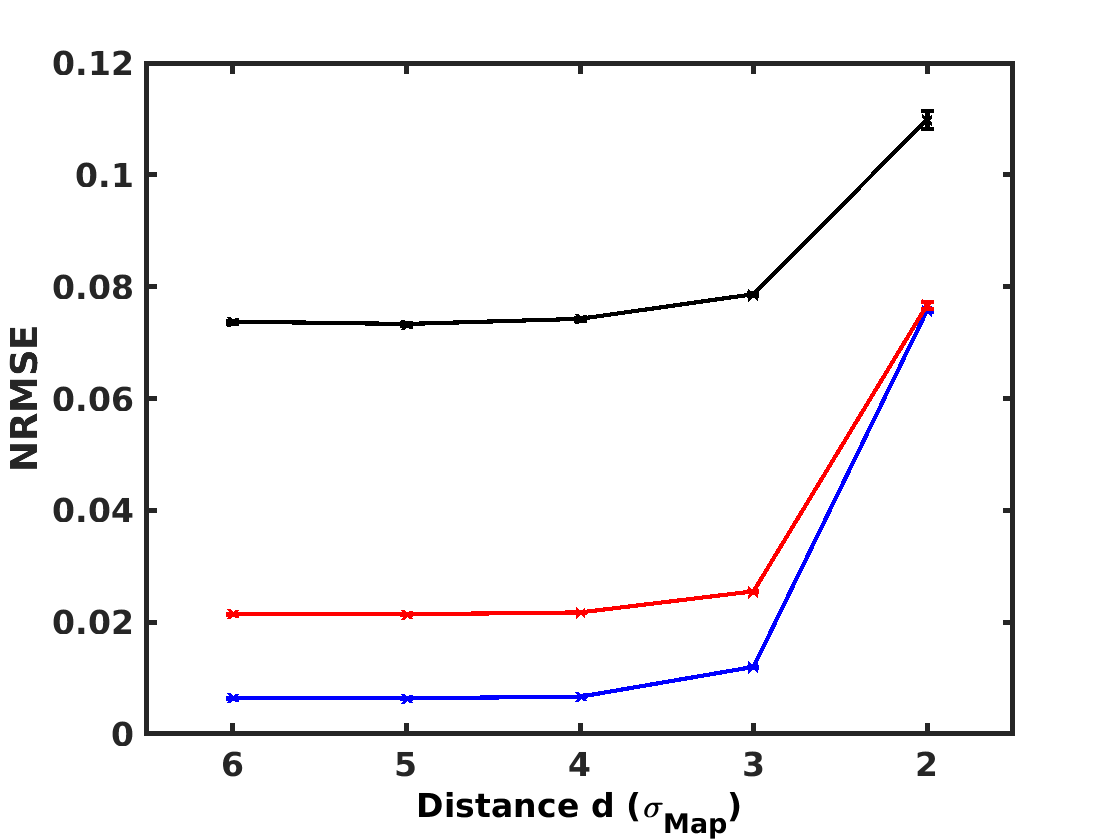}&
\includegraphics[width=0.3\textwidth]{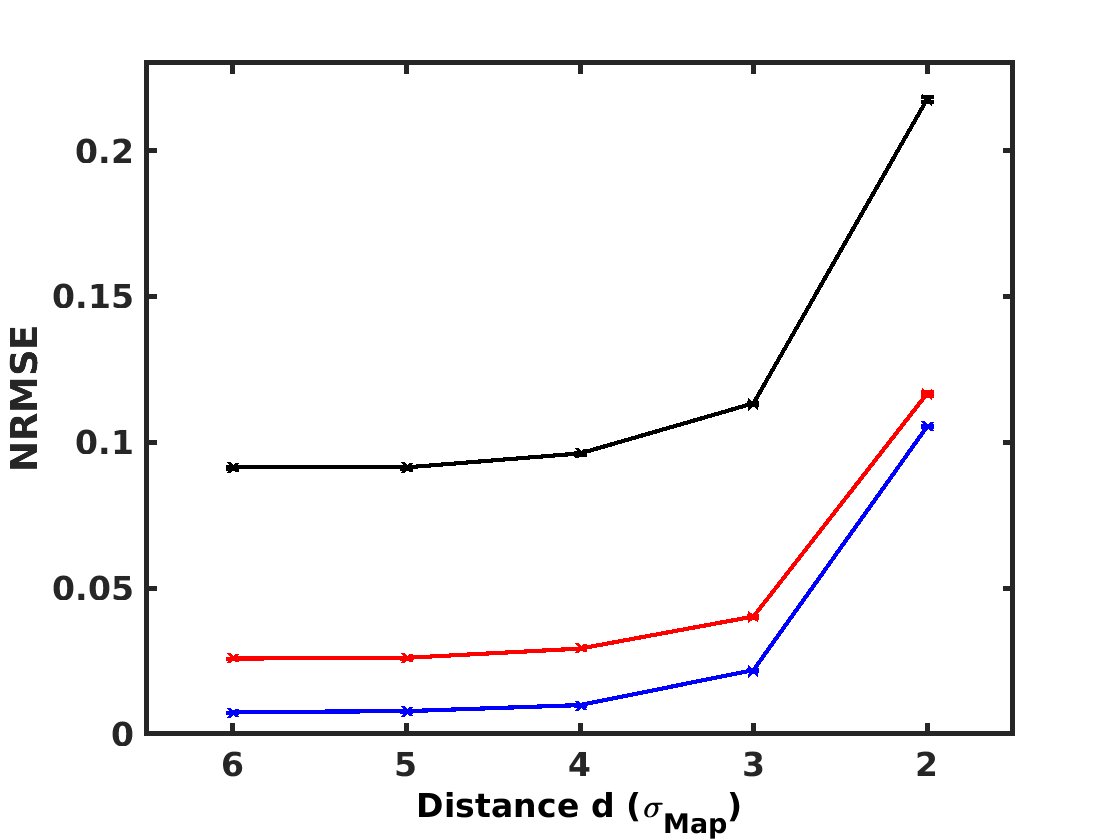}&
\includegraphics[width=0.3\textwidth]{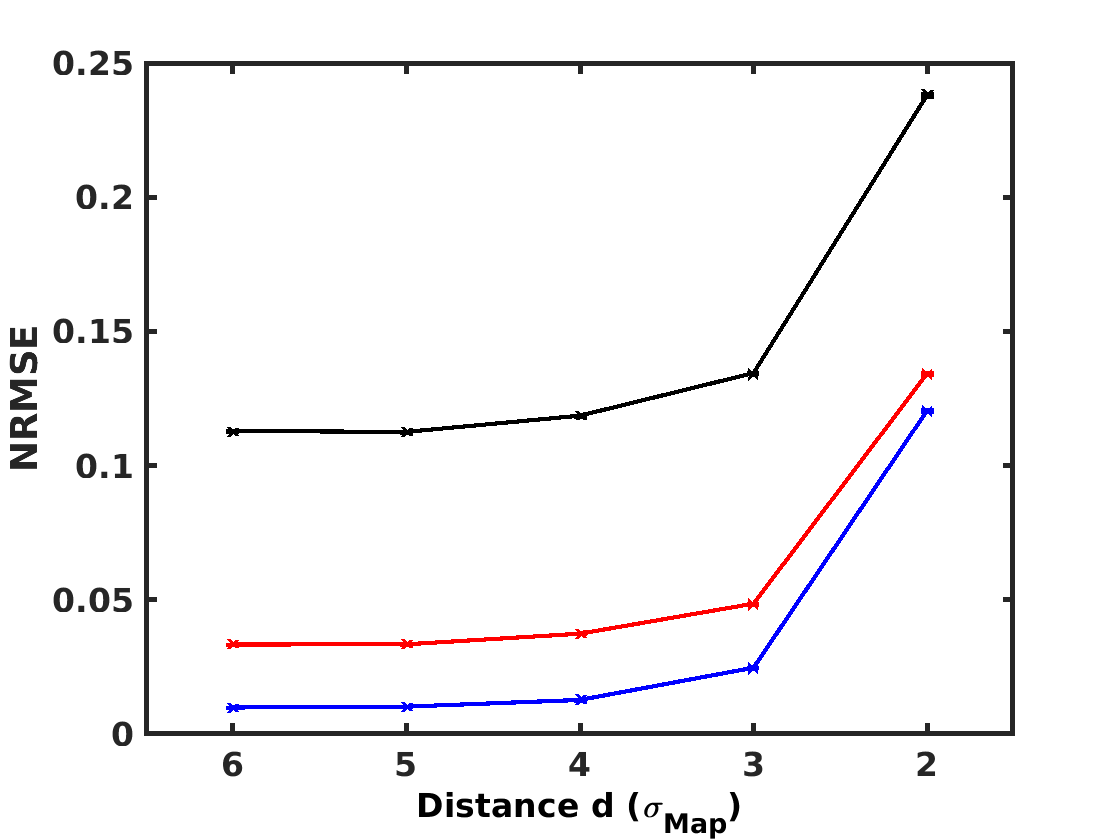}\\
2 sources & 4 sources & 6 sources
\end{tabular}
\end{center}
\caption{Performances achieved by 
\ytextmodifhershelvonestepfive{SC-NMF-Map}
on the 45 synthetic cubes for 100 
realizations of noise with 
\ytextmodifhershelvonestepfive{an}
SNR of 30 dB (in blue), 20 dB (in red), and 10 dB 
(in black). The error bars give the standard deviation 
\ytextmodifhershelvonestepfive{over}
the 100 
realizations of noise.} 
\label{fig_perf_HybridMap_chap3}
\end{figure}

\begin{figure}[h]
\begin{center}	
\begin{tabular}{ccc}
\includegraphics[width=0.3\textwidth]{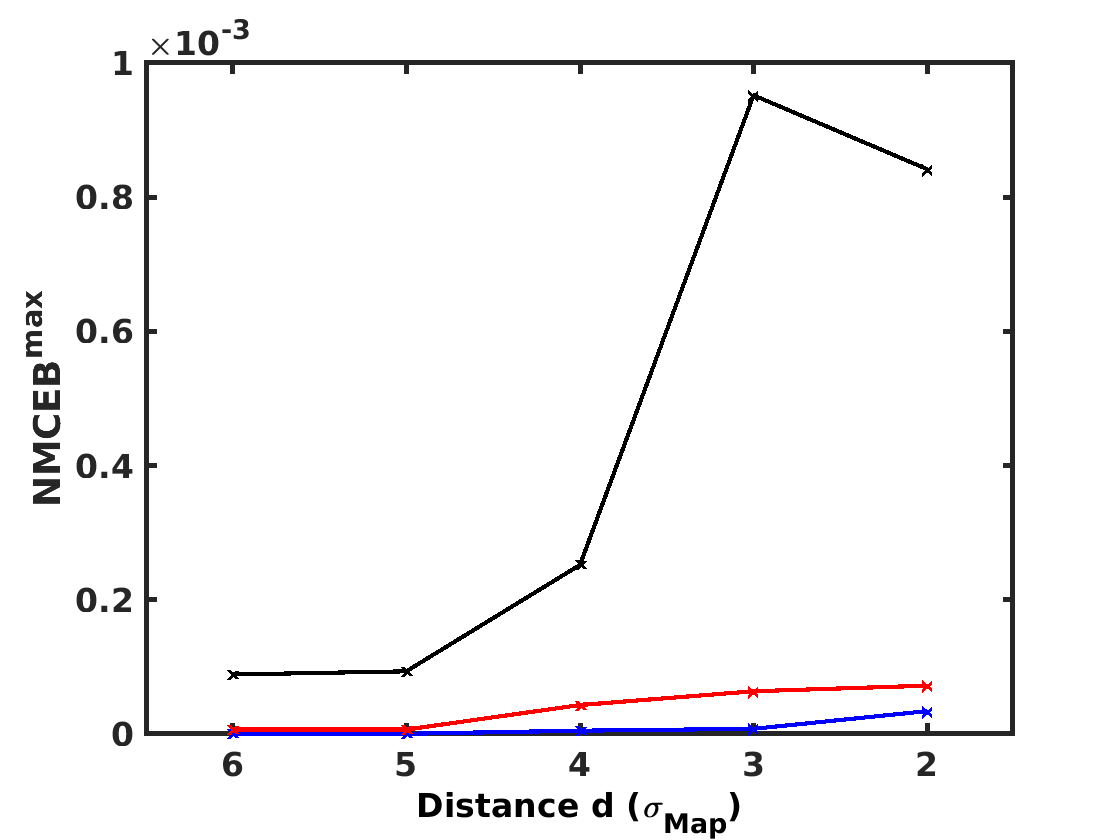}&
\includegraphics[width=0.3\textwidth]{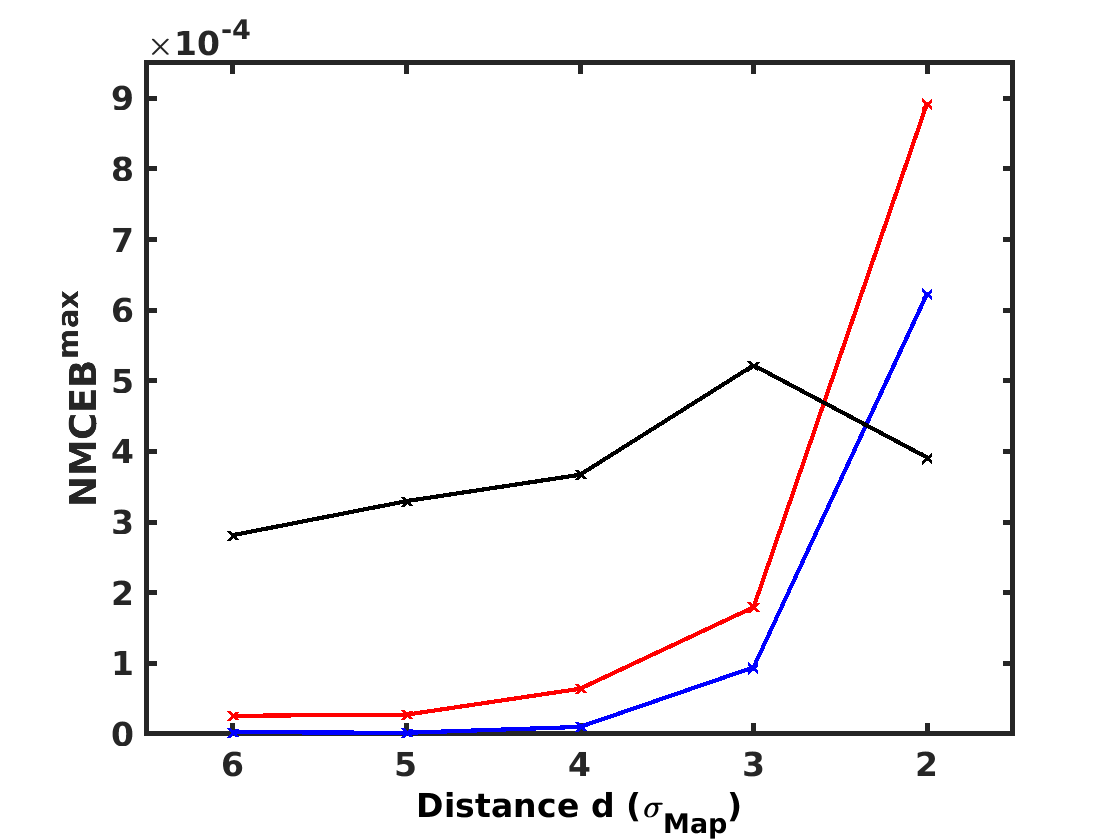}&
\includegraphics[width=0.3\textwidth]{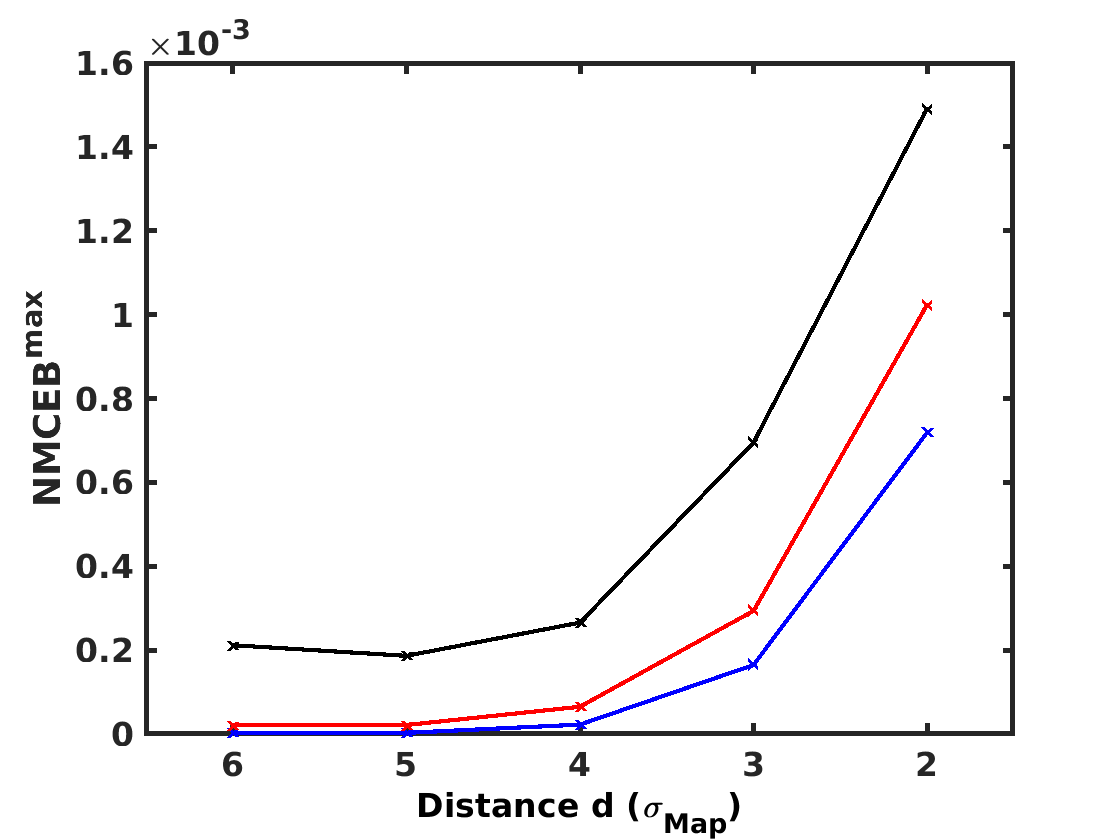}\\
2 sources & 4 sources & 6 sources
\end{tabular}
\end{center}
\caption{%
\ytextmodifhershelvonestepfive{Spread (NMCEB)}
of the solutions of 
\ytextmodifhershelvonestepfive{SC-NMF-Map}
obtained on the 45 synthetic 
cubes with 100 realizations of noise.} 
\label{fig_errbar_HybridMap_chap3}
\end{figure}

\ytextmodifhershelvonestepfive{The SC-NMF-Map
hybrid version 
(Fig.
\ref{fig_perf_HybridMap_chap3})
yields results with the same trend as that 
obtained by SpaceCorr used alone
(Fig. 
\ref{fig_perf_SpaceCORR_chap3}),
whatever the number of sources.
Moreover, 
\ytextmodifhershelvonestepfive{SC-NMF-Map}
results in significantly
lower estimation errors, especially
in difficult cases.}
The 
\ytextmodifhershelvonestepfive{spread}
of the results 
given by 
\ytextmodifhershelvonestepfive{SC-NMF-Map}
%
\ytextmodifhershelvonestepfive{(Fig.
\ref{fig_errbar_HybridMap_chap3})}
becomes negligible, with 
\ytextmodifhershelvonestepfive{an improvement}
of an order of magnitude 
over the amplitude of the error bars  
\ytextmodifhershelvonestepfive{of the SC-NMF-Spec}
hybrid version.

\subsection{\ytextmodifhershelvonestepfive{Results with MASS}}
%
\label{sec-appendix-Results-MASS}

MASS gives excellent results in the case of sufficiently sparse data and in the 
presence of a reasonable noise level (20 or 30 dB
\ytextmodifhershelvonestepfive{SNR}%
). It is further noted that for 
mixtures of two or four sources, the results are correct even under the least 
favorable conditions of sparsity, 
\ytextmodifhershelvonestepfive{again with an SNR}
of 20 or 30 dB.

The main disadvantage highlighted by the synthetic data tests is the sensitivity 
of the method to the high level of noise. Indeed, the performances for the 
mixtures 
\ytextmodifhershelvonestepfive{with a 10 dB SNR}
are weak, even under optimal conditions of 
sparsity. This sensitivity comes from the structure of the method that performs 
the estimation of the mixing matrix by directly selecting the columns from the 
observed vectors. The introduction of a tolerance angle in this selection has an 
effect 
\ytextmodifhershelvonestepfive{at}
reasonable noise levels but becomes less effective at higher noise 
levels. In addition, reducing the tolerance threshold would allow greater 
robustness to noise, to the detriment of component separation. Non single-source 
observations would then be used in the estimation of the columns of the mixing 
matrix. The sensitivity of MASS to the noise level can be mitigated by 
hybridization with MC-NMF.

\begin{figure}[H]
\begin{center}	
\begin{tabular}{ccc}
\includegraphics[width=0.3\textwidth]{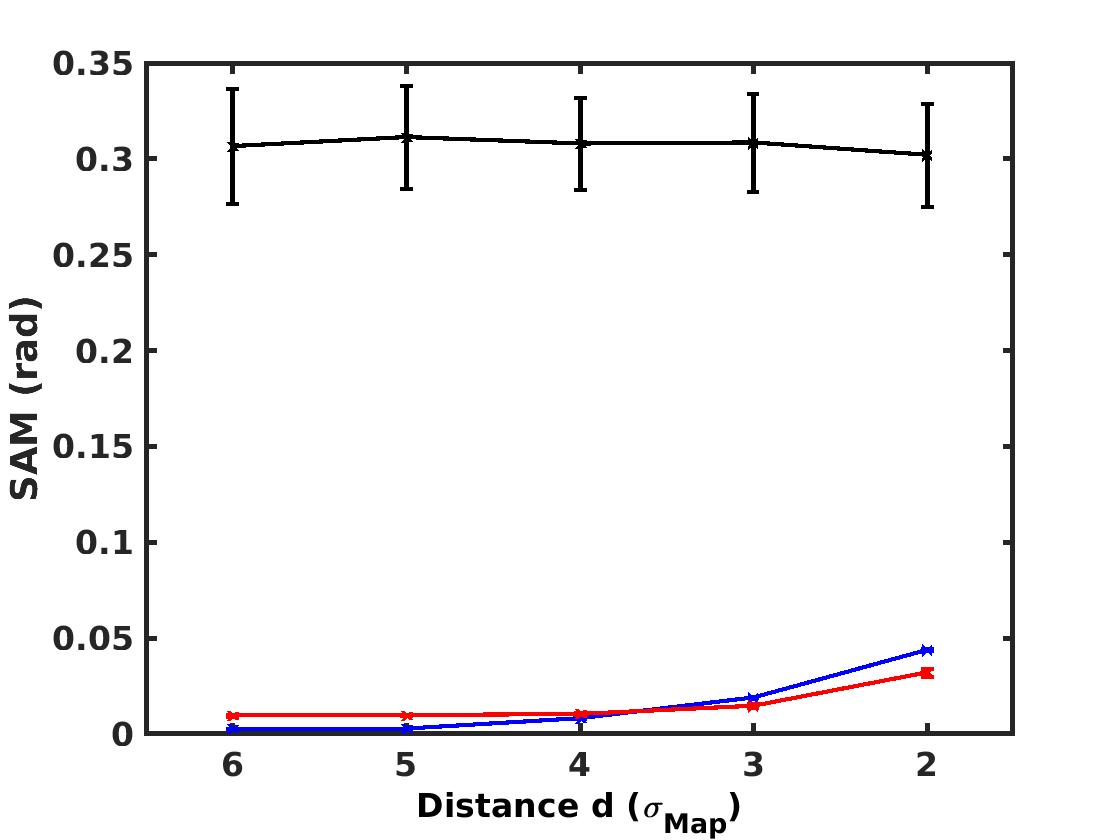}&
\includegraphics[width=0.3\textwidth]{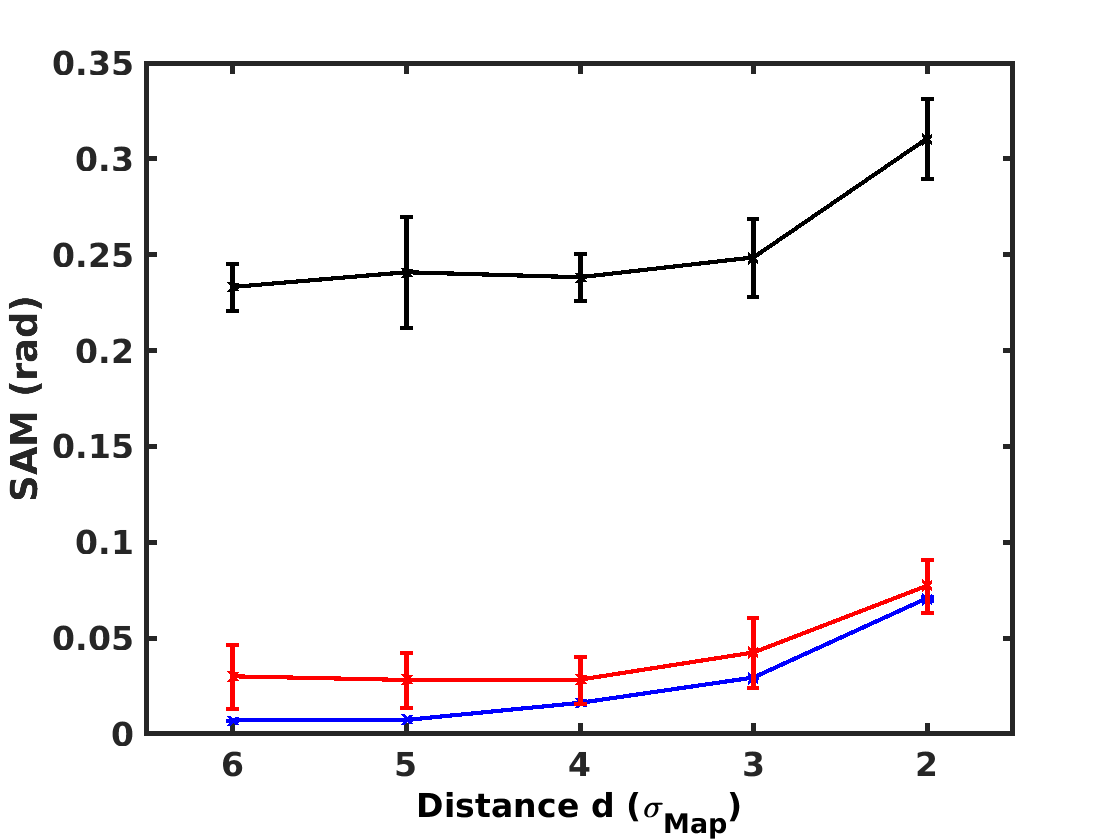}&
\includegraphics[width=0.3\textwidth]{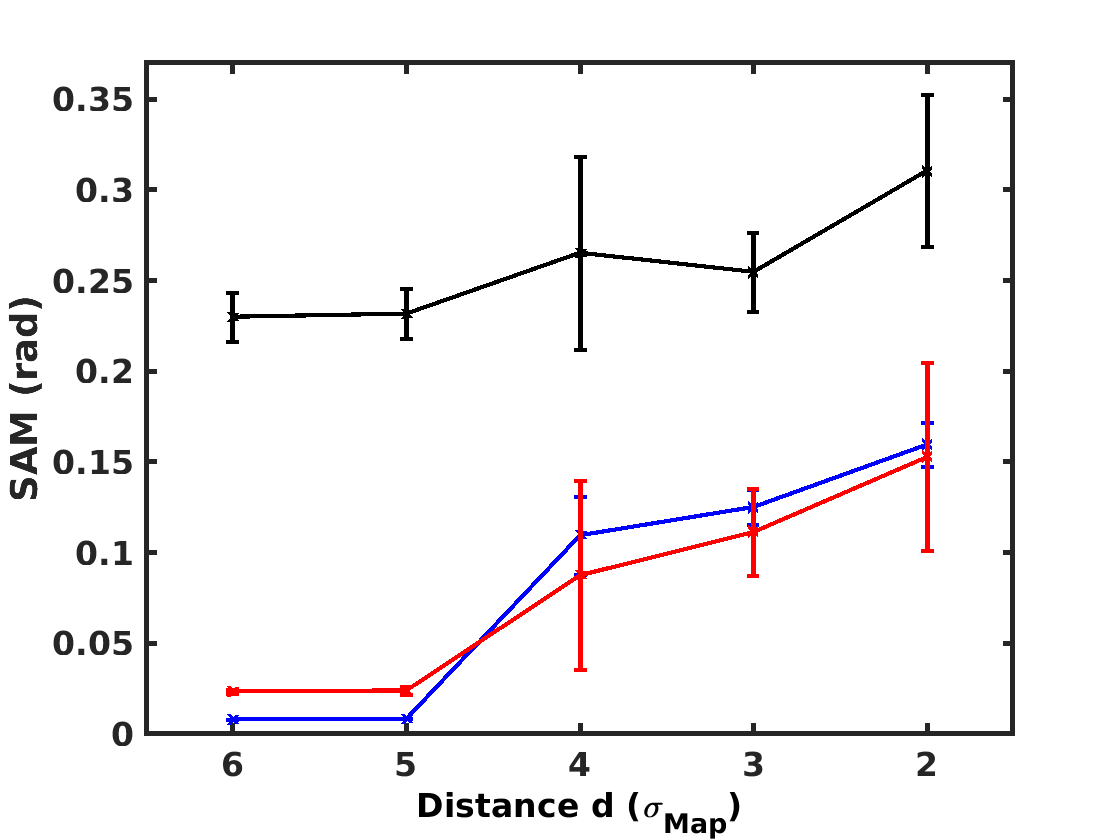}\\
\includegraphics[width=0.3\textwidth]{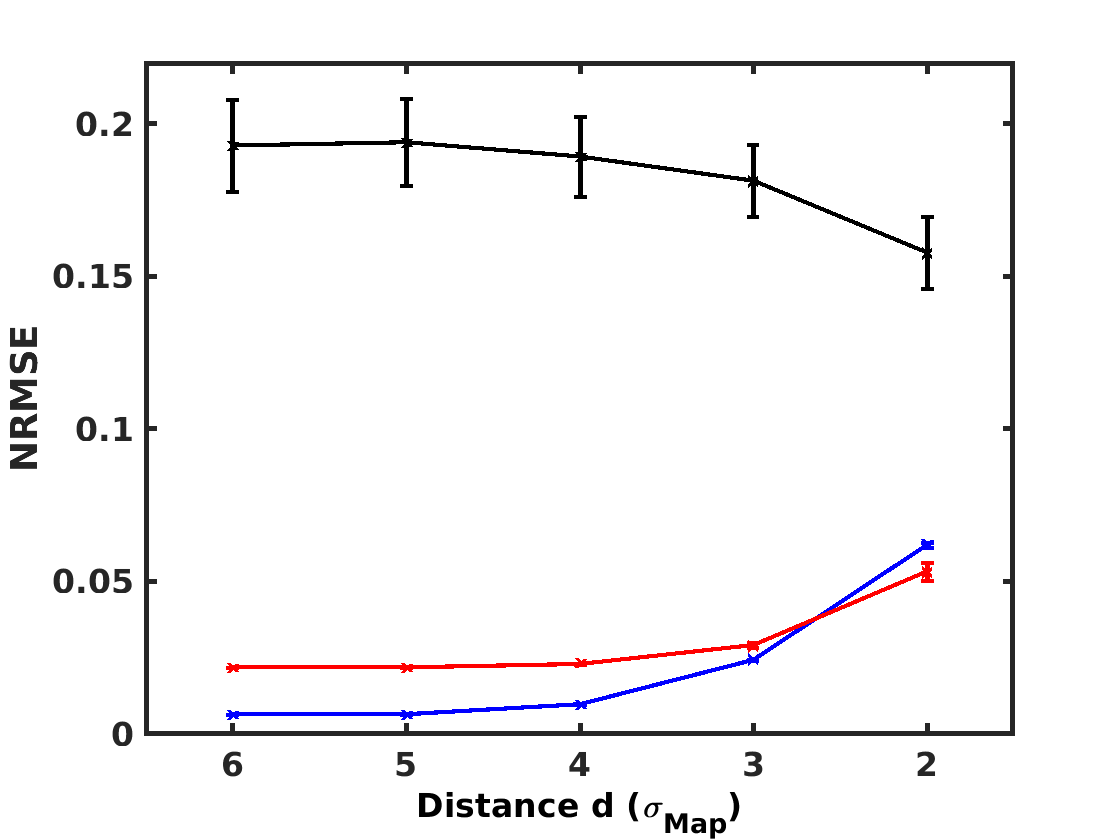}&
\includegraphics[width=0.3\textwidth]{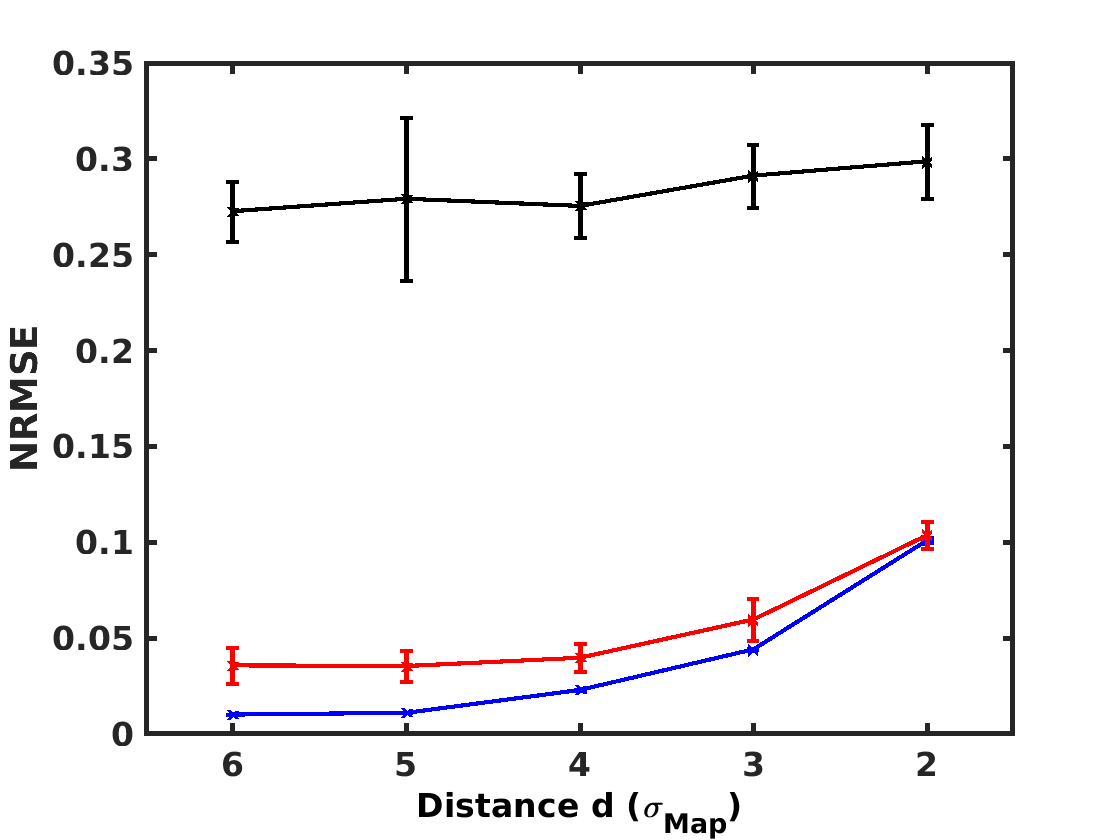}&
\includegraphics[width=0.3\textwidth]{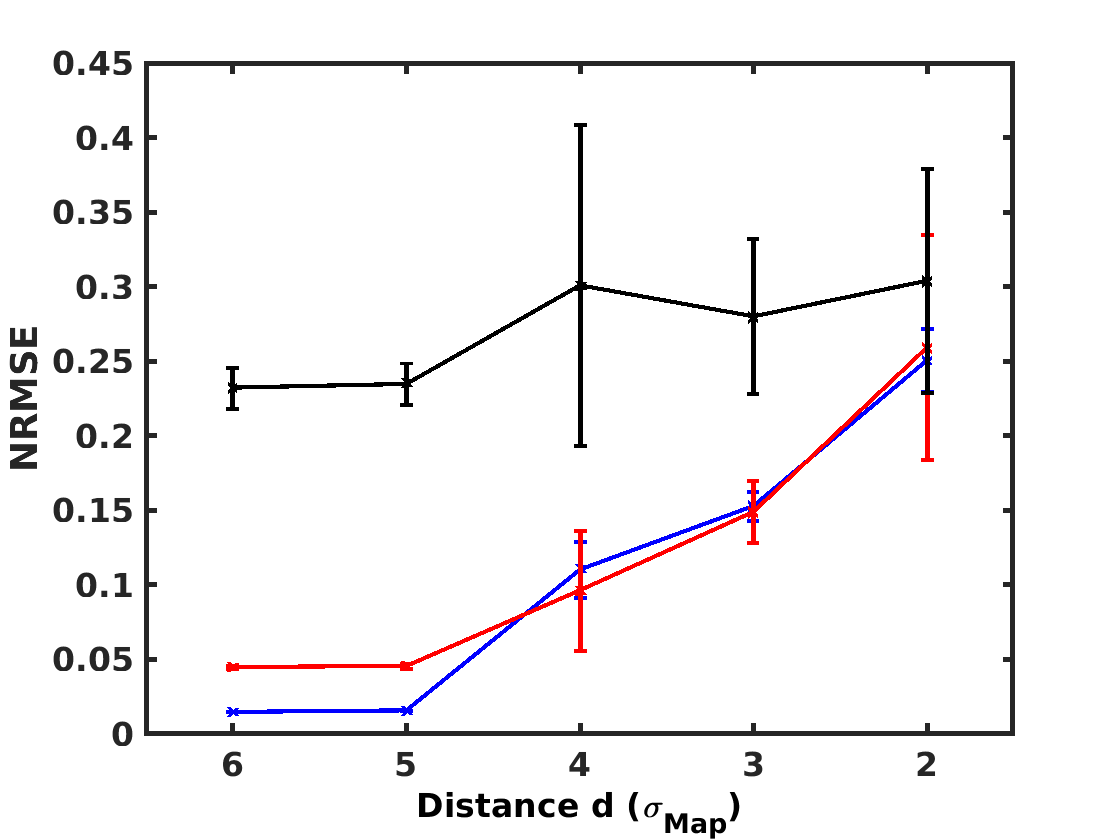}\\
2 sources & 4 sources & 6 sources
\end{tabular}
\end{center}
\caption{Performances achieved by MASS on the 45 synthetic cubes for 100 
realizations of noise with 
\ytextmodifhershelvonestepfive{an}
SNR of 30 dB (in blue), 20 dB (in red), and 10 dB 
(in black). The error bars give the standard deviation 
\ytextmodifhershelvonestepfive{over}
the 100 
realizations of noise.} 
\label{fig_perf_MASS_chap4}
\end{figure}

\subsection{\ytextmodifhershelvonestepfive{Results with MASS-NMF-Spec}}

\begin{figure}[H]
\begin{center}	
\begin{tabular}{ccc}
\includegraphics[width=0.3\textwidth]{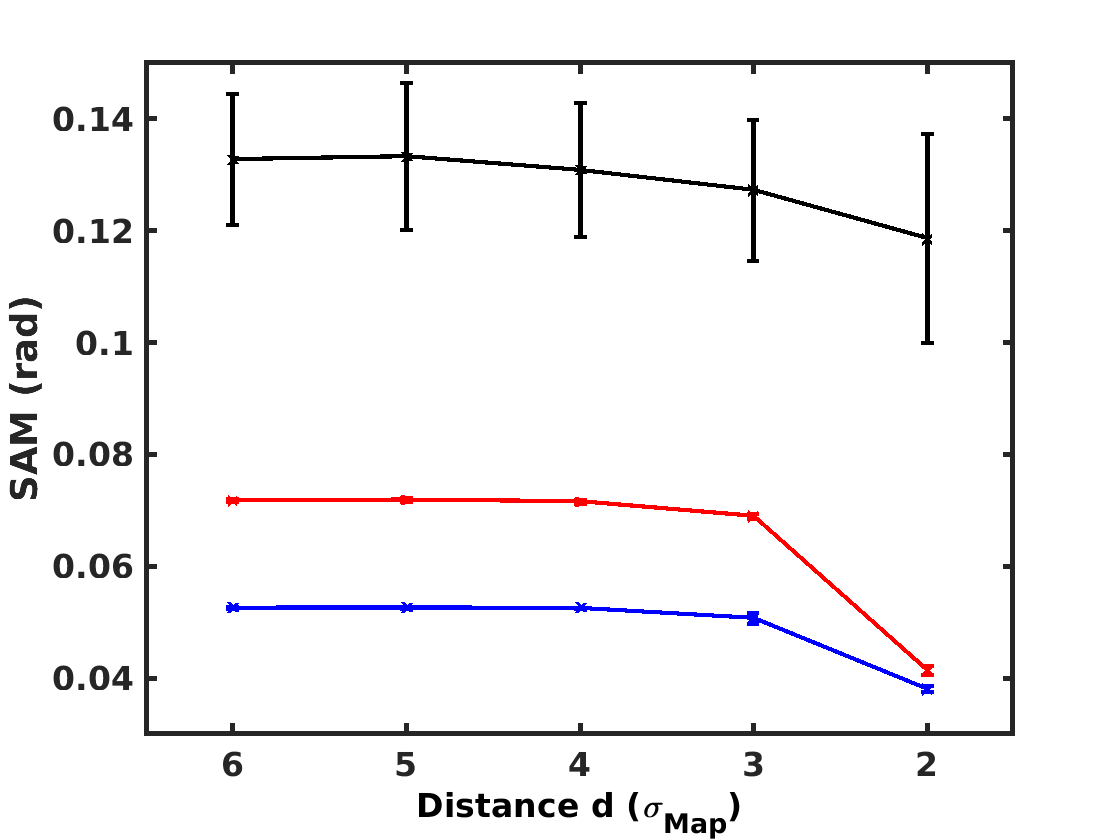}&
\includegraphics[width=0.3\textwidth]{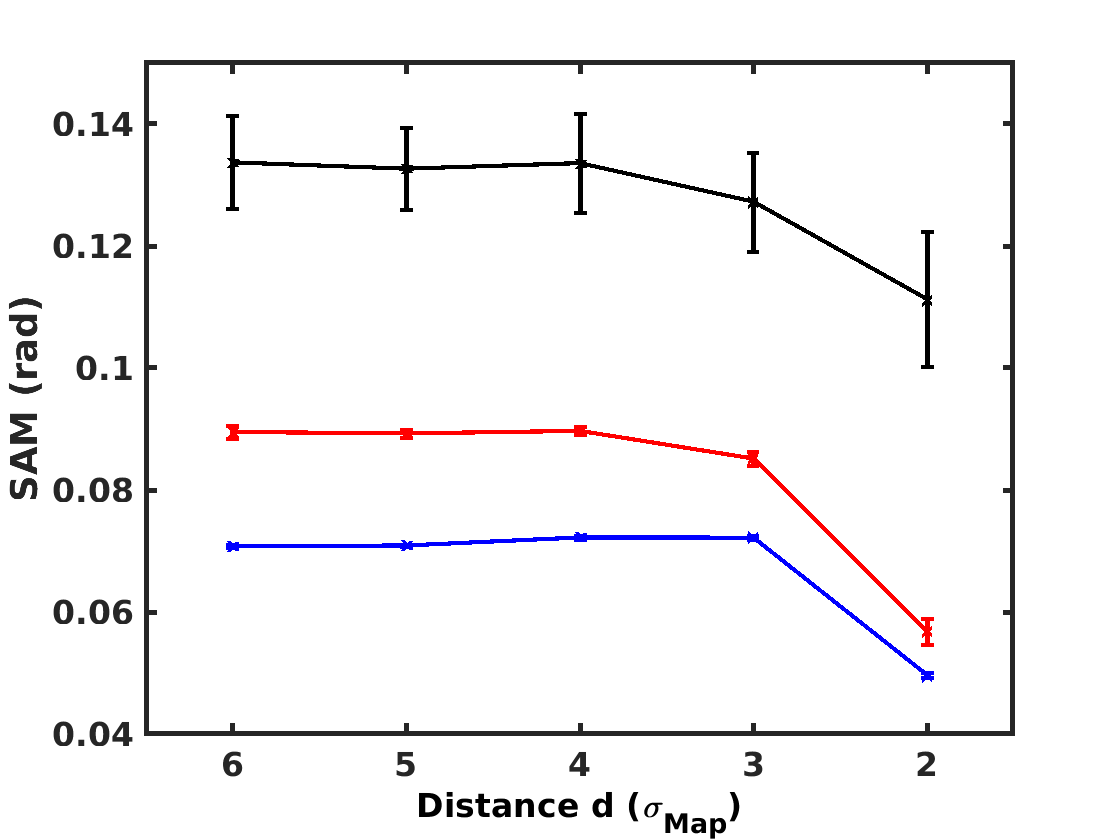}&
\includegraphics[width=0.3\textwidth]{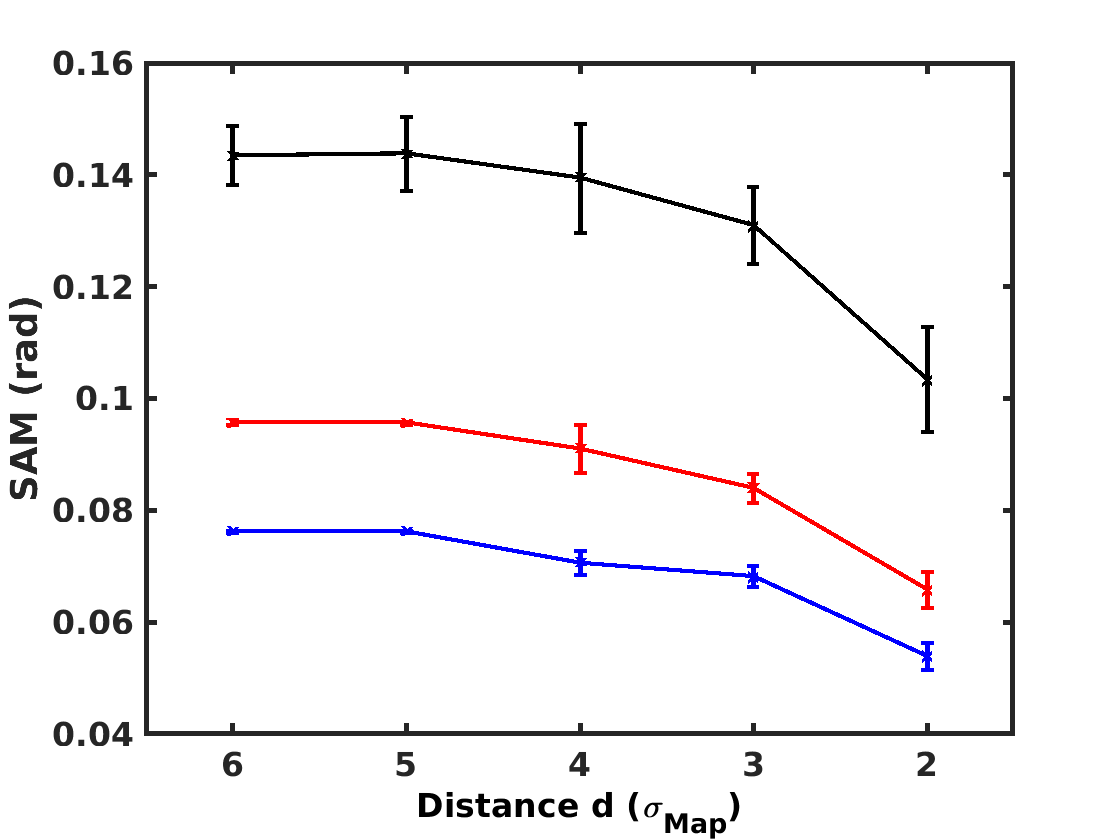}\\
\includegraphics[width=0.3\textwidth]{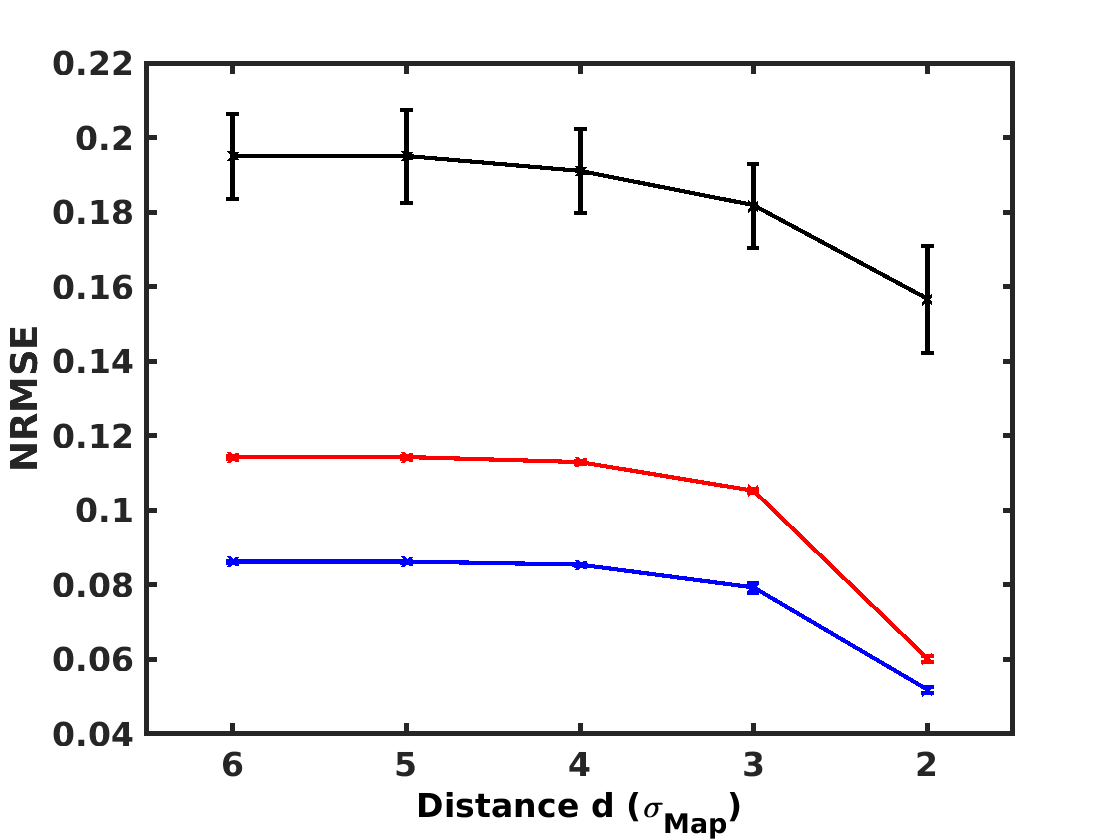}&
\includegraphics[width=0.3\textwidth]{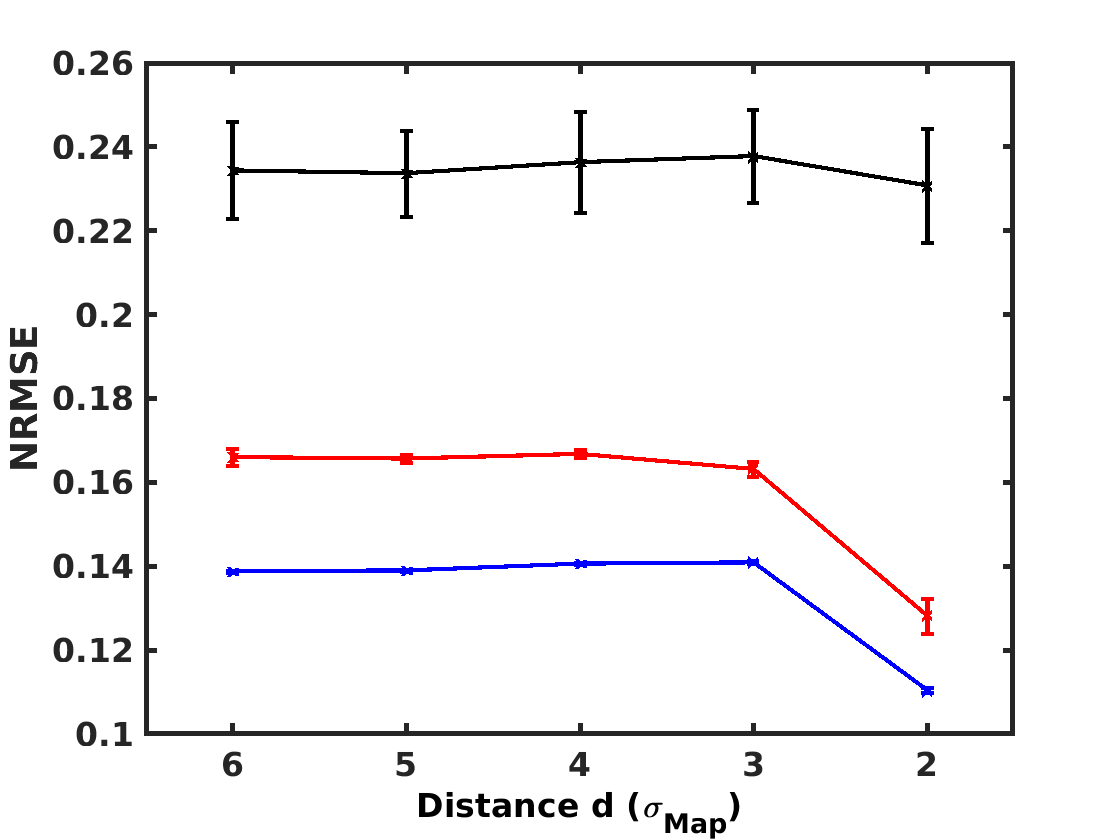}&
\includegraphics[width=0.3\textwidth]{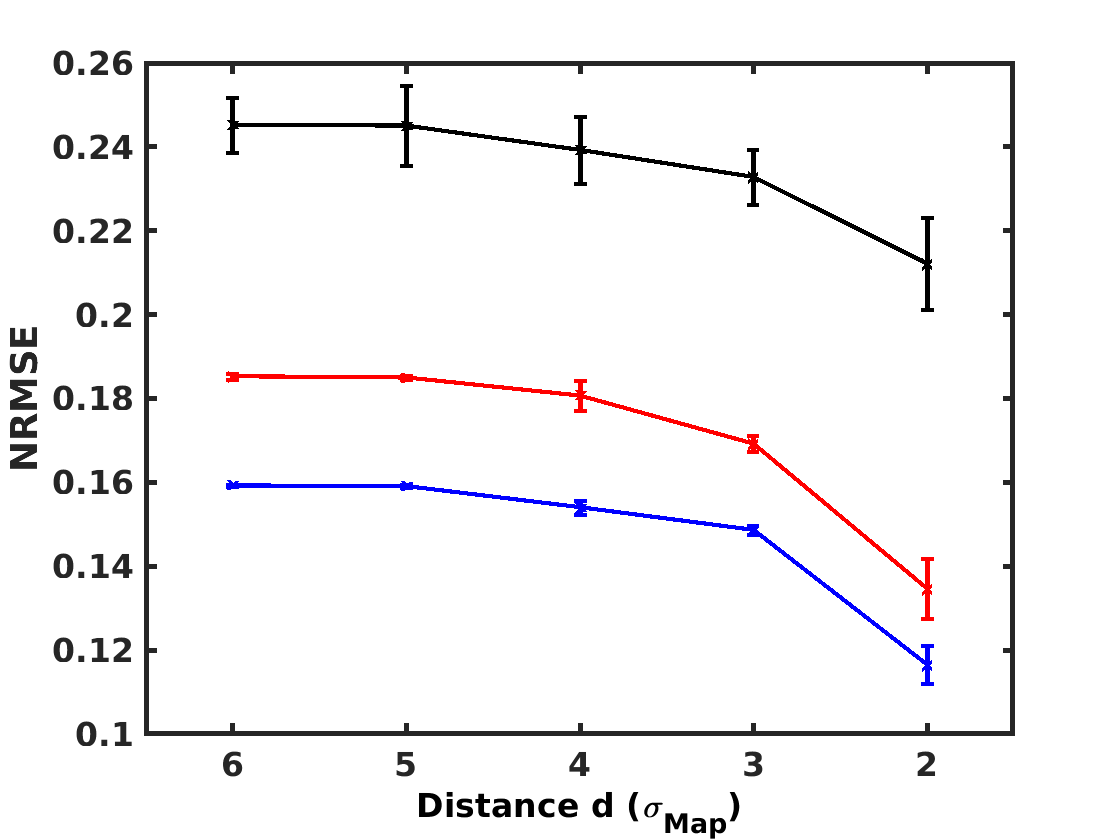}\\
2 sources & 4 sources & 6 sources
\end{tabular}
\end{center}
\caption{Performances achieved by MASS-NMF-Spec on the 45 synthetic cubes for 
100 realizations of noise with 
\ytextmodifhershelvonestepfive{an}
SNR of 30 dB (in blue), 20 dB (in red), and 10 
dB (in black). The error bars give the standard deviation 
\ytextmodifhershelvonestepfive{over}
the 100 
realizations of noise.} 
\label{fig_perf_HybridSpec_chap4}
\end{figure}

\begin{figure}[h]
\begin{center}	
\begin{tabular}{ccc}
\includegraphics[width=0.3\textwidth]{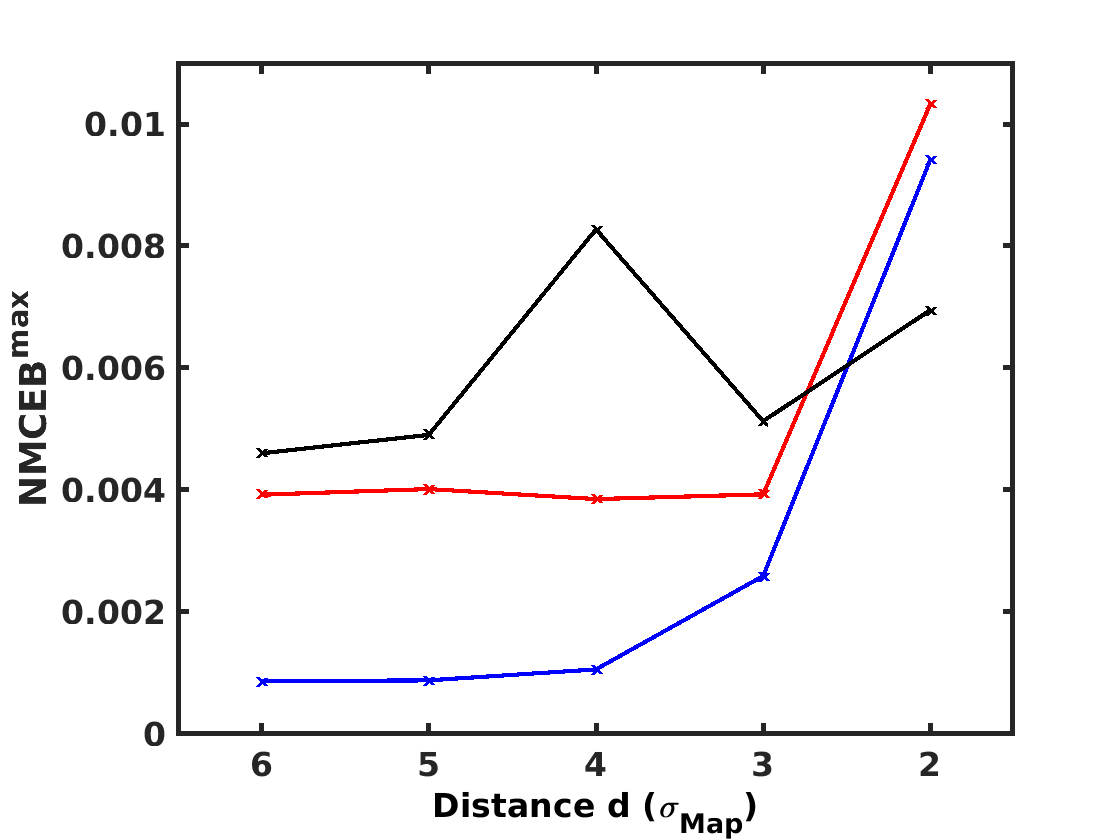}&
\includegraphics[width=0.3\textwidth]{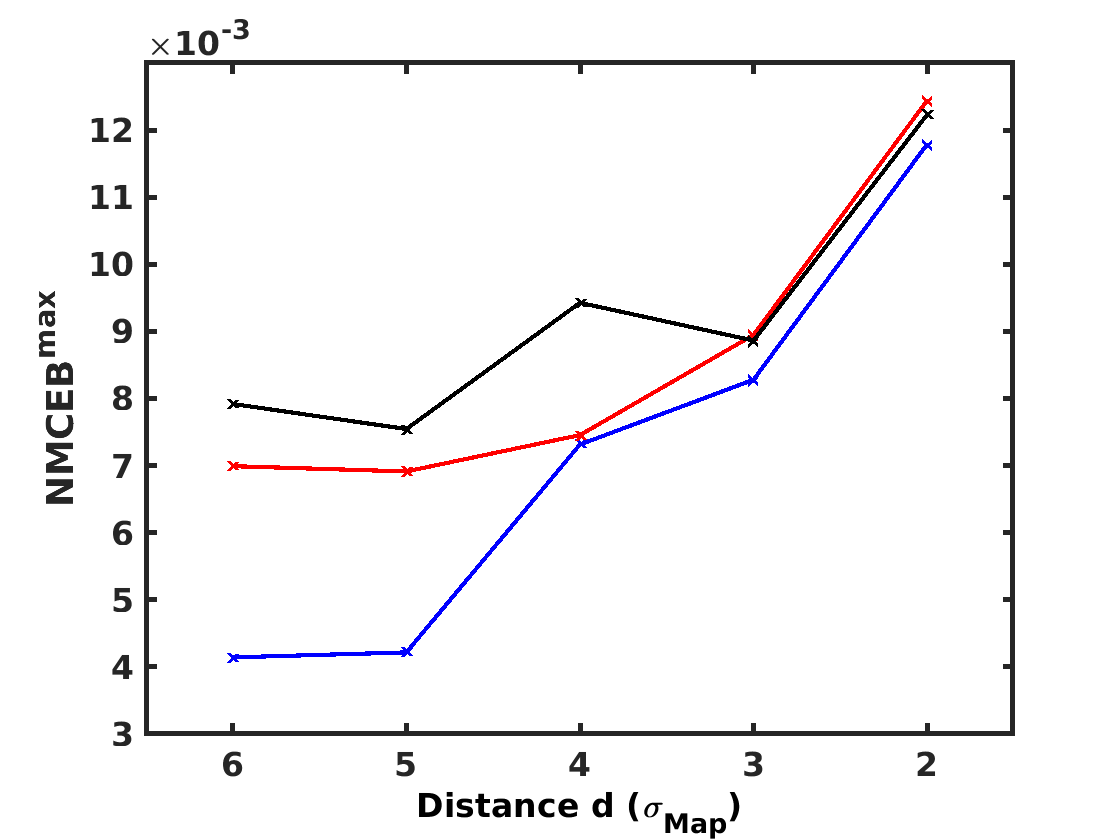}&
\includegraphics[width=0.3\textwidth]{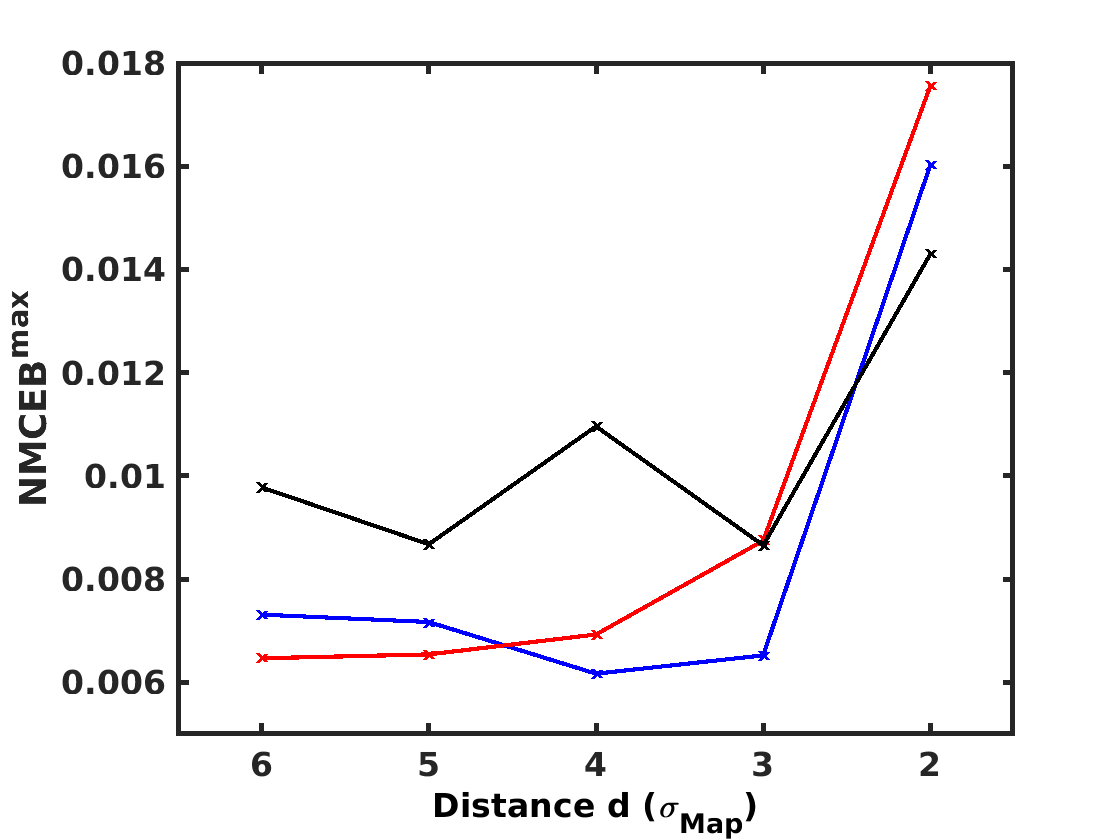}\\
2 sources & 4 sources & 6 sources
\end{tabular}
\end{center}
\caption{%
\ytextmodifhershelvonestepfive{%
\ytextmodifhershelvonestepfive{Spread (NMCEB)}
of the solutions of MASS-NMF-Spec}
obtained on the 45 
synthetic cubes with 100 realizations of noise.} 
\label{fig_errbar_HybridSpec_chap4}
\end{figure}

\ytextmodifhershelvonestepfive{The MASS-NMF-Spec 
hybrid version 
(Fig.
\ref{fig_perf_HybridSpec_chap4})
yields the same trends as those}
obtained by 
\ytextmodifhershelvonestepfive{SC-NMF-Spec}
%
(Fig. \ref{fig_perf_HybridSpec_chap3}). 
In comparison with the MASS method alone, the main advantage is an overall 
improvement of the performance criteria for mixtures with 
\ytextmodifhershelvonestepfive{10 dB SNR.}
The 
MASS method used alone in this configuration gave unsatisfactory results. They 
are markedly improved during hybridization with MC-NMF. Finally, we note the 
\ytextmodifhershelvonestepfive{major}
decrease in the 
\ytextmodifhershelvonestepfive{spread}
of the solutions given by MASS-NMF-Spec 
\ytextmodifhershelvonestepfive{(Fig.
\ref{fig_errbar_HybridSpec_chap4})}
compared to that encountered with the MC-NMF (Fig.
\ref{fig_errbar_NMF_chap3})%
\ytextmodifhershelvonestepfive{, especially for four or six sources.}
The error bars encountered here are of the same 
order of magnitude as those obtained with 
\ytextmodifhershelvonestepfive{SC-NMF-Spec}
(Fig.
\ref{fig_errbar_HybridSpec_chap3}).

\subsection{\ytextmodifhershelvonestepfive{Results with MASS-NMF-Map }}
%
\label{sec-appendix-Results-MASS-NMF-Map}

\begin{figure}[H]
\begin{center}	
\begin{tabular}{ccc}
\includegraphics[width=0.3\textwidth]{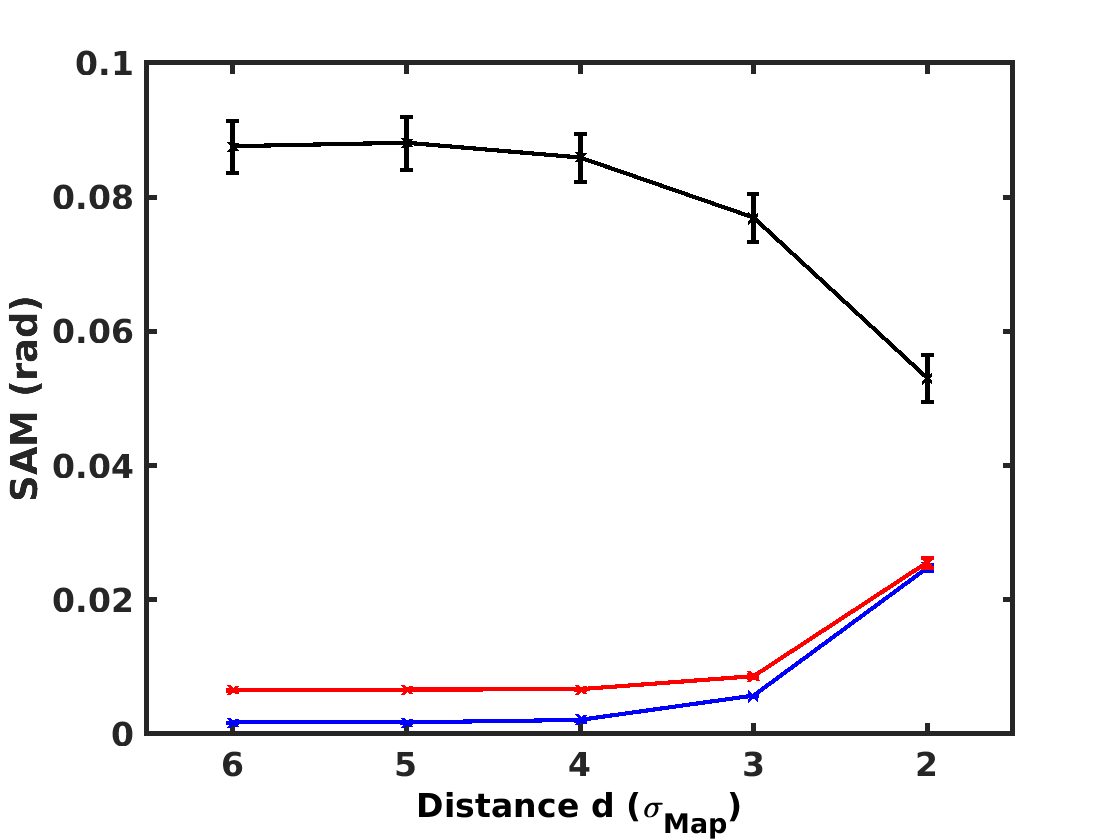}&
\includegraphics[width=0.3\textwidth]{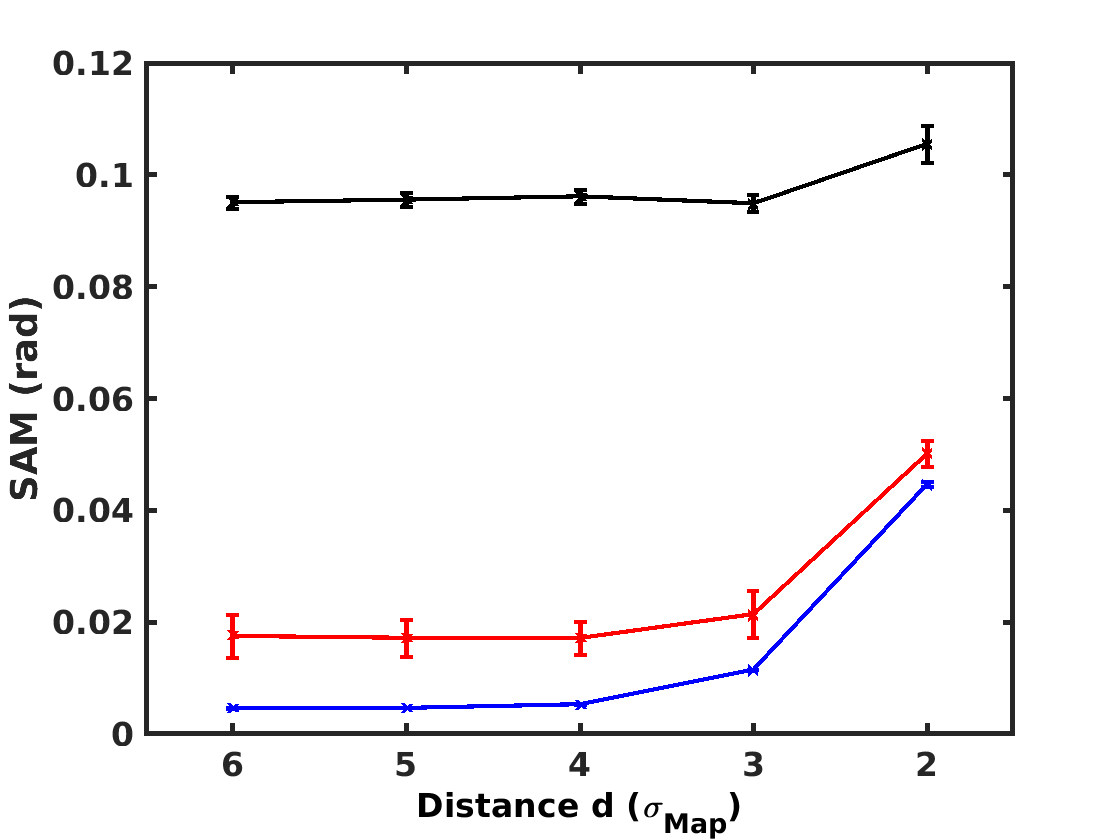}&
\includegraphics[width=0.3\textwidth]{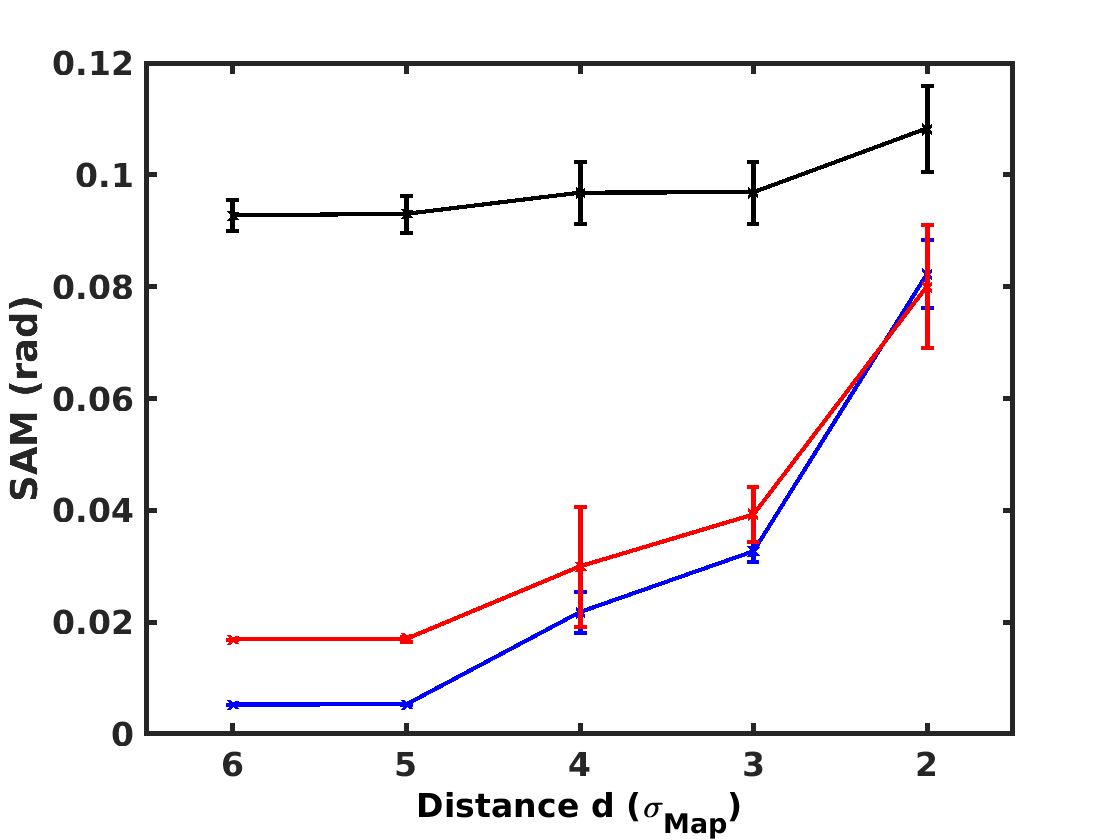}\\
\includegraphics[width=0.3\textwidth]{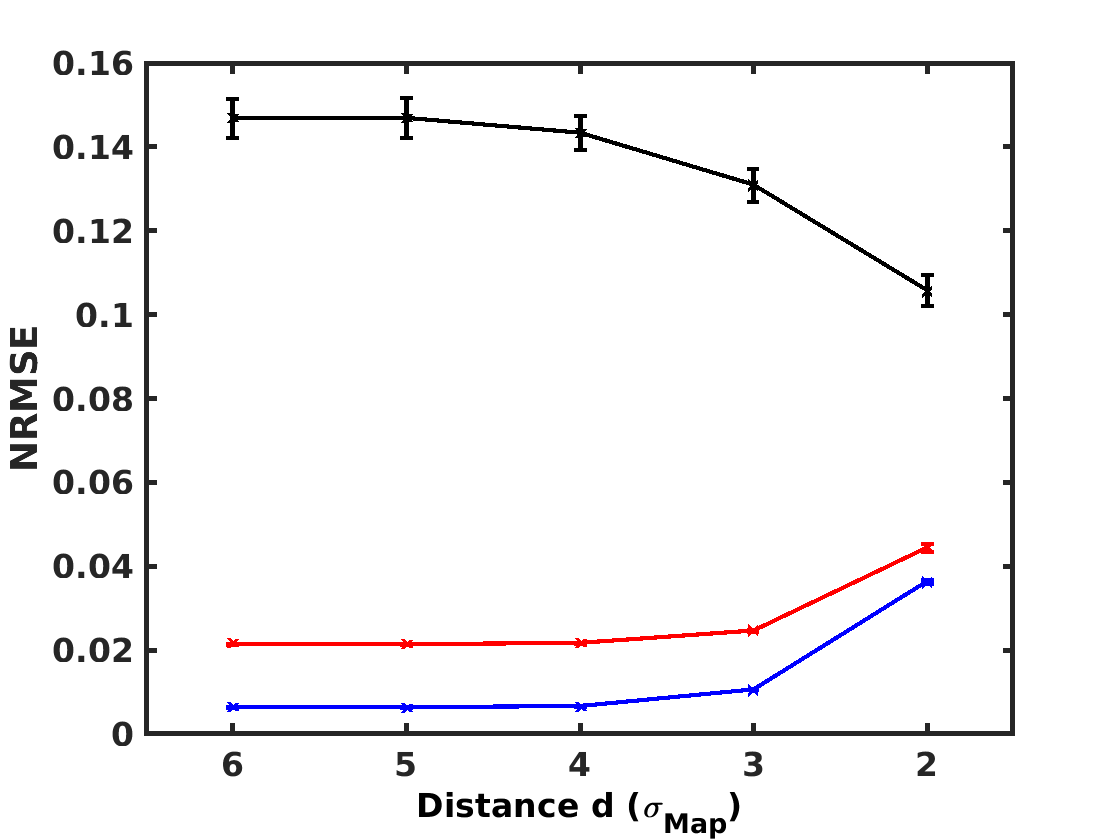}&
\includegraphics[width=0.3\textwidth]{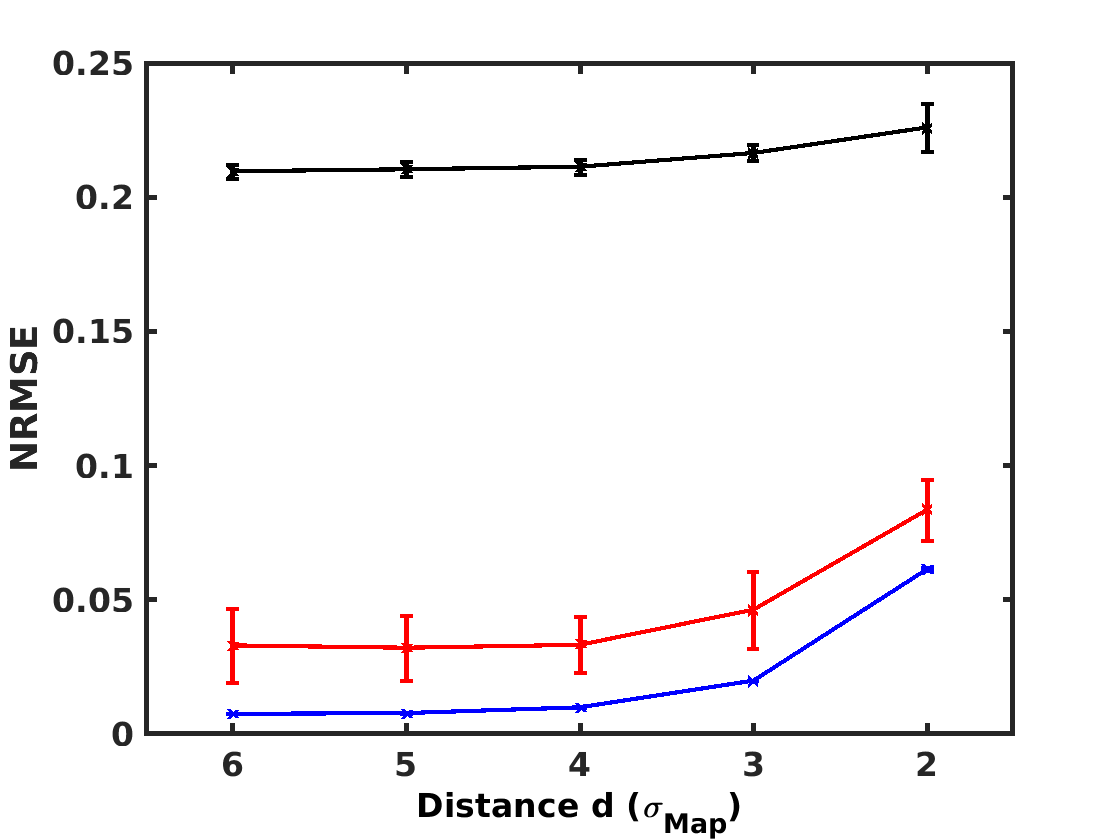}&
\includegraphics[width=0.3\textwidth]{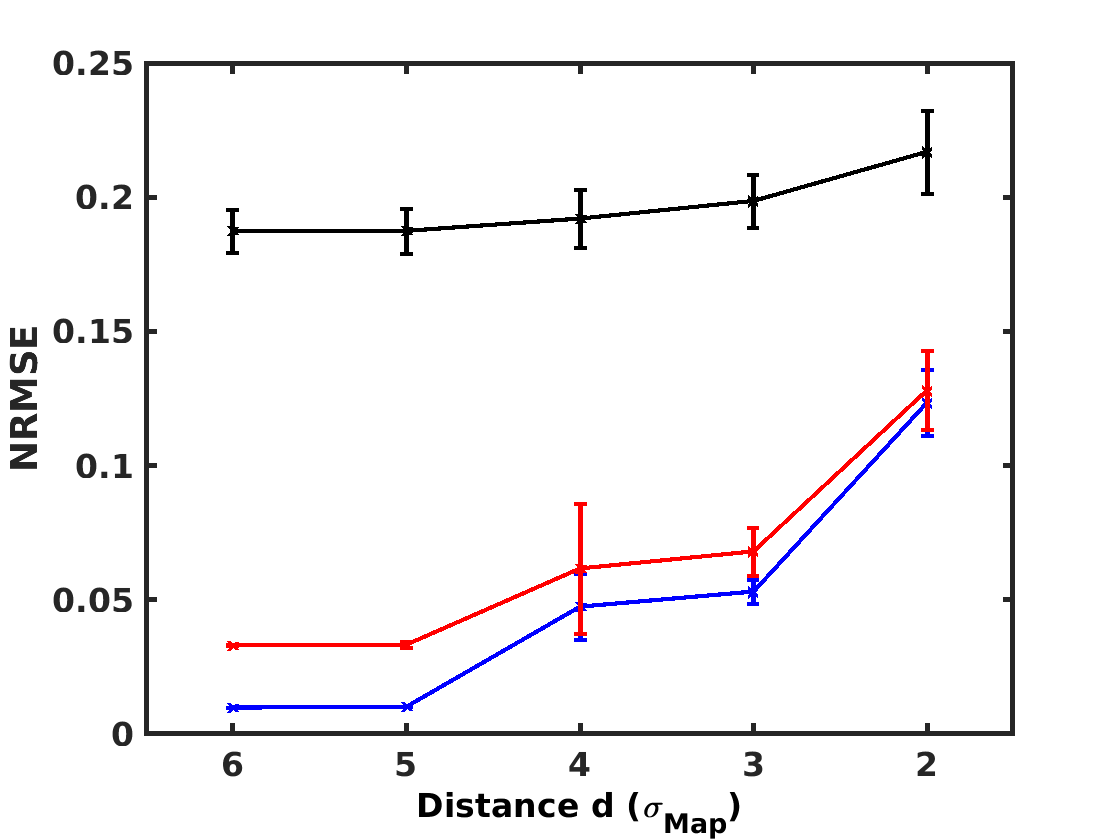}\\
2 sources & 4 sources & 6 sources
\end{tabular}
\end{center}
\caption{Performances achieved by MASS-NMF-Map on the 45 synthetic cubes for 100 
realizations of noise with 
\ytextmodifhershelvonestepfive{an}
SNR of 30 dB (in blue), 20 dB (in red), and 10 dB 
(in black). The error bars give the standard deviation 
\ytextmodifhershelvonestepfive{over}
the 100 
realizations of noise.} 
\label{fig_perf_HybridMap_chap4}
\end{figure}

\begin{figure}[h]
\begin{center}	
\begin{tabular}{ccc}
\includegraphics[width=0.3\textwidth]{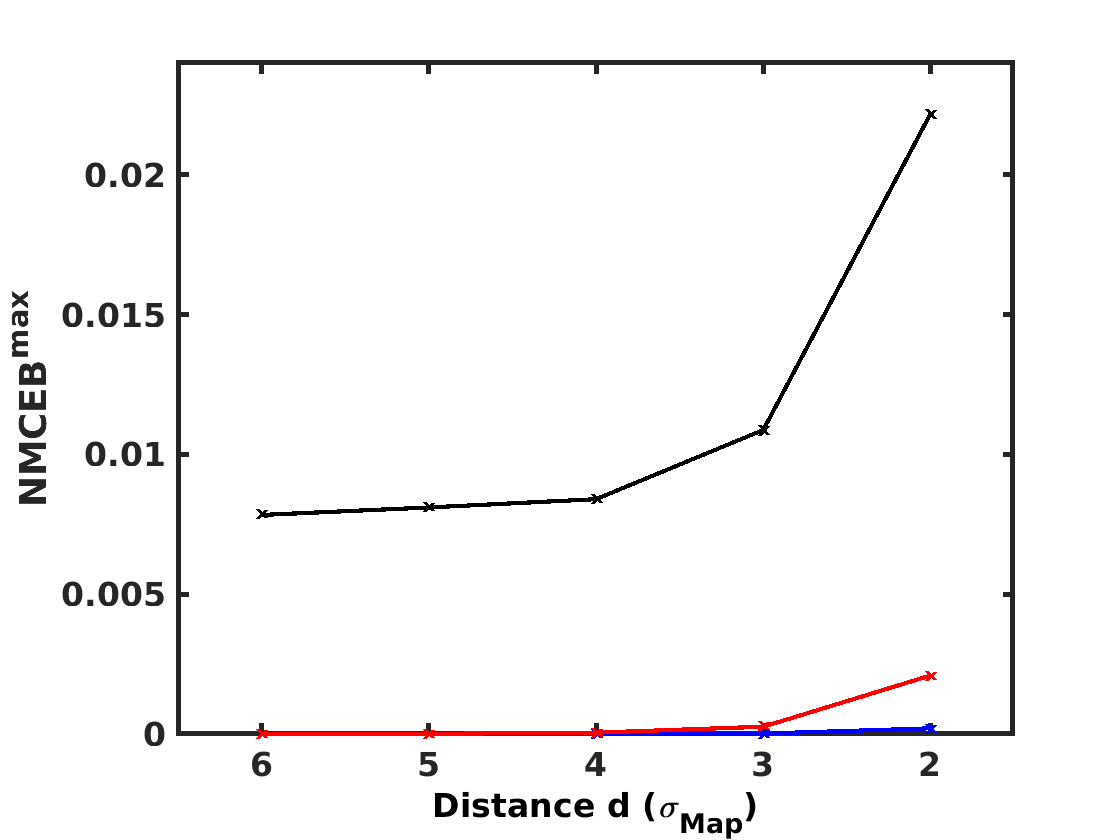}&
\includegraphics[width=0.3\textwidth]{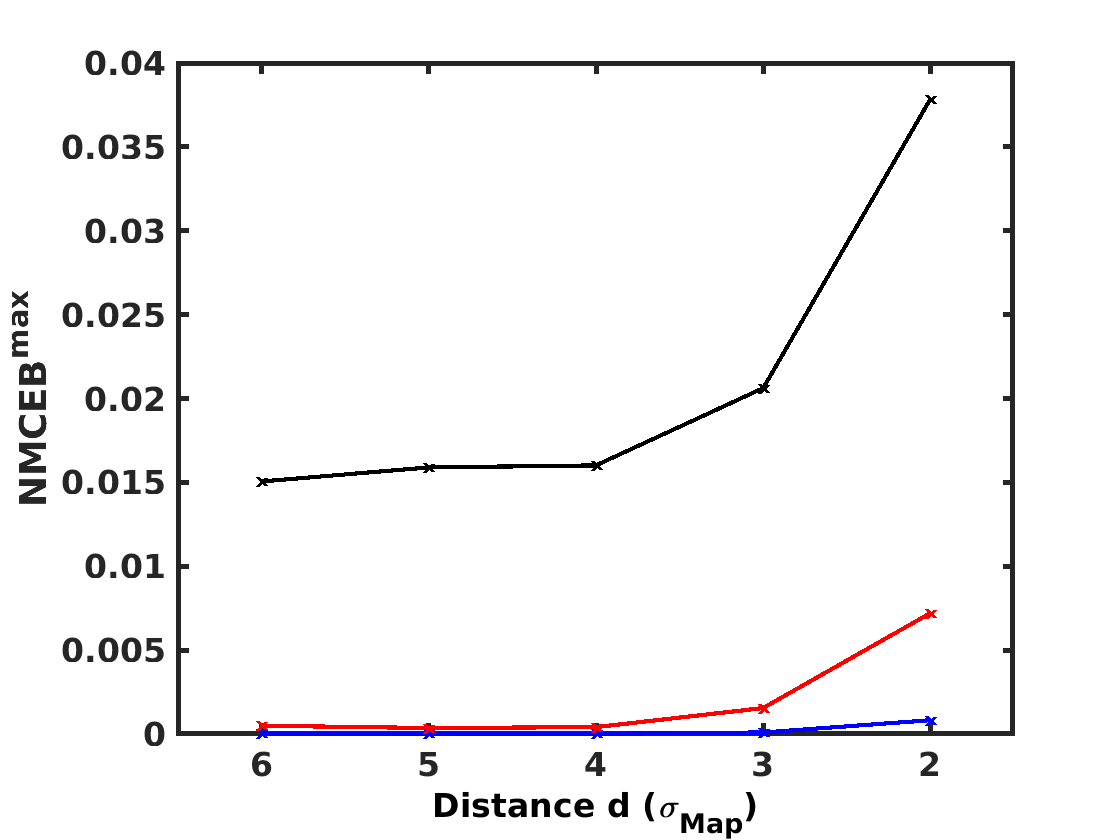}&
\includegraphics[width=0.3\textwidth]{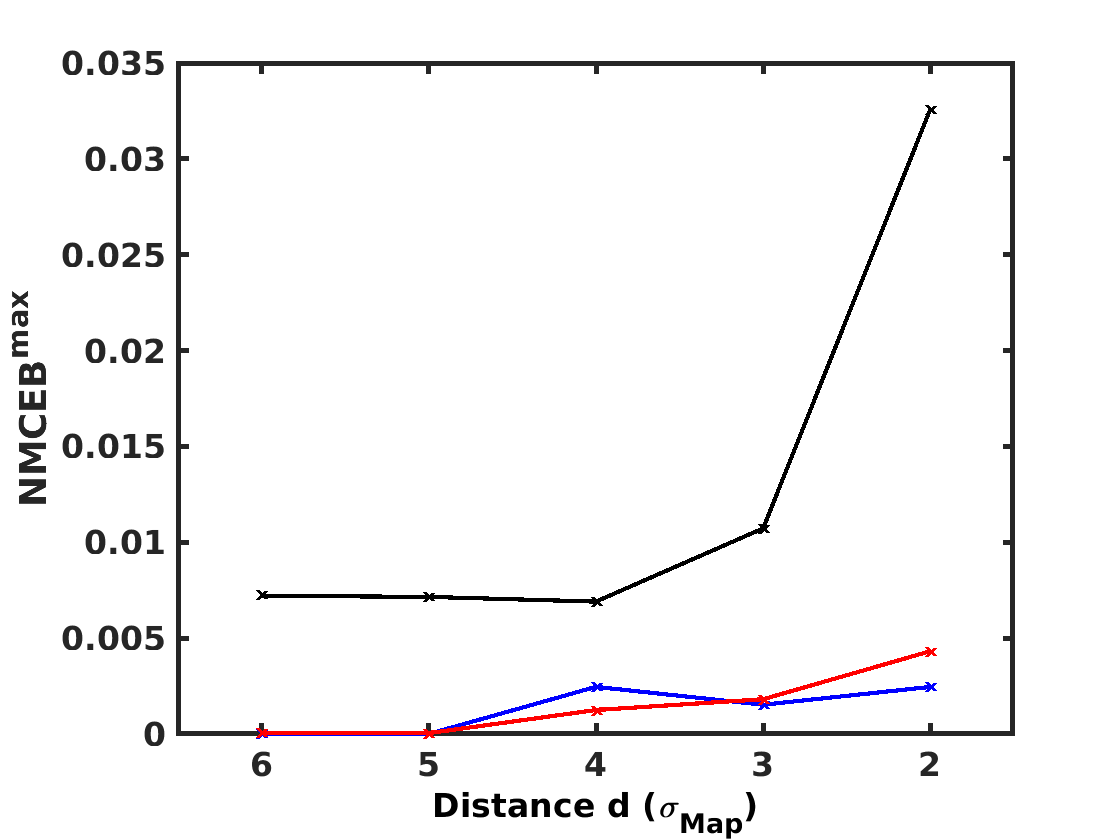}\\
2 sources & 4 sources & 6 sources
\end{tabular}
\end{center}
\caption{%
\ytextmodifhershelvonestepfive{%
\ytextmodifhershelvonestepfive{Spread (NMCEB)}
of the solutions of MASS-NMF-Map}
obtained on the 45 
synthetic cubes with 100 realizations of noise.} 
\label{fig_errbar_HybridMap_chap4}
\end{figure}

\ytextmodifhershelvonestepfive{The MASS-NMF-Map
hybrid version 
(Fig.
\ref{fig_perf_HybridMap_chap4})
yields the same trends as those
obtained by MASS used alone
(Fig.
\ref{fig_perf_MASS_chap4}).}
Again, there is a noticeable improvement in 
performance for mixtures 
\ytextmodifhershelvonestepfive{with a 10 dB SNR}
compared to MASS used alone%
\ytextmodifhershelvonestepfive{,}
as well as 
a clear improvement in the 
\ytextmodifhershelvonestepfive{spread}
of the solutions 
\ytextmodifhershelvonestepfive{(Fig.
\ref{fig_errbar_HybridMap_chap4})}
compared to that obtained 
with the MC-NMF method alone (Fig. \ref{fig_errbar_NMF_chap3})%
\ytextmodifhershelvonestepfive{, especially for 4 or 6 sources.}

%
%
%

\newpage

\subsection{\ytextmodifhershelvtwostepfiftyone{Results for
very low signal-to-noise ratios}}
\label{sec-appendix-Results-low-SNR}
\begin{figure}[H]
\begin{center}	
\begin{tabular}{ccc}
\includegraphics[width=0.3\textwidth]{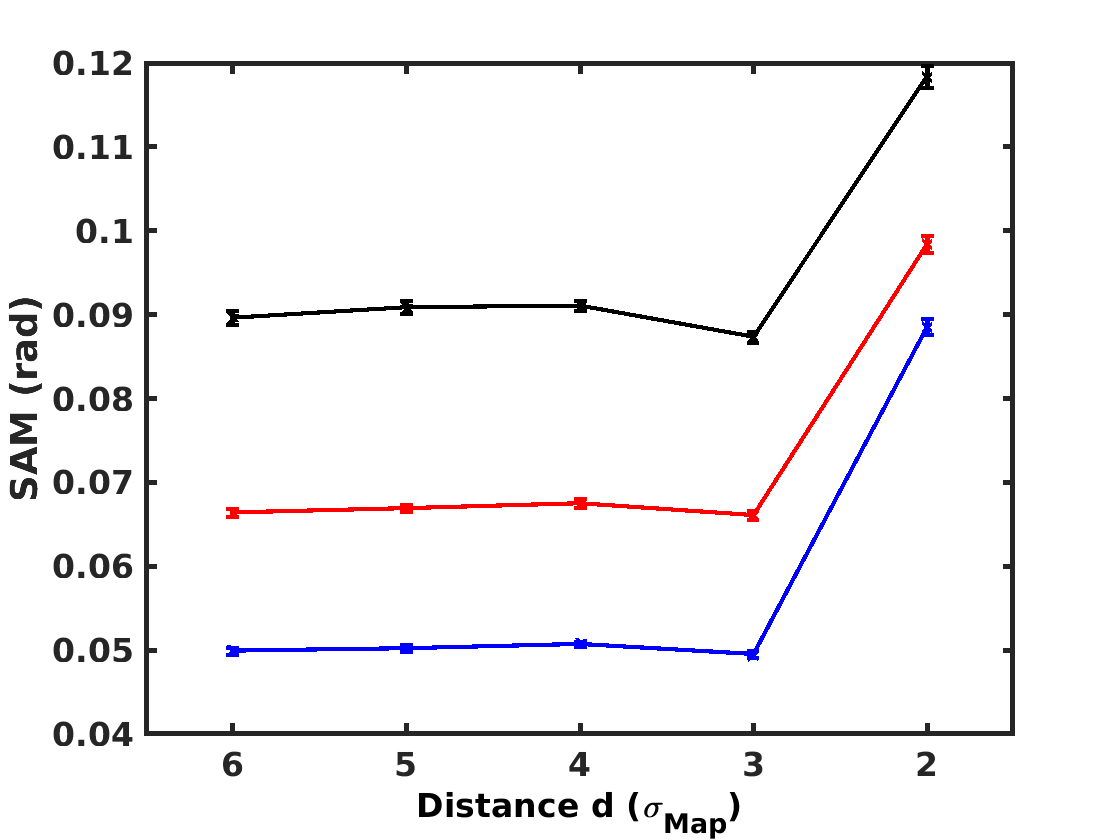}&
\includegraphics[width=0.3\textwidth]{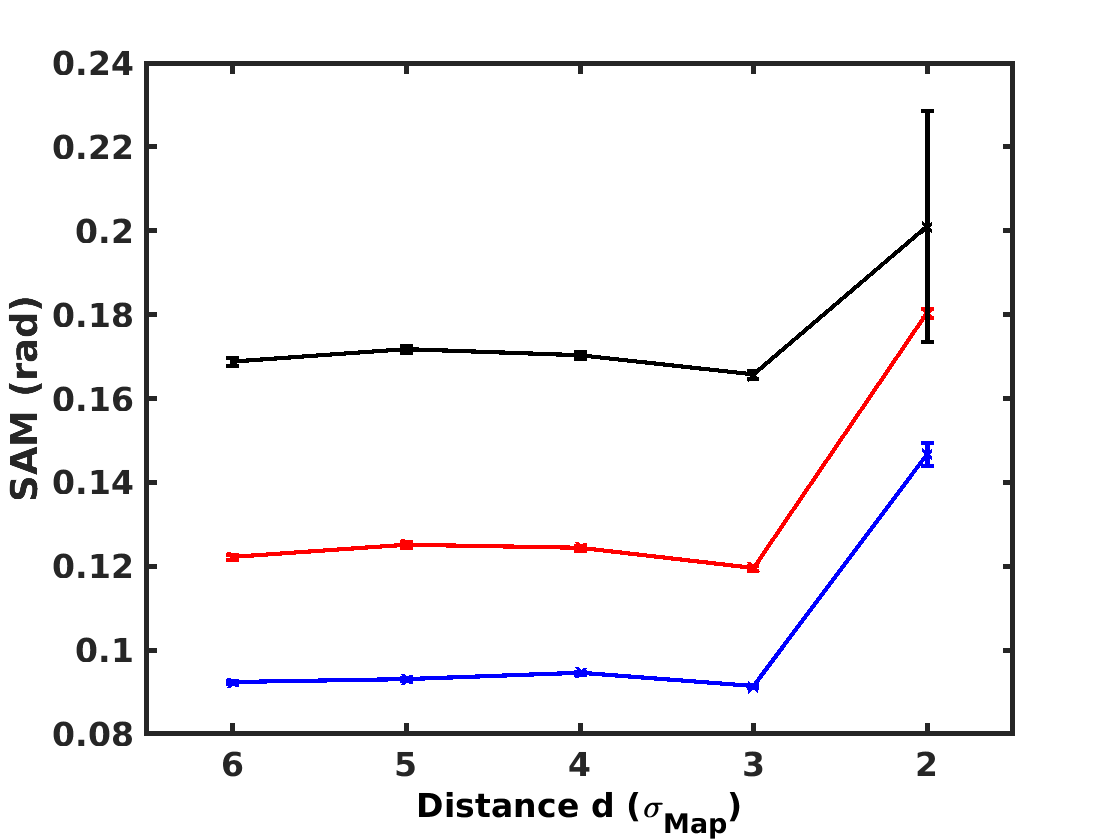}&
\includegraphics[width=0.3\textwidth]{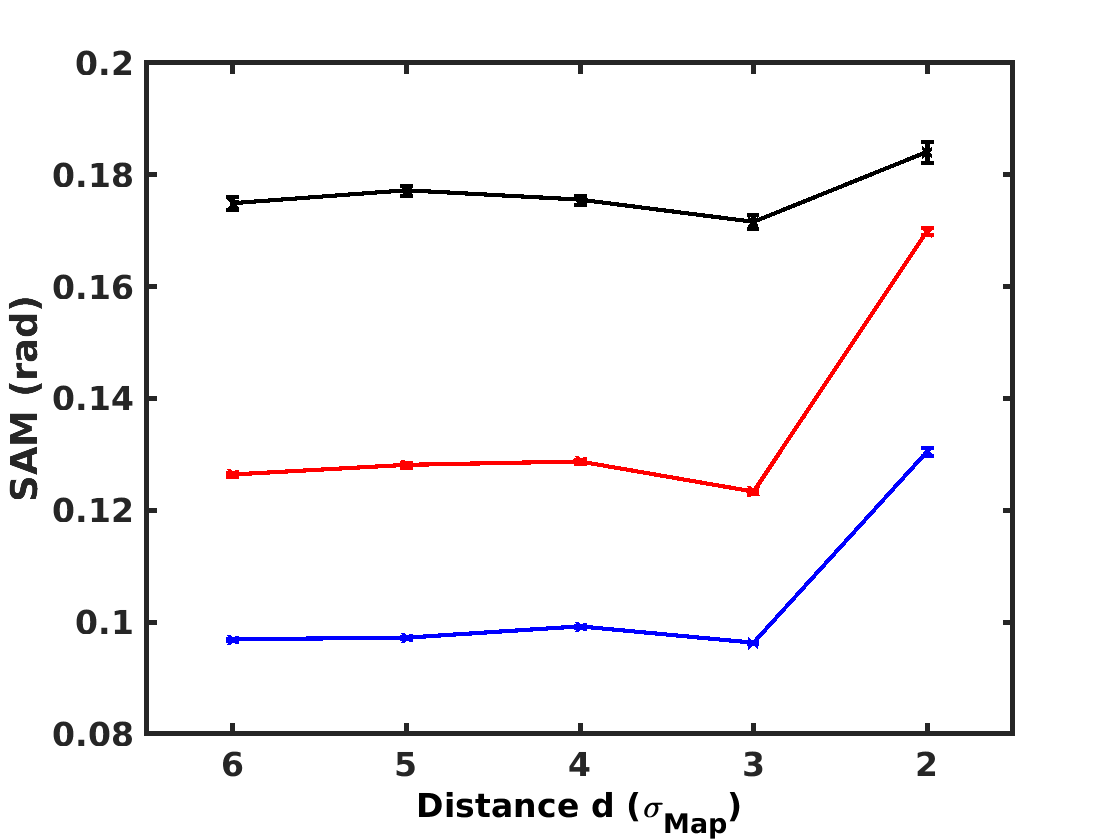}\\
\includegraphics[width=0.3\textwidth]{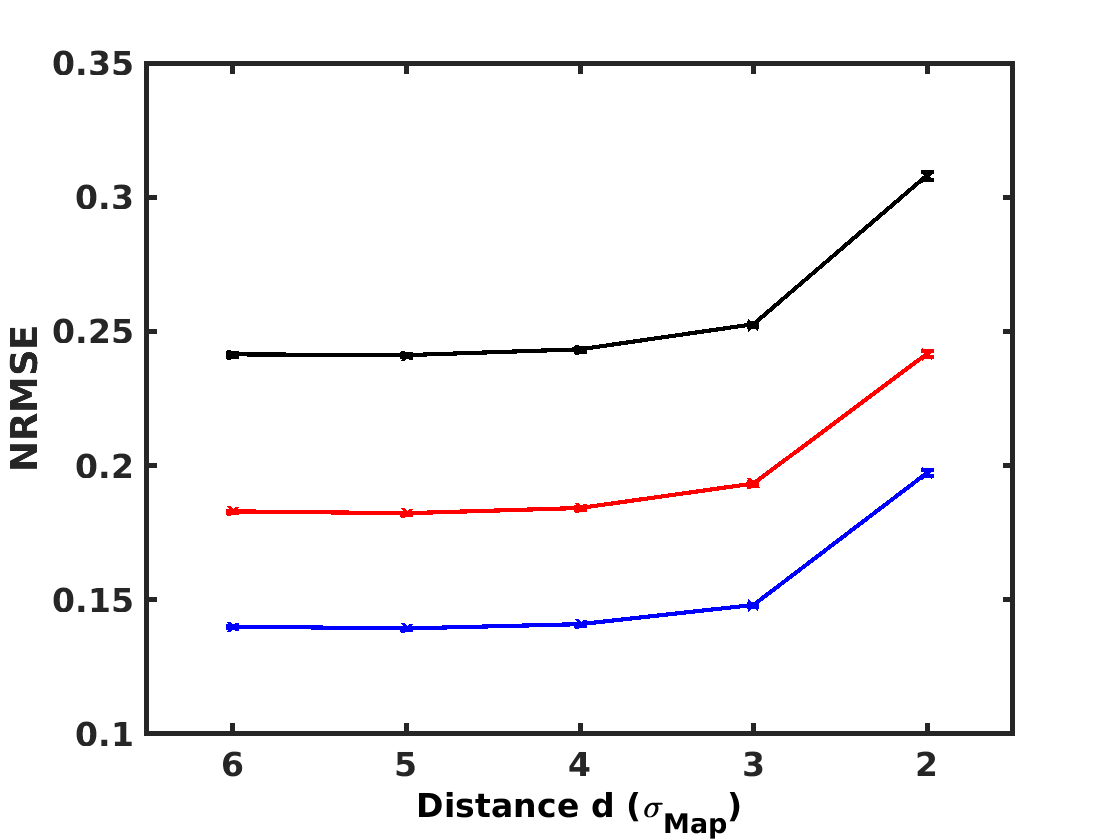}&
\includegraphics[width=0.3\textwidth]{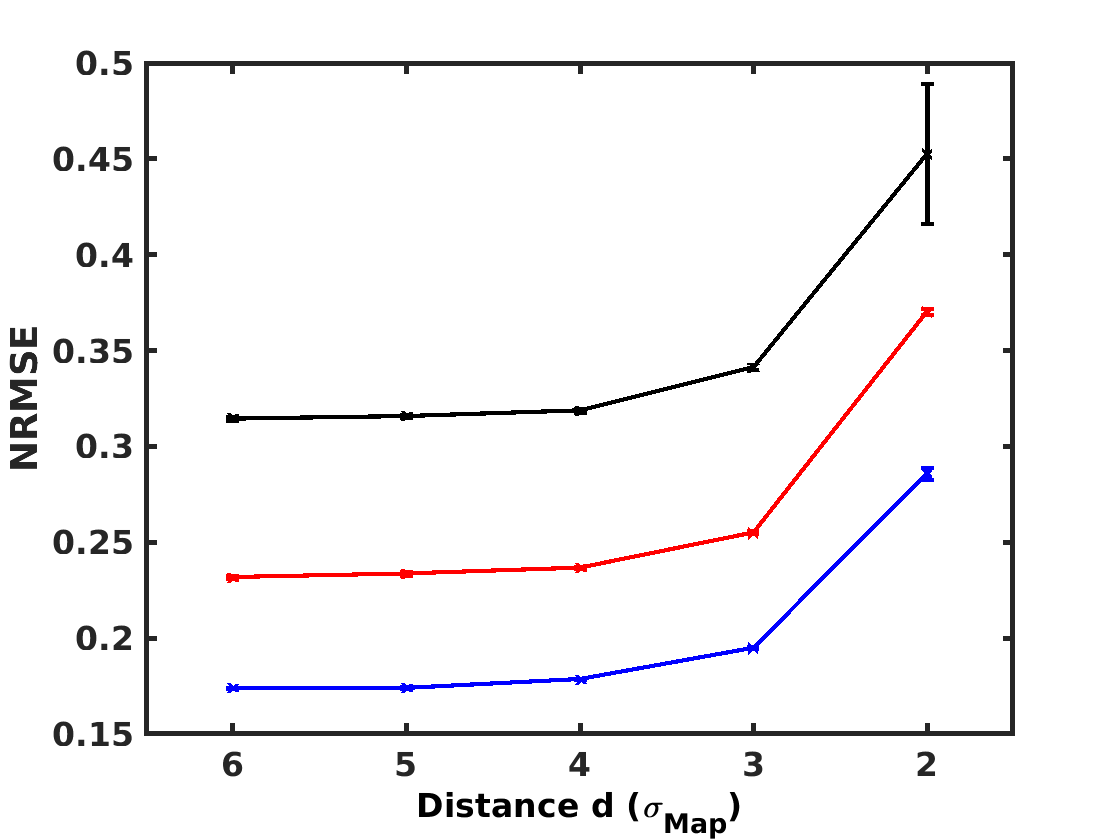}&
\includegraphics[width=0.3\textwidth]{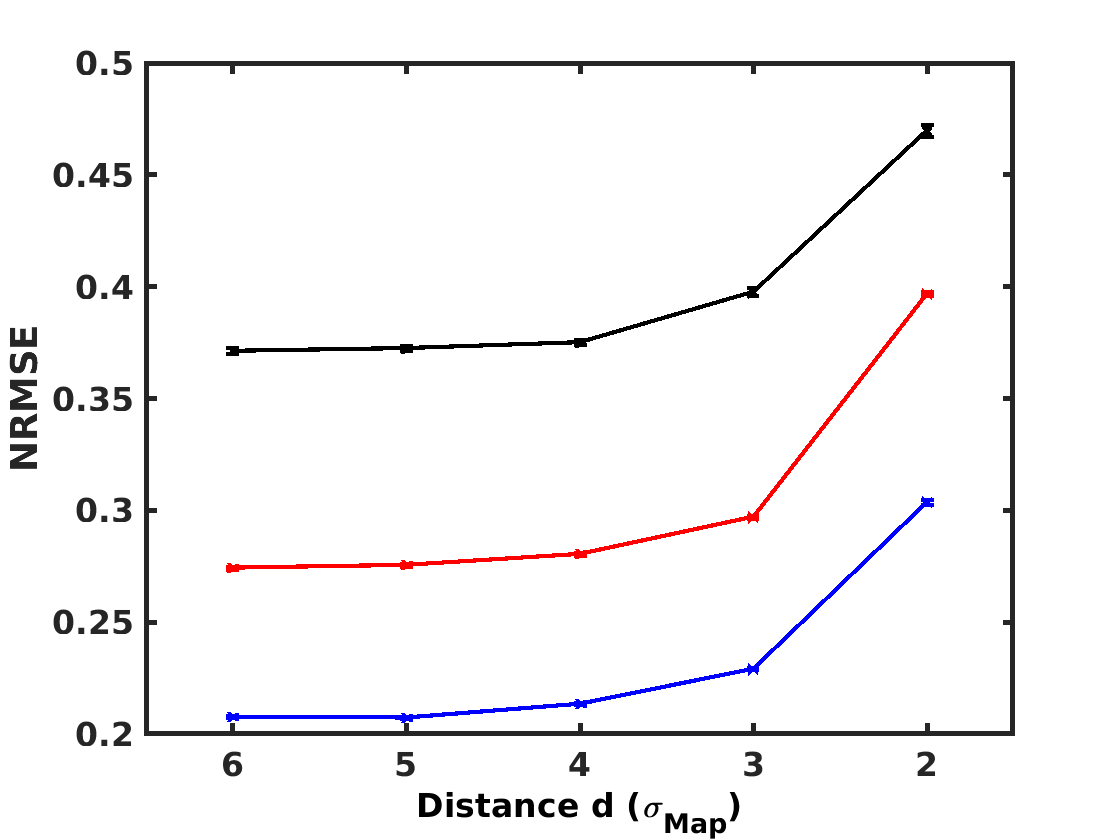}\\
2 sources & 4 sources & 6 sources
\end{tabular}
\end{center}
\caption{%
\ytextmodifhershelvtwostepfiftyone{%
Performances achieved by 
\ytextmodifhershelvonestepfive{SC-NMF-Map}
on the 45 synthetic cubes for 100 
realizations of noise with 
\ytextmodifhershelvonestepfive{an}
SNR of 5 dB (in blue), 3 dB (in red), and 1 dB 
(in black). The error bars give the standard deviation 
\ytextmodifhershelvonestepfive{over}
the 100 
realizations of noise.%
}
} 
\label{fig_perf_HybridMap_chap3_lsnr}
\end{figure}

\begin{figure}[H]
\begin{center}	
\begin{tabular}{ccc}
\includegraphics[width=0.3\textwidth]{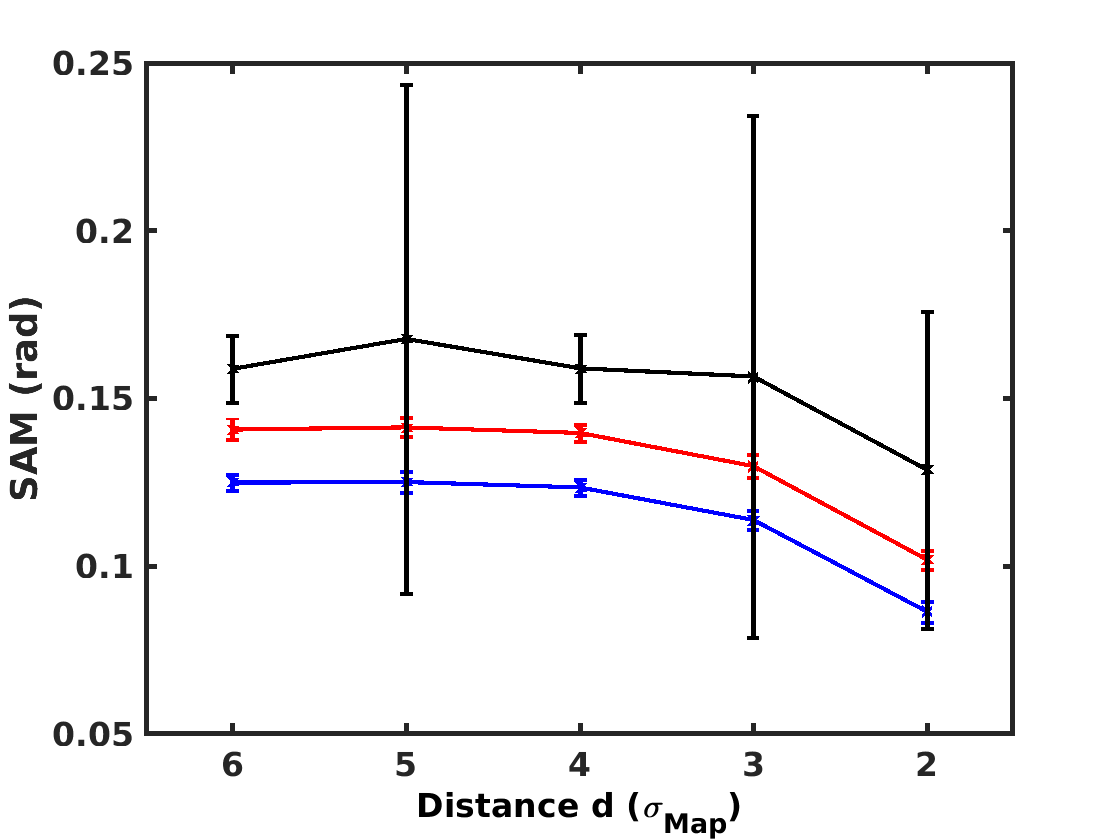}&
\includegraphics[width=0.3\textwidth]{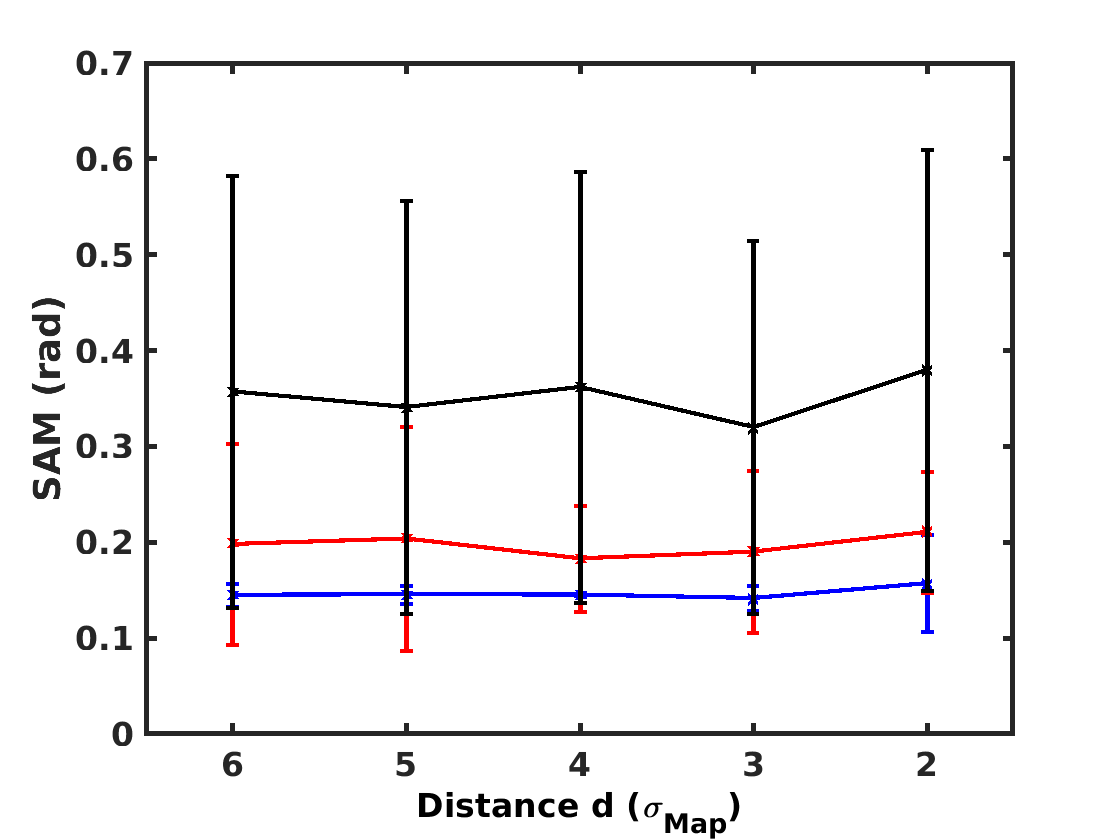}&
\includegraphics[width=0.3\textwidth]{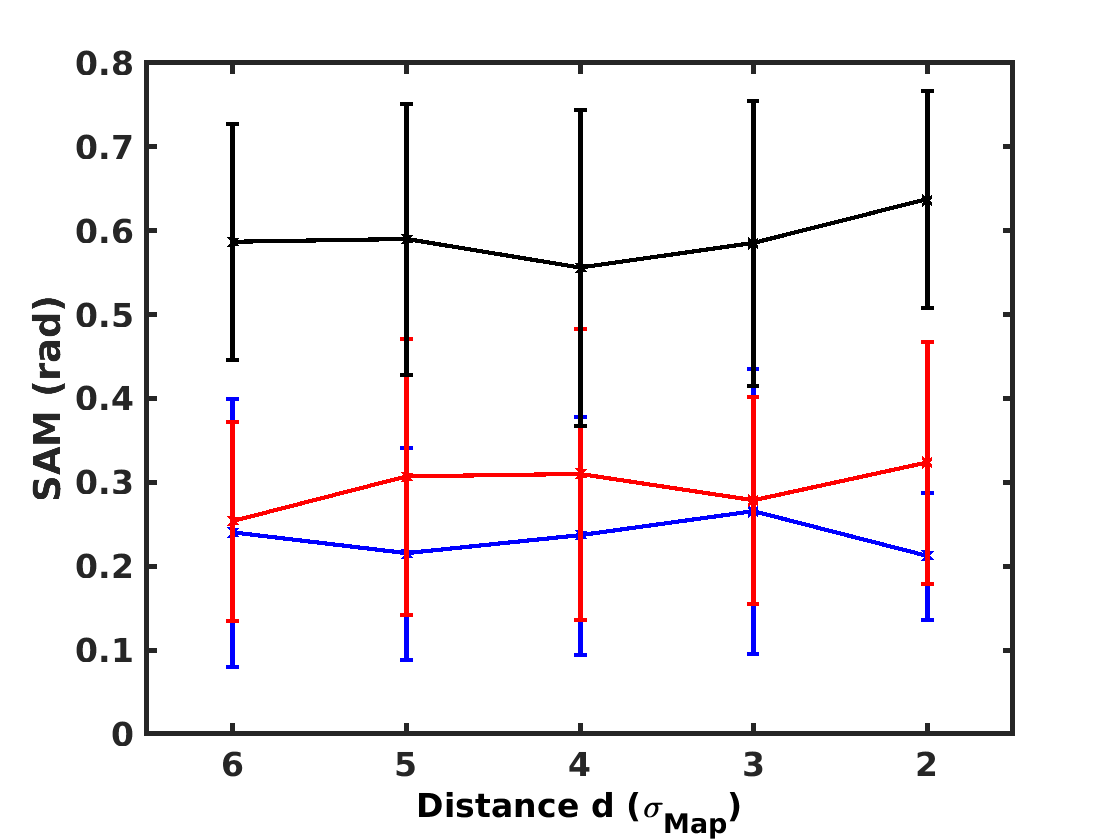}\\
\includegraphics[width=0.3\textwidth]{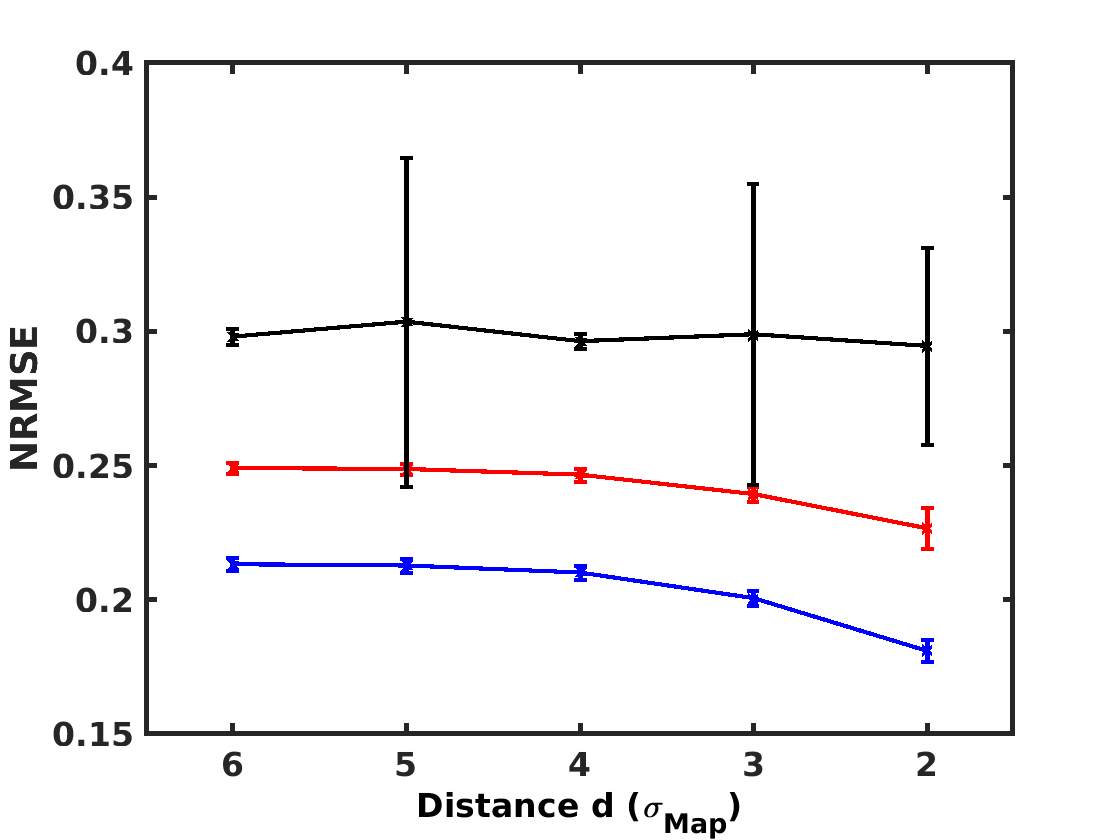}&
\includegraphics[width=0.3\textwidth]{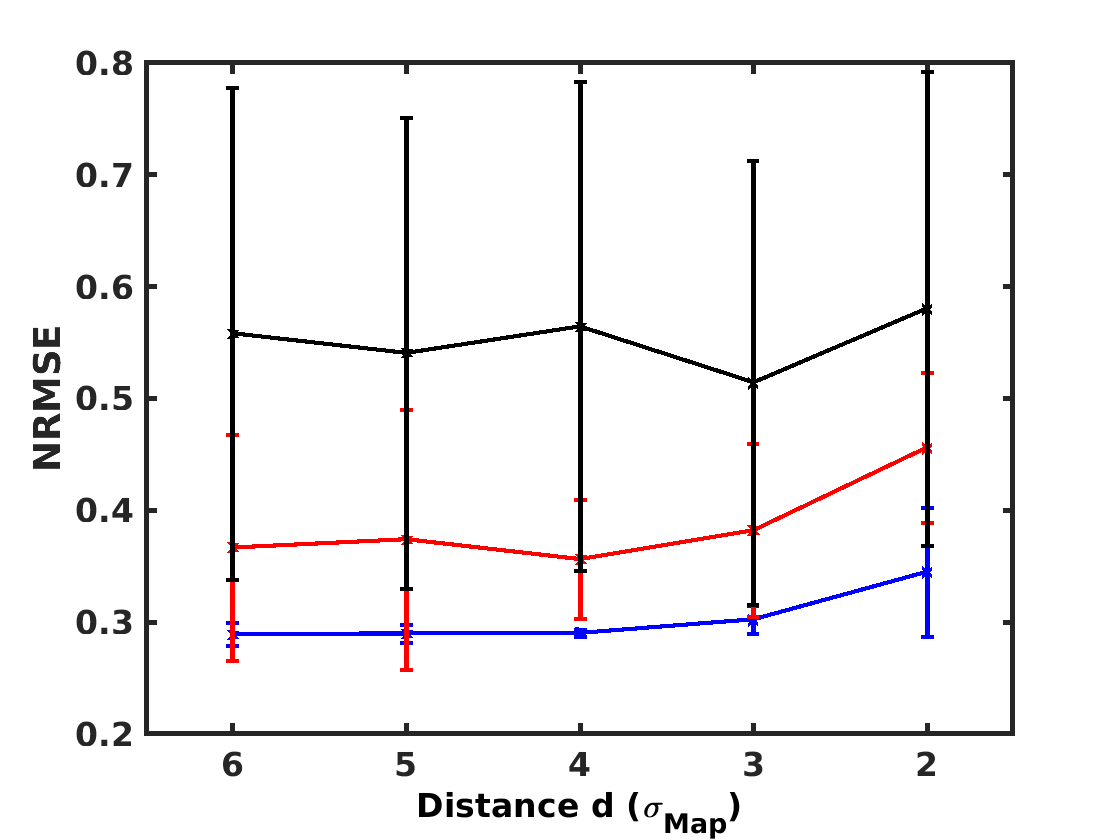}&
\includegraphics[width=0.3\textwidth]{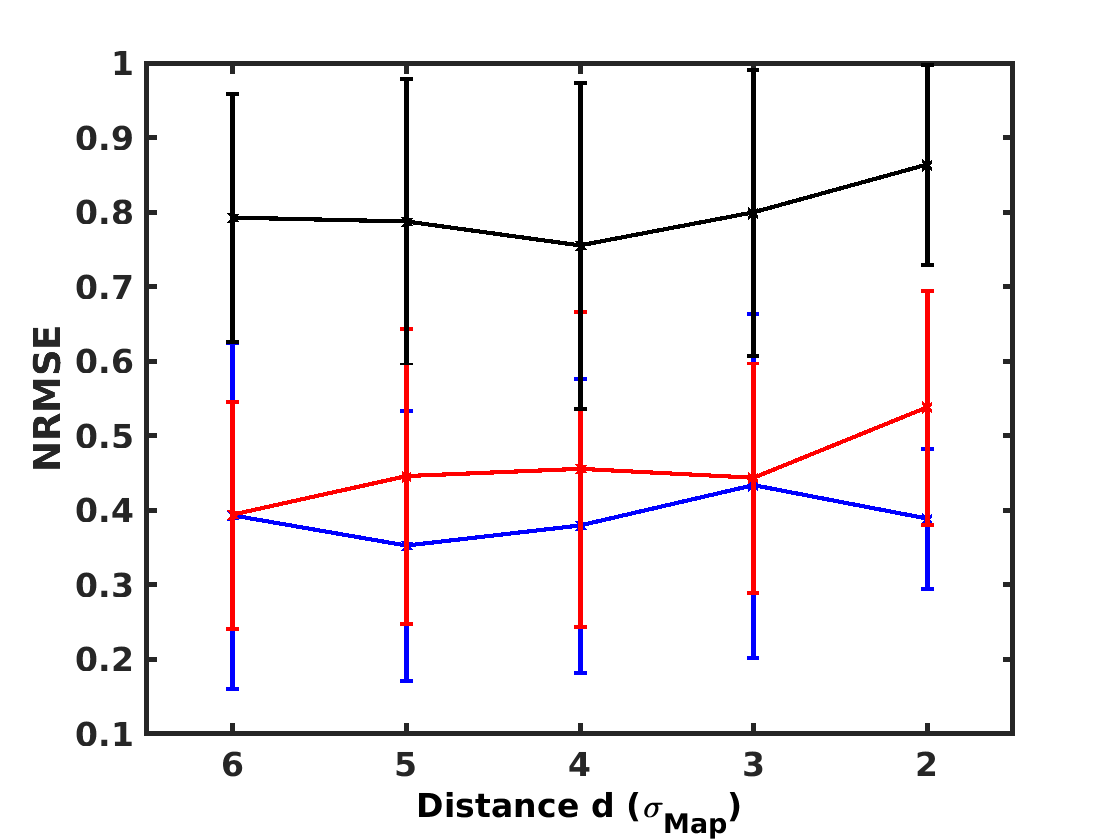}\\
2 sources & 4 sources & 6 sources
\end{tabular}
\end{center}
\caption{%
\ytextmodifhershelvtwostepfiftyone{%
Performances achieved by MASS-NMF-Map on the 45 synthetic cubes for 100 
realizations of noise with 
\ytextmodifhershelvonestepfive{an}
SNR of 5 dB (in blue), 3 dB (in red), and 1 dB 
(in black). The error bars give the standard deviation 
\ytextmodifhershelvonestepfive{over}
the 100 
realizations of noise.
}
} 
\label{fig_perf_HybridMap_chap4_lsnr}
\end{figure}

\ytextmodifhershelvtwostepfiftyone{%
Finally, we performed additional tests for the two methods
which appeared to be the most attractive ones in
Section
\ref{sec-results-summary},
namely
MASS-NMF-Map and
SC-NMF-Map.
These tests aim at further analyzing the behavior of these
preferred methods
for very low SNRs, that is five, three, and one dB.
The results thus obtained are shown in Fig.
\ref{fig_perf_HybridMap_chap3_lsnr}
and
\ref{fig_perf_HybridMap_chap4_lsnr}.
This shows that SC-NMF-Map here yields significantly
better performance than
MASS-NMF-Map, as opposed to the results obtained in Section
\ref{sec-results-summary} for significantly higher SNRs.
This difference is reasonable and coherent with the comments
that we provided in Section
\ref{sec-appendix-Results-MASS}:
although SC-NMF-Map is constraining in terms of sparsity
requirements, it has the advantage of averaging the data over an
analysis zone (instead of using a single point in the basic version
of
MASS-based methods), which reduces its sensitivity to noise.
For very noisy data, such as those considered here, this
feature is of utmost importance, so that
SC-NMF-Map yields better performance than
MASS-NMF-Map.}
%

%

%

\newpage

\subsection{\ytextmodifhershelvtwostepfiftyone{Results for asymmetric scenes}}
\label{sec-appendix-Results-asym}

\begin{figure}[H]
\begin{center}	
\includegraphics[width=1\textwidth]{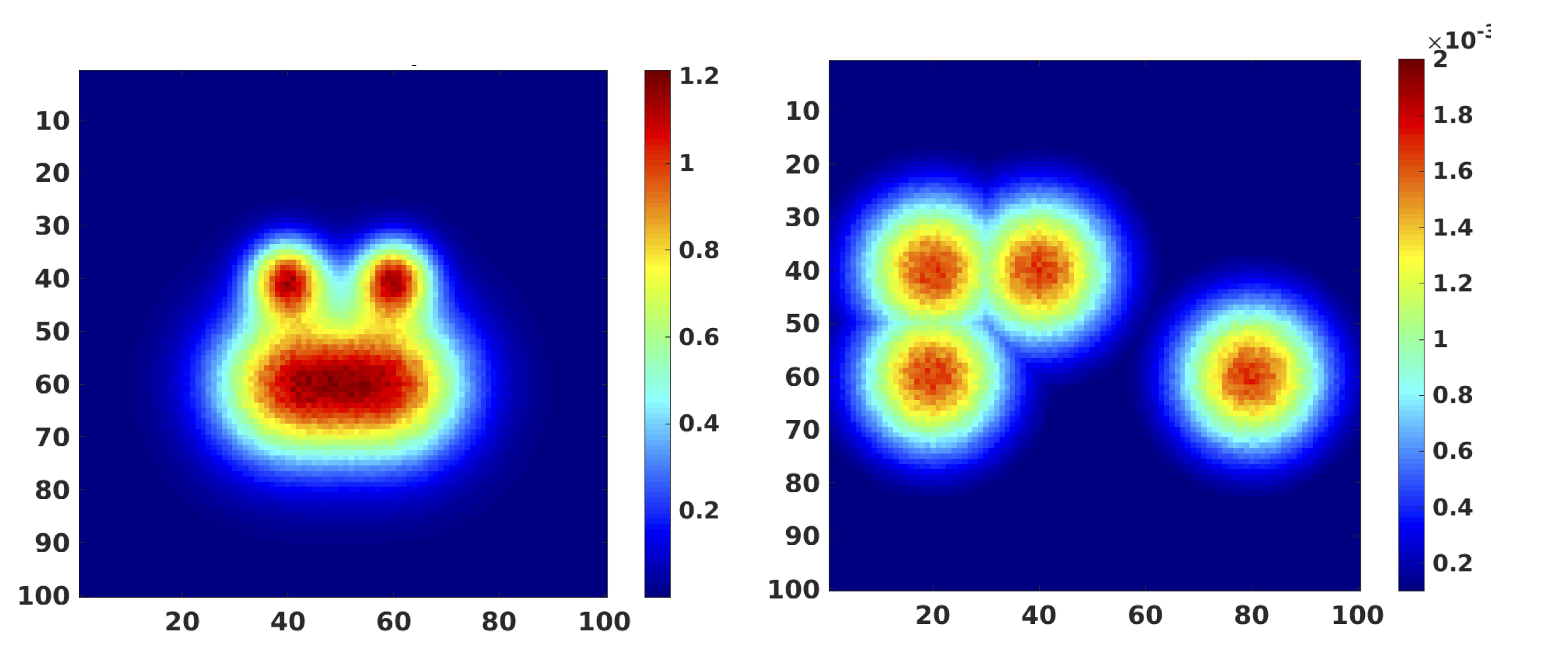}
\end{center}
\caption{Two cases of symmetry breaking for the maps : reduction of the size of two spatial sources (left) and 
displacement of one source (right).\label{Fig_assym_cropped}}
\end{figure}

{ Here we test the effect of breaking the symmetry of the spatial scene presented 
in Fig \ref{fig_Maps}. We do these by two means, 1) we decrease the size of some sources,
2) we displace one of the sources outside the square grid considered so far. These two cases 
are illustrated in Fig.~\ref{Fig_assym_cropped}. We have considered other variations as well
(e.g. smaller or larger displacement). Overall, the analysis of the results shows that the 
methods provide results with similar performances as for the standard cases described in the 
core of the paper.  Hence, the symmetry of the scene does not appear to be critical, as long 
as the important hypotheses required by the methods (positivity and sparsity) are verified.}

\end{document}